\documentclass[twocolumn]{aastex63} 
\usepackage{amsmath}
\usepackage{natbib}
\usepackage{float}

\graphicspath{{./}{figures/}}

\submitjournal{ApJS}

\shorttitle{Radial and Rotational Velocities of T Dwarfs}
\shortauthors{Hsu et al.}

\begin{document}

\title{The Brown Dwarf Kinematics Project (BDKP). V. Radial and Rotational Velocities of T Dwarfs from Keck/NIRSPEC High-Resolution Spectroscopy}

\correspondingauthor{Chih-Chun Hsu}
\email{chh194@ucsd.edu}

\author[0000-0002-5370-7494]{Chih-Chun Hsu}
\affil{Center for Astrophysics and Space Science, University of California San Diego, La Jolla, CA 92093, USA}

\author[0000-0002-6523-9536]{Adam J.\ Burgasser}
\affil{Center for Astrophysics and Space Science, University of California San Diego, La Jolla, CA 92093, USA}

\author[0000-0002-9807-5435]{Christopher A.\ Theissen}
\altaffiliation{NASA Sagan Fellow}
\affil{Center for Astrophysics and Space Science, University of California San Diego, La Jolla, CA 92093, USA}

\author{Christopher R.\ Gelino}
\affiliation{NASA Exoplanet Science Institute, Mail Code 100-22, California Institute of Technology, 770 South Wilson Avenue, Pasadena, CA 91125, USA}
\affiliation{Infrared Processing and Analysis Center, Mail Code 100-22, California Institute of Technology, 1200 E. California Boulevard, Pasadena, CA 91125, USA}

\author[0000-0002-7961-6881]{Jessica L.\ Birky}
\affil{Center for Astrophysics and Space Science, University of California San Diego, La Jolla, CA 92093, USA}
\affil{Department of Astronomy, University of Washington, Seattle, WA 98195, USA}

\author{Sharon J.\ M.\ Diamant}
\affiliation{Leiden Observatory, Leiden University, Leiden, the Netherlands}

\author[0000-0001-8170-7072]{Daniella C. Bardalez Gagliuffi}
\affiliation{Department of Astrophysics, American Museum of Natural History, NY, NY 10024}

\author[0000-0003-2094-9128]{Christian Aganze}
\affil{Center for Astrophysics and Space Science, University of California San Diego, La Jolla, CA 92093, USA}

\author[0000-0002-6096-1749]{Cullen H.\ Blake}
\affiliation{University of Pennsylvania Department of Physics and Astronomy, 209 S 33rd St, Philadelphia, PA 19104, USA}

\added{\author[0000-0001-6251-0573]{Jacqueline K. Faherty}}
\added{\affil{Department of Astrophysics, American Museum of Natural History, Central Park West at 79th Street, NY 10024, USA}}

\begin{abstract}
We report multi-epoch radial velocities, rotational velocities, and atmospheric parameters for 37 T-type brown dwarfs observed with Keck/NIRSPEC. Using a Markov Chain Monte Carlo forward-modeling method, we achieve median precisions of 0.5 km s$^{-1}$ and 0.9 km s$^{-1}$ for radial and rotational velocities, respectively. All of the T dwarfs in our sample are thin disk brown dwarfs. We confirm previously reported moving group associations for \replaced{two}{four} T dwarfs\deleted{, and identify \replaced{two additional}{a new} candidate kinematic member\deleted{s} of the \deleted{$\beta$ Pictoris and} Carina-Near moving group}\replaced{s, although the lack of spectral indicators of youth}{. However, the lack of spectral indicators of youth in two of these sources} suggests that these are chance alignments. We confirm two previously un-resolved binary candidates, \added{the} T0+T4.5 2MASS J11061197+2754225 and \added{the} L7+T3.5 2MASS J21265916+7617440, with orbital periods of 4~yr and 12~yr, respectively. We find a kinematic age of 3.5$\pm$0.3~Gyr for local T dwarfs, consistent with nearby late-M dwarfs (4.1$\pm$0.3~Gyr). Removal of thick disk L dwarfs in the local ultracool dwarf sample gives a similar age for L dwarfs (\replaced{4.1}{4.2}$\pm$0.3~Gyr), \added{largely} resolving the local L dwarf age anomaly. The kinematic ages of local late-M, L, and T dwarfs can be accurately reproduced with population simulations incorporating standard assumptions of the mass function, star formation rate, and brown dwarf evolutionary models. A kinematic dispersion break is found at \added{the} L4–L6 subtypes, likely reflecting the terminus of the stellar Main Sequence. We provide a compilation of precise radial velocities for \replaced{173}{172} late-M, L, and T dwarfs within $\sim$20~pc of the Sun.
\end{abstract}

\section{Introduction} \label{sec:intro}

Ultracool dwarfs (UCDs) are low-mass ($M < 0.1$~M$_{\odot}$) stellar and substellar objects with effective temperatures $T_{\text{eff}} \, \leq $ 3,000 K and spectral types M7 or later, encompassing late-M, L, T, and Y dwarfs 
\replaced{\citep{Kirkpatrick:2005aa}}{\citep{Kirkpatrick:2005aa, Cushing:2011aa}}. 
UCDs are abundant in the Galaxy (20--50\% of all stars; \citealt{Bochanski:2007aa,Kirkpatrick:2019aa}) and have lifetimes orders of magnitude longer than the current age of the Universe \citep{1997ApJ...482..420L}. Every UCD ever formed still exists. UCDs thus provide a means of investigating the low-mass star formation and chemical enrichment history of the Milky Way and its substructures \citep{Bochanski:2007aa,Burgasser:2009aa}. Substellar UCDs, or brown dwarfs \citep[M $<$ 0.072 M$_{\odot}$;][]{Kumar:1963aa,Hayashi:1963aa}, cannot sustain hydrogen fusion in their cores and cool over time. 
The evolutionary nature of brown dwarfs makes them useful tracers of age in various populations.
Brown dwarfs have been used to age-date young clusters \citep{1998ApJ...499L.199S,2018ApJ...856...40M} and 
binaries \citep{2002ApJ...581L..43S}, and could potentially  
be used to age-date other Galactic populations, such as globular clusters and the Galactic halo \citep{Burgasser:2009aa,2017arXiv170200091C}. 

In mixed populations such as the Galactic disk, the overlapping temperatures and luminosities of stellar and substellar UCDs results in an observational age-mass degeneracy, requiring additional diagnostics to disentangle these parameters.
High-resolution spectroscopy provides several ways of doing this. 
Precise radial velocities (RVs) can uncover binaries with periodically variable RVs from which masses can be inferred \citep{Basri:1999aa, Reid:2002aa, Joergens:2007aa, Blake:2008aa, Blake:2010aa, Konopacky:2010aa, Burgasser:2012aa, Burgasser:2015aa, Burgasser:2016aa}. 
Kinematics can identify UCD members of young clusters and moving groups, yielding an age by association \citep{Zuckerman:2004aa, Gagne:2015aa}.
Kinematic association with the Galactic thin disk, thick disk, or halo can also provide coarse age constraints \citep{Bensby:2003aa, Pinfield:2014aa}. 
As the velocity dispersion of field stars increases over time due to diffuse dynamical scattering with Galactic structures, 
kinematics can also provide a statistical measure of age for a given population \citep{Wielen:1977aa, Aumer:2009aa}.
Finally, the magnetized winds of low-mass stars reduces stellar rotation over time\replaced{, and g}{. G}yrochronology relations can be applied to rotational velocities ($v\sin{i}$) or periods to infer ages \citep{Barnes:2003aa,2019AJ....158..173A}, although it is unclear how low in mass these relations can be applied \replaced{\citep{Reiners:2009aa,2011ApJ...727...56I,2013MNRAS.432.1203M,2018AJ....156..217N}}{\citep{Reiners:2009aa,2011ApJ...727...56I,2013MNRAS.432.1203M,2018AJ....156..217N,Popinchalk:2021aa}}.

\replaced{Statistical methods of age-dating UCD populations from kinematics}{The study of UCD populations} must account for the evolution of brown dwarfs\replaced{, and t}{. T}ypical analyses make use of population simulations incorporating stellar and substellar evolution to model present-day observables \citep{Reid:1999aa,Burgasser:2004aa,Kirkpatrick:2019aa}. 
A consistent prediction of these simulations is that L-type UCDs, a mix of stars and rapidly-cooling brown dwarfs ($\tau$ $\approx$ 0.5--1~Gyr; \citealt{Burrows:1997aa,Baraffe:2003aa}), should be on average younger than late M-type UCDs, which are predominantly \added{long-lived} stars \citep{Reid:1999aa, Burgasser:2004aa, Allen:2005aa, Ryan:2017aa}.
\replaced{However, o}{O}bservations of statistical ages through kinematics have \replaced{not confirmed}{mostly contradicted} this prediction. 
In one of the first 3D kinematic studies of UCDs\replaced{ (}{, combining }RVs and tangential velocities\deleted{)}, \cite{Zapatero-Osorio:2007aa} did find that local L and T dwarfs had a lower velocity dispersion than local late-M dwarfs, although the sample for the former included only 21 sources. 
A 2D proper motion survey of 277 UCDs by \cite{Faherty:2009aa} found local\added{ 20~pc} L dwarfs to be marginally older than late-M and T dwarfs\added{, while their whole sample of 841 UCDs gave consistent ages.}\deleted{, while} \cite{Schmidt:2010aa} identified both young and old L dwarf populations in a 3D kinematic survey of over 300 sources using low-resolution spectroscopy.
The diagnostic power of these early studies was limited by incomplete samples and relatively low-precision measurements ($\sigma_\text{RV}$ $\gtrsim$ 5--10 km s$^{-1}$). 

Subsequent volume-limited 3D kinematic surveys of late-M and L dwarfs using high-resolution spectra and improved astrometry have consistently found the local L dwarf population to be significantly more dispersed and hence older than the late-M dwarfs, in disagreement with simulation predictions \citep{Reiners:2009aa, Seifahrt:2010aa, Blake:2010aa, Burgasser:2015ac}. 
The most recent study by \citet{Burgasser:2015ac} attempted to explain this discrepancy as an evolution of the mass function over time, with a greater abundance of brown dwarfs at earlier ages. While this ansatz was able to correctly predict the relative ages of late-M and L dwarfs, the actual ages and corresponding mass function were inconsistent with observations.  
Whether this persistent disagreement between simulated and observed ages of L dwarfs
arises from a more complex star and brown dwarf formation history, a problem with brown dwarf evolutionary models, 
or incompleteness or bias in the local UCD sample remains uncertain.

The kinematics of local T dwarfs \replaced{provides}{can provide} clarity on this issue. These low-temperature, evolved brown dwarfs having cooling ages \replaced{longer than the age of the Milky Way}{of several Gyr}, and are predicted to be kinematically older than L dwarfs \citep{Burgasser:2004aa}.
T dwarfs are also \replaced{much}{intrinsically} fainter than late-M and L dwarfs, limiting the number of sources with suitable high-precision RV measurements. While there are more than 500 T dwarfs currently known, there are fewer than 15 T dwarfs with precise RV ($\sigma_\text{RV} \leq 3$ km s$^{-1}$) and rotational velocity ($\sigma_{v\sin{i}} \leq 5$ km s$^{-1}$) measurements \added{reported} in the literature \citep{Zapatero-Osorio:2006aa, Zapatero-Osorio:2007aa, Prato:2015aa, Gagne:2017aa, Gagne:2018aa, Vos:2017aa, Vos:2018aa}.
This sample is insufficient to accurately measure kinematic dispersions and ages, or test population simulations. 
The lack of precise RV 

\begin{longrotatetable}
\begin{deluxetable*}{lllrrrrrccl}
\tablecaption{NIRSPEC T Dwarf Sample\label{tablesample}}
\tablewidth{700pt}
\tabletypesize{\tiny}
\tablehead{
\colhead{} & \colhead{} & \colhead{} & \colhead{} & \colhead{} & \colhead{} & \colhead{} & \colhead{} & \colhead{} & \colhead{} & \colhead{}\\
\colhead{} & \colhead{} & \colhead{} & \colhead{} & \colhead{} & \colhead{} & \colhead{} & \colhead{} & \colhead{} & \colhead{} & \colhead{}\\
\colhead{Source Name} & \colhead{Coordinates} & 
\colhead{SpT} & \colhead{2MASS $J$} & 
\colhead{$J \, - \, K$} & \colhead{$\mu_{\alpha}$} & 
\colhead{$\mu_{\delta}$} & \colhead{$d$} & 
\colhead{Published RV} & \colhead{Published $v\sin i$} & \colhead{References$^{a}$} \\ 
\colhead{} & \colhead{(J2000)} & \colhead{} & \colhead{(mag)} & 
\colhead{(mag)} & \colhead{(mas yr$^{-1}$)} & \colhead{(mas yr$^{-1}$)} &
\colhead{(pc)} & \colhead{(km s$^{-1}$)} & \colhead{(km s$^{-1}$)} & \colhead{}
} 
\startdata
J0000+2554 & 00 00 13.54 +25 54 18.0 & T4.5 & $15.06 \pm 0.04$ & $0.22 \pm 0.13$ & $-19 \pm 2$ & $127 \pm 1$ & $14.12 \pm 0.38$ & \nodata & \nodata & (18, 14, 13) \\ 
J0034+0523 & 00 34 51.58 +05 23 05.1 & T6.5 & $15.140 \pm 0.004$ & $-0.933 \pm 0.03$ & $674 \pm 1$ & $178 \pm 2$ & $8.42 \pm 0.19$ & \nodata & \nodata & (8, 10, 35) \\ 
J0136+0933$^{b}$ & 01 36 56.56 +09 33 47.3 & T2.5 & $13.46 \pm 0.03$ & $0.68 \pm 0.04$ & $1238 \pm 1$ & $-16 \pm 0$ & $6.12 \pm 0.02$ & $11.5 \pm 0.4$ & $50.9 \pm 0.8$ & (1, 14, 33, 39) \\ 
J0150+3827 & 01 50 09.97 +38 27 25.9 & T0 & $16.11 \pm 0.08$ & $1.63 \pm 0.1$ & $881 \pm 1$ & $-120 \pm 1$ & $22.42 \pm 1.61$ & \nodata & \nodata & (17, 35) \\ 
J0213+3648 & 02 13 19.86 +36 48 38.0 & T3 & $15.3 \pm 0.5$ & $0.5 \pm 0.5$ & $65 \pm 65$ & $0 \pm 0$ & $14.28 \pm 0.04$ & \nodata & \nodata & (12, 32) \\ 
J0243$-$2453 & 02 43 13.72 $-$24 53 29.8 & T6 & $15.38 \pm 0.05$ & $0.17 \pm 0.18$ & $-288 \pm 4$ & $-208 \pm 3$ & $10.68 \pm 0.41$ & \nodata & \nodata & (5, 14, 31) \\ 
J0415$-$0935 & 04 15 19.54 $-$09 35 06.7 & T8 & $15.7 \pm 0.1$ & $0.3 \pm 0.2$ & $2214 \pm 1$ & $536 \pm 1$ & $5.71 \pm 0.06$ & $49.6 \pm 1.2$ & $33.5 \pm 2.0$ & (5, 13, 43) \\ 
J0559$-$1404 & 05 59 19.19 $-$14 04 49.2 & T4.5 & $13.8 \pm 0.02$ & $0.23 \pm 0.06$ & $571 \pm 1$ & $-338 \pm 1$ & $10.5 \pm 0.08$ & $-9.0 \pm 3.0$ & $20.1 \pm 4.8$ & (5, 20, 33, 41) \\ 
J0627$-$1114 & 06 27 20.08 $-$11 14 24.1 & T6 & $15.49 \pm 0.05$ & $0.06 \pm 0.19$ & $-13 \pm 1$ & $-338 \pm 1$ & $13.37 \pm 0.64$ & \nodata & \nodata & (17, 34) \\ 
J0629+2418 & 06 29 05.12 +24 18 08.7 & T2sb$^{e}$ & $15.89 \pm 0.09$ & $0.72 \pm 0.18$ & $-35 \pm 4$ & $-368 \pm 4$ & $26.67 \pm 2.35$ & \nodata & \nodata & (23, 35) \\ 
J0755+2212 & 07 55 47.95 +22 12 16.9 & T5 & $15.73 \pm 0.06$ & $0.0 \pm 0.2$ & $-21 \pm 1$ & $-256 \pm 1$ & $14.84 \pm 0.7$ & \nodata & \nodata & (5, 35) \\ 
J0819$-$0335 & 08 19 58.21 $-$03 35 26.6 & T4 & $14.99 \pm 0.04$ & $0.41 \pm 0.11$ & $-199 \pm 3$ & $-166 \pm 2$ & $14.01 \pm 0.43$ & \nodata & \nodata & (17, 29, 35) \\ 
J0909+6525 & 09 09 00.86 +65 25 27.6 & T1.5+T2.5$^{e}$ & $16.03 \pm 0.09$ & $0.86 \pm 0.17$ & $-223 \pm 1$ & $-120 \pm 1$ & $15.65 \pm 0.96$ & \nodata & \nodata & (10, 9, 35) \\ 
J0937+2931 & 09 37 34.88 +29 31 41.0 & T6 & $14.65 \pm 0.04$ & $-0.62 \pm 0.13$ & $973 \pm 6$ & $-1298 \pm 5$ & $6.12 \pm 0.07$ & $-5.0 \pm 3.0$ & $60 \pm 10.0$ & (5, 14, 31, 41) \\ 
J1106+2754 & 11 06 11.92 +27 54 21.6 & T0+T4.5$^{f}$ & $14.82 \pm 0.04$ & $1.02 \pm 0.07$ & $-271 \pm 1$ & $-452 \pm 1$ & $20.3 \pm 0.5$ & \nodata & \nodata & (21, 9, 33) \\ 
J1217$-$0311 & 12 17 11.10 $-$03 11 13.2 & T7.5 & $15.86 \pm 0.06$ & $-0.03 \pm 0.07$ & $-1054 \pm 2$ & $76 \pm 2$ & $11.01 \pm 0.27$ & $5.0 \pm 1.6$ & $31.4 \pm 2.1$ & (3, 25, 14, 31, 43) \\ 
J1225$-$2739 & 12 25 54.32 $-$27 39 46.7 & T5.5+T8$^{f}$ & $15.26 \pm 0.05$ & $0.19 \pm 0.16$ & $385 \pm 3$ & $-628 \pm 2$ & $13.32 \pm 0.44$ & \nodata & \nodata & (3, 13, 14, 31) \\ 
J1254$-$0122 & 12 54 53.90 $-$01 22 47.3 & T2e & $14.89 \pm 0.04$ & $1.05 \pm 0.06$ & $-492 \pm 4$ & $111 \pm 3$ & $13.48 \pm 0.42$ & $4.0 \pm 3.0$ & $27.3 \pm 2.5$ & (19, 6, 32, 41) \\ 
J1324+6358$^{c}$ & 13 24 33.86 +63 58 30.7 & T2p & $15.6 \pm 0.07$ & $1.54 \pm 0.09$ & $-364 \pm 2$ & $-72 \pm 2$ & $10.03 \pm 0.56$ & $-23.7 \pm 0.4$ & $11.5 \pm 1.0$ & (21, 17, 35, 40) \\ 
J1331$-$0116 & 13 31 48.95 $-$01 16 50.1 & T0 & $15.46 \pm 0.04$ & $1.39 \pm 0.08$ & $-422 \pm 6$ & $-1039 \pm 5$ & $20.0 \pm 2.0$ & \nodata & \nodata & (15, 26, 37) \\ 
J1346$-$0031 & 13 46 46.35 $-$00 31 50.1 & T6.5 & $16.0 \pm 0.1$ & $0.23 \pm 0.29$ & $-503 \pm 3$ & $-114 \pm 2$ & $14.64 \pm 0.49$ & $-23.1 \pm 1.5$ & $<15$ & (3, 9, 14, 31, 43) \\ 
J1457$-$2122 & 14 57 14.96 $-$21 21 47.8 & T8 & $15.32 \pm 0.05$ & $0.08 \pm 0.16$ & $1034 \pm 2$ & $-1726 \pm 1$ & $5.91 \pm 0.06$ & $28.9 \pm 2.4$ & $28.6 \pm 2.4$ & (4, 38, 31, 43) \\ 
J1503+2525 & 15 03 19.61 +25 25 19.8 & T5.5 & $13.94 \pm 0.02$ & $-0.03 \pm 0.06$ & $87 \pm 1$ & $558 \pm 1$ & $6.42 \pm 0.03$ & $-40.5 \pm 2.1$ & $32.8 \pm 2.0$ & (7, 33, 43) \\ 
J1506+7027 & 15 06 52.44 +70 27 25.1 & T6 & $13.7 \pm 0.03$ & $-0.09 \pm 0.07$ & $-1194 \pm 1$ & $1042 \pm 1$ & $5.16 \pm 0.02$ & \nodata & \nodata & (16, 17, 33) \\ 
J1520+3546 & 15 20 39.75 +35 46 21.0 & T0 & $15.54 \pm 0.06$ & $1.54 \pm 0.08$ & $315 \pm 2$ & $-378 \pm 2$ & $13.59 \pm 1.05$ & \nodata & \nodata & (10, 14, 35) \\ 
J1553+1532 & 15 53 02.28 +15 32 36.9 & T6.5+T7.5$^{f}$ & $15.83 \pm 0.07$ & $0.32 \pm 0.2$ & $-386 \pm 1$ & $166 \pm 1$ & $13.32 \pm 0.16$ & $-32.9 \pm 3.0$ & $29.4 \pm 2.3$ & (5, 13, 43) \\ 
J1624+0029 & 16 24 14.37 +00 29 15.8 & T6 & $15.49 \pm 0.05$ & $-0.02 \pm 0.07$ & $-373 \pm 2$ & $-9 \pm 2$ & $11.0 \pm 0.15$ & $-30.7 \pm 3.0$ & $38.5 \pm 2.0$ & (27, 5, 14, 31, 43) \\ 
J1629+0335 & 16 29 18.41 +03 35 37.1 & T2 & $15.29 \pm 0.04$ & $1.11 \pm 0.07$ & $234 \pm 2$ & $-144 \pm 2$ & $12.99 \pm 0.67$ & \nodata & \nodata & (11, 24, 28, 29) \\ 
J1809$-$0448 & 18 09 52.56 $-$04 48 08.1 & T1 & $15.14 \pm 0.05$ & $1.18 \pm 0.08$ & $-54 \pm 2$ & $-402 \pm 5$ & $20.33 \pm 1.2$ & \nodata & \nodata & (2, 35) \\ 
J1928+2356 & 19 28 41.55 +23 56 01.6 & T6 & $14.34 \pm 0.06$ & $0.25 \pm 0.08$ & $-248 \pm 1$ & $239 \pm 1$ & $6.46 \pm 0.08$ & \nodata & \nodata & (24, 35) \\ 
J1952+7240$^{d}$ & 19 52 46.66 +72 40 00.8 & T4 & $15.09 \pm 0.05$ & $0.44 \pm 0.09$ & $-294 \pm 1$ & $-355 \pm 2$ & $12.1 \pm 0.3$ & \nodata & \nodata & (17, 28) \\ 
J2030+0749 & 20 30 42.33 +07 49 35.8 & T1.5 & $14.22 \pm 0.03$ & $0.91 \pm 0.05$ & $664 \pm 1$ & $-112 \pm 1$ & $9.73 \pm 0.08$ & \nodata & \nodata & (24, 33) \\ 
J2126+7617 & 21 26 59.14 +76 17 43.3 & T0p$^{f}$ & $14.34 \pm 0.03$ & $1.18 \pm 0.05$ & $756 \pm 1$ & $822 \pm 1$ & $16.35 \pm 0.16$ & \nodata & \nodata & (16, 17, 33) \\ 
HN Peg B & 21 44 31.33 +14 46 18.9 & T2.5 & $15.86 \pm 0.03$ & $0.46 \pm 0.03$ & $231 \pm 0$ & $-113 \pm 0$ & $18.13 \pm 0.01$ & \nodata & \nodata & (22, 33) \\ 
J2236+5105 & 22 36 16.86 +51 05 48.7 & T5 & $14.58 \pm 0.04$ & $0.13 \pm 0.1$ & $729 \pm 2$ & $324 \pm 2$ & $9.97 \pm 0.36$ & \nodata & \nodata & (2, 35) \\ 
J2254+3123 & 22 54 18.92 +31 23 49.8 & T5 & $15.26 \pm 0.05$ & $0.36 \pm 0.15$ & $60 \pm 3$ & $187 \pm 2$ & $13.89 \pm 0.58$ & $14.0 \pm 3.0$ & $15 \pm 5.0$ & (5, 28, 36, 41) \\ 
J2356$-$1553 & 23 56 54.77 $-$15 53 11.1 & T6 & $15.82 \pm 0.06$ & $0.05 \pm 0.19$ & $-423 \pm 4$ & $-616 \pm 4$ & $13.44 \pm 1.05$ & $19.0 \pm 3.0$ & $15 \pm 5.0$ & (5, 31, 41) \\
\enddata
\tablenotetext{a}{References are in the order of discovery, classification, astrometry, previously published RV and $v\sin{i}$ measurements.}
\tablenotetext{b}{Previously identified as a member of the Carina-Near moving group.}
\tablenotetext{c}{Previously identified as a member of the AB Doradus moving group.}
\tablenotetext{d}{Distance is estimated using $M_J$/spectral type relation in \cite{Dupuy:2012aa}.}
\tablenotetext{e}{Suspected binary based on blended light spectrum; the component types are estimated to be
L7+T5.5 for J0629+2418 \citep{Mace:2013aa} and T1.5+T2.5 for J0909+6525 \citep{Burgasser:2010aa}.}
\tablenotetext{f}{Confirmed binary as reported in \citet{2003ApJ...586..512B,2006ApJS..166..585B,Burgasser:2010aa,Kirkpatrick:2010aa,Dupuy:2012aa}; and this paper}
\tablerefs{Source discovery/classification: (1) \citet{Artigau:2006aa},  (2) \citet{Best:2013aa},  (3) \citet{Burgasser:1999aa},  (4) \citet{Burgasser:2000aa},  (5) \citet{Burgasser:2002aa},  (6) \citet{Burgasser:2003aa},  (7) \citet{Burgasser:2003ab},  (8) \citet{Burgasser:2004ab},  (9) \citet{Burgasser:2010aa},  (10) \citet{Chiu:2006aa},  (11) \citet{Deacon:2011aa},  (12) \citet{Deacon:2017aa},  (13) \citet{Dupuy:2012aa},  (14) \citet{Faherty:2009aa},  (15) \citet{Hawley:2002aa},  (16) \citet{Kirkpatrick:2010aa},  (17) \citet{Kirkpatrick:2011aa},  (18) \citet{Knapp:2004aa},  (19) \citet{Leggett:2000aa},  (20) \citet{Liu:2006aa},  (21) \citet{Looper:2007aa},  (22) \citet{Luhman:2007aa},  (23) \citet{Mace:2013aa},  (24) \citet{Mace:2014aa},  (25) \citet{Metchev:2008aa},  (26) \citet{Schneider:2014aa},  (27) \citet{Strauss:1999aa},  Source astrometry: (28) \citet{Best:2018aa},  (29) \citet{Best:2020aa},  (30) \citet{Dupuy:2017aa},  (31) \citet{Faherty:2012aa},  (32) \citet{Gaia-Collaboration:2018ab},  (33) \citet{Gaia-Collaboration:2020aa},  (34) \citet{Kirkpatrick:2019aa},  (35) \citet{Kirkpatrick:2021aa},  (36) \citet{Manjavacas:2013aa},  (37) \citet{Smart:2018aa},  (38) \citet{Weinberger:2016aa}, Source previously published RV and  $v\sin{i}$: (39) \citet{Gagne:2017aa},  (40) \citet{Gagne:2018aa},  (41) \citet{Prato:2015aa},  (42) \citet{Zapatero-Osorio:2006aa},  (43) \citet{Zapatero-Osorio:2007aa}}
\end{deluxetable*}
\end{longrotatetable}

\startlongtable
\begin{deluxetable*}{ccccccccccc}
\tablecaption{NIRSPEC T Dwarf Observing Log\label{tableobservinglog}}
\tablewidth{700pt}
\tabletypesize{\scriptsize}
\tablehead{
& & & & & & & & \colhead{Echelle}  & \colhead{Cross-disperser}  & \colhead{Barycentric } \\
\colhead{Source} & \colhead{Program PI} & 
\colhead{UT Date} & \colhead{UT Time$^{a}$} & \colhead{Integration$^{b}$} & 
\colhead{Airmass$^{a}$} & \colhead{Filter$^{d}$} & \colhead{Slit} & \colhead{Angle$^{a}$} & \colhead{Angle$^{a}$} & \colhead{Correction} \\
\colhead{} & \colhead{} & 
\colhead{} & \colhead{(hh:mm:ss)} & \colhead{(s)} & 
\colhead{} & \colhead{} & \colhead{} & \colhead{(deg)} & \colhead{(deg)} & \colhead{(km s$^{-1}$)} 
} 
\startdata
J0000+2554$^{c}$ & Burgasser & 2019-10-17 & 08:39:13 & 2 $\times$ 1500 & 1.01 & $N3$ & 0$\farcs$432$\times$12 & 62.97 & 34.09 &$-$6.329 \\ 
J0034+0523$^{c}$ & Burgasser & 2020-09-03 & 08:39:31 & 2 $\times$ 1400 & 1.62 & $N3$ & 0$\farcs$432$\times$12 & 62.98 & 34.09 & 14.315 \\ 
J0136+0933 & McLean & 2008-12-04 & 04:38:53 & 4 $\times$ 600 & 1.21 & $N3$ & 0$\farcs$432$\times$12 & 63.00 & 34.08 &$-$21.792 \\ 
\nodata & Burgasser & 2013-10-16 & 11:42:53 & 2 $\times$ 900 & 1.08 & $N7$ & 0$\farcs$432$\times$12 & 62.97 & 35.47 & 0.857 \\ 
\nodata & Burgasser & 2016-02-03 & 05:04:31 & 2 $\times$ 750 & 1.15 & $N7$ & 0$\farcs$432$\times$12 & 63.03 & 35.46 &$-$29.081 \\ 
J0150+3827$^{c}$ & Burgasser & 2020-08-05 & 13:56:36 & 2 $\times$ 1500 & 1.10 & $N7$ & 0$\farcs$432$\times$12 & 62.98 & 35.73 & 26.588 \\
J0213+3648$^{c}$ & Burgasser & 2020-09-03 & 10:08:18 & 1 $\times$ 750 & 1.54 & $N3$ & 0$\farcs$432$\times$12 & 62.98 & 34.09 & 24.419 \\
J0243$-$2453$^{c}$ & Burgasser & 2021-01-01 & 05:27:08 & 2 $\times$ 1800 & 1.45 & $N3$ & 0$\farcs$432$\times$12 & 62.95 & 34.06 &$-$22.462 \\ 
J0415$-$0935 & Mart\'{i}n & 2005-10-26 & 12:55:37 & 2 $\times$ 600 & 1.16 & $N3$ & 0$\farcs$432$\times$12 & 63.00 & 34.08 & 11.18 \\ 
\nodata & Wainscoat & 2006-01-18 & 09:01:02 & 2 $\times$ 600 & 1.39 & $N3$ & 0$\farcs$432$\times$12 & 63.00 & 34.08 &$-$22.389 \\ 
J0559$-$1404 & Engineering & 2000-10-10 & 13:53:03 & 4 $\times$ 600 & 1.27 & $N3$ & 0$\farcs$576$\times$12 & 63.00 & 34.08 & 22.527 \\ 
\nodata & McLean & 2001-10-09 & 15:06:35 & 2 $\times$ 600 & 1.20 & $N3$ & 0$\farcs$432$\times$12 & 63.00 & 34.08 & 22.564 \\ 
\nodata & Rayner & 2001-11-02 & 13:17:34 & 10 $\times$ 300 & 1.21 & $N3$ & 0$\farcs$432$\times$12 & 63.00 & 34.08 & 17.938 \\ 
\nodata & McLean & 2001-12-29 & 07:49:17 & 2 $\times$ 600 & 1.41 & $N3$ & 0$\farcs$432$\times$12 & 63.00 & 34.08 &$-$3.097 \\ 
\nodata & Mart\'{i}n & 2002-11-25 & 13:06:34 & 3 $\times$ 600 & 1.25 & $N3$ & 0$\farcs$576$\times$12 & 63.00 & 34.08 & 10.44 \\ 
\nodata & Wainscoat & 2004-12-05 & 13:03:11 & 2 $\times$ 300 & 1.34 & $N3$ & 0$\farcs$432$\times$12 & 63.00 & 34.08 & 6.31 \\ 
\nodata & Mart\'{i}n & 2005-10-26 & 13:37:56 & 3 $\times$ 480 & 1.21 & $N3$ & 0$\farcs$432$\times$12 & 63.00 & 34.08 & 19.706 \\ 
\nodata & Mart\'{i}n & 2005-10-27 & 12:35:45 & 3 $\times$ 480 & 1.29 & $N3$ & 0$\farcs$432$\times$12 & 63.00 & 34.08 & 19.593 \\ 
\nodata & Mart\'{i}n & 2005-10-28 & 12:49:07 & 3 $\times$ 480 & 1.26 & $N3$ & 0$\farcs$432$\times$12 & 63.00 & 34.08 & 19.326 \\ 
\nodata & McLean & 2006-01-11 & 07:48:54 & 26 $\times$ 300 & 1.27 & $N3$ & 0$\farcs$432$\times$12 & 63.00 & 34.08 &$-$8.445 \\ 
\nodata & McLean & 2008-03-19 & 05:55:54 & 7 $\times$ 600 & 1.30 & $N3$ & 0$\farcs$432$\times$12 & 63.00 & 34.08 &$-$23.852 \\ 
\nodata & Burgasser & 2015-12-29 & 11:17:25 & 2 $\times$ 1500 & 1.30 & $N7$ & 0$\farcs$432$\times$12 & 63.02 & 35.48 &$-$3.265 \\
\nodata$^{c}$ & Burgasser & 2021-01-01 & 07:58:37 & 2 $\times$ 1200 & 1.33 & $N3$ & 0$\farcs$432$\times$12 & 62.95 & 34.06 &$-$4.434 \\
J0627$-$1114$^{c}$ & Burgasser & 2021-01-01 & 09:08:11 & 2 $\times$ 1800 & 1.21 & $N3$ & 0$\farcs$432$\times$12 & 62.95 & 34.06 &$-$1.101 \\
J0629+2418 & Burgasser & 2012-11-28 & 12:50:54 & 2 $\times$ 1500 & 1.00 & $N7$ & 0$\farcs$432$\times$12 & 63.02 & 35.55 & 14.826 \\
J0755+2212$^{c}$ & Burgasser & 2021-01-01 & 10:16:22 & 1 $\times$ 1800 & 1.05 & $N3$ & 0$\farcs$432$\times$12 & 62.95 & 34.06 & 8.315 \\ 
J0819$-$0335$^{c}$ & Burgasser & 2021-01-01 & 11:36:35 & 2 $\times$ 1800 & 1.09 & $N3$ & 0$\farcs$432$\times$12 & 62.95 & 34.06 & 12.859 \\
J0909+6525 & Burgasser & 2010-12-26 & 12:12:07 & 4 $\times$ 1200 & 1.45 & $N7$ & 0$\farcs$432$\times$12 & 63.00 & 35.55 & 7.764 \\ 
J0937+2931 & McLean & 2002-04-23 & 05:15:35 & 4 $\times$ 300 & 1.03 & $N3$ & 0$\farcs$432$\times$12 & 63.00 & 34.08 &$-$27.634 \\ 
\nodata & McLean & 2003-03-24 & 08:12:11 & 6 $\times$ 600 & 1.02 & $N3$ & 0$\farcs$432$\times$12 & 63.00 & 34.08 &$-$20.594 \\ 
\nodata & Prato & 2003-05-12 & 05:30:39 & 13 $\times$ 300 & 1.03 & $N3$ & 0$\farcs$432$\times$12 & 63.00 & 34.08 &$-$28.564 \\ 
\nodata & McLean & 2006-01-10 & 13:10:31 & 24 $\times$ 300 & 1.02 & $N3$ & 0$\farcs$432$\times$12 & 63.00 & 34.08 & 13.261 \\ 
\nodata & McLean & 2006-05-19 & 07:15:48 & 8 $\times$ 300 & 1.35 & $N3$ & 0$\farcs$432$\times$12 & 63.00 & 34.08 &$-$28.282 \\ 
J1106+2754 & McLean & 2008-03-19 & 08:19:51 & 12 $\times$ 600 & 1.06 & $N3$ & 0$\farcs$432$\times$12 & 63.00 & 34.08 &$-$10.193 \\ 
\nodata & Burgasser & 2010-12-26 & 15:07:25 & 3 $\times$ 1000 & 1.01 & $N3$ & 0$\farcs$432$\times$12 & 62.95 & 34.08 & 25.007 \\ 
\nodata & Burgasser & 2012-04-02 & 11:07:11 & 3 $\times$ 1500 & 1.20 & $N7$ & 0$\farcs$432$\times$12 & 63.00 & 35.52 &$-$16.475 \\ 
\nodata & Burgasser & 2012-11-27 & 13:42:12 & 3 $\times$ 1200 & 1.43 & $N7$ & 0$\farcs$432$\times$12 & 63.01 & 35.52 & 28.583 \\ 
\nodata & Burgasser & 2013-02-05 & 14:15:17 & 3 $\times$ 1500 & 1.07 & $N7$ & 0$\farcs$432$\times$12 & 63.03 & 35.46 & 9.608 \\ 
\nodata & Burgasser & 2015-01-01 & 15:00:50 & 2 $\times$ 1500 & 1.01 & $N7$ & 0$\farcs$432$\times$12 & 63.03 & 35.46 & 23.434 \\ 
\nodata & Burgasser & 2016-01-18 & 14:03:05 & 2 $\times$ 1500 & 1.01 & $N7$ & 0$\farcs$432$\times$12 & 63.02 & 35.48 & 17.818 \\ 
\nodata & Burgasser & 2016-02-16 & 09:47:02 & 3 $\times$ 1400 & 1.13 & $N7$ & 0$\farcs$432$\times$12 & 63.03 & 35.48 & 5.18 \\ 
\nodata & Burgasser & 2016-04-22 & 08:43:42 & 3 $\times$ 1500 & 1.06 & $N7$ & 0$\farcs$432$\times$12 & 62.98 & 35.48 &$-$22.888 \\ 
\nodata & Burgasser & 2016-05-22 & 07:56:10 & 3 $\times$ 1500 & 1.22 & $N7$ & 0$\farcs$432$\times$12 & 62.97 & 35.48 &$-$27.661 \\ 
\nodata & Burgasser & 2017-03-22 & 13:00:58 & 2 $\times$ 900 & 1.02 & $N7$ & 0$\farcs$432$\times$12 & 63.02 & 35.52 & 9.678 \\ 
\nodata & Burgasser & 2017-05-06 & 07:14:02 & 2 $\times$ 1200 & 1.01 & $N7$ & 0$\farcs$432$\times$12 & 63.02 & 35.52 &$-$25.802 \\ 
\nodata & Burgasser & 2018-01-01 & 14:47:39 & 2 $\times$ 1500 & 1.01 & $N7$ & 0$\farcs$432$\times$12 & 63.01 & 35.48 & 23.406 \\ 
\nodata & Burgasser & 2018-06-03 & 06:19:27 & 2 $\times$ 1500 & 1.05 & $N7$ & 0$\farcs$432$\times$12 & 63.01 & 35.38 &$-$27.541 \\
\nodata$^{c}$ & Burgasser & 2021-01-01 & 12:57:30 & 2 $\times$ 1500 & 1.11 & $N3$ & 0$\farcs$432$\times$12 & 62.95 & 34.06 & 23.544 \\ 
\nodata$^{c}$ & Burgasser & 2021-01-01 & 14:20:08 & 2 $\times$ 1500 & 1.01 & $N7$ & 0$\farcs$432$\times$12 & 62.95 & 34.06 & 23.394 \\ 
J1217$-$0311 & Mart\'{i}n & 2001-06-15 & 06:08:18 & 2 $\times$ 900 & 1.11 & $N3$ & 0$\farcs$576$\times$12 & 63.00 & 34.08 &$-$28.873 \\ 
\nodata & Wainscoat & 2006-01-18 & 14:07:41 & 2 $\times$ 600 & 1.10 & $N3$ & 0$\farcs$432$\times$12 & 63.00 & 34.08 & 27.977 \\ 
\nodata & Wainscoat & 2006-01-19 & 12:40:31 & 3 $\times$ 1200 & 1.27 & $N3$ & 0$\farcs$432$\times$12 & 63.00 & 34.08 & 27.929 \\ 
J1225$-$2739 & McLean & 2002-04-23 & 08:37:15 & 6 $\times$ 300 & 1.48 & $N3$ & 0$\farcs$432$\times$12 & 63.00 & 34.08 &$-$6.921 \\ 
J1254$-$0122 & McLean & 2001-12-31 & 15:13:36 & 3 $\times$ 300 & 1.15 & $N3$ & 0$\farcs$432$\times$12 & 63.00 & 34.08 & 30.294 \\ 
\nodata & Basri & 2002-05-17 & 06:55:33 & 9 $\times$ 300 & 1.09 & $N3$ & 0$\farcs$432$\times$12 & 63.00 & 34.08 &$-$19.715 \\ 
\nodata & Prato & 2003-05-14 & 05:50:51 & 14 $\times$ 300 & 1.24 & $N3$ & 0$\farcs$432$\times$12 & 63.00 & 34.08 &$-$18.322 \\ 
\nodata & Wainscoat & 2006-01-19 & 13:55:21 & 9 $\times$ 600 & 1.15 & $N3$ & 0$\farcs$432$\times$12 & 63.00 & 34.08 & 29.206 \\ 
\nodata & McLean & 2007-05-31 & 06:51:50 & 8 $\times$ 600 & 1.15 & $N3$ & 0$\farcs$432$\times$12 & 63.00 & 34.08 &$-$24.187 \\ 
\nodata & Burgasser & 2011-06-10 & 05:46:19 & 2 $\times$ 900 & 1.07 & $N7$ & 0$\farcs$432$\times$12 & 63.00 & 35.53 &$-$26.549 \\ 
J1324+6358 & Burgasser & 2016-05-22 & 09:03:28 & 2 $\times$ 3000 & 1.44 & $N7$ & 0$\farcs$432$\times$12 & 62.97 & 35.48 &$-$13.606 \\ 
J1331$-$0116 & Burgasser & 2011-03-18 & 12:20:39 & 4 $\times$ 1200 & 1.07 & $N7$ & 0$\farcs$432$\times$12 & 63.00 & 35.47 & 12.621 \\ 
\nodata & Burgasser & 2011-07-06 & 06:32:30 & 2 $\times$ 1500 & 1.17 & $N7$ & 0$\farcs$432$\times$12 & 63.00 & 35.53 &$-$28.921 \\ 
\nodata & McLean & 2013-05-24 & 08:27:28 & 4 $\times$ 600 & 1.08 & $N3$ & 0$\farcs$432$\times$12 & 63.00 & 34.08 & 11.046 \\ 
J1346$-$0031 & Mart\'{i}n & 2001-06-15 & 07:10:43 & 3 $\times$ 900 & 1.08 & $N3$ & 0$\farcs$576$\times$12 & 63.00 & 34.08 &$-$24.832 \\ 
\nodata & Wainscoat & 2006-01-18 & 14:55:32 & 3 $\times$ 600 & 1.14 & $N3$ & 0$\farcs$432$\times$12 & 63.00 & 34.08 & 29.934 \\ 
J1457$-$2122 & Mart\'{i}n & 2001-06-15 & 08:58:36 & 2 $\times$ 900 & 1.41 & $N3$ & 0$\farcs$576$\times$12 & 63.00 & 34.08 &$-$17.283 \\ 
\nodata & Wainscoat & 2006-01-18 & 15:37:10 & 2 $\times$ 600 & 1.53 & $N3$ & 0$\farcs$432$\times$12 & 63.00 & 34.08 & 28.542 \\ 
\nodata & McLean & 2008-03-19 & 11:22:16 & 8 $\times$ 600 & 1.61 & $N3$ & 0$\farcs$432$\times$12 & 63.00 & 34.08 & 23.02 \\ 
J1503+2525 & Wainscoat & 2006-01-19 & 15:54:04 & 2 $\times$ 300 & 1.08 & $N3$ & 0$\farcs$432$\times$12 & 63.00 & 34.08 & 23.037 \\ 
\nodata & McLean & 2008-03-19 & 13:48:14 & 8 $\times$ 600 & 1.01 & $N3$ & 0$\farcs$432$\times$12 & 63.00 & 34.08 & 13.207 \\ 
J1506+7027$^{c}$ & Burgasser & 2020-09-03 & 05:17:38 & 2 $\times$ 1200 & 1.77 & $N3$ & 0$\farcs$432$\times$12 & 62.98 & 34.09 & 2.035 \\ 
J1520+3546 & Burgasser & 2012-04-02 & 14:19:49 & 2 $\times$ 1200 & 1.09 & $N7$ & 0$\farcs$432$\times$12 & 63.00 & 35.52 & 6.498 \\ 
\deleted{J1534$-$2952 & Ghez & 2008-06-01 & 08:22:35 & 4 $\times$ 1800 & 1.60 & $K$-$AO$ & 0.041x2.26 & 63.00 & 35.65 &$-$5.919 \\} 
J1553+1523 & Mart\'{i}n & 2001-06-15 & 09:38:29 & 2 $\times$ 900 & 1.03 & $N3$ & 0$\farcs$576$\times$12 & 63.00 & 34.08 &$-$12.915 \\ 
J1624+0029 & Mart\'{i}n & 2001-06-15 & 10:22:21 & 2 $\times$ 900 & 1.11 & $N3$ & 0$\farcs$576$\times$12 & 63.00 & 34.08 &$-$9.423 \\ 
\nodata & Basri & 2002-05-17 & 12:23:06 & 8 $\times$ 300 & 1.12 & $N3$ & 0$\farcs$432$\times$12 & 63.00 & 34.08 & 3.83 \\ 
\nodata & McLean & 2005-06-04 & 08:08:51 & 1 $\times$ 600 & 1.18 & $N3$ & 0$\farcs$432$\times$12 & 63.00 & 34.08 &$-$4.155 \\ 
J1629+0335 & Burgasser & 2011-08-11 & 06:55:17 & 2 $\times$ 1500 & 1.11 & $N7$ & 0$\farcs$432$\times$12 & 63.00 & 35.47 &$-$25.693 \\ 
\nodata & Burgasser & 2011-09-07 & 05:15:20 & 2 $\times$ 1200 & 1.12 & $N7$ & 0$\farcs$432$\times$12 & 63.00 & 35.46 &$-$26.499 \\ 
J1809$-$0448$^{c}$ & Burgasser & 2020-08-25 & 06:37:23 & 2 $\times$ 1500 & 1.10 & $N7$ & 0$\farcs$432$\times$12 & 62.98 & 35.75 &$-$24.312 \\ 
J1928+2356$^{c}$ & Burgasser & 2019-09-12 & 08:00:26 & 1 $\times$ 1500 & 1.04 & $N7$ & 0$\farcs$432$\times$12 & 63.00 & 35.76 &$-$16.283 \\ 
\nodata & Burgasser & 2019-10-17 & 06:31:42 & 2 $\times$ 1200 & 1.2 & $N3$ & 0$\farcs$432$\times$12 & 62.97 & 34.09 &$-$21.221 \\ 
J1952+7240$^{c}$ & Burgasser & 2019-10-17 & 07:31:51 & 2 $\times$ 1500 & 1.91 & $N3$ & 0$\farcs$432$\times$12 & 62.97 & 34.09 & 1.957 \\ 
J2030+0749$^{c}$ & Burgasser & 2020-07-10 & 09:55:16 & 2 $\times$ 900 & 1.13 & $N7$ & 0$\farcs$432$\times$12 & 63.00 & 35.73 & 10.869 \\
J2126+7617 & Burgasser & 2011-06-10 & 14:59:25 & 2 $\times$ 750 & 1.81 & $N7$ & 0$\farcs$432$\times$12 & 63.00 & 35.53 & 3.919 \\ 
\nodata & Burgasser & 2011-07-06 & 13:51:48 & 2 $\times$ 1200 & 1.83 & $N7$ & 0$\farcs$432$\times$12 & 63.00 & 35.53 & 7.054 \\ 
\nodata & Burgasser & 2011-08-11 & 11:38:20 & 2 $\times$ 1500 & 1.84 & $N7$ & 0$\farcs$432$\times$12 & 63.00 & 35.47 & 9.141 \\ 
\nodata & Burgasser & 2011-09-07 & 10:49:22 & 2 $\times$ 1200 & 1.92 & $N7$ & 0$\farcs$432$\times$12 & 63.00 & 35.46 & 8.549 \\ 
\nodata & Burgasser & 2013-09-17 & 10:46:37 & 2 $\times$ 1500 & 2.00 & $N7$ & 0$\farcs$432$\times$12 & 62.97 & 35.51 & 7.816 \\ 
\nodata & Burgasser & 2013-10-16 & 10:34:47 & 2 $\times$ 1500 & 2.38 & $N7$ & 0$\farcs$432$\times$12 & 62.97 & 35.47 & 4.528 \\ 
\nodata & Burgasser & 2014-09-02 & 09:25:11 & 2 $\times$ 1200 & 1.81 & $N7$ & 0$\farcs$288$\times$12 & 63.02 & 35.49 & 8.839 \\
\nodata$^{c}$ & Burgasser & 2020-08-25 & 08:52:17 & 2 $\times$ 1500 & 1.82 & $N7$ & 0$\farcs$432$\times$12 & 62.98 & 35.75 & 9.12 \\ 
\nodata$^{c}$ & Burgasser & 2020-09-03 & 06:46:14 & 2 $\times$ 1200 & 1.93 & $N3$ & 0$\farcs$432$\times$12 & 62.98 & 34.09 & 8.838 \\ 
HN Peg B & Skemer & 2017-06-09 & 11:21:04 & 20 $\times$ 13200 & 1.59 & $N7$ & 0$\farcs$432$\times$12 & 63.01 & 36.60 & 25.737 \\ 
\nodata & Skemer & 2017-10-08 & 05:36:07 & 10 $\times$ 9000 & 1.07 & $N7$ & 0$\farcs$432$\times$24 & 62.93 & 36.55 &$-$17.657 \\ 
J2236+5105$^{c}$ & Burgasser & 2020-09-03 & 07:44:53 & 2 $\times$ 1200 & 1.34 & $N3$ & 0$\farcs$432$\times$12 & 62.98 & 34.09 & 8.472 \\ 
J2254+3123 & Prato & 2003-08-10 & 09:20:13 & 10 $\times$ 300 & 1.32 & $N3$ & 0$\farcs$432$\times$12 & 63.00 & 34.08 & 16.077 \\ 
\nodata & McLean & 2005-07-19 & 12:17:11 & 4 $\times$ 600 & 1.06 & $N3$ & 0$\farcs$432$\times$12 & 63.00 & 34.08 & 21.277 \\ 
\nodata & McLean & 2004-11-21 & 05:38:26 & 1 $\times$ 300 & 1.03 & $N5$ & 0$\farcs$288$\times$24 & 62.61 & 36.90 &$-$21.596 \\ 
\nodata & McLean & 2007-06-26 & 14:15:19 & 4 $\times$ 600 & 1.04 & $K$ & 0$\farcs$432$\times$24 & 63.00 & 35.65 & 23.945 \\ 
J2356$-$1553 & McLean & 2003-07-20 & 12:50:03 & 4 $\times$ 600 & 1.36 & $N3$ & 0$\farcs$432$\times$12 & 63.00 & 34.08 & 23.527 \\ 
\nodata & McLean & 2005-07-19 & 13:52:18 & 3 $\times$ 600 & 1.25 & $N3$ & 0$\farcs$432$\times$12 & 63.00 & 34.08 & 23.519 \\ 
\nodata & McLean & 2005-12-10 & 05:21:03 & 6 $\times$ 600 & 1.23 & $N3$ & 0$\farcs$432$\times$12 & 63.00 & 34.08 &$-$29.26
\enddata
\tablenotetext{a}{Values are taken from the configuration of the first file within the same night.}
\tablenotetext{b}{The number of files times the individual integration time.
}
\tablenotetext{c}{Observed with the upgraded NIRSPEC.}
\tablenotetext{d}{Filters: $N3$(NIRSPEC-3), $N5$(NIRSPEC-5), $N7$(NIRSPEC-7)\deleted{, $K-AO$(NIRSPEC K with Adaptive Optics)}, and $K$(NIRSPEC K)}
\end{deluxetable*}

\noindent measurements for T dwarfs also hinders the identification of young moving group members and short-period binaries, both of which serve as benchmarks for disentangling the age/mass degeneracy.

To improve our sampling of brown dwarf space motions, rotation, and multiplicity, and to properly compare the kinematic distributions of T dwarfs to warmer UCDs in the vicinity of the Sun, we report RV and $v\sin{i}$ measurements, and stellar atmospheric parameters, for 37 T dwarfs with Keck/NIRSPEC high-resolution spectroscopic data. 
In Section \ref{sec:observations}, we describe the sample and observations. 
In Section \ref{sec:methods}, we discuss the data reduction methods, wavelength calibration, and forward-modeling routines, and analyze the outcomes in the context of prior measurements in the literature.
In Section \ref{sec:analysis}, we analyze individual RV and $v\sin{i}$ measurements, deriving space motions, examining cluster and Galactic population membership, characterizing Galactic orbits, and examining multi-epoch measurements for evidence of RV variability. 
In Section \ref{sec:discussion}, we examine the kinematics of local T dwarfs as a population, evaluating their statistical age from velocity and vertical action dispersions, and we compare these to the distributions of late-M and L dwarfs and to population simulations. We find local T dwarfs have kinematic ages similar to those of late-M dwarfs and simulation predictions, and both are younger than local early- and mid-L dwarfs. Removing the thick disk L dwarfs resolves this age discrepancy.
In Section \ref{sec:explain}, we evaluate sample incompleteness and bias, and conduct a more detailed analysis of L dwarf kinematics.
In Section \ref{sec:summary} we summarize our results. 
As Section \ref{sec:observations} and \ref{sec:methods} describe our analysis methods in detail, readers primarily interested in the scientific aspects of this study can start at Section \ref{sec:analysis} without loss of content. 

\section{Observations} \label{sec:observations}

Our T dwarf sample was compiled from sources observed with the Keck Near-Infrared Spectrometer \citep[NIRSPEC; ][]{1998SPIE.3354..566M, 2000SPIE.4008.1048M}. NIRSPEC is a high-resolution, near-infrared, cross-dispersed spectrograph, mounted on the Keck II telescope, spanning 0.95 to 5.5 $\micron$ with a spectral resolution of $\sim$25,000 (pre-upgrade) and $\sim$35,000 (post-upgrade) for a slit width of 0$\farcs$432. Table \ref{tablesample} lists the 37 T dwarfs in our sample, with photometry from the Two Micron All Sky Survey \citep[2MASS; ][]{Skrutskie:2006aa} and the UKIRT Infrared Deep Sky Survey \citep[UKIDSS; ][]{Lawrence:2013aa}; astrometry from 2MASS, the \textit{Wide-field Infrared Survey Explorer} \citep[\textit{WISE}; ][]{Cutri:2012aa}, the Panoramic Survey Telescope and Rapid Response System \citep[Pan-STARRS; ][]{Chambers:2017aa} and $Gaia$ \citep{Gaia-Collaboration:2018ab, Gaia-Collaboration:2020aa}; and classifications from the literature. Our sample encompasses observations taken between 2000 and 2021, including publicly available archival data obtained from the Keck Observatory Archive (KOA)\footnote{\url{https://koa.ipac.caltech.edu}}. All public data were downloaded directly from the KOA, with the exception of data from 2001 June 15 \citep{Zapatero-Osorio:2006aa, Zapatero-Osorio:2007aa} and 2005 July 19 \citep{Prato:2015aa} which were provided by the PIs of these programs.
Fourteen sources were observed after the NIRSPEC upgrade \citep{2018SPIE10702E..0AM}. Table \ref{tableobservinglog} lists the details for each observation, segregated by epoch. Most observations were accompanied by observation of an early type star (typically A0) for telluric absorption correction. In total, we analyzed 290 near-infrared spectra in order $N3$ (1.143--1.375 $\micron$) and 101 near-infrared spectra in order $N7$ (1.839--2.630 $\micron$).

\section{Methods} \label{sec:methods}

\subsection{Spectral Data Reduction} \label{sec:data_reduction}

\subsubsection{NSDRP Pipeline and Modifications} \label{sec:nsdrp}

All of the NIRSPEC data were reduced using a modified version of the NIRSPEC Data Reduction Pipeline \citep[NSDRP;][]{Tran:2016aa}. 
Our modifications were as follows: 

\begin{enumerate}
\item We loosened the criteria for determining the edges of each dispersion order. NSDRP determines these edges from flat field frames, but the threshold value for edge detection was too strict and cut off some science spectra in certain orders. Note that NSDRP only processes orders that are completely within the image. Orders that are cut off at the top or bottom of the detector are ignored.

\item NSDRP determines the wavelength calibration using only sky emission lines, so we added arc lamp and etalon lamp exposures as additional inputs for wavelength calibration. Arc lamp line identifications were drawn primarily from the National Institute of Standards and Technology \cite[NIST;][]{NIST_ASD}, the Atomic Line List \citep{VanHoof:2018aa}, 
and \cite{Outred:1978aa}. 
Etalon lamp lines are not tied to an absolute wavelength scale but provide a way to refine the relative wavelength calibration across an order. 

\item When spatially rectifying the tilted orders, NSDRP uses the top and bottom of the edges of the flat field dispersions. This is suitable for data with low S/N, such as our science spectra. For data with high S/N, such as our telluric star spectra, we found that this rectification can be offset by 1--2 pixels across the spectral order. For these data, we used the trace of the bright object spectrum, \replaced{using}{determined by} a Gaussian profile fit along each column.

\item We \deleted{also }added a 3-$\sigma$ clipping algorithm to find the optimal spectral tilt (spatial y-direction) from the emission line traces.

\item We added new coefficients for the grating equation, determined empirically, for $N3$ ($J$-band) and $N7$ ($K$-band) \deleted{for} data obtained with the upgraded NIRSPEC instrument.
\end{enumerate}

\noindent These modifications have been integrated into an updated version of the NSDRP.\footnote{\url{https://github.com/ctheissen/NIRSPEC-Data-Reduction-Pipeline}} 

In addition, we implemented an algorithm to correct for fringing features in the flat field images.
Interference from reflections between the echelle gratings and internal optics in NIRSPEC produce fringing patterns, easily visible in high signal-to-noise (S/N) data and flat field images. The interference patterns reduce both our ability to model the spectra and achieve high radial velocity precision. We therefore added a defringing algorithm for the raw flat field files using the wavelet analysis described in \citet{Rojo:2006aa} and the \textit{wavelets}\footnote{\url{https://github.com/aaren/wavelets}} package described in \citet{Torrence:1998aa}. Briefly, each pixel in the flat field data is substituted for the median-average value of the nearest $10$ pixels in the vertical direction, making the horizontal fringe patterns more prominent. A continuum profile is determined from the binned data using a low-order cubic spline. Subtracting this continuum, we fit the difference using a wavelet analysis. The modeled fringe pattern for the flat images was then subtracted from the original flat field frame. This algorithm was only applied to flat field images, so fringing remains in the science frames. As our data were typically low S/N, we did not attempt to correct the science frames for fringing, and defer this to a future study.

\subsubsection{Telluric Wavelength Calibration} \label{wavelength calibration}

The wavelength calibration from NSDRP is a second-order polynomial fit to the sky, arc, and etalon lamp emission lines in the science and calibration data. We found this default calibration to be insufficient for precise RV measurement. 
We therefore adopted the wavelength calibration method described in \citet{Burgasser:2016aa}, cross-correlating the associated telluric standard star spectrum with a telluric absorption model \citep{Moehler:2014aa} over discrete wavebands of width 15~{\AA}. 
This fit was performed after first modeling out the continuum of the A0~V with a second-order polynomial and a Voigt absorption profile for orders containing H~\textsc{i} absorption lines\added{\footnote{Order 59 in $J$-band and order 35 in $K$-band are fit with a Voigt profile, multiplied by a second-order polynomial function, using \textit{scipy}'s \texttt{special.wofz} function. The Voigt profile parameters are optimized by a least squares fit.}}.
We used a fit to the residual shift to update the wavelength solution, both represented as fourth-order polynomial functions. This process was iterated until the wavelength solution residuals reached a minimum.
The calibrated wavelength solution of each telluric standard star spectrum was applied to the corresponding science spectrum with a wavelength offset determined in the forward-modeling (Section \ref{subsec:mcmc}). The typical standard deviation of velocity residuals in the calibrated telluric spectra ranges from 0.1 to 0.6 km s$^{-1}$, with a median residual of 0.3 km s$^{-1}$. Our baseline $N3$ order 58 has a median residual of 0.2 km s$^{-1}$, while our baseline $N7$ order 33 has a median residual of 0.4 km s$^{-1}$. We also determined wavelength calibrations for orders 32, 34, 37, and 38 ($N7$); and 57, 63, 64, 65, and 66 ($N3$). Orders 35 and 36 ($N7$), and 59, 60, 61, 62 ($N3$) do not have sufficient telluric absorption features to apply this method, while orders 55 and 56 have excessively strong telluric absorption and do not provide sufficient signal for our targets. \\

\subsection{Forward Modeling} \label{subsec:mcmc}

\subsubsection{Overview} \label{subsec:overview}

The typical approach to spectral reduction is to correct for instrumental and telluric atmospheric effects to infer the target's emitted spectrum. Here, we model these effects explicitly using a forward-modeling approach, following \cite{Blake:2010aa} and \cite{Burgasser:2016aa} \citep[see also][]{Tanner:2012aa, 2016ApJ...819..133A, Vos:2017aa, Cale:2019aa}. The stellar parameters (effective temperature, surface gravity, rotational velocity, and radial velocity), and calibration factors (continuum and wavelength corrections, instrumental line-spread function, and strength of telluric absorption) are determined using a Markov Chain Monte Carlo (MCMC) algorithm \citep{Goodman:2010aa} using the package \textit{emcee} \citep{Foreman-Mackey:2013aa}. Our forward-modeling method is optimal in those spectral orders with both strong telluric absorption features for accurate wavelength calibration, and sufficient structure in the stellar spectrum to distinguish it from the telluric absorption. We found order 33 in the $K$-band ($N7$; 22690--23410 {\AA}) and order 58 in the $J$-band ($N3$; 12990--13290 {\AA}) to be the ideal orders for the T dwarfs, similar to prior studies \citep{Blake:2010aa, Konopacky:2010aa, Burgasser:2012aa, Tanner:2012aa}. Our forward-modeling routine \textit{Spectral Modeling Analysis and RV Tool} \replaced{(SMART)\citep{Hsu:2021ab}}{(SMART; \citealt{Hsu:2021ab})} is open source and available online.\footnote{\url{https://github.com/chihchunhsu/smart}}

To summarize, there are three main steps in our MCMC forward-modeling scheme: 
\begin{enumerate}
\item An MCMC fit of the telluric spectrum is performed to determine the parameters for the instrumental line-spread function (LSF) and the strength of telluric absorption, which are used to initialize the MCMC of the science spectrum.
\item An initial MCMC fit of the science spectrum is conducted to estimate the stellar parameters for effective temperature ($T_{\text{eff}}$), surface gravity ($\log {g}$), rotational velocity ($v\sin{i}$), and radial velocity (RV), as well as calibration and nuisance parameters. 
\item The residuals between the best-fit model spectrum and data are used to generate a mask array to identify discrepant pixels, and a final MCMC fit of the masked science spectrum is run to obtain the best estimates of the fit parameters.
\end{enumerate}

\subsubsection{Telluric Star Modeling} \label{subsec:telluric_mcmc}

Each telluric standard star was forward-modeled to obtain initial estimates of the LSF and strength of telluric absorption. The data ($D$) are modeled as:

\begin{equation}
D[p] = C[p(\lambda)] \times \Big[ T \big[ p^*(\lambda) \big]^{\alpha} \otimes \kappa_G (\Delta \nu_\text{inst}(p)) \Big] + C_\text{flux}.
\label{eqn:forwardmodeltelluric}
\end{equation}

\noindent Here, $p$ is the pixel coordinate, $p^*(\lambda) = p(\lambda) + C_{\lambda}$ is the mapping of wavelength to pixel with a small constant offset, $C[p]$ is a second-order polynomial representing the continuum correction\deleted{ (including H~\textsc{i} absorption)}, $\kappa_G (\Delta \nu_\text{inst})$ is the intrumental LSF, assumed to be Gaussian of velocity width $\Delta \nu_\text{inst}$ ($\otimes$ represents convolution), $T \big[ p\big]$ is a model for telluric absorption from \citet{Moehler:2014aa}, $\alpha$ is a constant that scales with the airmass and precipitable water vapor, and $C_{\mathrm{flux}}$ is an additive offset for the overall flux. 
The default precipitable water vapor (pwv) for the \citet{Moehler:2014aa} models used was 0.5 mm. For some data, a telluric model with pwv = 1.5 mm produced a better fit and these were used instead.
We fit the parameters, $\Delta \nu_\text{inst}$, $C_{\mathrm{flux}}$, $C_{\lambda}$, and $\alpha$ assuming uniform priors (Table \ref{table:mcmc_parameter}), while the continuum was determined after each iteration through a least-squares fit of the ratio of the data and model to a second-order polynomial \deleted{and Voigt profile }(i.e., the continuum fit was done outside of the MCMC). 
The likelihood function was computed by assuming the noise follows a normal distribution:
\begin{equation}
\ln \, \mathcal{L} = -0.5 \times \left[ \sum \chi^2 + \sum \ln ( 2 \pi \sigma^2 ) \right],
\end{equation}

\noindent where $\chi = \frac{Data[p] - D[p]}{\sigma [p]}$, and $Data$ and $\sigma$ are the observed spectrum and noise. 
We used 50 walkers of 400 steps each and a burn-in of 300 steps, with these parameters chosen\footnote{See Section~\ref{subsec:science_mcmc} for discussions on convergence.} 
to optimize convergence. The convergence of each fit was checked visually and quantified using both the Gelman-Rubin scale reduction factor \citep{Gelman:1992aa} and integrated autocorrelation time statistics \citep{Goodman:2010aa}.
For each fit, we inferred the best-fit parameter values and their uncertainties by computing the 50$^\text{th}$, 16$^\text{th}$, and 84$^\text{th}$ percentiles of the marginalized posterior distributions from the residual MCMC chains.

\subsubsection{T Dwarf Modeling} \label{subsec:science_mcmc}

The T dwarf spectra were forward-modeled as:

\begin{equation}
\begin{split}
D[p] & = C[p] \times \Bigg[ \bigg(M \Big[p^* \big(\lambda \big[ 1 + \frac{RV^*}{c}\big] \big) , T_{\text{eff}}, \log \, g \Big]  \\ 
& \otimes \kappa_R (v\sin{i}) \bigg) \times T \big[ p^*(\lambda) \big]^{\alpha} \Bigg] \otimes \kappa_G (\Delta \nu_\text{inst}) + C_\text{flux} \, ,    
\end{split}
\label{eqn:forwardmodelscience}
\end{equation}

where the additional terms compared to equation~(\ref{eqn:forwardmodeltelluric}) include the solar-metallicity stellar atmosphere model\footnote{See Section \ref{sec:atmmodel} for discussions of model selection} $M[p]$ drawn from the BT-Settl \citep{Allard:2012aa} and Sonora \citep{Marley:2018aa} model grids, parameterized by effective temperature ($T_{\text{eff}}$) and surface gravity ($\log {g}$);
$RV^* = RV + v_\text{bary}$ is the radial velocity of the source plus barycentric motion of the Earth at the observed epoch; $c$ is the speed of light; $\kappa_R$ is the rotational broadening profile defined in \citet{Gray:1992aa} assuming a constant limb-darkening coefficient of $\epsilon = 0.6$ \citep{Claret:2000aa}; and $v\sin{i}$ is the projected rotational velocity. 
Atmosphere model log fluxes were linearly interpolated between grid points to approximate a continuous distribution of $T_{\text{eff}}$ and $\log {g}$ values.
The likelihood function is:
\begin{equation}
\ln \, \mathcal{L} = -0.5 \times \left[ \sum \chi^2/C_\text{noise}^2 + \sum{\ln ( 2 \pi (C_\text{noise} \sigma)^2 ) } \right],
\end{equation}
where $C_{\mathrm{noise}}$ is a constant scaling factor for the noise
($\sigma$) to take into account underestimates or overestimates of observational noise in computing $\chi^2$; as well as systematic errors between the model and the spectrum, such as missing line features. We performed an initial MCMC fit for the parameters $T_{\text{eff}}$, $\log {g}$, $RV^*$, $v\sin{i}$, $\alpha$, $C_{\lambda}$, $C_{\mathrm{flux}}$\added{\footnote{The values are determined based on the percentage of median flux $C_{\mathrm{flux}}$ = $C_{\mathrm{flux}^*} \times F$, where $F$ is the median flux.}}, and $C_{\mathrm{noise}}$, modeling the continuum in the same manner as the telluric standard. 
The nuisance parameter $C_{\lambda}$ takes into account the small shift in instrument alignment between the telluric and science integrations. RVs inferred between subsequent nods are more consistent when this nuisance parameter in included in the forward model. 
Stellar model parameter prior ranges were chosen to encompass the typical properties of T dwarfs, with $T_{\text{eff}}$ = 600 to 1300~K, $\log{g} = 3.5$ to $5.5$~dex (in units of cm\,s$^{-2}$), $v\sin{i} = 0$ to $100$\,km\,s$^{-1}$, and $RV^* = -200$ to $+200$\,km\,s$^{-1}$.
The MCMC bounds of $T_{\text{eff}}$ are set for the whole range of the available model sets (BT-Settl = 500 to 3500~K and Sonora = 200 to 2400~K) (Table \ref{table:mcmc_parameter}). The initial MCMC used 50 walkers of 600 steps each and burn-in of 300 steps, and convergence was verified by inspection of parameter chains \replaced{and evaluation of the integrated auto-correlation time, with a median auto-correlation time of 40 steps.}{and a requirement of the Gelman-Rubin scale reduction factor \citep{Gelman:1992aa} of less than 1.32. Typically convergence occurred after the first 100--200 steps.}

The Gelman-Rubin scale reduction factor and autocorrelation time suggested by \cite{Goodman:2010aa} were used to test convergence for a representative set of telluric and science spectra, and to set the number of walkers and steps for all fits. We tested the MCMC runs with 50 chains of 600 to 8,000 steps each. The $\chi^2$ values were similar\replaced{, but longer chains reduced the scale reduction from $\sim$ 1.5 to $\sim$ 1.03.
The autocorrelation time was typically 40 steps, and the chains converged after 10 autocorrelation times; hence, longer chains did not significantly improve the fits.}{, and the best-fit parameters were fully consistent within the uncertainties, but longer chains reduced the scale reduction factor from 1.2 to 1.01. The longest autocorrelation time was 120 steps, estimated from the runs with 8000 steps. Among these select sets of fits, we found that longer chains did not significantly improve the fits, which only have lower Gelman-Rubin scale reduction factor and converged autocorrelation times, but not change to the parameter values or uncertainties. Due to our limited computational resources, we chose to run only 600 steps, where the convergence was checked visually and confirmed by the requirement that the Gelman-Rubin scale reduction factors were all less than 1.32 (50$^\text{th}$, 16$^\text{th}$, and 84$^\text{th}$ percentiles of R = 1.05, 1.03, and 1.13).}

Residuals between the best-fit model and data from this first pass were used to generate a pixel mask rejecting $2.5\sigma$ outliers\footnote{Typically, $\sim$3\% of the pixels are masked.}, typically cosmic rays and bad pixels in the detector. A second MCMC was then run on the masked data using the same MCMC fit parameters and initializing model parameters from the first MCMC fit plus a random offset drawn from uniform parameter ranges spanning $\Delta T_{\text{eff}} =  \pm 20$ K, $\Delta \log{g} = \pm 0.1$ dex, $\Delta v\sin{i}  = \pm 1$ km s$^{-1}$, and $\Delta RV =\pm  1$ km s$^{-1}$. The masking step considerably improved the RV and $v\sin{i}$ uncertainties and the overall spectral fit. The derived RVs were corrected to the heliocentric frame using the \textit{astropy} function \texttt{radial\char`_velocity\char`_correction} to compute $v_\text{bary}$.

Each individual spectrum was forward-modeled, and measurements both within an epoch and across epochs (for multi-epoch data) were averaged using uncertainty weighting, with weight $W \propto 1/(\sigma_\text{lower}^2 + \sigma_\text{upper}^2)$, where $\sigma_\text{lower}$, $\sigma_\text{upper}$ are the uncertainties associated with the 16$^\text{th}$ and 84$^\text{th}$ percentiles of each marginalized parameter distribution. In a few cases where the S/N of an individual spectrum is lower than 10, all spectra in an epoch were coadded before forward-modeling. Table \ref{tableindmearurement} lists the RV, $v\sin{i}$, $T_{\text{eff}}$ and $\log {g}$ values inferred for each source and epoch, along with previously published values from the literature.

\begin{deluxetable}{lccc}
\tablecaption{Modeling Parameter Ranges \label{table:mcmc_parameter}}
\tablecolumns{4}
\tablehead{
\colhead{}  & \colhead{} & \colhead{} & \colhead{} \\
\colhead{Description} &  \colhead{Symbol (unit)}  & \colhead{Priors\tablenotemark{a}} & \colhead{Bounds}
}
\startdata
\multicolumn{4}{c}{Telluric Standard Star}\\
\hline
Line Spread Func. & $\Delta \nu_\text{inst}$ (km s$^{-1}$) & (3.0, 6.0) & (2.0, 10.0) \\
Flux Offset & $C_{\mathrm{flux}}$ & ($-$1.0, +1.0) & ($-$500, +500) \\
Wavelength Offset & $C_{\lambda}$ ($\AA$) &  ($-$0.02, +0.02) & ($-$0.04, +0.04)\added{\tablenotemark{b}} \\
Telluric Scaling & $\alpha$ & (0.3, 3.0) & (0.3, 10.0) \\
\hline
\multicolumn{4}{c}{T Dwarf} \\
\hline
Effective Temp. & $T_{\mathrm{eff}}$ (K) & (600, 1300) & B\tablenotemark{c}=(500, 3500) \\
\nodata & \nodata & \nodata & S\tablenotemark{d}=(200, 2400) \\
Surface Gravity & $\log{g}$ (cm s$^{-2}$) & (3.5, 5.5) & (3.5, 5.5) \\
Rotational Velocity & $v\sin{i}$ (km s$^{-1}$) & (0, 100) & (0, 100) \\
Radial Velocity & RV (km s$^{-1}$) & ($-$200, +200) & ($-$200, +200) \\
 \replaced{Telluric Scaling & $\alpha$ & (0.9, 1.1) & (0.1, 10.0) \\
Wavelength Offset & $C_{\lambda}$ ($\AA$) & ($-$0.6, $+$0.6) & ($-$0.6, $+$0.6) \\
Flux Offset & $C_{\mathrm{flux}^*}$ & ($-$0.01, +0.01) & ($-$0.05, +0.05) \\}
{Flux Offset & $C_{\mathrm{flux}^*}$ & ($-$0.01, +0.01) & ($-$0.05, +0.05)\tablenotemark{e} \\
Wavelength Offset & $C_{\lambda}$ ($\AA$) & ($-$0.6, $+$0.6) & ($-$0.6, $+$0.6)\tablenotemark{b} \\
Telluric Scaling & $\alpha$ & (0.9, 1.1) & (0.1, 10.0) \\}
Noise Factor & $C_{\mathrm{noise}}$ & (0.99, 1.01) & (0.1, 5.0)
\enddata
\tablenotetext{a}{Uniform priors}
\added{\tablenotetext{b}{The Telluric Standard Star wavelength offset bounds range is much smaller than the T Dwarf offset bounds range as the data for the former are used to formally derive the wavelength solution while the data for the latter incur pixel shifts due to instrumental flexure between pointings. See Section~\ref{subsec:science_mcmc} for more details.}}
\tablenotetext{c}{BT-Settl model}
\tablenotetext{d}{Sonora model}
\added{\tablenotetext{e}{The values are determined based on the percentage of median flux $C_{\mathrm{flux}}$ = $C_{\mathrm{flux}^*} \times F$, where $F$ is the median flux.}}
\end{deluxetable}

\subsection{Evaluating the Fits} \label{subsec:fits}

\subsubsection{Fit Quality and Parameter Correlations \label{subsec:example_fits}}

\begin{figure*}[!htb]
\includegraphics[width=\linewidth]{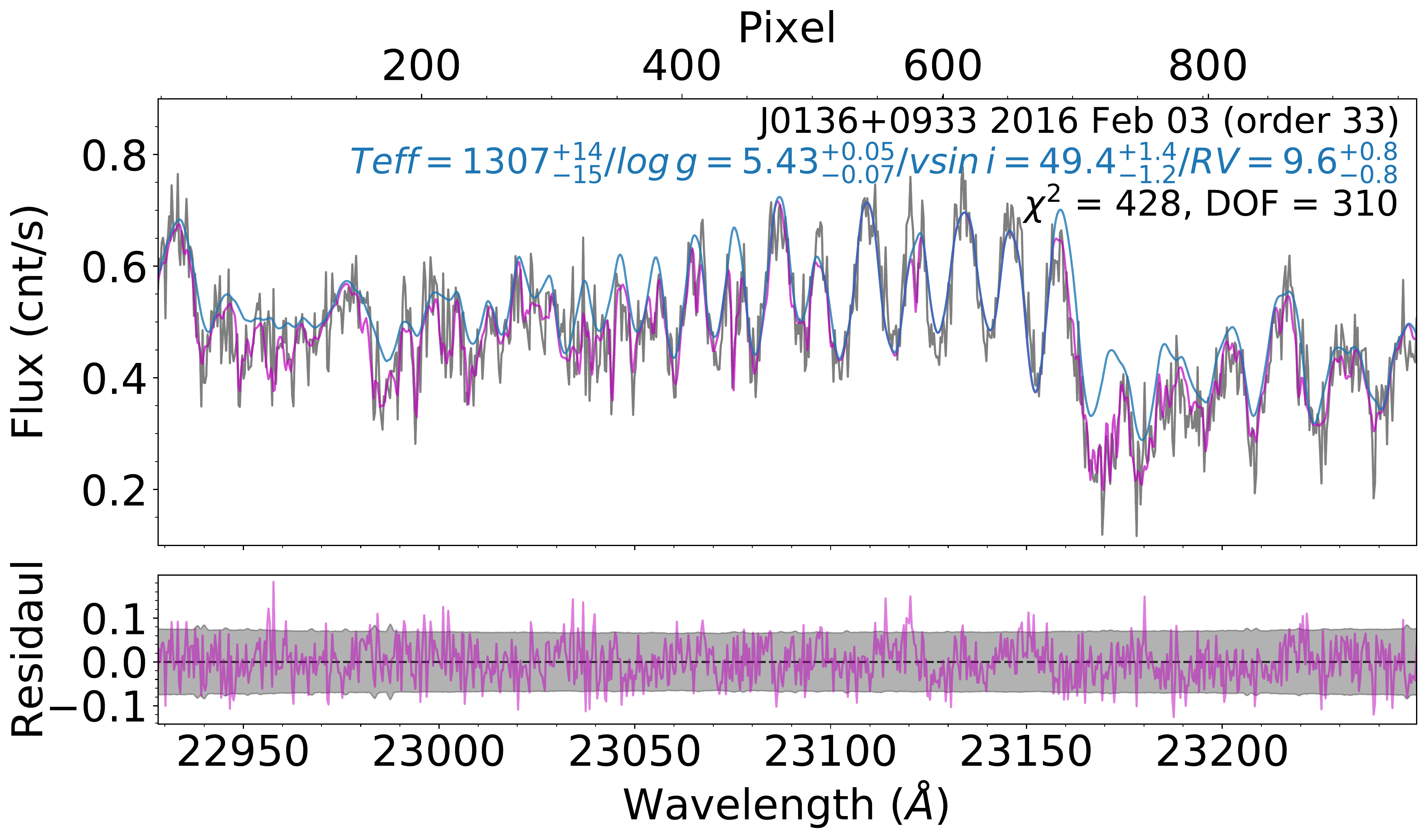}
\caption{BT-Settl model fit of the order 33 spectrum of the T2.5 J0136+0933, observed on 2016 February 3 (UT). The horizontal axis displays both pixel position (top axis) and wavelength (bottom axis). Upper panel: the grey line is the observed spectra; the magenta and blue lines are the stellar model with and without telluric absorption, respectively. Lower panel: difference of the data minus model (magenta) with $\pm$1$\sigma$ data uncertainty shaded in grey. The best-fit parameters are listed at the upper right corner of the top panel, with effective temperature ($T_{\text{eff}}$) in K, surface gravity ($\log {g}$) in cm s$^{-2}$, rotational velocity ($v\sin{i}$) in km s$^{-1}$, and radial velocity (RV) in km s$^{-1}$.
\label{fig:J0136fitspectrum}}
\end{figure*}

\begin{figure*}[!htb]
\includegraphics[width=\linewidth]{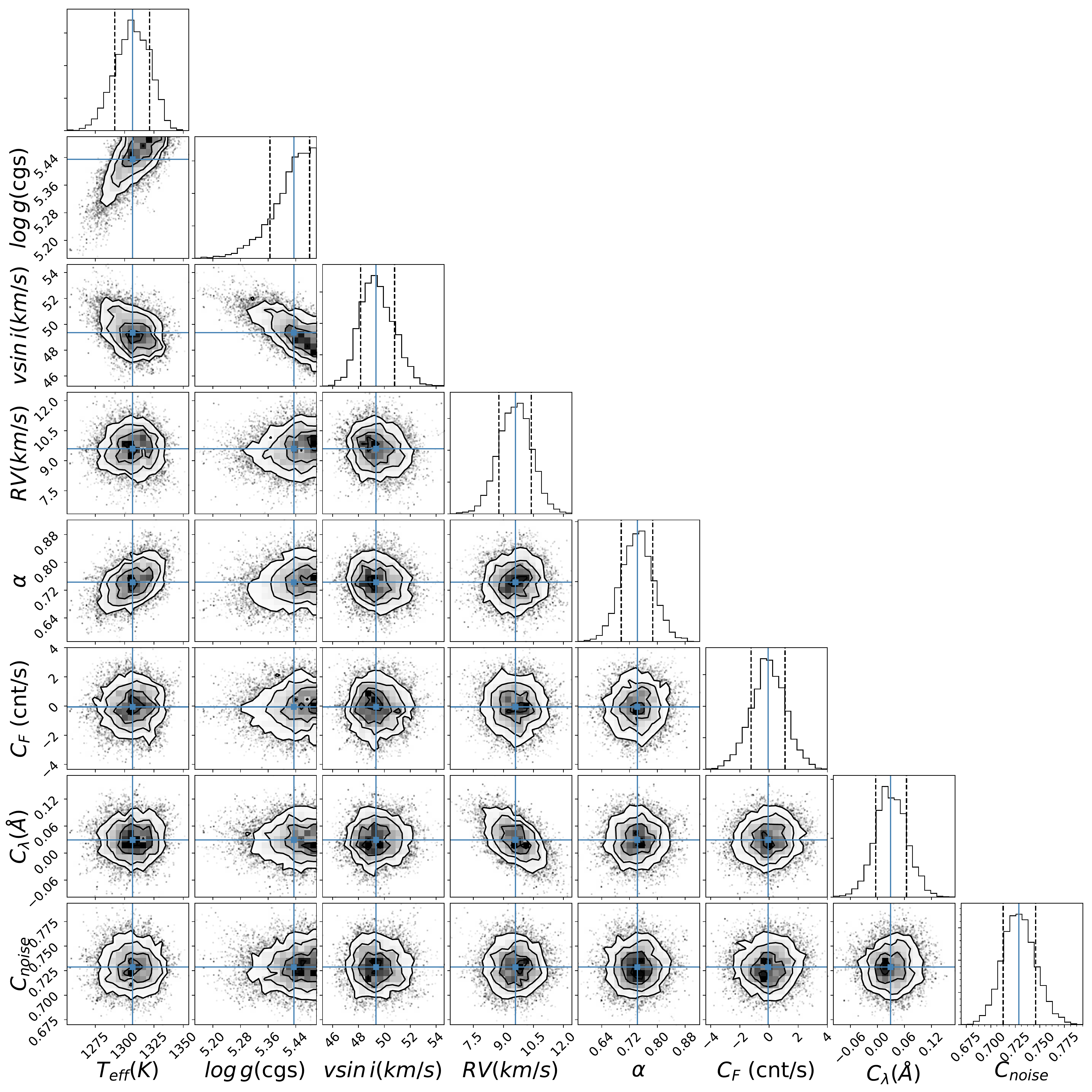}
\caption{The posterior probability distribution of fits to the order 33 spectrum of the T2.5 J0136+0933 observed on 2016 February 3 (UT). The parameters shown are effective temperatures ($T_{\text{eff}}$) in K, surface gravity ($\log {g}$) in cm s$^{-2}$,  projected rotational velocity ($v\sin{i}$) in km s$^{-1}$, radial velocity (RV) in km s$^{-1}$, telluric scale factor ($\alpha$), nuisance flux parameter ($C_{F}$) in count/s, nuisance wavelength parameter ($C_{\lambda}$) in \AA, and noise scale factor ($C_{\mathrm{noise}}$). The black dash lines are the 16$^\text{th}$ and 84$^\text{th}$ percentiles in the marginalized distributions (diagonal plots), and the blue lines denote the median values in both marginalized distributions and interior parameter correlation plots. 
\label{fig:J0136fitposterior}}
\end{figure*}

\begin{figure*}[!htb]
\includegraphics[width=\linewidth]{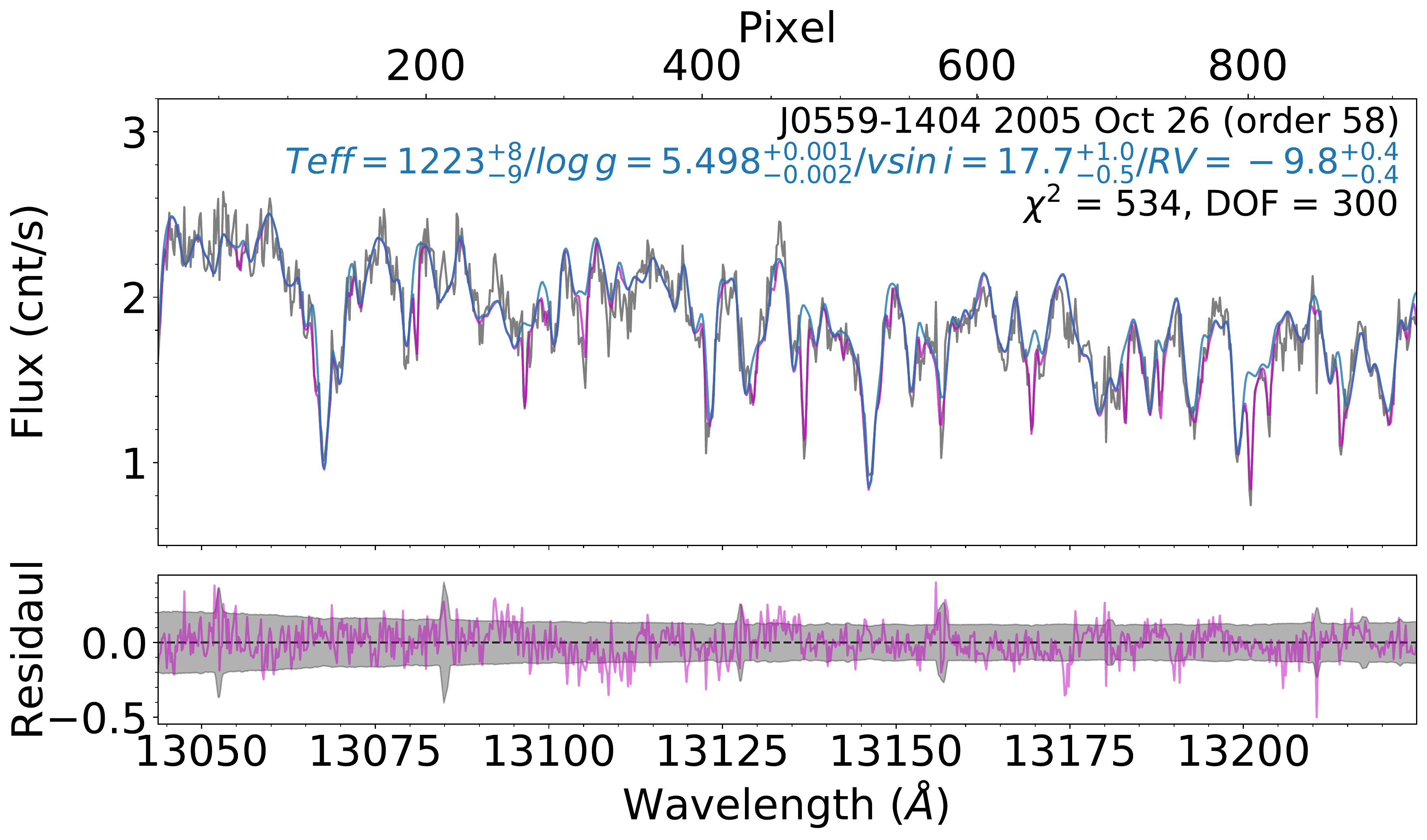}
\caption{Sonora model fit of the order 58 spectrum of the T4.5 J0559$-$1404, observed on 2005 October 26 (UT). Notation is identical to Figure \ref{fig:J0136fitspectrum}.
\label{fig:J0559fitspectrum}}
\end{figure*}

\begin{figure*}[!htb]
\includegraphics[width=\linewidth]{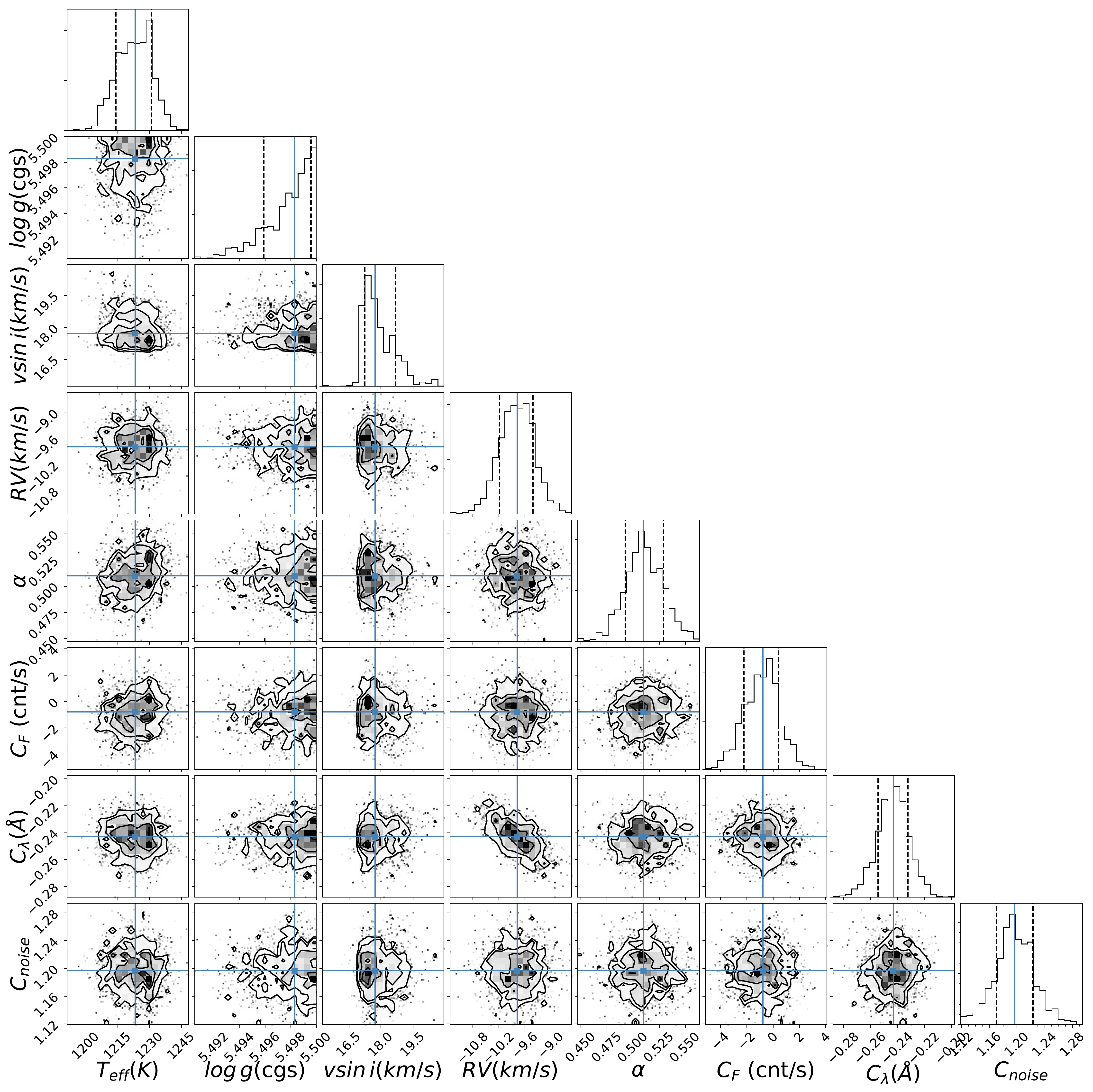}
\caption{The posterior probability distribution of fits to the order 58 spectrum of the T4.5 J0559$-$1404, observed on 2005 October 26 (UT). 
Notation is identical to Figure \ref{fig:J0136fitposterior}. \label{fig:J0559fitposterior}}
\end{figure*}

Figures \ref{fig:J0136fitspectrum}--\ref{fig:J0559fitposterior} illustrate representative fits to science data in orders 33 (T2.5 J0136+0933) and 58 (T4.5 J0559$-$1404). 

For the order 33 fit, residuals between the data and the best-fit BT-Settl model are on par with the scaled noise ($\chi^2_r$ = 1.3,\footnote{$\chi^2_r$ is the reduced chi-square statistic computed as $\chi^2_r = \frac{1}{N_\mathrm{DOF}}\sum{Mask[p]}\left( \frac{Data[p] - D[p]}{C_\text{noise}\sigma [p]}\right)^2$, where $Mask[p]$ is the pixel mask ($Mask$ = 1 for good data, $Mask$ = 0 for bad data) and $N_\mathrm{DOF}$ is the number of degrees of freedom computed as $N_\mathrm{DOF}$ = [number of unmasked data pixels]/3 $-$ [number of fit parameters]. 
The factor of 1/3 takes into account the pixel-to-pixel correlations caused by the finite slit width, typically 3 pixels ($0\farcs432$). We consider fits with $\chi^2_r < 2.5$ to be consistent within uncertainties.} 
for $C_{\mathrm{noise}}$ = 0.7), and all of the marginalized distributions show normal distributions, modulo parameter limits (e.g., $\log {g}$).
This fit exemplifies parameter correlations found in some (but not all) of the order 33 fits. 
First, we find a negative correlation between $v\sin{i}$ ($\Delta \, v\sin{i}$ = 3 km s$^{-1}$) and $\log{g}$ ($\Delta \log{g}$ = 0.2 dex) that we attribute to a degeneracy between rotational and pressure broadening. A larger $\log {g}$ results in greater pressure broadening,  which is compensated for by a smaller $v\sin{i}$, and vice-versa.
Disentangling this correlation in the line spread shape would require higher resolution and higher signal-to-noise data than is available with the current dataset.
Second, we find a positive correlation between $T_{\text{eff}}$ ($\Delta \, T_{\text{eff}}$ = 50 K) and $\log {g}$ ($\Delta \log{g}$ = 0.2 dex) that we attribute to temperature and pressure effects in the primary carbon reduction reaction, CO + 3H$_2$ $\Leftrightarrow$ CH$_4$ + H$_2$O. This reaction is driven toward the right (weaker CO and stronger CH$_4$) at low temperatures and high pressures. 
Disentangling \replaced{this}{the $T_{\text{eff}}$-$\log {g}$} correlation could be achieved with an accurate measure of the surface flux, and hence the luminosity and radius of each source, which is beyond the scope of this work. 
Finally, we find a positive correlation between RV ($\Delta{RV}$ = 0.4~km s$^{-1}$) and $C_{\lambda}$ ($\Delta{C_{\lambda}}$ = 0.03~{\AA}) 
which is inherent to the simultaneous fitting of the Doppler shift of the source and instrumental shift of the wavelength calibration. As noted above, the $C_{\lambda}$ term is necessary to enforce agreement of RVs measured within a single epoch, which far exceed the slight increase to our marginalized RV uncertainties.

For the order 58 fit, residuals between the data and the best-fit 
Sonora model are again consistent with uncertainty
($\chi^2_r$ = 1.7 for $C_{\mathrm{noise}}$ = 1.2), and
marginalized distributions for most parameters reflect normal distributions with the exception of $\log {g}$ (parameter limit) and $v\sin{i}$, the latter of which shows a sharp lower cutoff at 17 km s$^{-1}$. 
We find only a slight positive correlation between RV ($\Delta{RV}$ = 0.2~km s$^{-1}$) and $C_{\lambda}$ ($\Delta{C_{\lambda}}$ = 0.01~{\AA}). 

\subsubsection{Substellar Atmosphere Model Selection} \label{sec:atmmodel}

Both BT-Settl and Sonora model sets were used for all forward-modeling analyses, and we determined the best choice between these sets for each source and order through a combination of visual inspection, $\chi^2$, F-test, and Bayesian information criterion\footnote{BIC = $\chi^2_\text{min}$ + k log$_{10}N$, where $\chi^2$ is the chi-square statistic for the best-fitting model, $k$ is the number of parameters, and $N$ is the number of data points \citep{schwarz1978}. Statistical significance for ruling out the null hypothesis (in this case, that the model sets provide equivalent fits) is assessed following \cite{Kass:1995aa}, in which the $\Delta$BIC ranges of 0--2, 2--6, 6--10, and $>$ 10 are categorized as insignificant, positive, strong, and very strong evidence against the null hypothesis, respectively.}  (BIC).
In general, the Sonora models provide significantly better fits to order 58 ($J$-band) spectra of mid- and late-T dwarfs, while
the BT-Settl models provide marginally better fits for order 33 ($K$-band) spectra of early-T dwarfs.
We attribute the significant improvement in Sonora model fits to the $J$-band data to the updated CH$_4$ opacities in these models, a particularly important factor for the coldest brown dwarfs. 
The slightly better fits for the BT-Settl models at $K$-band
may be due to the inclusion of cloud opacity in these models, which are absent in the Sonora grid, although such opacity should have relatively modest influence in the 2~$\micron$ region.

The choice of model does influence the physical parameters inferred for each source.
Comparing the best-fit parameters between the two models 
across all sources, orders, and epochs, we found that measured RV and $v\sin{i}$ values are relatively 
robust to model choice, with median model discrepancies (Sonora minus BT-Settl) of
$\Delta{RV}$ = 0.6 $\pm$ 1.5~km s$^{-1}$ and
$\Delta{v\sin{i}}$ = $-$2 $\pm$ 4~km s$^{-1}$.
The RV offset between the models is dominated by order 58 fits of mid- and late- T dwarfs, for which $\Delta{RV}$ = 1.2~km s$^{-1}$.
Order 33 fits of early-T dwarfs have $\Delta{RV}$ = 0.4~km s$^{-1}$.
The ${v\sin{i}}$ offsets are again dominated by order 58 fits, where the velocity kernel must be broadened for the BT-Settl models to compensate for missing CH$_4$ opacities. The median $\Delta{v\sin{i}}$ for order 58 is $-$2.9~km s$^{-1}$ while the median $\Delta{v\sin{i}}$ for order 33 is 0.0~km s$^{-1}$.
The most discrepant $v\sin{i}$ measurements are among the order 58 fits for mid-T dwarfs, where the Sonora models are far more robust.

\begin{figure*}[!htbp]
\gridline{\fig{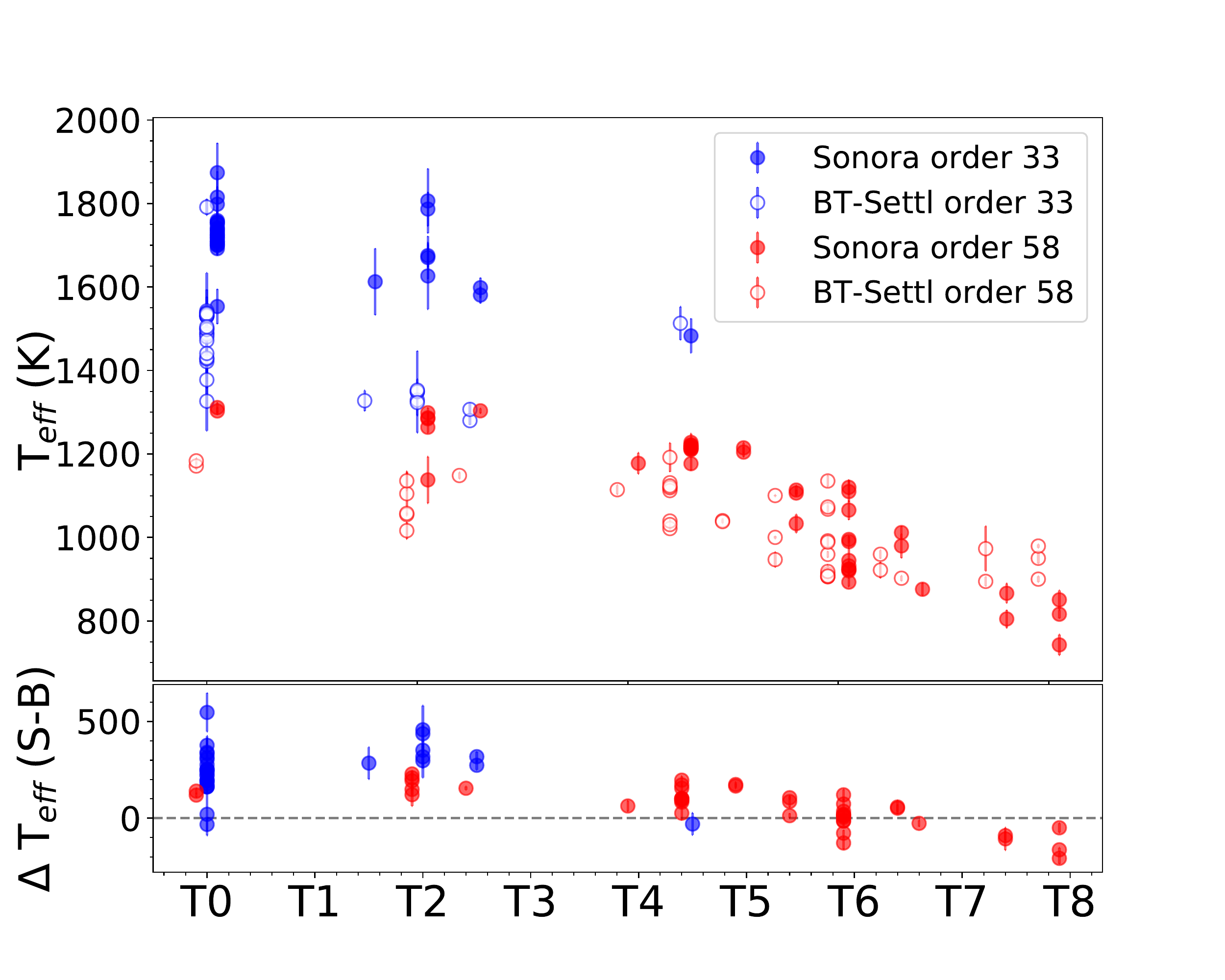}{0.5\textwidth}{}
\fig{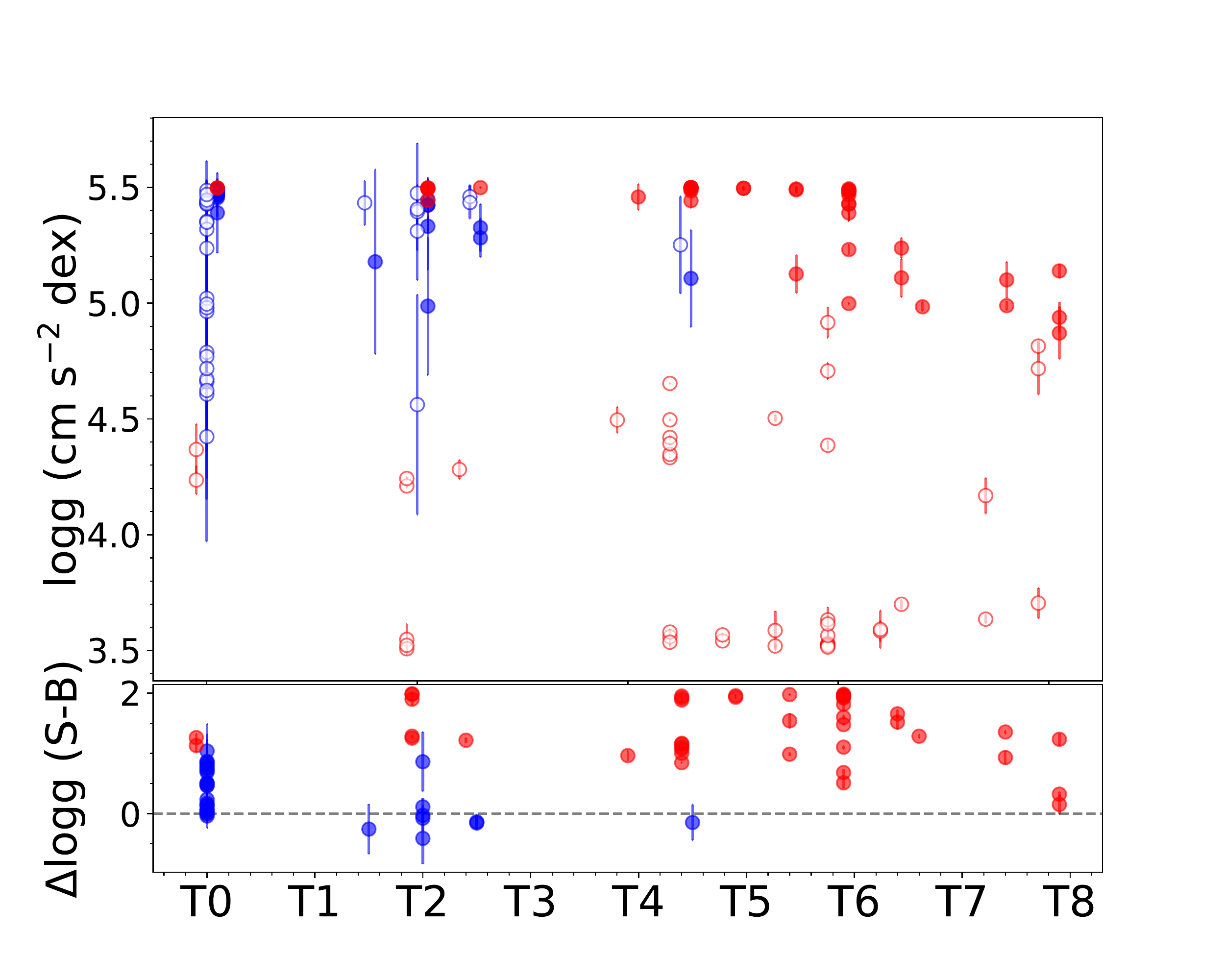}{0.5\textwidth}{}
}
\caption{Comparison of inferred $T_{\text{eff}}$ (left) and $\log {g}$ (right) parameters between the BT-Settl (open circles) and Sonora (filled circles) models as a function of spectral type, in orders 33 (blue) and 58 (red). Differences (Sonora$-$BT-Settl) are shown in the lower panels.
\label{fig:model_comparison}}
\end{figure*}

The atmosphere parameters $T_{\text{eff}}$ and $\log {g}$ show considerably more variance between the models (Figure \ref{fig:model_comparison}).
$T_{\text{eff}}$s inferred from the Sonora model fits are consistently hotter than those inferred from the BT-Settl model fits for early- to mid-T dwarfs in both orders 33 and 58, with a median offset of 240 $\pm$ 110~K\replaced{; f}{. F}or late-T dwarfs, the Sonora model fits are cooler.
More striking is the difference in $\log {g}$ values inferred for order 58 ($J$-band) data, 
for which BT-Settl models typically converge to $\log {g}$ $\approx$ 3.5 dex, the minimum of the model parameter range. 
In contrast, $\log {g}$ values inferred from fits to order 33 ($K$-band) data 
are generally consistent between the models, with a median difference of only 0.14 $\pm$ 0.40 dex,  
although both models often converge to the maximum model parameter value of 5.5 dex.

\subsubsection{Examination of Fits Across Different Orders 
\label{sec:compare_diff_order}}

For the sources with measurements in both bands, we examined the consistency of $T_{\text{eff}}$ and $\log {g}$ measurements between orders for six sources with measurements with both $J$- and $K$-band spectra. These sources are J0136+0933, J0559$-$1404, J1106+2754, J1254$-$0122 (orders 33 and 58), J1928+2356 (orders 37 and 58), and J2126+7617 (orders 33 and 57). 
$T_{\text{eff}}$ values inferred from $K$-band data are consistently higher than those inferred from $J$-band data \replaced{(average ${\Delta}T_{\text{eff}}$ = 187~K, range $-$182~K to 341~K; average ${\Delta}T_{\text{eff}}$ = 260~K, range 140~K to 341~K excluding J2126+7617).}
{(average ${\Delta}T_{\text{eff}}$ = 187~K, range $-$182~K to 341~K; excluding J2126+7617 yields an average ${\Delta}T_{\text{eff}}$ = 260~K, range 140~K to 341~K ).}
$\log {g}$ values differed by up to 1.20~dex between orders but with no clear trend (average $\log {g}$ = 0.27~dex, range $-$0.39~dex to 1.20~dex).
For J1928+2356 and J2126+7617, the average differences in $T_{\text{eff}}$ and $\log {g}$ between orders
are more than 150~K and 0.08 dex, respectively.
RV and $v\sin{i}$ values are generally consistent across the orders, except for sources with low S/N data (J1254$-$0122 on 2011 Jun 10 and J1928+2356 on 2019 Sep 12\footnote{RV values are consistent, but $v\sin{i}$ values are not.}) and binaries (J1106+2754 and J2126+7617).\added{

}
The best-fit parameters for sources observed in multiple orders were determined generally by spectral type: $N7$ for early T dwarfs and $N3$ for mid- and late-T dwarfs. \replaced{In some cases}{For J1254$-$0122}, we used the order with the higher S/N data. 

\subsubsection{Minimum $v\sin{i}$ \label{sec:min_vsini}}

The finite resolution of NIRSPEC data places a fundamental limit on our ability to measure rotational broadening for the slowest rotators, which scales with the width of the instrumental LSF. \citet{Blake:2010aa} found a minimum detectable $v\sin{i}$ of 9 km s$^{-1}$ for their NIRSPEC sample of M8--L6 dwarfs. Our T dwarf \replaced{sample contains}{spectra contain} a higher density of molecular features, with overlapping CO and CH$_4$ absorption bands, \replaced{our spectra have}{and} lower S/N.
We empirically determined the minimum detectable $v\sin{i}$ limits for order 33 and 58 data by analyzing simulated NIRSPEC data derived from the model grids, using the same forward-modeling method as the science data. We evaluated a representative set of models with $T_{\text{eff}}$ = 900, 1200, 1500 K; $\log {g}$ = 5.0 dex; RV = 0 km s$^{-1}$; pwv = 1.5 mm; airmass = 1.0; instrumental LSF = 4.8 km s$^{-1}$ (our typical value); $v\sin{i}$ = 1--15 km s$^{-1}$ in steps of 1 km s$^{-1}$, and 15--25 km s$^{-1}$ in steps of 5 km s$^{-1}$; and S/N = 1--10 in steps of 1, and 10--25 in steps of 5.
Gaussian noise was applied using pre-upgrade NIRSPEC values for detector gain, read noise, and dark current, and we assumed an integration time of 1500 s. 
We defined a robust measurement to be the difference between the true $v\sin{i}$ and measured $v\sin{i}$ of less than 1 km s$^{-1}$. With this benchmark, we determined the minimum robust $v\sin{i}$ to be 9 km s$^{-1}$ for S/N $\geq$ 5 for both orders 33 and 58, equivalent to the minimum $v\sin{i}$ for late-M and L dwarfs determined by \citet{Blake:2010aa}. Discrepancies generally increase from high to low S/N\footnote{See the diagnostic plots in Appendix \ref{appendix:minvsini}}. 
For S/N $<$ 5 data, $v\sin{i}$ fits become much less robust and a more conservative $v\sin{i}$ lower limit of 15 km s$^{-1}$ is adopted.
Hence, the slowest rotators in our sample, 
J0000+2554 ($v\sin{i}$ = 4 $\pm$ 2 km s$^{-1}$, S/N = 5),
J0627$-$1114 ($v\sin{i}$ = 5.4$^{+2.5}_{-2.0}$ km s$^{-1}$, S/N = 6), 
J0819$-$0335 ($v\sin{i}$ = 8.5$^{+1.4}_{-2.4}$ km s$^{-1}$, S/N = 9), and 
J2236+5105 ($v\sin{i}$ = 6.6 $\pm$ 0.9 km s$^{-1}$, S/N = 22) are assigned limits of $<$ 9 km s$^{-1}$; 
while J2030+0749 ($v\sin{i}$ = 14.0$^{+0.9}_{-1.0}$, S/N = 4), is assigned a limit of $<$15 km s$^{-1}$.

\subsection{Analysis of Fit Parameters} \label{sec:analysis}

\subsubsection{Radial Velocities} \label{sec:rv}

\begin{figure*}[!htbp]
\gridline{\fig{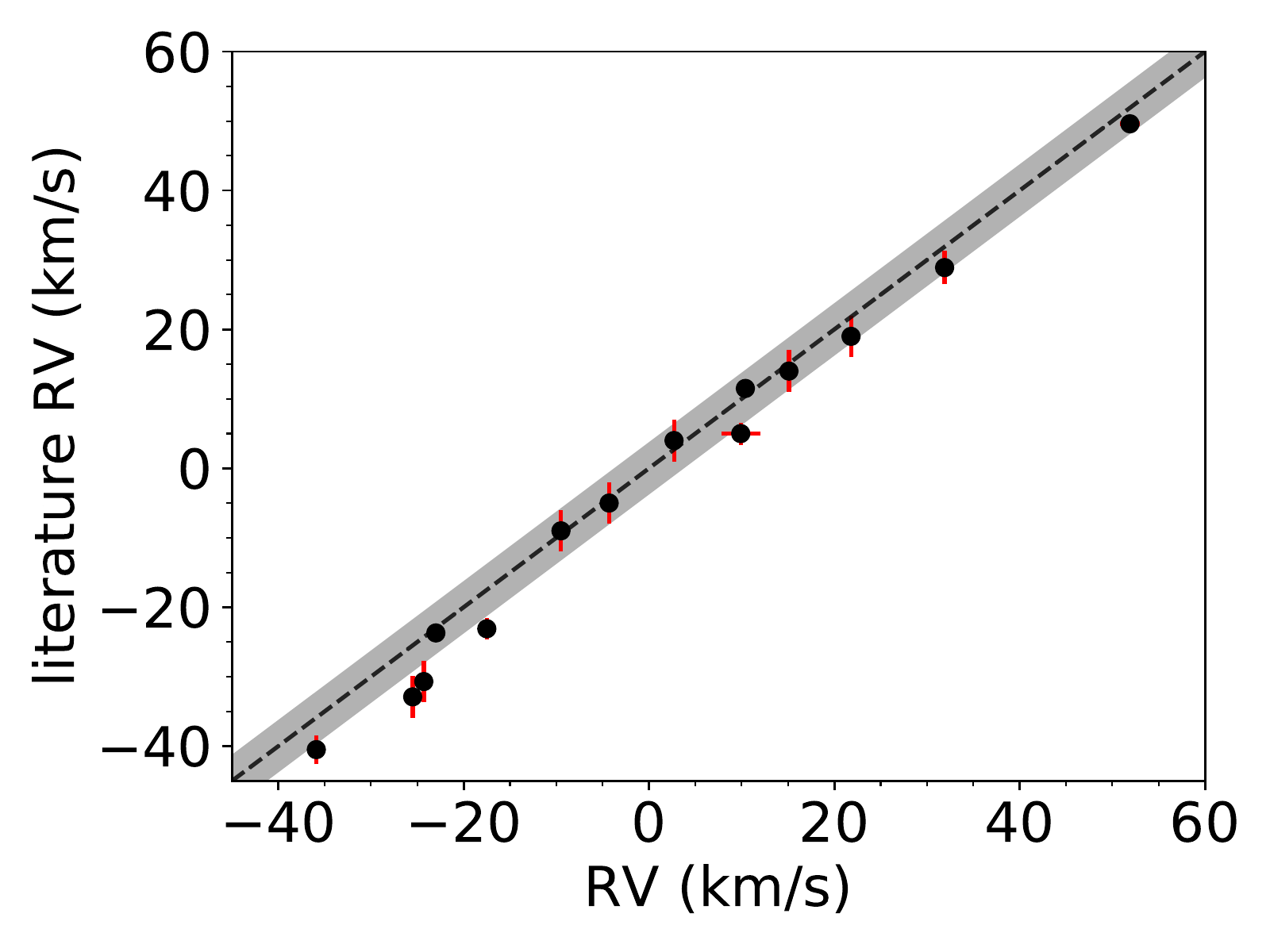}{0.5\textwidth}{}
\fig{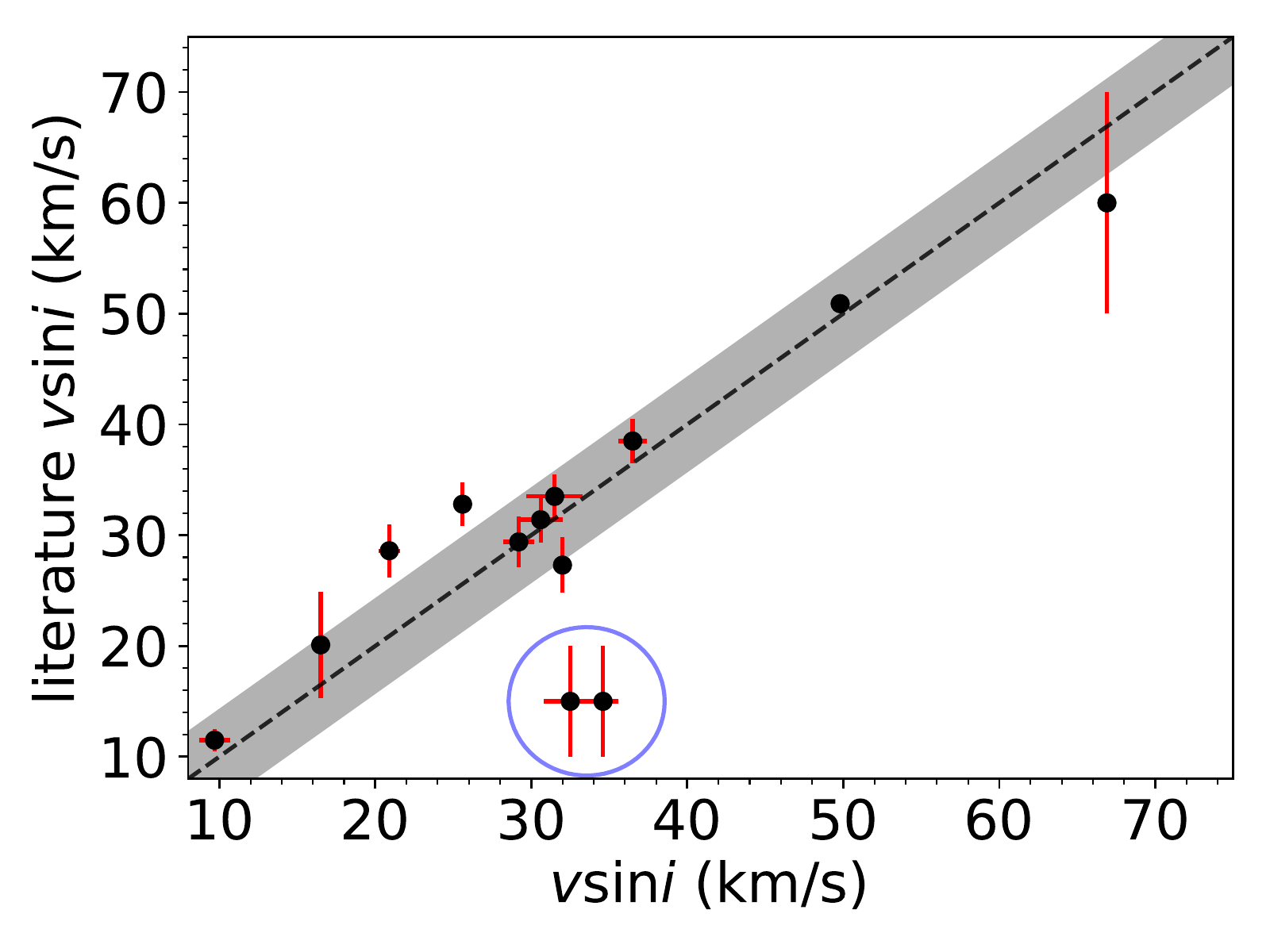}{0.5\textwidth}{}}
\caption{Comparison of RV (left) and $v\sin{i}$ (right) measurements from our NIRSPEC data to previously reported values for the same sources in the literature (see Table \ref{tableindmearurement}). The black dashed line delineates perfect agreement. The shaded region indicates the $\pm$1$\sigma$ scatter (3.8 km s$^{-1}$ for RV and 4.3 km s$^{-1}$ for $v\sin{i}$) between observed and previously reported values. Two outliers in the $v\sin{i}$ plot, noted by a blue circle, are excluded in computation of the $v\sin{i}$ scatter and are discussed in Section~\ref{sec:vsini}.} \label{fig:rvvsinicompub}
\end{figure*}

Our RV measurements range over $-$43 to +56 km s$^{-1}$ 
with a median value of 1.2 km s$^{-1}$ and a median uncertainty of 0.6 km s$^{-1}$.  
These measurements are consistent with a population drawn from the local disk, as verified in further detail below.
These measurements are also largely consistent with \replaced{those}{14 sources} previously reported in the literature (Figure \ref{fig:rvvsinicompub}\added{; Table~\ref{tableindmearurement}}), but with a median factor of 5.5 improvement in uncertainty.
The only significant ($> 3 \sigma$) RV outlier is J1346$-$0031, for which our measurement of $-$17.5$^{+0.6}_{-0.5}$ km s$^{-1}$ is 5.1 km s$^{-1}$ higher than that reported in \citet{Zapatero-Osorio:2007aa}.
For this source (and others), we found our RVs to be consistent between nod pairs in individual epochs and across 2 epochs, albeit with a 
relatively large $C_{\lambda}$ correction of 4 km s$^{-1}$, which may explain the difference with the previously reported value.
We found no correlation between RV discrepancies and spectral type.
The spatial kinematics of our sources are described further in Section~\ref{sec:discussion}.

\begin{figure*}[!htbp]
\includegraphics[width=1.0\textwidth, trim=0 0 0 0]{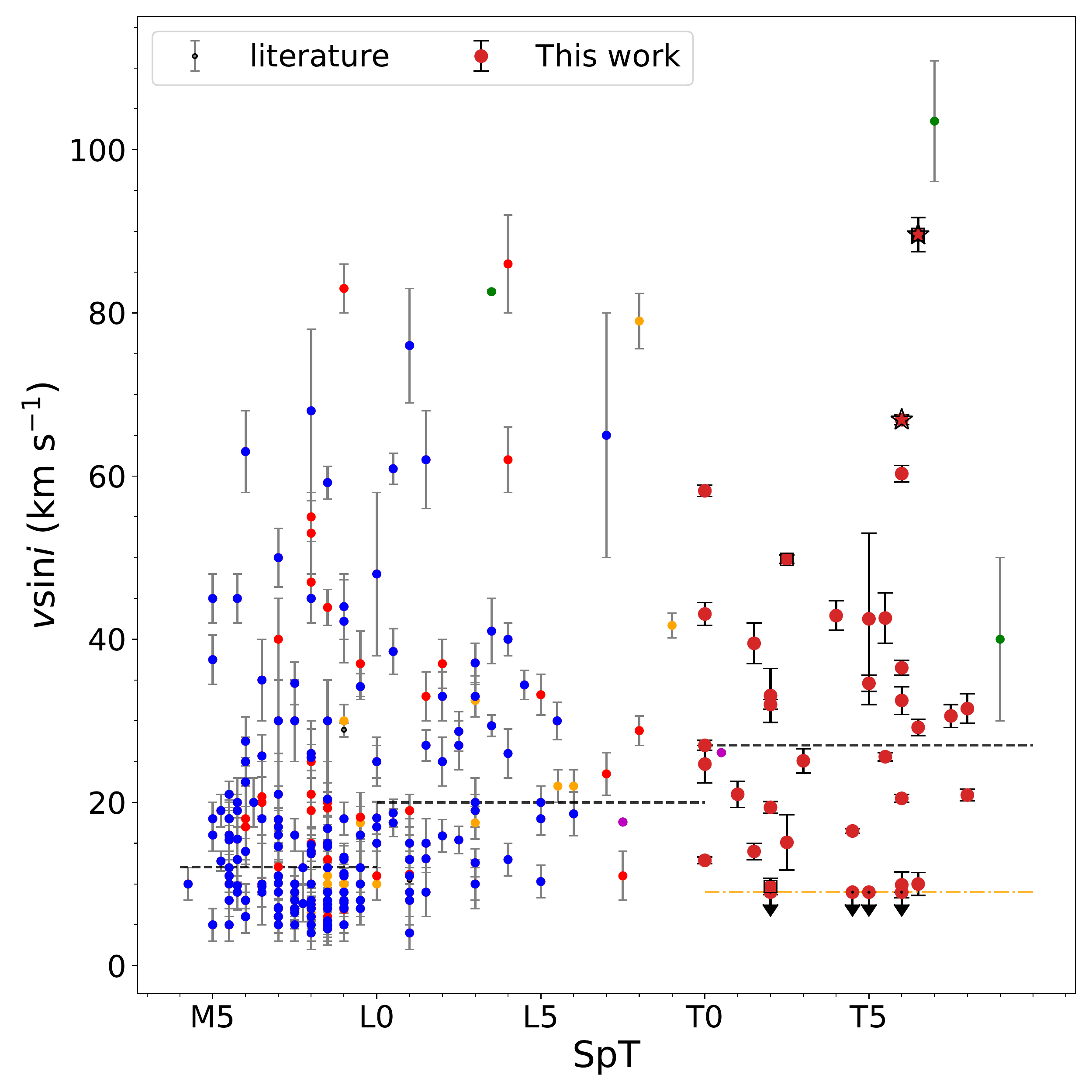}
\caption{$v\sin{i}$ measurements as a function of spectral type for a compilation of M4--T9 dwarfs from this work (large symbols) and the literature (small symbols). 
All symbols are color-coded by instrument resolutions (green for $R<10,000$, orange for $10,000 \leq R < 20,000$, red for $20,000 \leq R < 30,000$, blue for $30,000 \leq R < 50,000$, magenta for $R \geq 50,000$). 
Our T dwarf measurements are further segregated into normal sources (circles), young cluster members (squares), and spectrally peculiar (stars). 
The horizontal grey dashed lines indicate the median rotational velocities for late-M, L, and T dwarfs, respectively, in this sample. The horizontal orange dashed line indicates the NIRSPEC minimum $v\sin{i}$ floor.
\label{fig:vsini_comparison_literature}}
\end{figure*}

\subsubsection{Rotational Velocities} \label{sec:vsini}

Our $v\sin{i}$ measurements range over 4\added{\footnote{In Section~\ref{sec:min_vsini}, we determined the minimum $v\sin{i}$ to be 9~km s$^{-1}$, with four sources of our sample under such limit.}} to 90~km s$^{-1}$, 
with a median value of 27 km s$^{-1}$ and a median uncertainty of 0.9 km s$^{-1}$. 
These velocities correspond to maximum rotational periods of 1.5~hr to 28~hr for spheres of radius $R = 1~R_\mathrm{Jup}$ observed at an inclination of 90$\degr$ (equator-on), and indicate a population of rapid rotators as previously
inferred from other studies \citep{Zapatero-Osorio:2006aa,Prato:2015aa,2012ApJ...750..105R,Metchev:2015aa, Vos:2017aa, Tannock:2021aa}.
Our $v\sin{i}$ measurements are generally consistent with those previously reported in the literature (Figure \ref{fig:rvvsinicompub}), with a median factor of 2.7 improvement in uncertainties.
Again, we found internal agreement between nod pairs and across multiple epochs for $v\sin{i}$ measurements of these sources. 
The largest discrepancies were values reported in \citet{Prato:2015aa} for J2254+3123 and J2356$-$1553. 
That study visually compared their spectra with spectral templates convolved with a rotational broadening profile for different $v\sin{i}$ values. Forward-model fits to these spectra with $v\sin{i}$ values fixed to the \citet{Prato:2015aa} measurements result in  
significantly worse fits, with p-values\footnote{Throughout this paper, we adopt the convention that a p-value $\leq$ 0.1 is marginally significant, a p-value $\leq$ 0.05 is significant, and a p-value $\leq$ 0.01 is highly significant \citep{Nuzzo:2014aa}.} of $<$0.001 and 0.003, respectively.

Figure \ref{fig:vsini_comparison_literature} compares the distribution of our T dwarf $v\sin{i}$ measurements \replaced{in comparison}{along} with values reported in the literature \replaced{\cite[][and references therein]{Crossfield:2014aa}}{(\citealp{Crossfield:2014aa}, and references therein; \citealt{Tannock:2021aa})} between M4 and T9 dwarfs.
The median $v\sin{i}$ for T dwarfs in our sample is greater than those of both mid/late M dwarfs (12 km s$^{-1}$) and L dwarfs (20 km s$^{-1}$), continuing a previously identified trend of increasing rotation rate with decreasing mass and later spectral type \citep{Mohanty:2003aa, Reiners:2010aa,2011ApJ...727...56I}. Such a correlation is expected for both the smaller radii of lower mass stars and brown dwarfs, and the reduced angular momentum loss from weakened magnetic winds among lower-temperature dwarfs 
\citep{Mohanty:2003aa,2008ApJ...684.1390R}.
It should be noted that many of the M and L dwarfs shown in this panel were observed with higher-resolution spectrometers than NIRSPEC, and therefore have a lower $v\sin{i}$ floor. However, the observed trend persists even when a minimum cutoff of $v\sin{i}$ = 9 km s$^{-1}$ is applied. 
Among the T dwarfs, there is no significant correlation between $v\sin{i}$ and spectral type (\textit{R} = 0.10, p-value = 0.56).

\subsubsection{$T_\mathrm{eff}$ and $\log {g}$} \label{sec:tefflogg}

\begin{figure}[!htp]
\includegraphics[width=0.5\textwidth, trim=20 0 0 0]{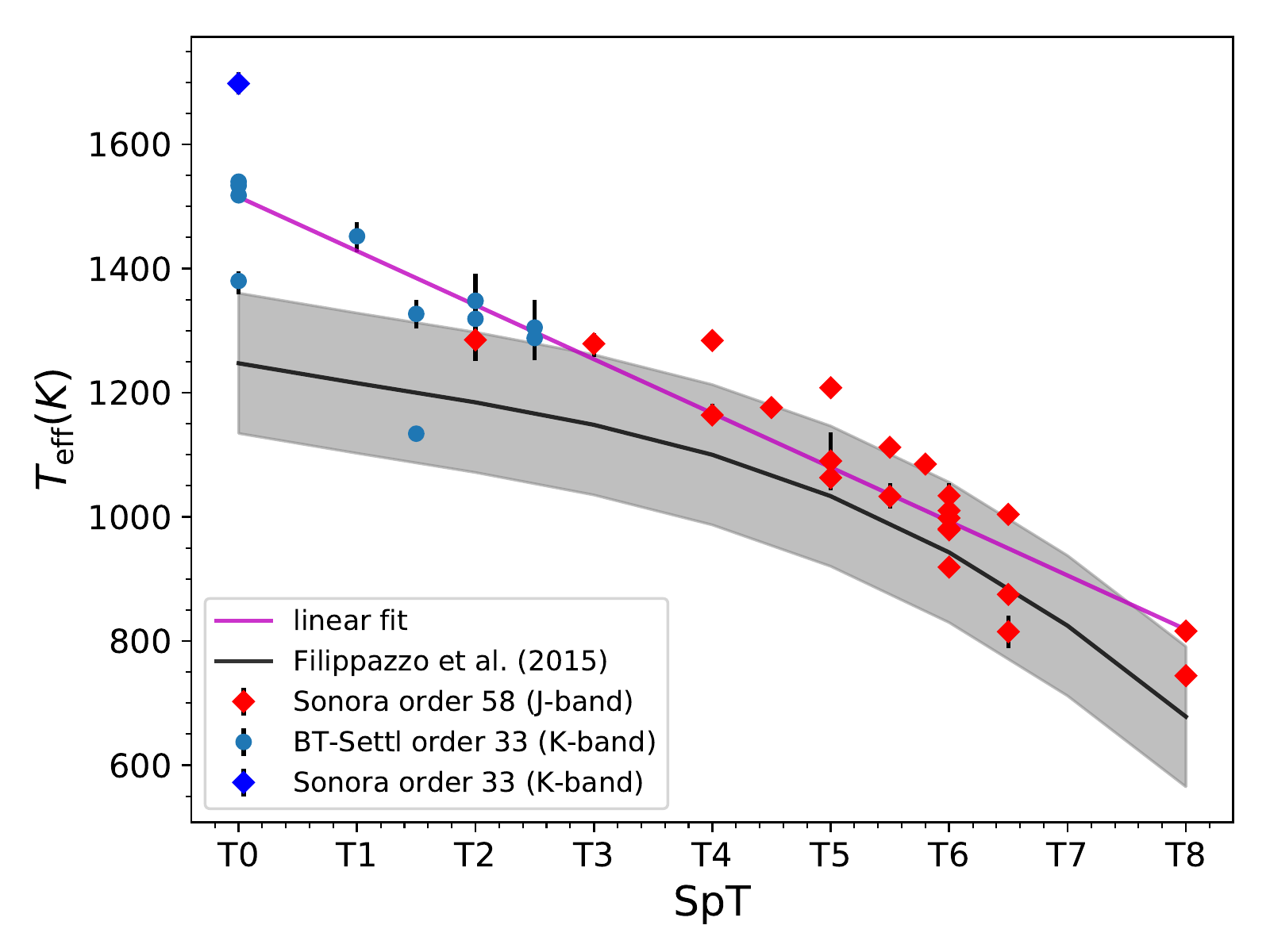}
\caption{Measured $T_{\text{eff}}$ values as a function of spectral type based on fits to orders 33 (blue) and 58 (red) data,
compared to the $T_{\text{eff}}$/spectral type relation of \cite{Filippazzo:2015aa} (black line with $\pm$1$\sigma$ uncertainty shaded in grey). 
The BT-Settl and Sonora model fits are marked as circles and diamonds, respectively. 
A linear fit to all of the measurements is indicated by the magenta line. 
} \label{fig:teff_spt}
\end{figure}

We \replaced{examined $T_{\text{eff}}$ and $\log {g}$ distribution and correlations between stellar parameters including}{evaluated our $T_{\text{eff}}$ and $\log {g}$ values by examining spectral type correlations with} RV and $v\sin{i}$, and \replaced{the}{comparing to} literature measurements. Table \ref{tableindmearurement} lists the best-fit $T_{\text{eff}}$ and $\log {g}$ for each source (744~K to 1700~K and 4.2 dex to 5.5~dex, respectively)\footnote{The observations of J1520+3546 on 2012 Apr 2 and J2126+7617 on 2020 Sep 3 have $T_{\text{eff}}$ $\sim$1700~K, which are determined with a wider $T_{\text{eff}}$ prior range to ensure the values are robust. The MCMC bounds are well above these temperatures. See Section \ref{subsec:science_mcmc} for more details.}.
These are based on \deleted{our }order 33 \replaced{and}{or} 58 fits, averaged over all spectra with a specific order.
Note that the uncertainties reported here for $T_{\text{eff}}$ and $\log {g}$ are just the Monte Carlo uncertainty, and do not reflect systematic error that may be present due to the narrow spectral bands we chose to model.
\deleted{We also examined correlations between spectral type and inferred $T_{\text{eff}}$ and $\log {g}$.} As expected, \replaced{the spectral type and $T_{\text{eff}}$ are strongly correlated}{$T_{\text{eff}}$ is strongly correlated with spectral type}, \deleted{with} decreasing \deleted{$T_{\text{eff}}$} to later types (Figure \ref{fig:teff_spt}; \textit{R}=$-$0.94, p-value $<$ 10$^{-3}$), although
the scatter about a linear trend can be as high as $\pm$300 K (standard deviation = 77~K).
Differences in $T_{\text{eff}}$ among equivalently-classified sources can be related to other physical properties (e.g., metallicity, $\log{g}$), but may also reflect systematic offsets between orders and \added{between} models, as discussed in Section \ref{sec:atmmodel}. 
We also find \added{that} our $\log {g}$ measurements based on Sonora model fits to order 58 data are strongly correlated with spectral type 
(\textit{R}=$-$0.62, p-value $<$ 0.01). Up to spectral type T5, $\log {g}$ is consistently around 5.5 dex, and then drops to lower values for later spectral types. This could reflect a systematic error or a \added{real} physical shift to lower average masses for cooler brown dwarfs.

\begin{figure*}[!htp]
\includegraphics[width=1.0\textwidth, trim=15 0 0 0]{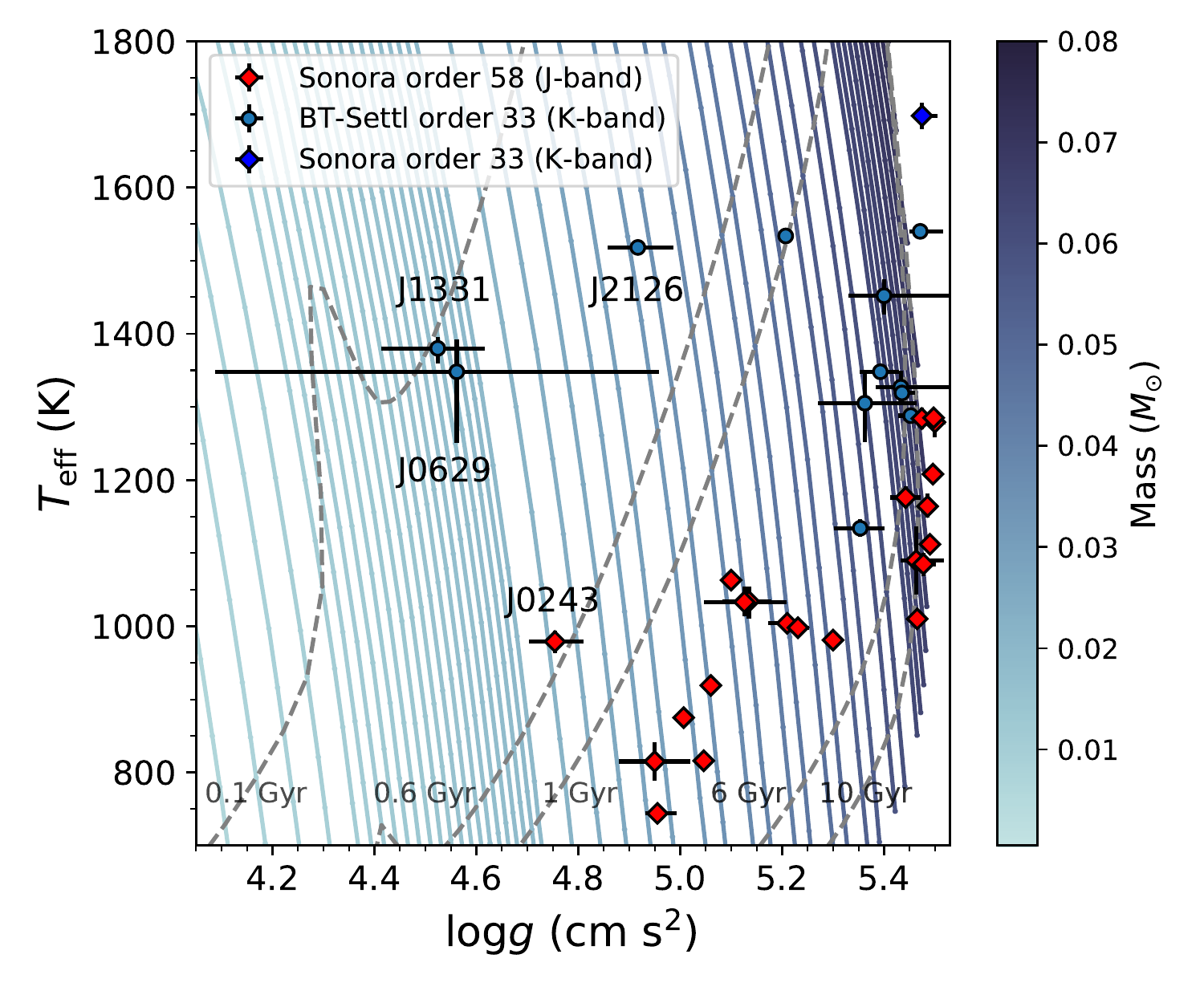}
\caption{Comparison of measured $T_{\text{eff}}$ and $\log {g}$ values based on fits to orders 33 (blue) and 58 (red) data. 
The BT-Settl and Sonora model fits are marked as circles and diamonds, respectively.
The \citet{Marley:2018aa}
evolutionary models are plotted, \replaced{color-coded with masses}{with lines of constant mass indicated by solid color-coded lines and lines of constant age (0.1, 0.6, 1, 6, and 10~Gyr) indicated by labeled dashed grey lines}. There are clear positive trends between $T_{\text{eff}}$ and $\log {g}$ for both BT-Settl and Sonora model fits \added{that run parallel to isoage lines}.  Note that earlier T dwarfs are mostly observed in $K$-band (orders 33), while later T dwarfs are mostly observed in $J$-band (order 58).
} \label{fig:teff_logg_order}
\end{figure*}

Next, we evaluated correlations \replaced{between}{among} $T_{\text{eff}}$, $\log {g}$, RV, and $v\sin{i}$. We find our inferred $T_{\text{eff}}$ and $\log {g}$ values to be positively correlated (Figure \ref{fig:teff_logg_order}) for Sonora model fits because of the temperature and pressure dependence of the CO$\rightarrow$CH$_4$ reduction reaction as described above (see Figure \ref{fig:J0136fitposterior}). Such a trend is not found in BT-Settl model fits, however, for which the lowest $\log {g}$ values correspond to L+T binaries (J0629+2418 and J2126+7617) and the blue L dwarf J1331$-$0116 (see Section \ref{sec:individual_source}). Excluding these three sources, the $\log {g}$ values are greater than 5.2~cm s$^{-2}$ dex for all fits, close to the $\log {g}$ ceiling of the models. \added{There are clear positive trends between $T_{\text{eff}}$ and $\log {g}$ paralleled to isoage lines (Figure \ref{fig:teff_logg_order}).}
We found no significant correlations between $\log {g}$ and $v\sin{i}$, $\log {g}$ and RV, $T_{\text{eff}}$ and $v\sin{i}$, or $T_{\text{eff}}$ and RV.

\begin{figure*}[!htbp]
\centering
\includegraphics[width=0.90\textwidth, trim=130 0 150 0]{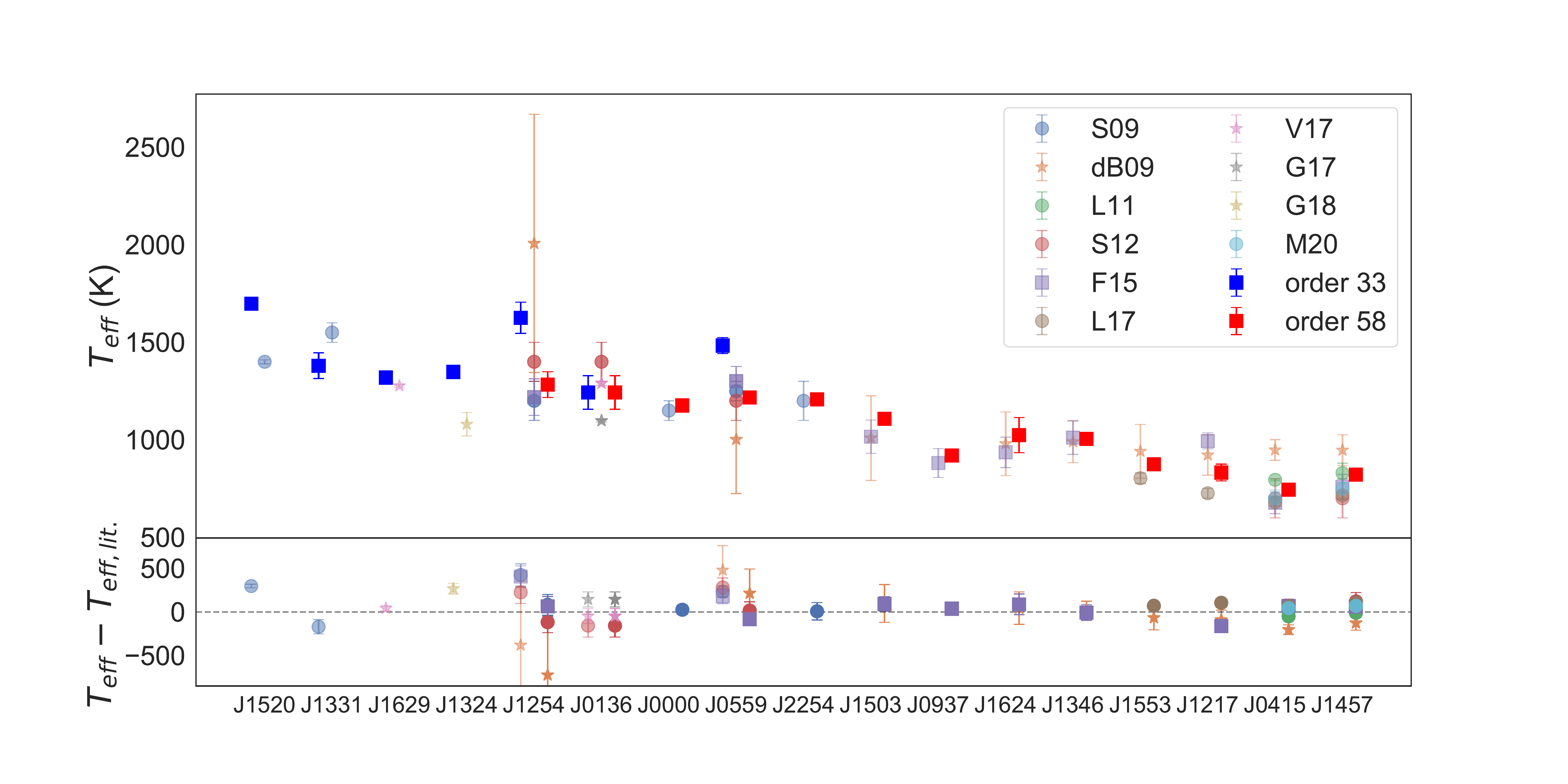}
\includegraphics[width=0.90\textwidth, trim=130 0 150 0]{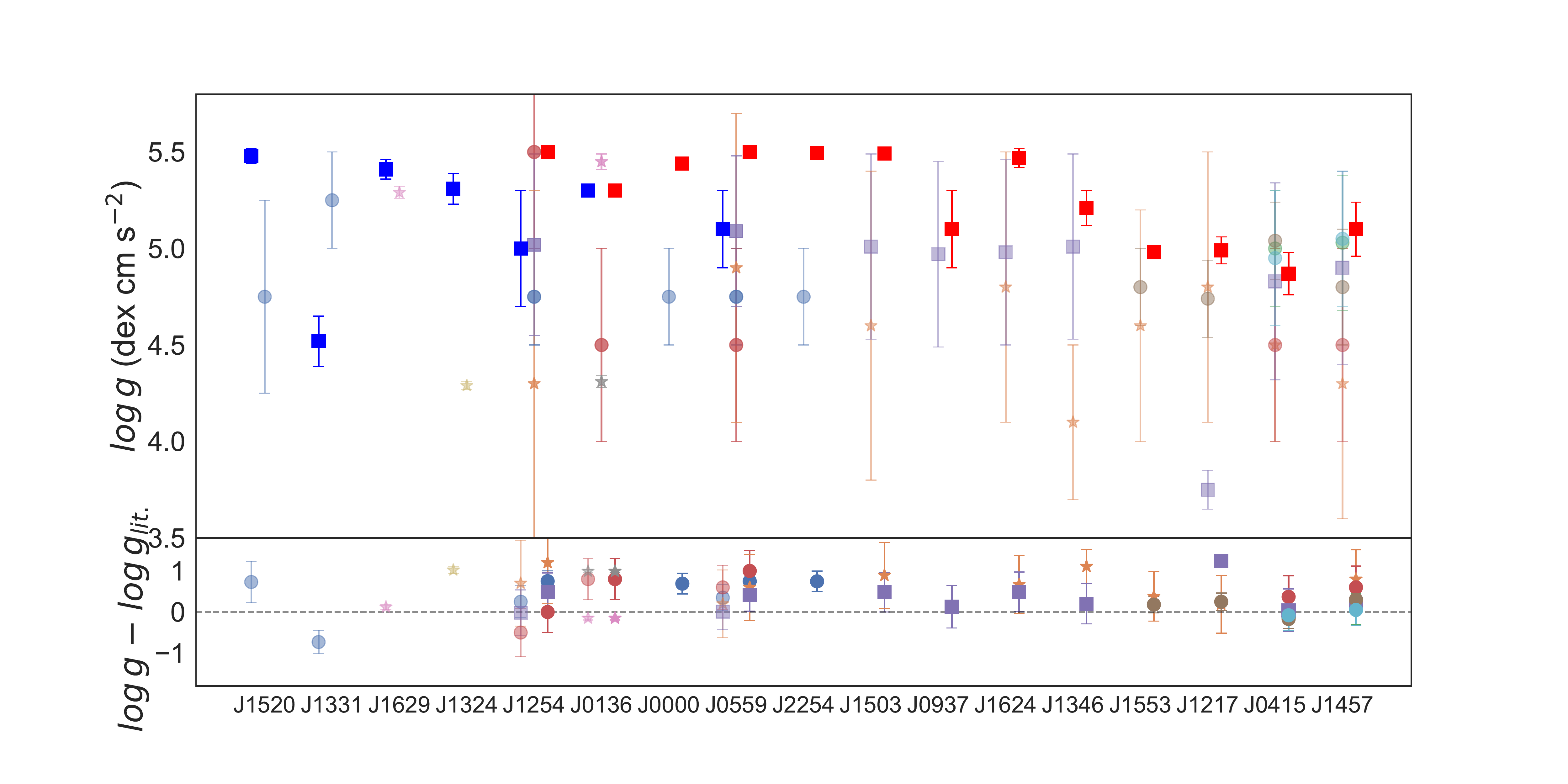}
\caption{Comparison of $T_{\text{eff}}$ measurements (top) and $\log {g}$ measurements (bottom) 
from our analysis with measurements from the literature. For each figure, the top panels show the measurements, with literature values slightly offset horizontally for clarity; the lower panels show the difference between measurement and literature values for each source. Measurements for orders 33 and 58 fits are labeled as blue and red squares, respectively. 
Literature references (from top to bottom in the legend) are: 
(S09): \citet{Stephens:2009aa} (light blue),  
(dB09): \citet{Del-Burgo:2009aa} (orange), 
(L11): \citet{Liu:2011aa} (green), 
(S12): \citet{Sorahana:2012aa} (red), 
(F15): \citet{Filippazzo:2015aa} (purple), 
(L17): \citet{Line:2017aa} (brown), 
(V17): \citet{Vos:2017aa} (pink), 
(G17): \citet{Gagne:2017aa} (grey), 
(G18): \citet{Gagne:2018aa} (olive), and
(M20): \citet{Miles:2020aa} (cyan). 
The stars, circles, and squares denote literature measurements based on high-resolution spectroscopy, medium-/low-resolution spectroscopy, and spectral energy distribution, respectively. \label{fig:teff_lit}}
\end{figure*}

Several T dwarfs in our sample have $T_{\text{eff}}$ and $\log {g}$ values inferred from other analysis, including high-resolution spectra, medium-/low-resolution spectra, and spectral energy distribution (SED) measurements. Figure \ref{fig:teff_lit} \replaced{show}{shows} our measurements of $T_{\text{eff}}$ and $\log {g}$\deleted{, respectively,} compared to the literature measurements.\footnote{The literature $T_{\text{eff}}$ and $\log {g}$ are drawn from
\citet[][medium-/low-resolution spectroscopy]{Stephens:2009aa}; 
\citet[][high-resolution spectroscopy]{Del-Burgo:2009aa}; 
\citet[][medium-resolution spectroscopy]{Liu:2011aa};
\citet[][low- and medium-resolution spectroscopy]{Sorahana:2012aa};
\citet[][SEDs]{Filippazzo:2015aa};
\citet[][low-resolution spectroscopy]{Line:2017aa};
\citet[][high-resolution spectroscopy]{Vos:2017aa};
\citet[][high-resolution spectroscopy]{Gagne:2017aa};
\citet[][high-resolution spectroscopy]{Gagne:2018aa}; and
\citet[][low-resolution spectroscopy]{Miles:2020aa}.}
Comparing against high-resolution spectra, 
our $T_{\text{eff}}$ values are on average 22$\pm$280~K lower and $\log {g}$ values on average are 0.6$\pm$0.4~dex higher than literature values, indicating overall consistency but with large scatter. 
Large discrepancies in $T_{\text{eff}}$ values inferred from high-resolution spectra \replaced{are in fact common}{have been previously reported} in the literature. 
\citet{Gagne:2017aa} and \citet{Vos:2017aa} analyzed NIRSPEC data for J0136+0933, both using forward-modeling techniques with the same model set, and
report $T_{\text{eff}}$ and $\log {g}$ values that differ by 192~K and 1.14~dex cm s$^{-2}$, respectively. 
\cite{Del-Burgo:2009aa} found a comparable degree of scatter in $T_{\text{eff}}$ $\approx$ 200~K and $\log {g}$ $\approx$ 0.7~dex in fits of PHOENIX AMES-COND cloudless models \citep{Allard:2001aa} across multiple  orders of T dwarf NIRSPEC spectra. Outdated methane opacities in the PHOENIX AMES-COND may be responsible for this scatter, as discussed in Section \ref{sec:atmmodel}. 
Comparing against low- and medium-resolution spectra, we find smaller differences and scatter, with our
$T_{\text{eff}}$ values on average 63$\pm$150~K higher and $\log {g}$ values on average 0.4$\pm$0.4~dex higher than literature values.
Similarly, our
$T_{\text{eff}}$ values are on average 68$\pm$145~K higher and $\log {g}$ values on average 0.3$\pm$0.4~dex higher than SED measurements from \citet{Filippazzo:2015aa}.
Again, such discrepancies are common in the literature, and reflect ongoing challenges in accurately modeling brown dwarf spectra.
In accord with our analysis, \cite{Logsdon:2018aa} found that $\log {g}$ values inferred from medium-resolution spectra of late-T dwarfs were highly dependent on the spectral band used, with the best-fit values of 3.0--3.5~dex in the $Y$-band ($\sim$0.95$-$1.12 $\micron$) and 5.0--5.5~dex in the $H$-band ($\sim$1.5$-$1.68 $\micron$) using the BT-Settl models. 
Different atmosphere models also yielded significantly different $\log {g}$ values, as discussed in Section~\ref{sec:atmmodel} 
\citep[see Table 3 in][]{Logsdon:2018aa}. 

Given the differences in $T_{\text{eff}}$ and $\log {g}$ values inferred between orders and in comparison with prior results, we urge caution in interpreting these quantities as they may not be accurate. Nevertheless, we have found that they are weakly correlated or uncorrelated with RV and $v\sin{i}$ and will not significantly influence the subsequent kinematic analysis. 

\begin{longrotatetable}
\begin{deluxetable*}{clllrrrrrrrrrrrll}
\tablecaption{Spectral Model Fit Parameters\label{tableindmearurement}}
\tablewidth{700pt}
\tabletypesize{\tiny}
\tablehead{
\colhead{} & \colhead{} & \colhead{} & \colhead{} & \colhead{} & \colhead{} & \colhead{} & \colhead{} &
\colhead{} & \colhead{} & \colhead{} & \colhead{} & \colhead{} & \colhead{} & \colhead{} & \colhead{} & \colhead{} \\
\colhead{Source} & 
\colhead{SpT} & \colhead{UT Date} & 
\colhead{O\tablenotemark{a}} & 
\colhead{S/N} & \colhead{RV} & \colhead{$v\sin i$} & 
\colhead{$\langle$RV$\rangle$\tablenotemark{b}} & \colhead{RV$_\text{LIT}$\tablenotemark{c}} & \colhead{$\langle$$v\sin i$$\rangle$\tablenotemark{b}} & \colhead{$v\sin i_\text{LIT}$\tablenotemark{c}} & \colhead{$T_\mathrm{eff}$} & \colhead{$\langle$$T_\mathrm{eff}$$\rangle$\tablenotemark{a}} &  \colhead{$\log \, g$} & \colhead{$\langle$$\log \, g$$\rangle$\tablenotemark{b}} & \colhead{M\tablenotemark{d}} & \colhead{Ref.\tablenotemark{e}} \\ 
\colhead{} & \colhead{} & \colhead{} & \colhead{} & \colhead{} & \colhead{(km s$^{-1}$)} & \colhead{(km s$^{-1}$)} & \colhead{(km s$^{-1}$)} &
\colhead{(km s$^{-1}$)} & \colhead{(km s$^{-1}$)} & \colhead{(km s$^{-1}$)} & \colhead{(K)} & \colhead{(K)} & \colhead{(cm s$^{-2}$)} & \colhead{(cm s$^{-2}$)} & \colhead{} & \colhead{}
} 
\startdata
J0000+2554 & T4.5 & 2019 Oct 17 & 58 & 5 & $6.4^{+0.4}_{-0.5}$ & $4.4^{+2.0}_{-2.5}$ & $6.4^{+0.4}_{-0.5}$ & \nodata & $<9$\tablenotemark{k} & \nodata & $1176^{+13}_{-15}$ & $1176^{+13}_{-15}$ & $5.44^{+0.03}_{-0.03}$ & $5.442^{+0.03}_{-0.03}$ & S & \\ 
J0034+0523 & T6.5 & 2020 Sep 3 & 58 & 14 & $16.6^{+1.4}_{-1.6}$ & $89.6^{+2.1}_{-2.1}$ & $16.6^{+1.4}_{-1.6}$ & \nodata & $89.6^{+2.1}_{-2.1}$ & \nodata & $815^{+26}_{-26}$ & $815^{+26}_{-26}$ & $4.95^{+0.07}_{-0.07}$ & $4.95^{+0.07}_{-0.07}$ & S & \\ 
J0136+0933\tablenotemark{h} & T2.5 & 2008 Dec 4 & 58 & 78 & $10.4^{+0.3}_{-0.4}$ & $49.9^{+0.9}_{-1.2}$ & $10.4^{+0.3}_{-0.3}$ & 11.5 $\pm$ 0.4 & $49.8^{+0.5}_{-0.5}$ & 50.9 $\pm$ 0.8 & $1148^{+6}_{-7}$ & $1288^{+8}_{-8}$ & $4.28^{+0.04}_{-0.04}$ & $5.45^{+0.02}_{-0.04}$ & B & (4) \\ 
\nodata & \nodata & 2013 Oct 16 & 33 & 10 & $11.6^{+0.8}_{-0.8}$ & $48.1^{+1.3}_{-1.3}$ & \nodata & \nodata & \nodata & \nodata & $1291^{+7}_{-8}$ & \nodata & $5.46^{+0.03}_{-0.05}$ & \nodata & B & \\ 
\nodata & \nodata & 2016 Feb 3 & 33 & 7 & $9.6^{+0.8}_{-0.8}$ & $49.4^{+1.4}_{-1.2}$ & \nodata & \nodata & \nodata & \nodata & $1303^{+6}_{-7}$ & \nodata & $5.43^{+0.05}_{-0.07}$ & \nodata & B & \\ 
J0150+3827 & T0 & 2020 Aug 5 & 33 & 4 & $56.2^{+0.9}_{-0.9}$ & $24.7^{+2.3}_{-2.3}$ & $56.2^{+0.9}_{-0.9}$ & \nodata & $24.7^{+2.3}_{-2.3}$ & \nodata & $1540^{+7}_{-7}$ & $1540^{+7}_{-7}$ & $5.47^{+0.02}_{-0.04}$ & $5.47^{+0.02}_{-0.04}$ & B & \\
J0213+3648 & T3 & 2020 Sep 3 & 58 & 4 & $-3.8^{+1.0}_{-0.9}$ & $25.1^{+1.1}_{-1.5}$ & $-3.8^{+1.0}_{-0.9}$ & \nodata & $25.1^{+1.1}_{-1.5}$ & \nodata & $1279^{+21}_{-17}$ & $1279^{+21}_{-17}$ & $5.5^{+0.01}_{-0.02}$ & $5.5^{+0.01}_{-0.02}$ & S & \\
J0243$-$2453 & T6 & 2021 Jan 1 & 58 & 5 & $-0.6^{+0.6}_{-0.7}$ & $9.9^{+1.6}_{-1.6}$ & $-0.6^{+0.6}_{-0.7}$ & \nodata & $9.9^{+1.6}_{-1.6}$ & \nodata & $979^{+16}_{-15}$ & $979^{+16}_{-15}$ & $4.75^{+0.05}_{-0.06}$ & $4.754^{+0.051}_{-0.057}$ & S & \\
\enddata
\tablerefs{(1) \citet{Zapatero-Osorio:2006aa}, (2) \citet{Zapatero-Osorio:2007aa}, (3) \citet{Prato:2015aa}, (4) \citet{Gagne:2017aa}, (5) \cite{Gagne:2018aa}}
\tablenotetext{a}{Spectral order of data.}
\tablenotetext{b}{Weighted average over all epochs.}
\tablenotetext{c}{Previously reported values in the literature.}
\tablenotetext{d}{Models used: S = Sonora 2018 \citep{Marley:2018aa}; B = BT-Settl \citep{Allard:2012aa}.}
\tablenotetext{e}{References for prior RV and $v\sin{i}$ measurements}
\tablenotetext{f}{No telluric observations available for this epoch.}
\tablenotetext{g}{Measurements are considered unreliable due to low S/N.}
\tablenotetext{h}{Adopted $T_{\text{eff}}$ and $\log {g}$ are determined using order 33 measurements only; see Section \ref{sec:tefflogg}.}
\tablenotetext{i}{Adopted $T_{\text{eff}}$ and $\log {g}$ are determined using order 58 measurements only; see Section \ref{sec:tefflogg}.}
\tablenotetext{j}{The spectra of order 57 are modeled in this case, as order 58 spectra are almost featureless due to large large $v\sin{i}$.}
\tablenotetext{k}{The $v\sin{i}$ detection floor for NIRSPEC is 9~km s$^{-1}$ for S/N $\geq$ 5, and 15~km s$^{-1}$ for S/N $<$ 5 data; see Section \ref{sec:min_vsini}.}
\tablecomments{Measurements from individual spectra over individual or multiple epochs are combined using inverse uncertainty weighting (weight = $1/(\sigma_\text{upper}^2 + \sigma_\text{lower}^2)$); upper and lower uncertainties are also combined using inverse uncertainty weighting. In cases where individual spectra have S/N $<$ 10, spectral data are combined first, the modeled.}
\end{deluxetable*}
\end{longrotatetable}

\section{Analysis} \label{sec:analysis}

\subsection{Galactic \textit{UVW} Space Motions and Kinematic Populations} \label{sec:uvwvel}

We combined astrometry and our measured RVs to compute Galactic \textit{UVW} space motions for our sample following the prescription of \citet{Johnson:1987aa}. \textit{UVW} velocities are defined here in a right-handed rectangular coordinate system centered on the Sun, with \textit{U} in the direction toward the Galactic center, \textit{V} in the direction of Galactic rotation, and \textit{W} in the direction toward the Galactic North pole (opposite the Galactic angular velocity vector). Uncertainties were propagated from the input quantities using the Monte Carlo method assuming Gaussian noise. We adopted a correction from the heliocentric frame to the local standard of rest (LSR) of $(U_{\odot}, V_{\odot}, W_{\odot}) = (11.1,	12.24, 7.25)$ km s$^{-1}$ from \citet{Schonrich:2010aa}. The T dwarf J1952+7240 does not have a parallax measurement, 
so we estimated its distance and uncertainty using the absolute magnitude/spectral type relations in \cite{Dupuy:2012aa}. Results are tabulated in Table \ref{tableuvw}. 

\begin{figure*}[!htbp]
\centering
\gridline{\fig{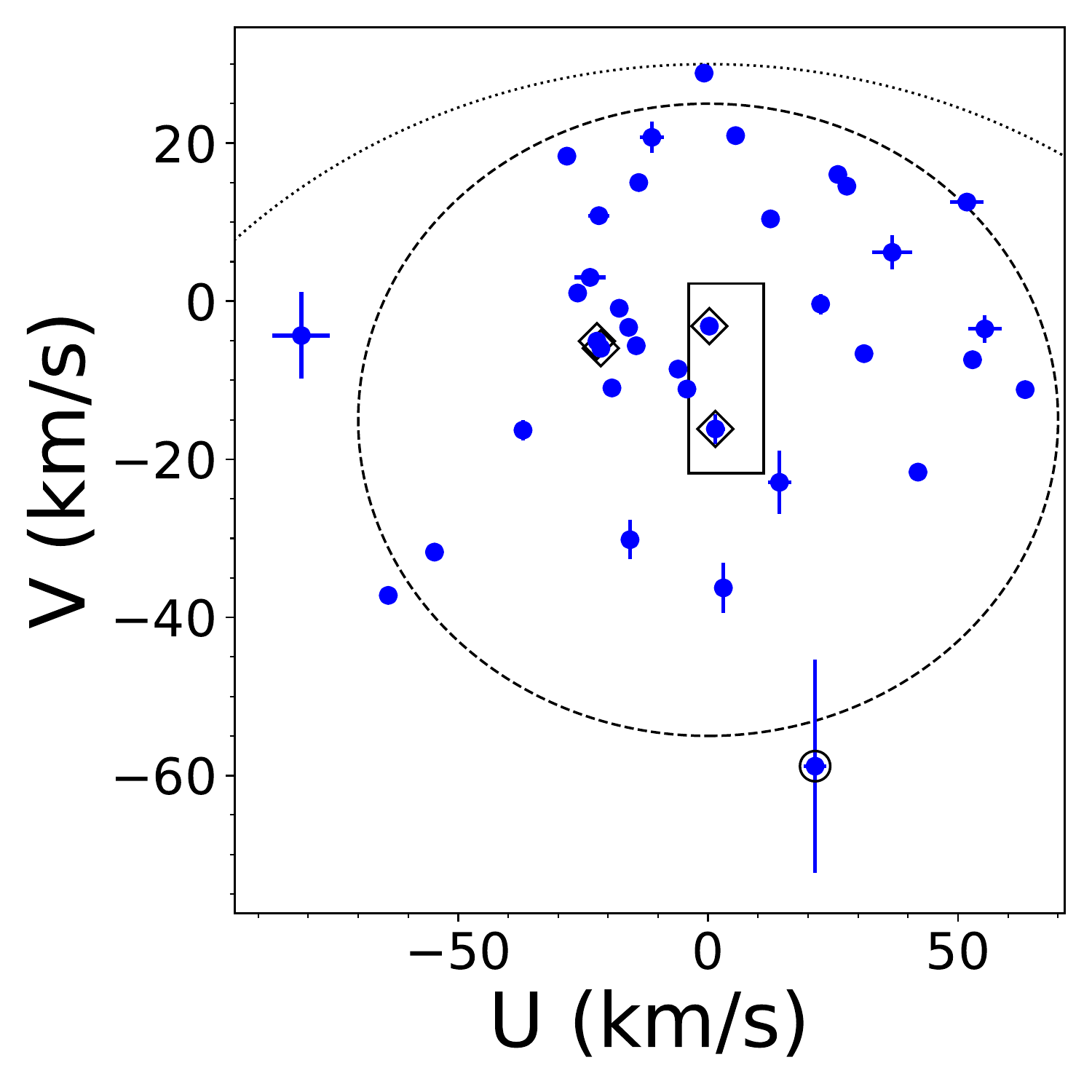}{0.45\textwidth}{}
\fig{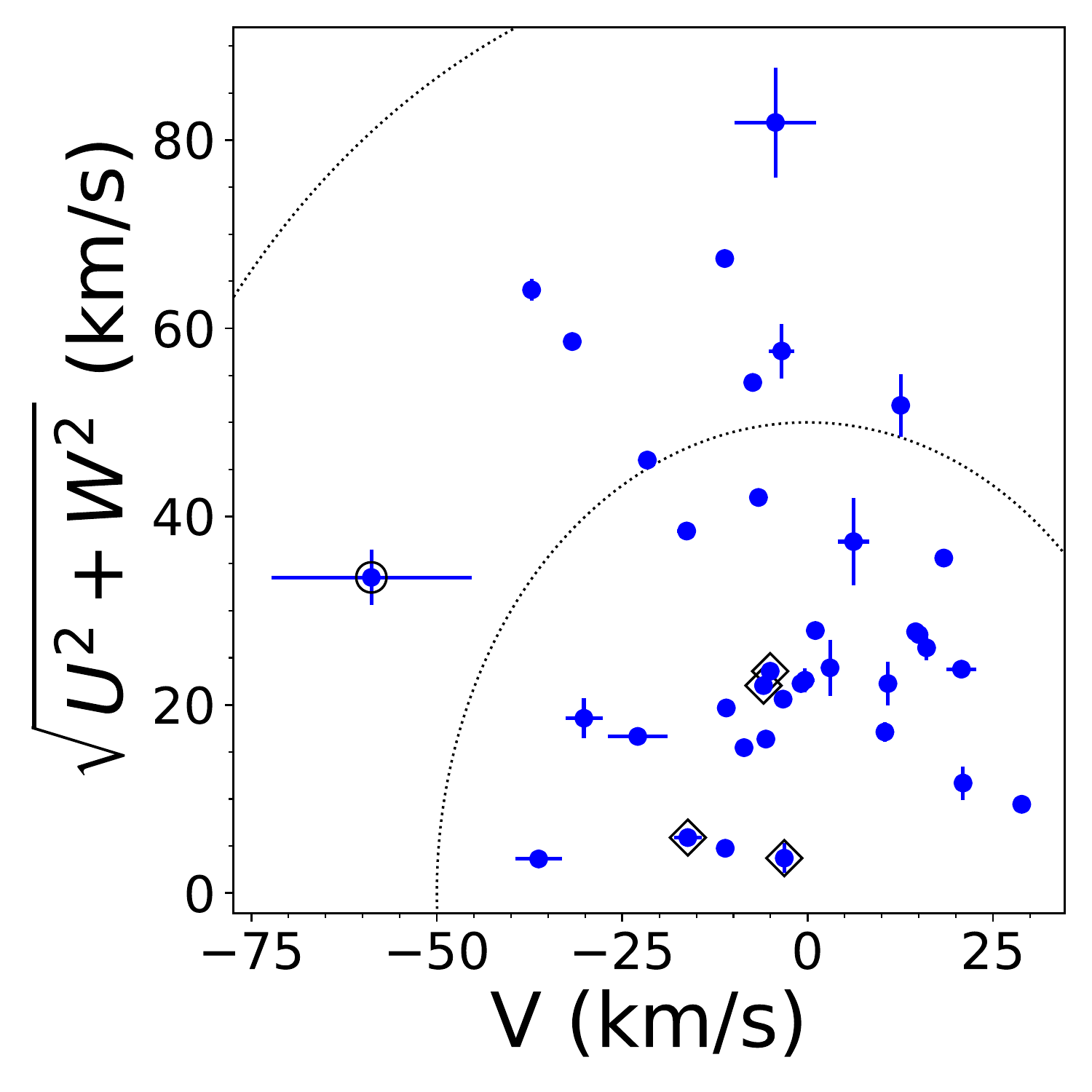}{0.45\textwidth}{}
}
\vspace{-1cm}
\gridline{\fig{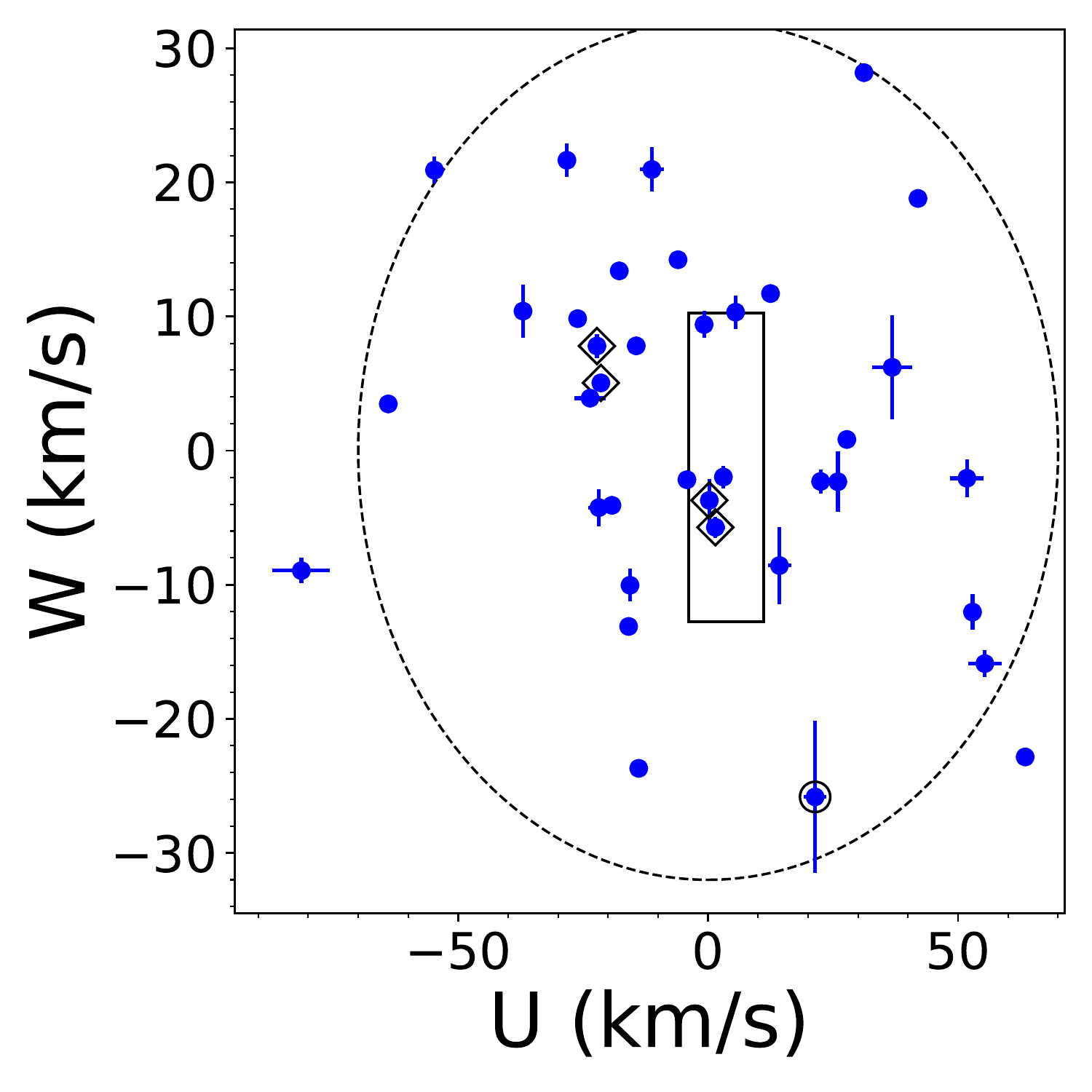}{0.45\textwidth}{}
\fig{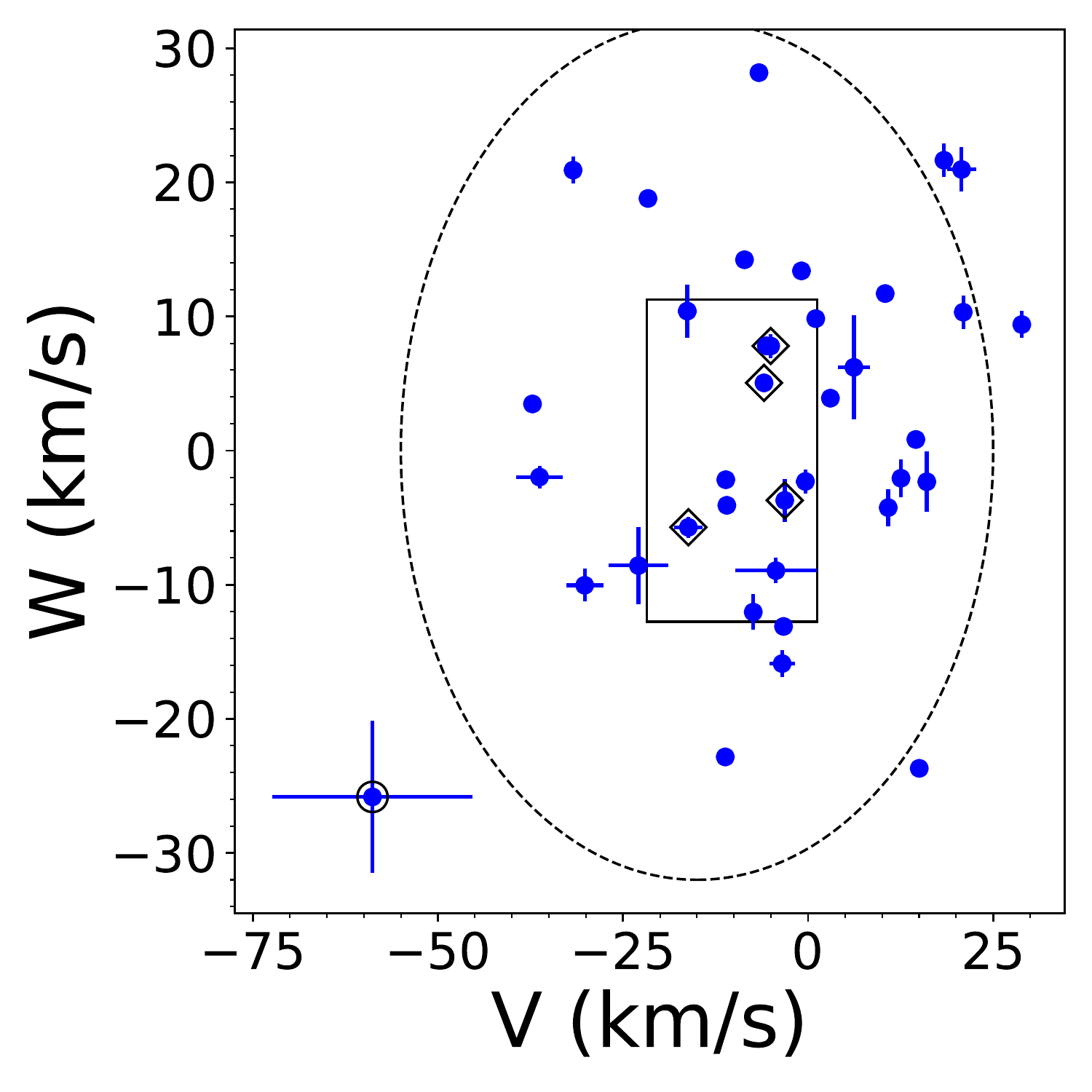}{0.45\textwidth}{}
}
\vspace{-1cm}
\caption{\textit{UVW} space motions of the T dwarf sample in the Local Standard of Rest \citep{Schonrich:2010aa}. The \textit{UV}, \textit{UW}, and \textit{VW} velocities are shown along with the 2$\sigma$ uncertainty spheres for the thin disk (dashed lines) and thick disk (dotted lines) populations from \citet{Bensby:2003aa}. The ``good box" from \citet{Zuckerman:2004aa} that segregates members of young moving groups is also labeled. The upper-right corner is a Toomre plot, with total velocities $v_\text{tot} = \sqrt{U^2 + V^2 +W^2}$ indicated in steps of $50$ km s$^{-1}$. Young sources and intermediate thin/thick disk sources ($0.1 < P(\text{TD})/P(\text{D}) < 10$) are highlighted with open diamonds and open circles, respectively. \label{fig:figUVWvelocity}}
\end{figure*}

Figure \ref{fig:figUVWvelocity} compares the distribution of \textit{UVW} velocities to the 2$\sigma$ velocity dispersion volumes of local thin and thick disk populations from \citet{Bensby:2003aa}. 
The average $U$ and $W$ velocities of the T dwarfs are consistent with zero, 
while a \replaced{slightly}{marginally significant net} negative average \textit{V} velocity ($\langle V \rangle = -3.6 \pm 2.7$ km s$^{-1}$) can be attributed to asymmetric drift
\citep{Stromberg:1924aa}.
We do not find any significant correlation between \textit{UV}, \textit{UW}, or \textit{VW} velocities.

Following \cite{Bensby:2003aa}, we computed relative probabilities of membership in the thick disk versus the thin disk ($P(\text{TD})/P(\text{D})$) and halo versus the thin disk ($P(\text{H})/P(\text{D})$) using the threshold criteria defined in \citet{Burgasser:2015ac}: thin disk membership is assigned for $P(\text{TD})/P(\text{D}) < 0.1$, thick disk membership is assigned for $ P(\text{TD})/P(\text{D}) >  10$, and intermediate population membership is assigned for $0.1 < P(\text{TD})/P(\text{D}) < 10$. All but one of our
sources are thin disk members, with J1331$-$0116 being identified as an intermediate thin disk/thick disk member and \deleted{as} an unusually blue L dwarf (see Section~\ref{sec:individual_source}).
None of the sources in our sample have a significant probability of halo membership.

\subsection{Galactic Orbits} \label{galacticorbits}

Additional insight into our sample's kinematic properties can be inferred by computing their Galactic orbits and orbital parameters. We used the package \textit{galpy} \citep{Bovy:2015aa} to compute the orbits, which is an ordinary differential equation solver that satisfies conservation of energy and angular momentum. We assumed an axisymmetric Galactic potential in a galactocentric cylindrical coordinate system ($R,\phi,Z$) using the parameters of \cite{Miyamoto:1975aa}, a Solar azimuthal velocity $v_{\phi}$ = 220 km s$^{-1}$ \citep{Bovy:2012aa}, and a Solar coordinate of ($R_{\odot}$, $Z_{\odot}$) = ($8.43, 0.027$) kpc at $\phi_{\odot}$  = 0 \citep{Chen:2001aa, Reid:2014aa}. 
Orbits were sampled over the period $-$5 to +5~Gyr. Uncertainties in the present-day position and velocity of each source were propagated using Monte Carlo sampling assuming Gaussian noise, resulting in 1,000 orbits per source from which we computed 
minimum and maximum Galactic cylindrical radius ($R_\text{max}, R_\text{min}$), maximum absolute Galactic vertical height ($\left| Z \right|$), median orbital eccentricity ($e \equiv \langle R_\text{max} - R_\text{min} \rangle / \langle R_\text{max} + R_\text{min} \rangle$), and median orbital inclination ($\tan{i} \equiv |Z / \sqrt{X^2 + Y^2}|$), with uncertainties determined from the distribution of simulated orbits. 

The majority of our sample possess circular and planar orbits ($e \leq 0.20$, $ i \leq 2^{\circ}$) as expected for a thin disk population. The intermediate thin/thick disk star J1331$-$0116 has the largest inclination and eccentricity in the sample ($i = 2.5^{\circ} \pm 0.8^{\circ}, \, e = 0.29 \pm 0.06, \, R_\text{min} = 4.6 \pm 0.6$ kpc). The median orbital parameters for the sample, $R_\text{min} = 7.5$ kpc, $R_\text{max} = 9.2$ kpc, $e = 0.12$, and $i = 0.69^{\circ}$, are consistent with the orbital parameters of local late-M and L dwarfs reported in \citet{Burgasser:2015ac}.\footnote{These values are $R_\text{min}$ = 8.0~kpc, $R_\text{max}$ = 9.5~kpc, and $e$ = 0.11 for late-M dwarfs, and $R_\text{min}$ = 7.8~kpc, $R_\text{max}$ = 10~kpc, and $e$ = 0.16 for late L dwarfs, with $i \leq 2^\circ$ for both populations.}

\subsection{Cluster Membership} \label{clustermembership}

UCDs, including T dwarfs, have been found to be members of nearby association and clusters, which provide independent age determinations and can potentially break the age-mass-$L_\text{bol}$ degeneracy. Cluster membership probability can be determined by the alignment in 6D configuration space (heliocentric \textit{XYZ} spatial and \textit{UVW} velocity coordinates) with other association members. 
\textit{XYZ} spatial coordinates are defined in the same direction as \textit{UVW}.
We used the BANYAN $\Sigma$ web tool\footnote{\url{http://www.exoplanetes.umontreal.ca/banyan/}} \citep{Gagne:2018ab} to compare the astrometry and radial velocities of our sources to 27 young clusters within 150~pc of the Sun. 
We confirmed that J0136+0933 \citep{Gagne:2017aa} and J1324+6358 \citep{Gagne:2018aa} are probable members (99$\%$) of the $\sim$200~Myr Carina-Near and 130~Myr AB Doradus moving groups, respectively \citep{2006ApJ...649L.115Z,2018ApJ...861L..13G}. 
\replaced{J0819$-$0335 is a newly identified candidate kinematic member of the $\beta$ Pictoris moving group \citep[age $\tau$ = 24$\pm$3~Myr,][]{Bell:2015aa}, with a 95\% probability of membership and a 5\% probability of being a field dwarf (candidate member reported in \citealp{Zhang:2021aa}).}
{We also confirmed that J0819$-$0335 is a candidate kinematic member of the $\beta$ Pictoris moving group \citep[age $\tau$ = 24$\pm$3~Myr,][]{Bell:2015aa}, with a 95\% probability of membership and a 5\% probability of being a field dwarf \citep{Zhang:2021aa}.}
J1553+1532 is \replaced{a newly identified}{also confirmed as a} kinematic member of Carina-Near moving group, with a 98\% probability of membership (2\% field object\added{; \citealp{Zhang:2021aa}}). We discuss these two sources in further detail below.
We are able to rule out three young moving group candidates reported in \citet{Zhang:2021aa}\added{, identified on the basis of spatial coordinates, proper motion, and the BANYAN $\Sigma$ tool. These are} J0627$-$1114 (99\% field object), J1624+0029 (27\% Carina-Near; 73\% field object), and J2236+5105 (99\% field object)\replaced{, which}{. Our ability to excluded these candidates} highlights the importance of precise RV measurements in assessing cluster membership.

\begin{deluxetable*}{llrccccc}[h]
\tablecaption{Radial Velocities and Heliocentric Space Motions \label{tableuvw}}
\tablewidth{700pt}
\tabletypesize{\scriptsize}
\tablehead{
\colhead{} & \colhead{} & \colhead{} & \colhead{} & \colhead{} & \colhead{} & \colhead{} & \colhead{} \\ 
\colhead{Source Name} & \colhead{SpT} & \colhead{Adpoted RV} & \colhead{$U$} & \colhead{$V$} & \colhead{$W$} & \colhead{P[TD]/P[D]\tablenotemark{a}} & \colhead{Population\tablenotemark{a}} \\ 
\colhead{} & \colhead{} & \colhead{(km s$^{-1}$)} & \colhead{(km s$^{-1}$)} & \colhead{(km s$^{-1}$)} & \colhead{(km s$^{-1}$)} & \colhead{} & \colhead{}
} 
\startdata
J0000+2554 & T4.5 & $6.4^{+0.4}_{-0.5}$ & 7.1$\pm$0.2 & 21.6$\pm$0.4 & 10.5$\pm$0.4 & 0.01 & D \\ 
J0034+0523 & T6.5 & $16.6^{+1.4}_{-1.6}$ & $-$19.0$\pm$0.7 & 11.5$\pm$0.8 & $-$4.6$\pm$1.4 & 0.01 & D \\ 
J0136+0933 & T2.5 & $10.4^{+0.3}_{-0.3}$ & $-$21.5$\pm$0.2 & $-$6.0$\pm$0.1 & 5.1$\pm$0.2 & 0.01 & D \\ 
J0150+3827 & T0 & $56.2^{+0.9}_{-0.9}$ & $-$94.6$\pm$5.0 & $-$15.5$\pm$4.6 & $-$5.2$\pm$0.8 & 0.09 & D \\ 
J0213+3648 & T3 & $-3.8^{+1.0}_{-0.9}$ & 10.7$\pm$3.2 & 7.2$\pm$3.6 & 10.1$\pm$4.1 & 0.01 & D \\ 
J0243$-$2453 & T6 & $-0.6^{+0.6}_{-0.7}$ & 27.8$\pm$0.7 & 14.6$\pm$0.3 & 0.8$\pm$0.7 & 0.01 & D \\ 
J0415$-$0935 & T8 & $51.9^{+1.1}_{-1.1}$ & $-$54.9$\pm$0.8 & $-$31.8$\pm$0.4 & 21.2$\pm$0.8 & 0.05 & D \\ 
J0559$-$1404 & T4.5 & $-9.5^{+0.1}_{-0.1}$ & 31.1$\pm$0.1 & $-$6.4$\pm$0.2 & 28.0$\pm$0.1 & 0.02 & D \\ 
J0627$-$1114 & T6 & $1.2^{+0.9}_{-1.1}$ & 23.8$\pm$1.0 & $-$1.7$\pm$1.0 & $-$3.0$\pm$0.5 & 0.01 & D \\ 
J0629+2418 & L7+T5.5\tablenotemark{b} & $0.5^{+2.3}_{-2.1}$ & 13.9$\pm$2.3 & $-$27.1$\pm$3.5 & $-$17.8$\pm$2.3 & 0.01 & D \\ 
J0755+2212 & T5 & $22.3^{+2.7}_{-1.7}$ & $-$5.9$\pm$2.3 & $-$10.8$\pm$1.1 & 9.0$\pm$1.1 & 0.01 & D \\ 
J0819$-$0335 & T4 & $14.4^{+0.8}_{-0.9}$ & 0.3$\pm$0.6 & $-$3.5$\pm$0.7 & $-$4.5$\pm$0.6 & 0.01 & D \\ 
J0909+6525 & T1.5+T2.5\tablenotemark{b} & $37.8^{+1.6}_{-1.6}$ & $-$28.1$\pm$1.4 & 19.6$\pm$0.8 & 21.0$\pm$1.2 & 0.02 & D \\ 
J0937+2931 & T6 & $-4.3^{+0.4}_{-0.4}$ & 42.0$\pm$0.4 & $-$21.6$\pm$0.4 & 18.8$\pm$0.4 & 0.02 & D \\ 
J1106+2754 & T0+T4.5\tablenotemark{b} & $2.5^{+0.2}_{-0.2}$ & 3.2$\pm$0.2 & $-$37.0$\pm$1.2 & $-$2.0$\pm$0.3 & 0.01 & D \\ 
J1217$-$0311 & T7.5 & $9.9^{+2.3}_{-1.9}$ & $-$37.0$\pm$1.2 & $-$16.3$\pm$1.3 & 10.4$\pm$2.0 & 0.01 & D \\ 
J1225$-$2739 & T5.5+T8\tablenotemark{b} & $18.5^{+1.5}_{-1.2}$ & 52.9$\pm$1.3 & $-$7.4$\pm$1.1 & $-$12.0$\pm$1.3 & 0.01 & D \\ 
J1254$-$0122 & T2e & $2.7^{+0.4}_{-0.5}$ & $-$17.8$\pm$1.0 & $-$0.9$\pm$0.5 & 13.4$\pm$0.5 & 0.01 & D \\ 
J1324+6358 & T2p & $-23.0^{+0.5}_{-0.4}$ & 5.5$\pm$0.7 & $-$12.5$\pm$0.8 & $-$6.8$\pm$0.5 & 0.01 & D \\ 
J1331$-$0116 & T0 & $-3.3^{+0.4}_{-0.4}$ & 25.4$\pm$1.6 & $-$83.8$\pm$9.7 & $-$36.3$\pm$4.1 & \replaced{16.69}{0.27} & \added{D/}TD \\ 
J1346$-$0031 & T6.5 & $-17.5^{+0.6}_{-0.5}$ & $-$19.2$\pm$0.8 & $-$11.0$\pm$1.0 & $-$4.1$\pm$0.5 & 0.01 & D \\ 
J1457$-$2122 & T8 & $31.9^{+0.3}_{-0.3}$ & 63.4$\pm$0.4 & $-$11.2$\pm$0.2 & $-$22.8$\pm$0.5 & 0.04 & D \\ 
J1503+2525 & T5.5 & $-35.9^{+0.4}_{-0.3}$ & $-$13.8$\pm$0.2 & 14.9$\pm$0.1 & $-$23.7$\pm$0.4 & 0.02 & D \\ 
J1506+7027 & T6 & $2.5^{+0.4}_{-0.5}$ & $-$26.0$\pm$0.2 & 1.0$\pm$0.4 & 9.8$\pm$0.3 & 0.01 & D \\ 
J1520+3546 & T0 & $4.1^{+1.0}_{-1.0}$ & 42.4$\pm$2.4 & 13.3$\pm$0.5 & 0.8$\pm$1.1 & 0.01 & D \\ 
J1553+1532 & T6.5+T7.5\tablenotemark{b} & $-25.5^{+0.6}_{-0.5}$ & $-$21.6$\pm$0.4 & $-$4.6$\pm$0.2 & 7.1$\pm$0.5 & 0.01 & D \\ 
J1624+0029 & T6 & $-24.3^{+0.5}_{-0.5}$ & $-$14.4$\pm$0.4 & $-$5.6$\pm$0.2 & 7.8$\pm$0.3 & 0.01 & D \\ 
J1629+0335 & T2 & $7.6^{+0.6}_{-0.6}$ & 26.0$\pm$0.7 & 16.9$\pm$0.2 & $-$2.8$\pm$0.8 & 0.01 & D \\ 
J1809$-$0448 & T1 & $-43.4^{+1.3}_{-1.3}$ & $-$12.0$\pm$1.5 & $-$38.0$\pm$2.0 & $-$11.8$\pm$0.9 & 0.02 & D \\ 
J1928+2356 & T6 & $-26.3^{+0.2}_{-0.3}$ & $-$5.3$\pm$0.2 & $-$9.1$\pm$0.3 & 16.0$\pm$0.1 & 0.01 & D \\ 
J1952+7240 & T4 & $-12.0^{+1.3}_{-0.7}$ & 39.2$\pm$0.7 & 5.4$\pm$1.2 & 9.7$\pm$0.5 & 0.01 & D \\ 
J2030+0749 & T1.5 & $-21.2^{+0.6}_{-0.7}$ & $-$16.2$\pm$0.4 & $-$3.3$\pm$0.5 & $-$13.5$\pm$0.3 & 0.01 & D \\ 
J2126+7617 & L7+T3.5\tablenotemark{b} & $-18.4^{+0.4}_{-0.4}$ & $-$62.6$\pm$0.8 & $-$36.7$\pm$0.5 & 3.5$\pm$0.1 & 0.05 & D \\ 
HN Peg B & T2.5 & $-19.5^{+1.3}_{-1.3}$ & $-$4.2$\pm$0.4 & $-$11.1$\pm$1.1 & $-$2.2$\pm$0.6 & 0.01 & D \\ 
J2236+5105 & T5 & $-1.2^{+0.3}_{-0.3}$ & $-$25.2$\pm$1.3 & 2.6$\pm$0.4 & 3.6$\pm$0.2 & 0.01 & D \\ 
J2254+3123 & T5 & $15.1^{+0.5}_{-0.6}$ & 0.1$\pm$0.4 & 28.7$\pm$0.6 & 8.9$\pm$0.5 & 0.01 & D \\ 
J2356$-$1553 & T6 & $21.8^{+0.9}_{-1.0}$ & 55.4$\pm$3.3 & $-$3.5$\pm$1.7 & $-$15.9$\pm$1.0 & 0.02 & D
\enddata
\tablenotetext{a}{Galactic thin disk (D), thick disk (TD), intermediate populations (D/TD) are assigned according to probability ratios 
$P(\mathrm{TD})/P(\mathrm{D}) < 0.1$, $P(\mathrm{TD})/P(\mathrm{D}) > 10$, and $0.1 < P(\mathrm{TD})/P(\mathrm{D}) < 10$, respectively, following \citet{Bensby:2003aa}.}
\tablenotetext{b}{Known or candidate binary.}
\end{deluxetable*}

\subsection{Individual Sources of Interest} \label{sec:individual_source}

\textit{2MASS J00345157+0523050} is a peculiar T6.5/T7 \citep{Chiu:2006aa, Burgasser:2004ab}. It has the largest $v\sin{i}$ in our sample ($v\sin{i}$ = 90 $\pm$ 2 km s$^{-1}$), making it one of the fastest rotating brown dwarfs found to date\replaced{(cf.\ 
2MASS~J04070752+1546457: 82.6$\pm$0.2~ km s$^{-1}$
LP 349-25B: 83 $\pm$ 3 km s$^{-1}$, 
HD 130948BC: 86 $\pm$ 6 km s$^{-1}$,
and 2MASS~J03480772$-$6022270: 104 $\pm$ 7 km s$^{-1}$,
\citealt{Konopacky:2012aa,2021arXiv210301990T}).}
{\footnote{Other unusually fast rotators plotted in Figure~\ref{fig:vsini_comparison_literature} include 2MASS~J04070752+1546457 (82.6 $\pm$ 0.2~km s$^{-1}$),
LP 349-25B (83 $\pm$ 3 km s$^{-1}$), 
HD 130948BC (86 $\pm$ 6 km s$^{-1}$),
and 2MASS~J03480772$-$6022270 (104 $\pm$ 7 km s$^{-1}$); see
\citet{Konopacky:2012aa, 2021arXiv210301990T}.}.}
Assuming a radius of 1~$R_\mathrm{Jup}$ and edge-on rotation, this corresponds to a {\it maximum} rotational period of 1.4~hr. 
Assuming a mass of 0.05 M$_{\odot}$ (which corresponds to a 900~K brown dwarf at 5~Gyr; \citealt{Baraffe:2003aa}),
the observed velocity corresponds to 30\% of the break-up rotational velocity ($v_\mathrm{break} = \sqrt{GM/R} \sim$ 300 km s$^{-1}$).
The source exhibits a very blue near-infrared color $J-K = -0.93 \pm 0.03$ \citep{Lawrence:2013aa} and other spectral peculiarities that have been attributed to enhanced H$_2$ collision-induced absorption (CIA) in the high-pressure atmosphere of a relatively massive (high surface gravity) and/or low-metallicity (low opacity) brown dwarf \citep{1969ApJ...156..989L,Burgasser:2004ab}.
The absence of the K~\textsc{i} doublet absorption \replaced{$\sim$}{at} 1.25~$\micron$, which is generally present in the spectra of T6--T7 dwarfs \citep{Martin:2017aa}, 
may be an indicator of low metallicity effects or shallowing of the line features due to the object's fast rotation.
Either trait would imply that J0034+0523 is a relatively old brown dwarf with a high mass and \deleted{more }compact radius which has not undergone significant angular momentum loss. We note that our forward-modeling fit utilizes solar-metallicity atmosphere models, and the potential subsolar metalicity of this peculiar T dwarf may influence the derived parameters, including $v\sin{i}$.

\added{\textit{2MASS~J02431371$-$2453298} \citep{Burgasser:1999aa} stands out in Figure~\ref{fig:teff_logg_order} as the only non-binary or UBL with a gravity-based model age significantly less than 1~Gyr. This T6 dwarf had previously been identified as a candidate member of the $\sim$400~Myr Ursa Majoris moving group based on its distance and proper motion \citep{2007MNRAS.378L..24B,2008MNRAS.385.1771J}, and prior low-resolution spectral analyses have also indicated evidence of low surface gravity \citep{Burgasser:2006ab}. 
This source thus exhibits both kinematic and spectral indicators of relative youth, and is a potential benchmark for spectral age indicators for mid- and late-type T dwarfs. However,  it should be noted that two other mid- to late-type T dwarfs identified by \citep{2007MNRAS.378L..24B} as potential members of the $\sim$650~Myr Hyades moving group, the T7 J1217$-$0311 and the T6 J1624+0029, are not identified as low surface gravity objects in this analysis.} 

\begin{figure*}[!htbp]
\includegraphics[width=1.0\textwidth, trim = 20 0 70 0 ]{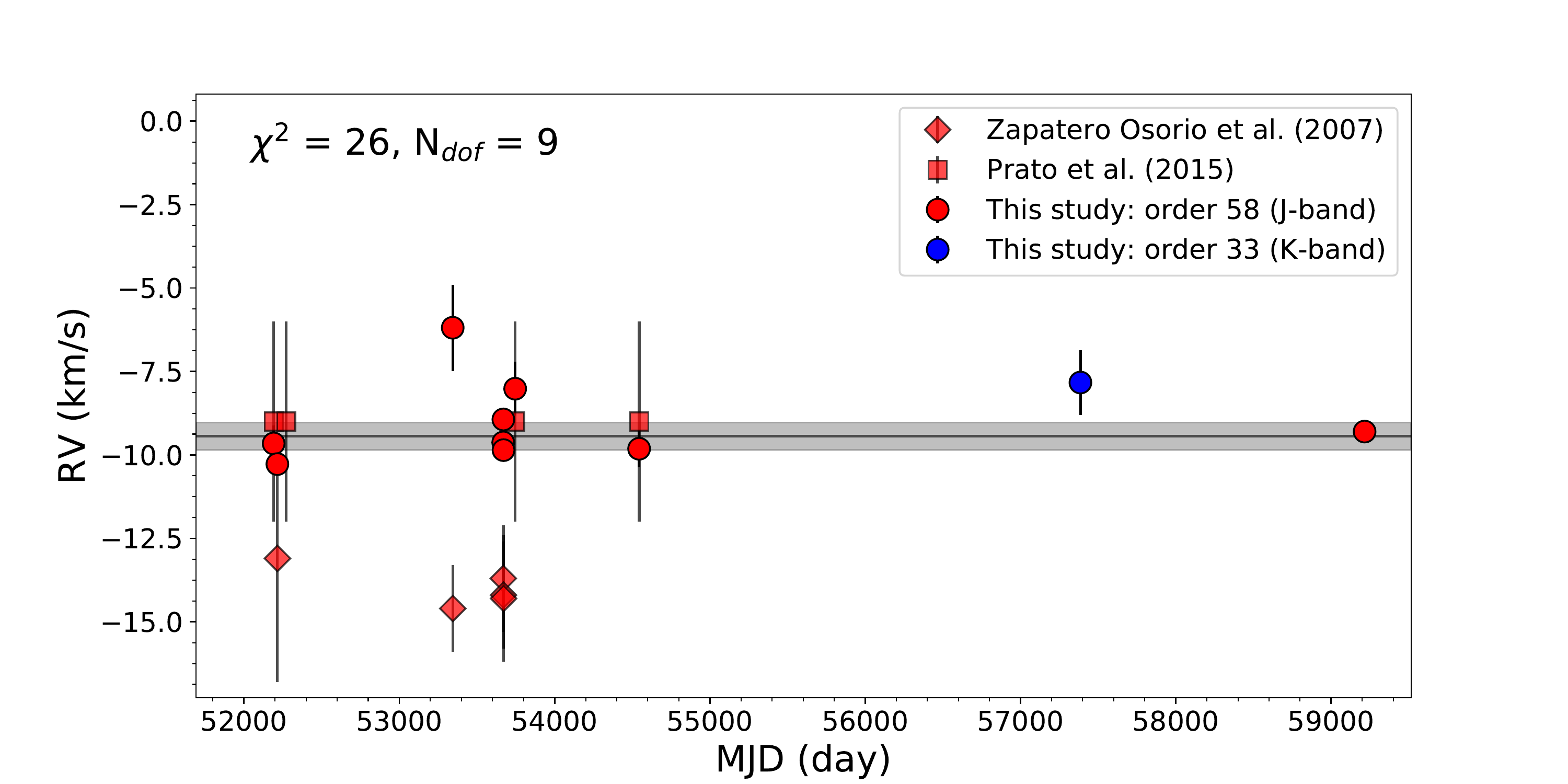}
\caption{RV time series for all of the NIRSPEC measurement epochs for 2MASS J0559$-$1404.  
Measurements made in order 33 and 58 are shown as blue and red circles, respectively.
The weighted average of our measurements and uncertainty are indicated by the horizontal line and grey shading. 
Prior measurements reported by \citet{Zapatero-Osorio:2007aa} and \citet{Prato:2015aa} based on NIRSPEC data are indicated by red diamonds and red squares, respectively. 
\label{fig:figJ0559variation}}
\end{figure*}

\textit{2MASS J05591914$-$1404488} \citep{Burgasser:2000ab} is a \added{relatively bright and} seemingly overluminous T4.5 \added{($M_J$ = 13.7)}, but has \replaced{never}{not yet} been confirmed as a binary system \citep{Burgasser:2003aa, Golimowski:2004aa, Liu:2006aa, Stephens:2009aa, Burgasser:2010aa, Dupuy:2017aa, Manjavacas:2019aa}. It is one of five sources in our sample that has multi-epoch observations\replaced{ (N$_{\text{obs}}>$4) sufficient to evaluate RV variability. We find that 10 epochs of observations}{, with 10 observations} spanning 2001 Oct 9 to 2021 Jan 1\replaced{ do}{. These measurements} show significant evidence of RV variation, differing from a constant velocity model by $\chi^2$ = 26, (Degrees of freedom DOF = 9, p-value $\leq$ 0.005; Figure \ref{fig:figJ0559variation}). 
This is in contrast to the conclusions of \citet{Zapatero-Osorio:2007aa} and \citet{Prato:2015aa} who report constant---but differing---RVs based on subsets of the same data. Indeed, the standard deviation of our re-analyzed measurements ($\sigma_\text{RV}$ = 1.2 km s$^{-1}$\added{)} is smaller than the difference between the previously published RVs ($RV_\mathrm{Zap}-RV_\mathrm{Pra}$ = $-$5.0 km s$^{-1}$).
The sparse sampling of the data prevents a robust assessment of potential orbital motion, 
and follow-up RV measurements are warranted to validate this RV variation and assess the multiplicity of this unresolved source. 

\begin{figure}[!htbp]
\includegraphics[width=0.5\textwidth, trim=20 0 0 0]{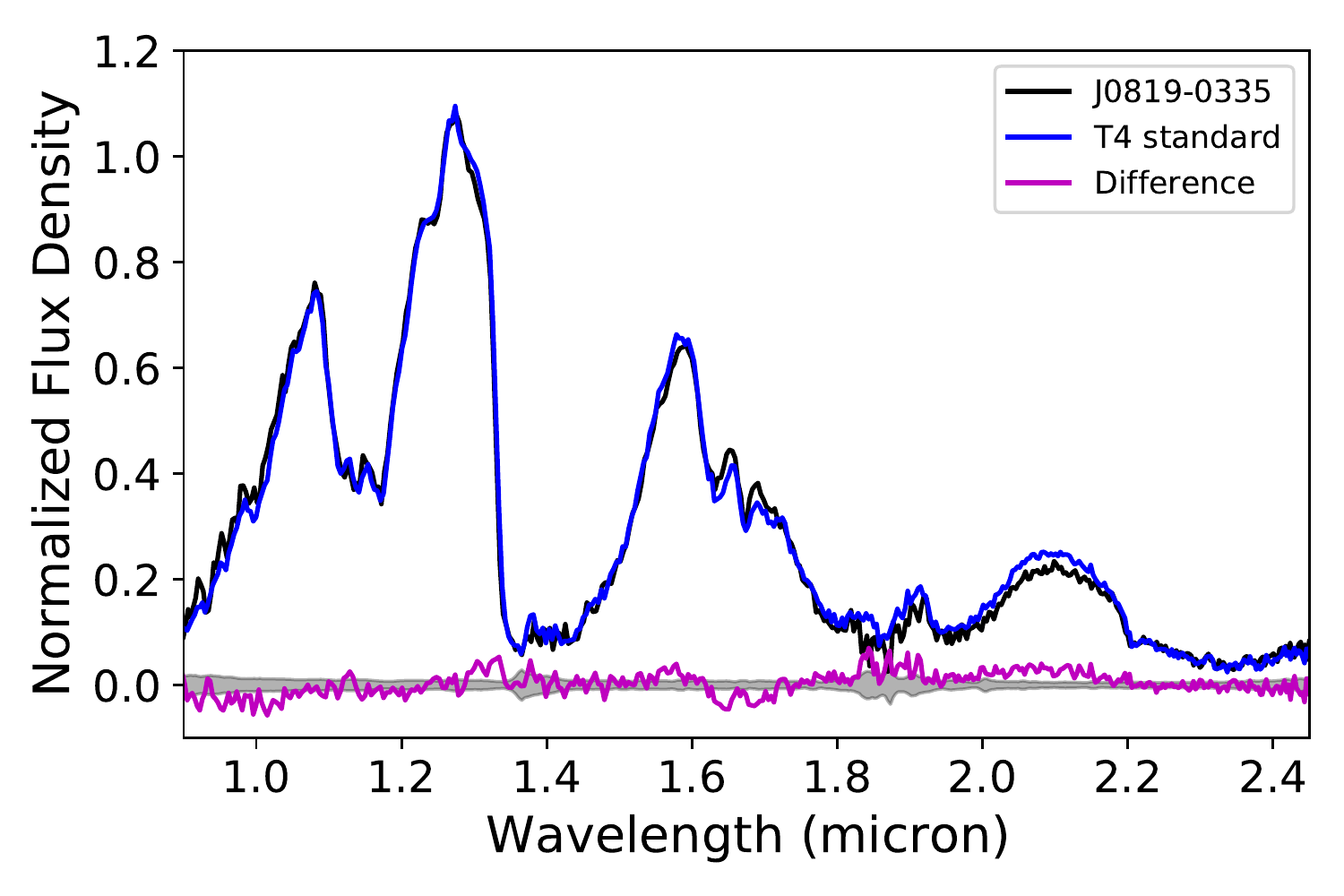}
\caption{Low-resolution near-infrared spectrum of the T dwarf 2MASS J0819$-$0335 (black lines; uncertainty in grey shading centered at zero flux) from \citet{Burgasser:2004ab} compared to (blue lines) the T4 near-infrared spectral standard 2MASSI~J2254188+312349 (data from \citealt{Kirkpatrick:2011aa}). The spectra are normalized to align in the 0.90--1.25~$\micron$ region, and the magenta lines show the difference spectra (template minus J0819$-$0335).}
\label{fig:J0819lowres}
\end{figure}

\textit{2MASS J08195820$-$0335266} \citep{Kirkpatrick:2011aa} is a T4 dwarf with spatial and kinematic evidence of membership in the 24~Myr-old $\beta$ Pictoris moving group (see Section \ref{clustermembership})\added{, consistent with \citet{Zhang:2021aa} on the basis of 5D kinematics}. 
At this age, and assuming $T_\mathrm{eff}$ = 1100~K based on \citet{Filippazzo:2015aa}, evolutionary models predict a mass of only 7 Jupiter masses, well below the deuterium burning mass limit
\citep{Baraffe:2003aa}.
However, spectral evidence of youth for this \replaced{souce}{source} is not clear.
Its near-infrared low-resolution spectrum is fully consistent with the T4 standard 2MASSI J2254188+312349 \citep[Figure \ref{fig:J0819lowres};][]{Burgasser:2004ab}.
\cite{Pineda:2016aa} \replaced{measured}{reported} the red optical (0.7--1.0~$\micron$) spectrum of this source, recommending it as the T4 optical standard with features that naturally transition between T2 and T5 optical standards \citep{Burgasser:2003aa}. They found no evidence of H$\alpha$ emission to a limit of $\log_{10}L_\mathrm{H\alpha}/L_\text{bol} < -5.7$. 
The lack of activity is consistent with its small $v\sin{i}$, for which we are only able to determine an upper limit of $v\sin{i}$ $<$ 9 km s$^{-1}$.
\cite{Heinze:2015aa} report the possible detection of variability ($>3.6\%$ in amplitude) at red optical wavelengths (0.7--0.95~$\micron$), which is common for both young brown dwarfs and objects spanning the L dwarf/T dwarf transition \citep{2012ApJ...750..105R,Metchev:2015aa}. Taken together, we suspect that this source is a field brown dwarf with a chance kinematic alignment with the $\beta$ Pictoris moving group.

\textit{2MASSI J0937347+293142} \citep{Burgasser:2002aa} is a peculiar T6 dwarf with the second largest rotational velocity in our sample, $v \sin {i}$ = 66.9$^{+0.5}_{-0.6}$ km s$^{-1}$. Assuming a radius of 1~R$_\mathrm{Jup}$, this rotational speed corresponds to a maximum rotation period of 2~hours.
Like J0034+0523, J0937+2931 has an unusually blue near-infrared color $J-K = -1.10 \pm 0.06$ \citep{2010ApJ...710.1627L}.
This and other spectral peculiarities, including the weak or absent K~\textsc{i} lines at 1.25~$\micron$, have been cited as evidence of high surface gravity and/or subsolar metallicity for this source \citep{Burgasser:2002aa, Burgasser:2006ab, McLean:2007aa, Prato:2015aa, Martin:2017aa, Zhang:2019aa}. The rapid rotation of J0937+2931 may also be partly responsible for the weakened atomic features.
Again, these physical traits imply an old age, indicating that J0937+2931 is a massive and compact brown dwarf that has not had appreciable angular momentum loss in its late evolution.

\begin{figure*}[!htbp]
\includegraphics[width=1.0\textwidth]{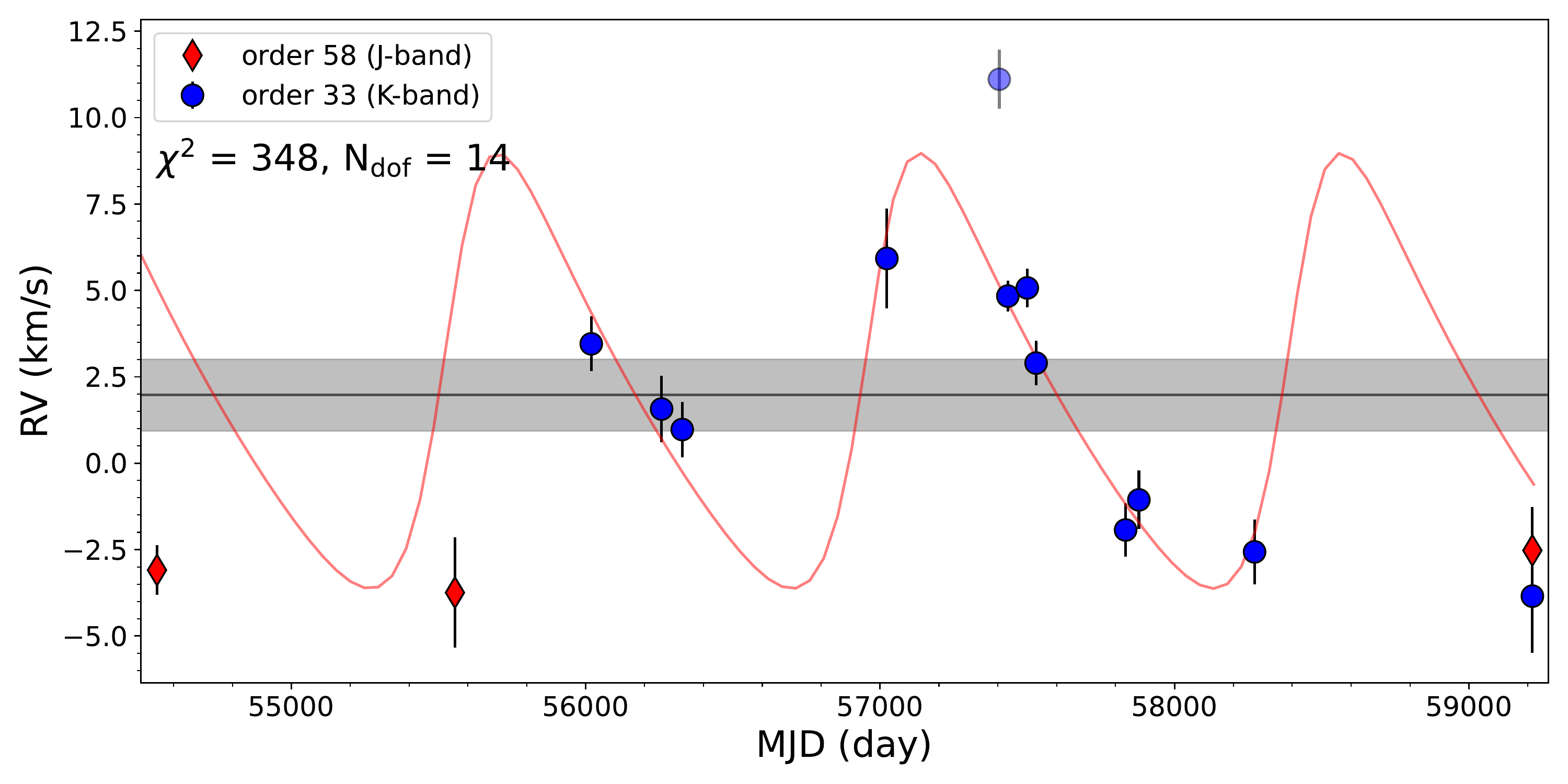}
\caption{RV time series for all of the NIRSPEC measurement epochs of 2MASS J1106+2754. Measurements of orders 33 ($K$-band) and 58 ($J$-band) are labeled as blue \added{circles} and red \added{diamonds}, respectively.
The horizontal line is the weighted average of our measurements, with the uncertainty shaded in grey. 
Also shown in a best-fit RV orbital curve with a semi-amplitude of $K_1$ = 6.30\deleted{$^{+0.05}_{-0.05}$} km s$^{-1}$ and a period of $P \replaced{\sim}{=}$ \replaced{4}{3.92}~yr, based on \replaced{ten}{eleven} epochs in order 33 with a removal of a bad observation on 2016 Jan 18 (light blue) \added{and 2018 Jan 1 (outside of the range presented here)}. The data in order 58 were not fit because the secondary component is \replaced{the dominant flux in}{brighter at} $J$-band. See Section \ref{sec:individual_source} for details.
\label{fig:figJ1106variation}}
\end{figure*}

\textit{2MASS J11061197+2754225} \citep{Looper:2007aa} is a previously reported candidate binary system with hypothesized T0.0+T4.5 components based on analysis of its low-resolution near-infrared spectrum \citep{Looper:2007aa, Burgasser:2010aa, Bardalez-Gagliuffi:2014aa}. It is also highly overluminous
\citep{Manjavacas:2013aa}, but
has not been resolved by direct imaging \citep{Looper:2008aa}. Our measurements show highly significant RV variations over \replaced{14}{15} epochs spanning 2008 Mar 19 to 2021 Jan 1, deviating from a constant velocity model by $\chi^2=318$ (DOF = 15, p-value $<$ 0.001; Figure \ref{fig:figJ1106variation}).
\replaced{We performed a preliminary orbit fit to the RV data using the package \textit{RadVel} \citep{Fulton:2018aa}, and converged on a set of solutions with primary semi-amplitude $K_1 = 6.30\pm0.05$ km s$^{-1}$, period $P = 3.92^{+0.07}_{-0.09}$~yr, eccentricity $e = 0.33^{+0.04}_{-0.02}$, and center of mass radial velocity V$_\text{COM} = 2.0\pm0.1$ km s$^{-1}$. Only eleven epochs of order 33 data were selected to fit. The data of order 58 were not used for the fit due to the secondary component contributing significantly to the $J$-band flux of the combined light system (see \citealt{Burgasser:2010aa}). We also excluded data from 2016 Jan 18 and 2018 Jan 1 in the fit due to the slit width $0\farcs432$ being wider than the unusually small seeing on these nights. This system is examined in further detail in a companion paper (Burgasser et al., in prep).}
{We performed a preliminary orbit fit on 11 epochs of these data\added{ ($\chi^2=155$, p-value $<$ 0.001)}, rejecting order 33 data from 2016 Jan 18 and 2018 Jan 1 due to the slit width $0\farcs432$ being wider than the seeing on these nights; and order 58 data on 2008 Mar 19 and 2010 Dec 26 which are likely contaminated by light from the secondary, which is brighter than the primary in the $J$-band (see \citealt{Burgasser:2010aa}). 
We used the package \textit{RadVel} \citep{Fulton:2018aa}, and converged on a set of solutions with primary semi-amplitude $K_1 = 6.30\pm0.05$ km s$^{-1}$, period $P = 3.92^{+0.07}_{-0.09}$~yr, eccentricity $e = 0.33^{+0.04}_{-0.02}$, and center of mass radial velocity V$_\text{COM} = 2.0\pm0.1$ km s$^{-1}$. This system is examined in further detail in a companion paper (Burgasser et al., in prep).}

\begin{figure}[!htbp]
\centering
\includegraphics[width=\linewidth]{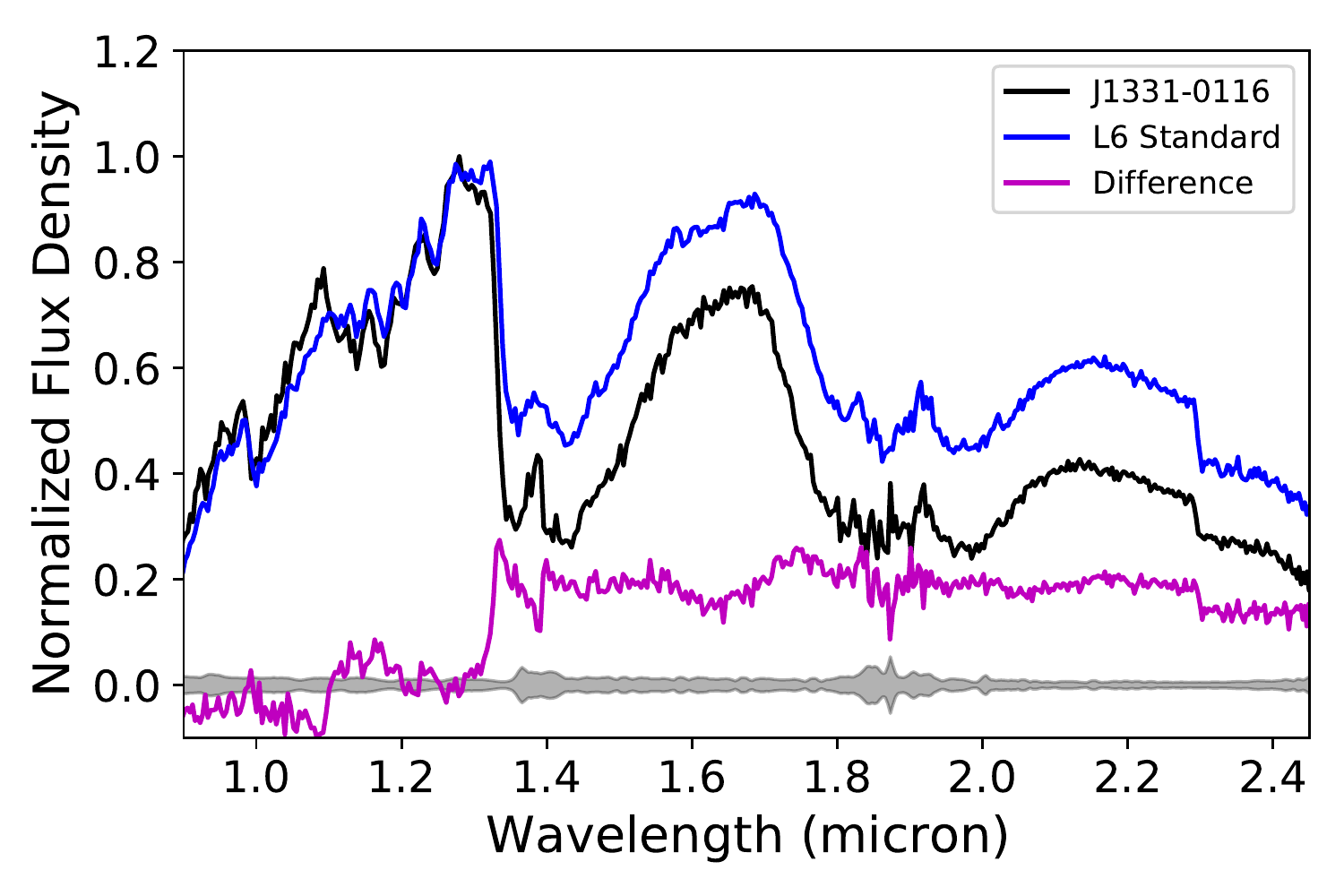}
\includegraphics[width=\linewidth]{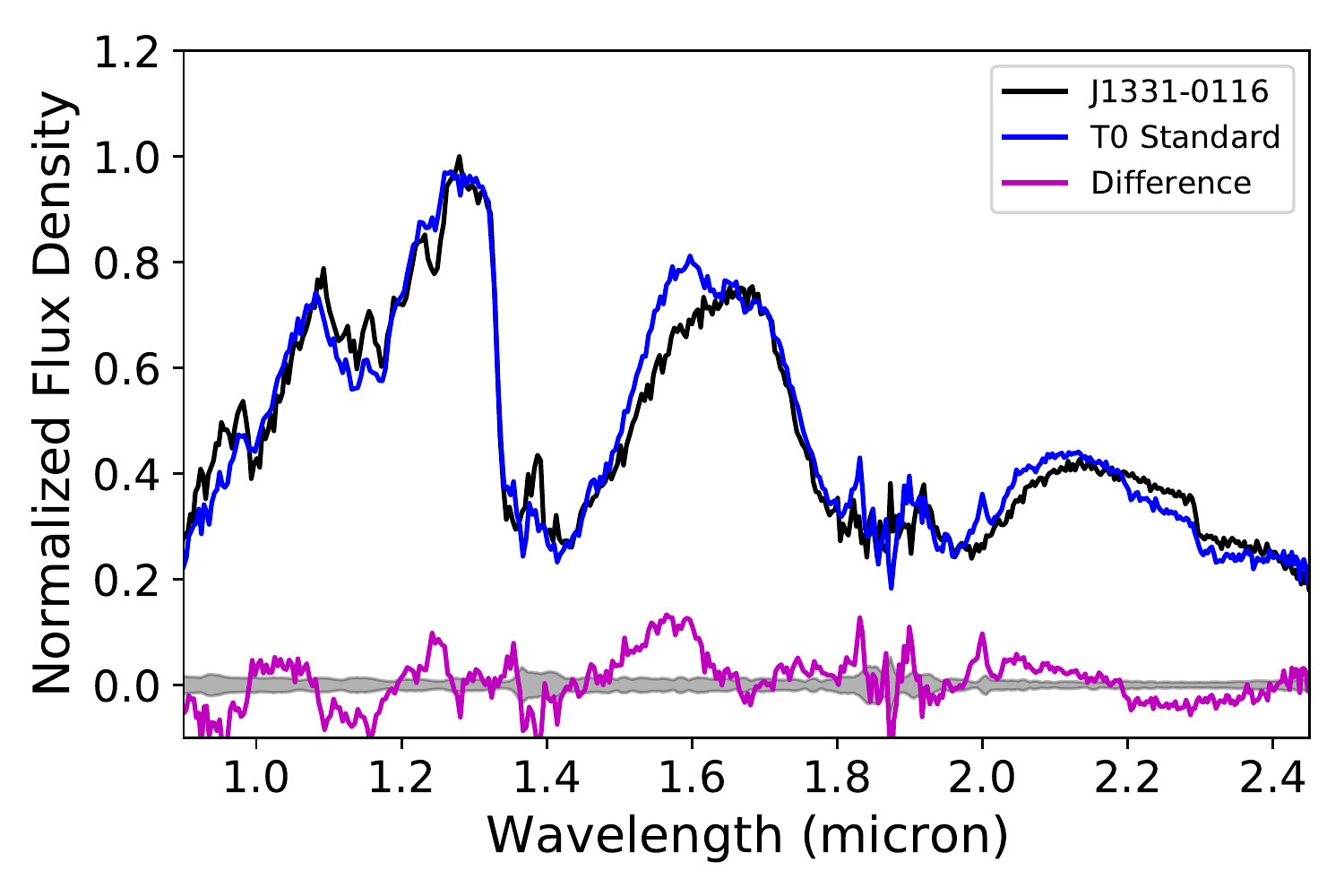}
\includegraphics[width=\linewidth]{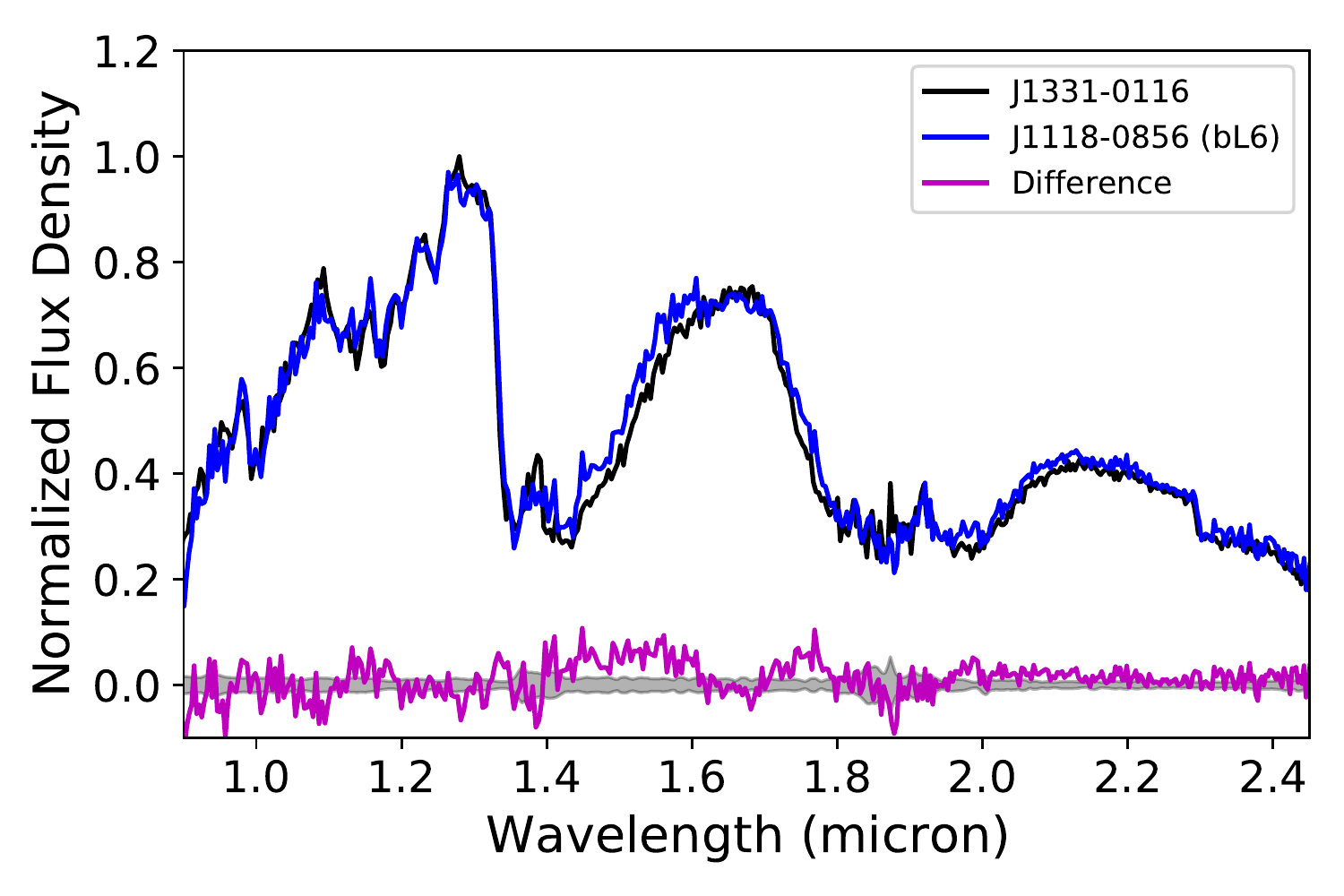}
\caption{Low-resolution near-infrared spectrum of the peculiar L dwarf J1331$-$0116 (black lines; uncertainty in grey shading centered at zero flux) from \citet{Bardalez-Gagliuffi:2014aa} compared to (blue lines) the L6 near-infrared spectral standard 2MASSI~J1010148$-$040649 (top; data from \citealt{Reid:2006aa}), the T0 spectral standard SDSS~J120747.17+024424.8 (middle; data from \citealt{Looper:2007aa}), and the unusually blue L6 dwarf 2MASS~J11181292$-$0856106 (bottom; data from \citealt{Kirkpatrick:2010aa}). All spectra are normalized to align in the 0.90--1.25~$\micron$ region, and the magenta lines show the difference spectra (template minus J1331$-$0116).
\label{fig:j1331spectrum}}
\end{figure}

\textit{SDSS J133148.92$-$011651.4} \citep{Hawley:2002aa} is the only source in our sample identified as an intermediate member in the Galactic thin and thick disk populations.
Like J0937+2931, this source has an unusually blue near-infrared color for an early-T dwarf ($J-K = 1.279\pm0.008$; \citealt{2007MNRAS.379.1599L}) and a peculiar spectrum that has challenged classification. Optical spectral classifications have ranged from L1 pec \citep{Marocco:2013aa} to L6 \citep{Hawley:2002aa} while near-infrared spectral classifications have ranged from L6 \citep{Bardalez-Gagliuffi:2014aa} to T0 \citep{Schneider:2014aa}.
There are also conflicting determinations of this source being either metal-poor \citep{Marocco:2013aa} or lacking in subdwarf spectral features \citep{Kirkpatrick:2016aa}, and model fits indicate unusually thin clouds for a late L dwarf \citep{Stephens:2009aa}. Figure~\ref{fig:j1331spectrum} shows the low-resolution near-infrared spectrum of this source from \citet{Bardalez-Gagliuffi:2014aa} compared to L6 and T0 spectral standards, and the near-infrared spectrum of the L subdwarf 2MASS~J11181292$-$0856106 \citep{Kirkpatrick:2010aa}. 
The L subdwarf is the best match of the three, meaning that J1331$-$0116 is likely an old, slightly metal-poor L dwarf whose spectrum is shaped by enhanced H$_2$ CIA and possibly atmospheric condensates. Given its distinct classification from the rest of the T sample, we exclude this source from the T dwarf kinematic analysis presented below.

\begin{figure}[!htbp]
\includegraphics[width=\linewidth]{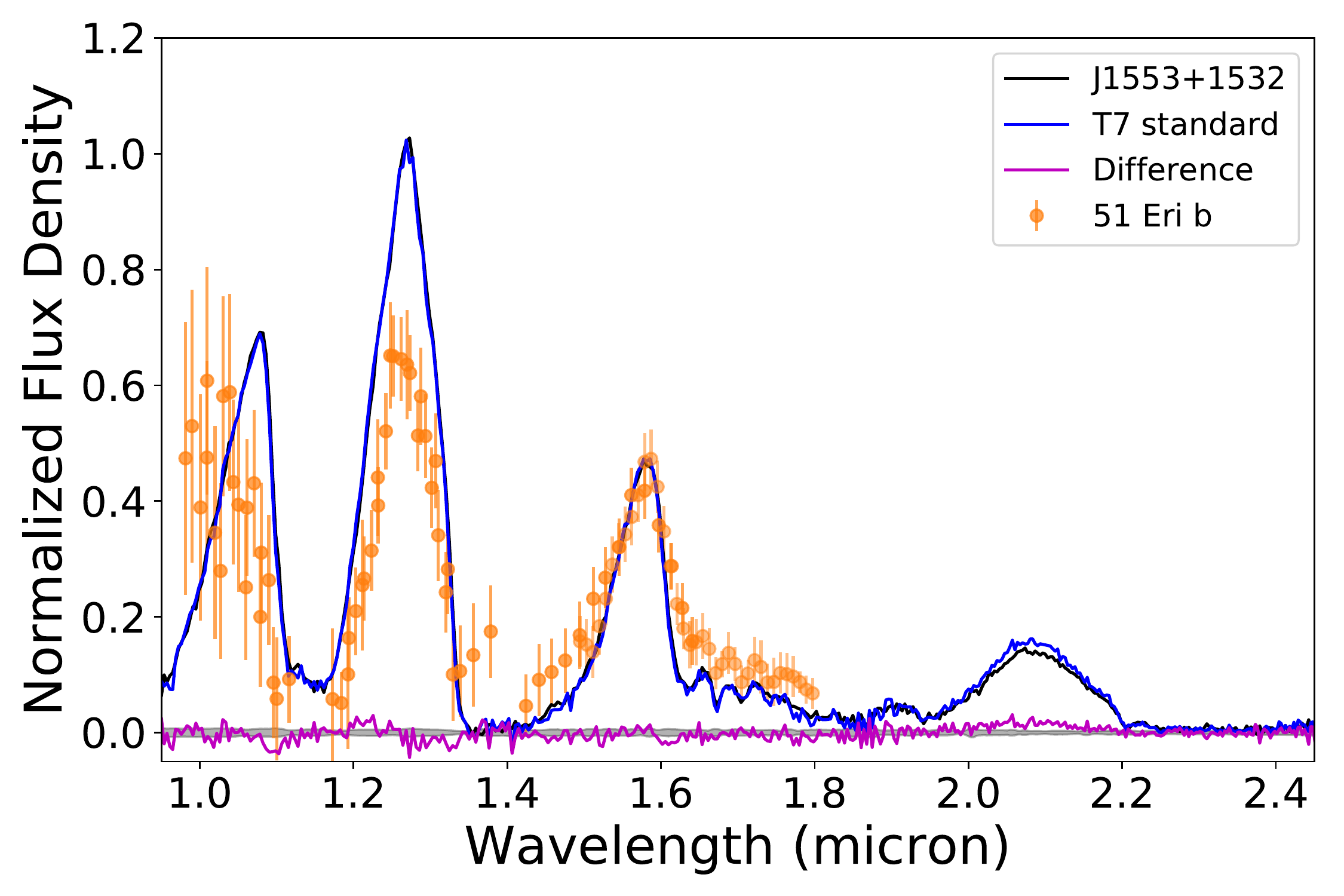}
\caption{Low-resolution near-infrared spectrum of the T dwarf 2MASS J1553+1532 (black line; uncertainty in grey shading centered at zero flux) from \citet{Burgasser:2010aa} compared to \replaced{(blue lines) the T7 near-infrared spectral standard 2MASS J07271824+1710012 (data from \citealt{Burgasser:2006ab})}{the T7 near-infrared spectral standard 2MASS J07271824+1710012 (blue line; data from \citealt{Burgasser:2006ab})}. The spectra are normalized to align in the 0.90--1.25~$\micron$ region, and the magenta \replaced{lines show the difference spectra}{line shows the difference spectrum} (template minus J1553+1532). Also shown is the 51 Eri b spectrum from \citet{Samland:2017aa}, normalized to $H$ band peak.}
\label{fig:J1553lowres}
\end{figure}

\textit{2MASS J15530228+1532369AB} \citep{Burgasser:2002aa} is a resolved T6.5+T7.5 binary with a separation of 0$\farcs$349 $\pm$ 0$\farcs$005, $\Delta M_\mathrm{bol}$ = 0.31 $\pm$ 0.12, and mass ratio $q$ = 0.90 $\pm$ 0.02. Here, we find spatial and kinematic evidence of membership in the $\sim$200~Myr-old Carina-Near moving group (see Section \ref{clustermembership})\added{, consistent with \citet{Zhang:2021aa} on the basis of 5D kinematics}. 
At this age, and assuming component T$_\mathrm{eff}$s = 750~K \replaced{adn}{and} 890~K based on \citet{Filippazzo:2015aa}, evolutionary models predict masses of \replaced{9}{10$^{+0.8}_{-1.6}$} and 11\added{.3$^{+0.4}_{-0.5}$} Jupiter masses, below the deuterium burning mass limit \citep{Baraffe:2003aa}. \added{The kinematically young late-type T binary could join an exclusive club of AB Doradus T3.5 GU Psc b \citep{Naud:2014aa}, AB Doradus T5.5 SDSS J111010.01+011613.1 \citep{Gagne:2015ab}, and candidate AB Doradus L+T binary WISE J135501.90$-$825838.9 \citep{Bardalez-Gagliuffi:2018aa}, and TW Hydrae L7 binary 2MASS~J11193254$-$1137466 \citep{Best:2017aa}.}
However, spectral evidence of youth for this source is lacking. Its near-infrared low-resolution spectrum is fully consistent with \added{the }T7 standard 2MASS J07271824+1710012 (Figure \ref{fig:J1553lowres}; \citealt{Burgasser:2010aa}) and inconsistent with the \replaced{young}{20~Myr} \replaced{T7}{T} dwarf 51 Eri b\added{\footnote{The spectral features might not be the same between 20~Myr and 200~Myr T7 dwarf, as L dwarfs have different spectral features between intermediate and very-low gravity in \citet{Allers:2013aa}. T7 dwarfs at the ages of 20~Myr and 200~Myr correspond to $\log{g}$=3.5--4.0~dex and 4.2--4.5~dex, respectively, using \citet{Baraffe:2003aa} models.}} (\added{$T_\text{eff}$=760$\pm$20~K; }data from VLT/SPHERE and Gemini Planet Imager; \citealp{Macintosh:2015aa, Samland:2017aa}). \citet{Line:2017aa} measured a $\log{g}$ of 4.8$^{+0.1}_{-0.2}$~dex \added{for J1553+1532} after correcting for binarity, but with subsolar metallicity ($[M/\mathrm{H}]$ = $-0.19^{+0.04}_{-0.06}$~dex and supersolar $\log{\mathrm{C}/\mathrm{O}}$ = $-0.11^{+0.09}_{-0.09}$). 
This source does not exhibit significant variability in $JHK_\mathrm{S}$ or $J_\mathrm{S}$ \citep{Koen:2004aa, Wilson:2014aa}. 
Taken together, we suspect that this source is also a field brown dwarf with a chance kinematic alignment with the Carina-Near moving group.

\begin{figure*}[!htbp]
\includegraphics[width=1.0\textwidth, trim=20 0 0 0]{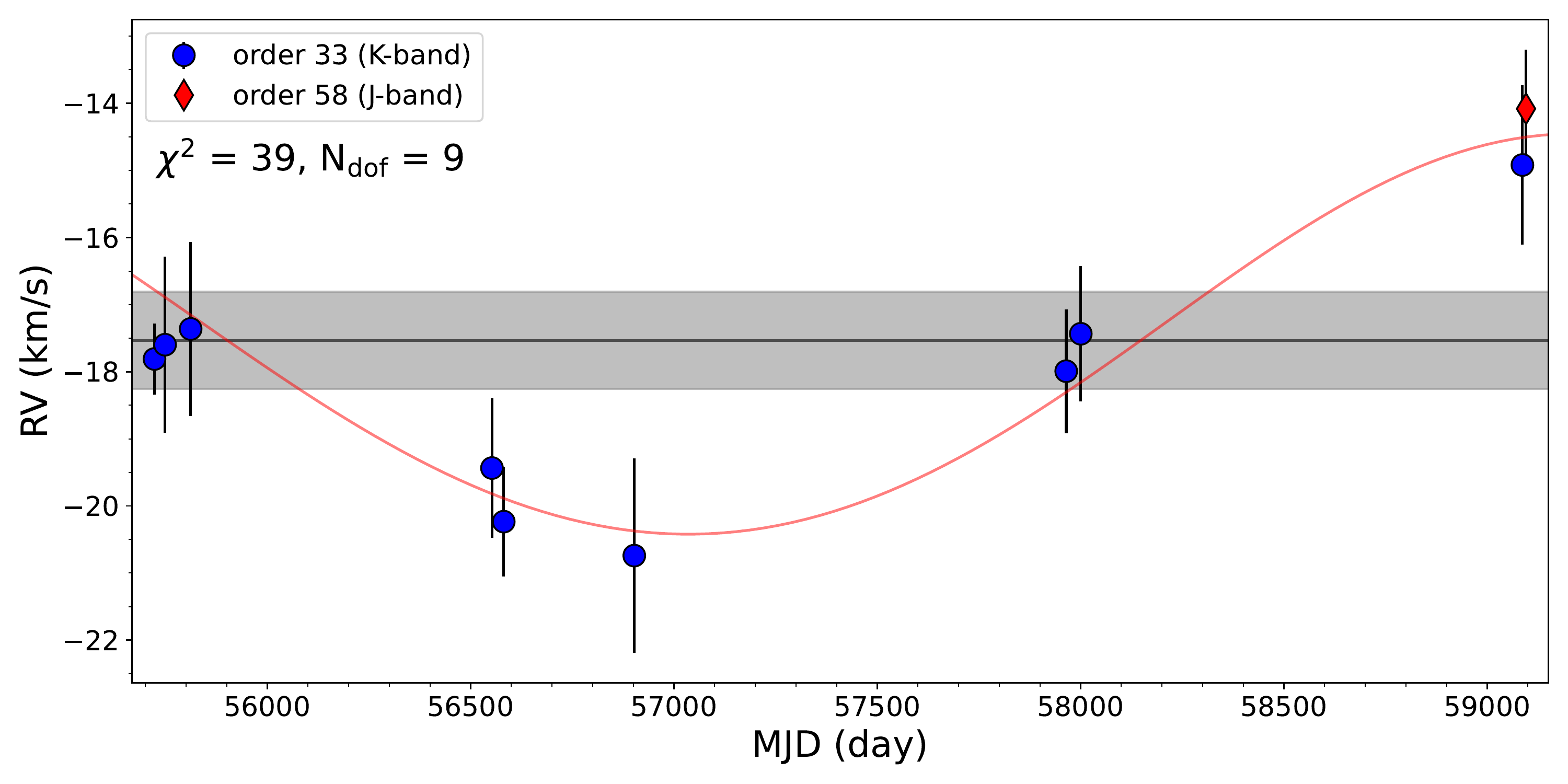}
\caption{RV time series for all of the NIRSPEC measurement epochs of 2MASS J2126+7617. Measurements of orders 33 ($K$-band) and 57 ($J$-band) are labeled as \replaced{orange}{blue circles} and \replaced{light blue}{red diamonds}, respectively. 
The horizontal line is the weighted average of our measurements, with the uncertainty shaded in grey. 
Also shown is a best-fit RV orbital curve with a semi-amplitude $K_1$ = 3.0\deleted{$^{+0.7}_{-0.6}$} km s$^{-1}$ and period $P \replaced{\approx}{=}$ 12\added{.0}~yr.
\label{fig:figJ2126variation}}
\end{figure*}

\textit{2MASS J21265916+7617440} is a T0p dwarf proposed to be an L7+T3.5 blended-light binary based on medium-resolution spectroscopy \citep{Kirkpatrick:2010aa}. High-resolution imaging by \citet{Bardalez-Gagliuffi:2015aa} failed to resolve the system and constrained its angular separation to $<$106 mas or $<$1.3 au. 
Our measurements show significant RV variations over 10 epochs spanning 2011 Jun 10 to 2020 Sep 3, deviating from a constant velocity model by $\chi^2$=39 (DOF = 9, p-value $<$ 0.001; Figure \ref{fig:figJ2126variation}).
Following the analysis of J1106+2754, we performed a preliminary orbit fit to the RV data using \textit{RadVel}, and converged on a set of solutions with primary semi-amplitude $K_1 = 3.0^{+0.7}_{-0.6}$ km s$^{-1}$, period $P$ = 12.0$^{+1.5}_{-1.2}$~yr, and center of mass radial velocity V$_\text{COM} = -17.5\pm0.4$ km s$^{-1}$. 
\replaced{The}{This} system is examined in detail in a companion paper (Hsu et al., in prep).

\textit{HN Peg B} \citep{Luhman:2007aa} is a T2.5 dwarf that is a wide \replaced{(782 au) companion}{companion (782 au)} to the young ($\sim$300 Myr) G0V star HN Peg. It is an important benchmark for testing brown dwarf evolutionary models near the L dwarf/T dwarf transition \citep{2008ApJ...682.1256L}. Our best-fit models of HN Peg B for data from 2017 Jun 9 yield RV = $-19.5\pm1.3$ km s$^{-1}$, consistent with the primary star's RV = $-17.18\pm0.17$ km s$^{-1}$ \citep{Gaia-Collaboration:2018ab}. HN Peg B is also a photometrically variable source \citep{Metchev:2015aa, Zhou:2018aa, Vos:2019aa}. Our $v\sin{i} = 15\pm3$ km s$^{-1}$ combined with the period measurement from \cite{Zhou:2018aa} of $15.4\pm0.5$ hr yields a rotation axis inclination angle of 63$^\circ\pm7^\circ$ assuming a model radius of $0.108^{+0.014}_{-0.006}~R_{\odot}$ using the theoretical evolutionary models of \citet{Burrows:1997aa} and \citet{Baraffe:2003aa} \citep{Luhman:2007aa}.
This is consistent with the current hypothesis that the frequency and amplitude of variability are higher for brown dwarfs observed at intermediate-to-high viewing angles ($i > 60^{\circ}$; \citealt{Heinze:2013aa, Metchev:2015aa, Vos:2017aa}).

\section{Kinematics of Ultracool Dwarfs} \label{sec:discussion}

\subsection{Velocity Dispersions and the Kinematic Age of T Dwarfs} \label{ages}

The velocity dispersion,
\begin{equation}
\label{eq:vel_dispersion}
\sigma_\text{tot}^2 = \sigma_U^2 + \sigma_V^2 + \sigma_W^2
\end{equation}
of a stellar population evolves as the population ages, as gravitational interactions with Galactic structures perturb orbits, increasing Galactic scale heights and relative velocities over time \citep{Wielen:1977aa}. Velocity dispersions thus provide a means of measuring the ages of stellar populations, including UCDs, which depend on the mass function, star formation history, and thermal evolution of brown dwarfs \citep{Burgasser:2004aa}. 

We measured a total velocity dispersion for our T dwarfs to be $\sigma_\text{tot} = 39.0 \pm 1.0$ km s$^{-1}$. To convert this into an age, we evaluated two age-dispersion relations. The first is the exponential decay law from \citet{Wielen:1977aa} based on the $|W|$-weighted total velocity dispersion:
\begin{equation}
\label{eq:Wielen1977}
\bar{\sigma}_\text{tot}(\tau)^3 = \sigma_{\text{tot}, 0}^3 + 1.5 \, \gamma_{\nu, p} \, T_{\gamma} \, (e^{\tau/T_{\gamma}} - 1), 
\end{equation}
where $\tau$ is the statistical age in Gyr, $\sigma_{\text{tot}, 0} = 10$ km s$^{-1}$, $\gamma_{\nu, p} = 1.1 \times 10^4$ (km s$^{-1}$)$^3$ Gyr$^{-1}$, $T_{\gamma} = 5$ Gyr, and  $\bar{\sigma}_\text{tot}$ is the $|W|$-weighted total velocity dispersion in km s$^{-1}$:
\begin{equation}
\begin{split}
\bar{\sigma}_\text{tot}^2 &= \frac{ \Sigma_i |W_i| (U_i - \bar{U})^2}{\Sigma_i |W_i| } + \frac{ \Sigma_i |W_i| (V_i - \bar{V})^2}{\Sigma_i |W_i| } \\
& + 0.5 \frac{ \Sigma_i |W_i| (W_i - \bar{W})^2}{\Sigma_i |W_i| },
\end{split}
\end{equation}
We also examined the age-dispersion law from \citet{Aumer:2009aa}:
\begin{equation}
\label{eq:AumerBinney2009}
\sigma_\text{tot} = \nu_{1, 0} \Big( \frac{\tau + \tau_1}{10 \, \mbox{Gyr} + \tau_1} \Big)^{\beta},
\end{equation}
where 
$\nu_{1, 0}$ $\in$ (55.179, 57.975) km s$^{-1}$, $\tau_1$ $\in$ (0.148, 0.261) Gyr, and $\beta$ $\in$ (0.349, 0.385) are model parameters drawn from Table 2 of \citet{Aumer:2009aa}. \added{It is noted that \cite{Bird:2019aa} measured a very similar $\beta$ = 0.389 $\pm$ 0.018 using \textit{Gaia} and APOGEE red clump stars.}

Kinematic ages were inferred by inverting these relations to solve for $\tau$. The age uncertainty was estimated statistically through Monte-Carlo sampling, assuming that \textit{UVW} velocity uncertainties follow normal distributions. To account for sample bias, we applied a Jackknife test by iteratively removing one source from our sample and recomputing dispersions and kinematic ages for each of the resulting subsamples. The resulting \textit{UVW} dispersions are summarized in Table \ref{tablekinematicage}.

Using the \citet{Aumer:2009aa} relation (averaged over all model parameters), we inferred a kinematic age of $3.5 \pm 0.3$ Gyr for our full sample of \added{36 }T dwarfs (excluding J1331$-$0116). The Wielen relation yields a smaller but statistically consistent age of $3.0 \pm 0.1$ Gyr. 
To mitigate possible velocity biases, we also examined kinematic ages for the sample excluding the resolved and candidate binaries J0629+2418, J0909+6525, J1106+2754, J1225$-$2739, J1553+1532, and J2126+7617 \citep{2006ApJS..166..585B,Dupuy:2012aa,Manjavacas:2013aa,Bardalez-Gagliuffi:2014aa}; and separately the young T dwarfs J0136+0933, J0819$-$0335, J1324+6538, J1553+1532, and HN Peg b \citep{Gagne:2017aa, Gagne:2018aa, Leggett:2008aa, Zhang:2021aa}, and found equivalent ages as the full sample (Table \ref{tablekinematicage}). 
We also found consistent kinematic ages between early-T (T0--T4; $3.3\pm0.6$~Gyr) and late-T (T5--T8; $3.6\pm0.4$~Gyr) subgroups.
The age of local T dwarfs from the \citet{Aumer:2009aa} relation is consistent with the age of local M dwarfs, but is younger than local L dwarfs \citep{Reid:2002aa, Reiners:2009aa, Blake:2010aa, Seifahrt:2010aa, Burgasser:2015ac}, as discussed in further detail below.

\begin{figure*}[!htbp]
\includegraphics[width=\textwidth, trim=10 10 0 0]{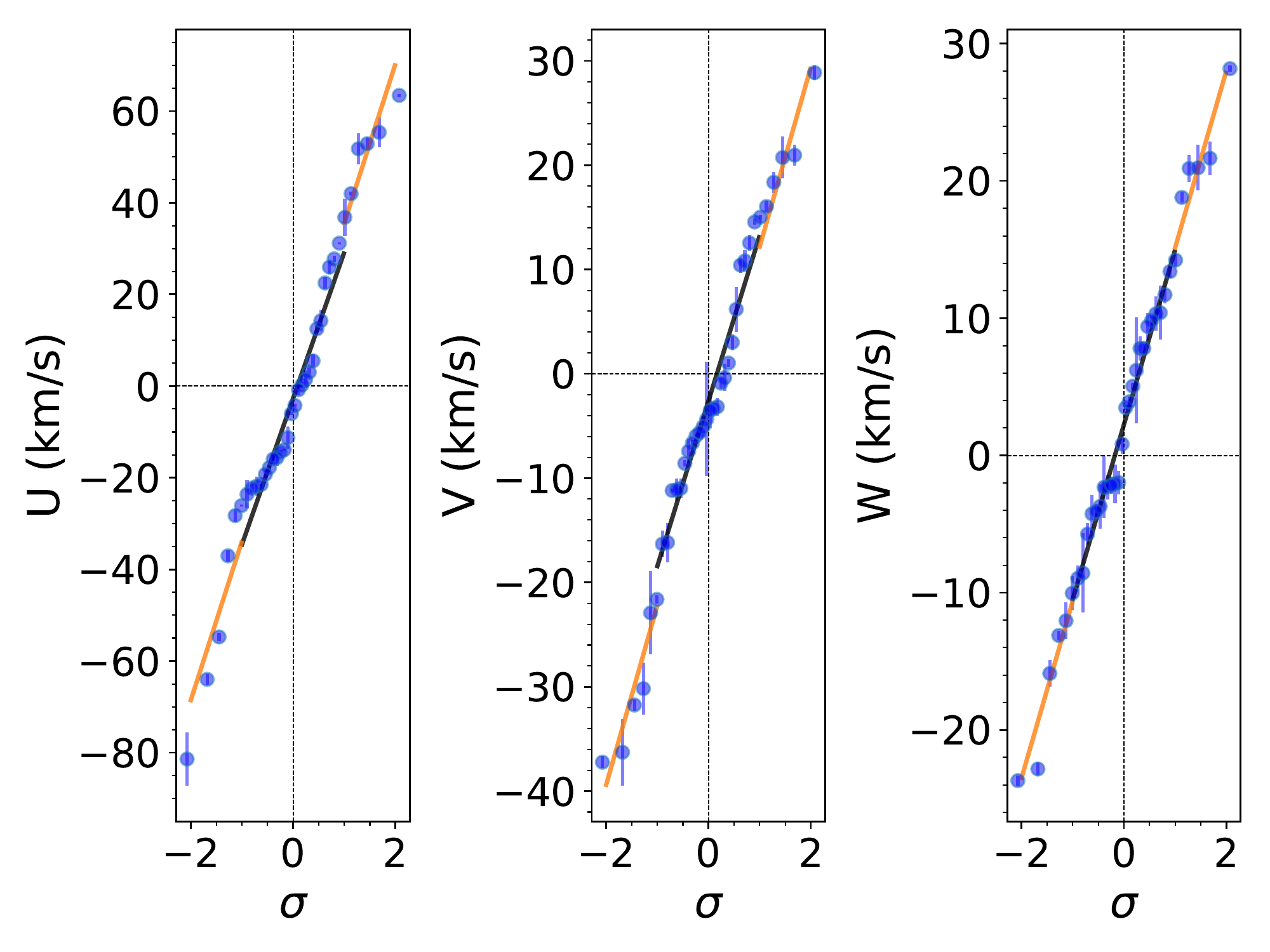}
\caption{Space velocity probability plots (probit plots) of the T dwarf sample. Individual velocities are indicated by blue circles, while a piece-wise linear fit broken at $\pm$1$\sigma$ for each velocity component (black and orange dashed lines, respectively) are shown. Note that the unusually blue L dwarf J1331$-$0116 is not included here. 
\label{fig:figvelocitydispersion}}
\end{figure*}

Figure \ref{fig:figvelocitydispersion} shows \textit{UVW} velocity probability plots, or probit plots, following \cite{Lutz:1980aa} and based on the percent point function defined in \citet{Filliben:1975aa}; see also \citet{Reid:2002aa} and \citet{Bochanski:2007aa}. Probit plots are 
linear if the velocities are drawn from a single normal distribution, with the slope equal to the distribution width ($\sigma$).
Both \textit{U} and \textit{V} velocities show two populations, with an inner ``shallow" population in the $\pm$1$\sigma$ region and a steeper ``wide" population. To quantify the difference in kinematic ages, we performed a piece-wise linear fit to each component and used the slopes to compute ages with the \citet{Aumer:2009aa} relations and Monte Carlo uncertainties \citep{Bochanski:2007aa, Burgasser:2015ac}. The ages for the inner and outer components are $3.2\pm0.5$~Gyr and $4.0\pm0.3$~Gyr, respectively, and hence marginally consistent.

\subsection{Velocity Dispersions and the Kinematic Ages of Late-M and L Dwarfs\label{ages_lateML}}

\begin{figure*}[!htbp]
\includegraphics[width=1.1\textwidth, trim=60 0 0 0]{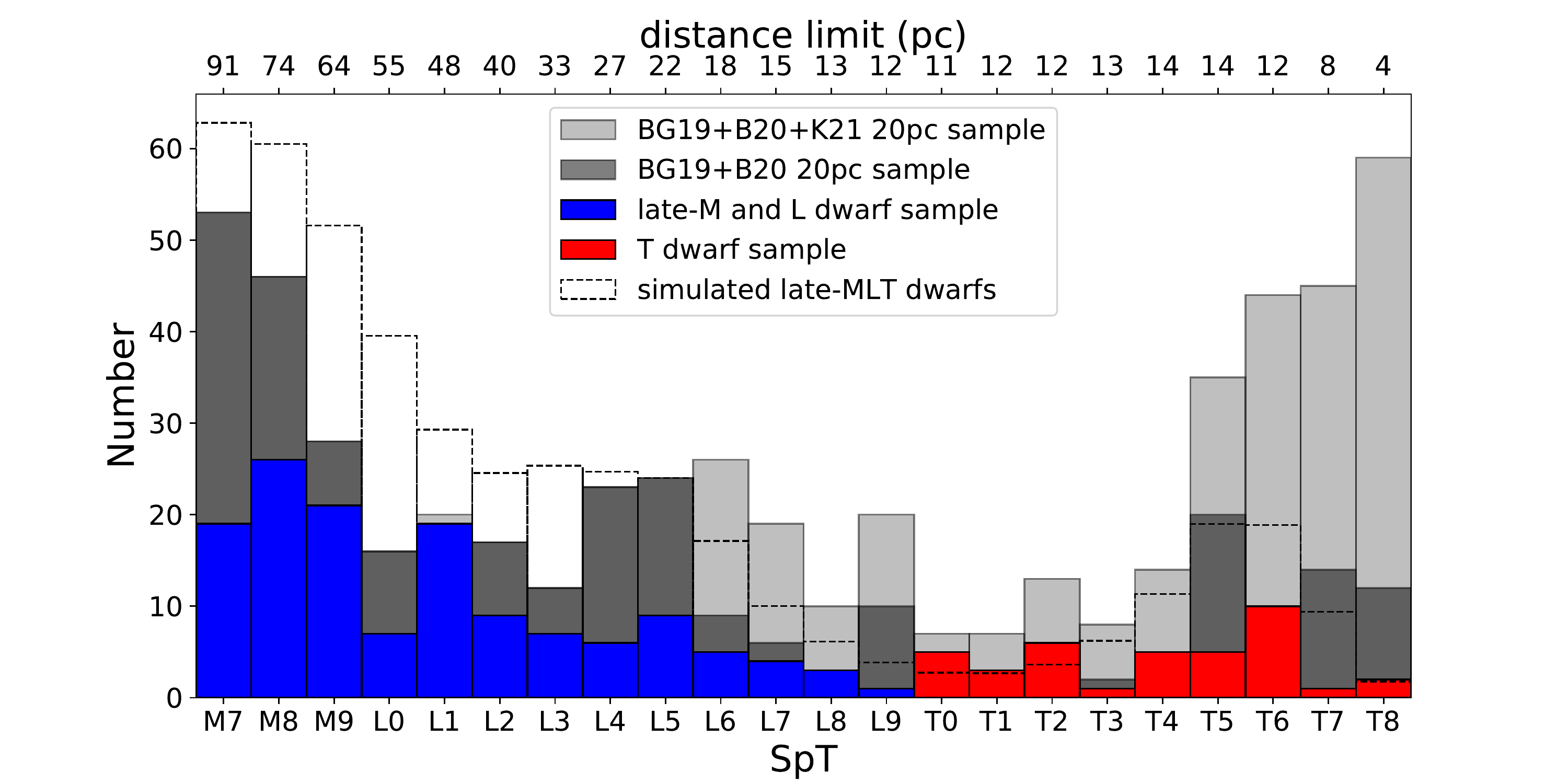}
\caption{Spectral type distribution of our 20 pc late-M and L dwarf kinematic sample with RV uncertainty of $\leq$ 3 km s$^{-1}$ (blue histogram), and our NIRSPEC T dwarf sample (red histogram). Also shown are the combined volume-limited samples of M7$-$L5 dwarfs from \citet[BG19;][]{Bardalez-Gagliuffi:2019aa} and L0$-$T8 dwarfs from \cite[B20;][]{2020AJ....159..257B} (dark grey), a 20 pc full sample from BG19, B20, and \citet[K21;][]{Kirkpatrick:2021aa} (light grey), and a simulated population (dashed histogram; Section \ref{sec:compare_simulated_ages}) normalized to agree with the observed sample at spectral type L5.
Distance limits assuming an apparent magnitude limit of $J$, $K$ $<$ 15.5 are shown along the top x-axis. 
\label{fig:SpT_lateML_sample_with_simulated_population}}
\end{figure*}

To place the T dwarf velocity dispersions and kinematic ages in context, we compiled all late-M and L dwarfs within 20 pc with published RV measurements that have uncertainties of $\leq$ 3 km s$^{-1}$, based on medium and high-resolution spectroscopic measurements (Table \ref{table:lateML_sample}). There are 65 late-M dwarfs (M7--M9) and \replaced{72}{71} L dwarfs (L0--L9) that match these criteria\added{ (J1331$-$0116 is included here; see Section~\ref{sec:individual_source} for more details)}.
Figure \ref{fig:SpT_lateML_sample_with_simulated_population} shows the spectral type distribution of the full late-M, L, and T dwarf RV sample, overplotted with a simulated local population and local 20~pc UCDs found to date (see Section \ref{sec:compare_simulated_ages} for further discussion). 
We combined RV measurements with \textit{Gaia} DR2 and eDR3 parallaxes and proper motions\added{ where available} to improve the \textit{UVW} precisions over prior studies, following the same analysis as that done for the T dwarfs.

\begin{figure*}[!htbp]
\centering
\gridline{\fig{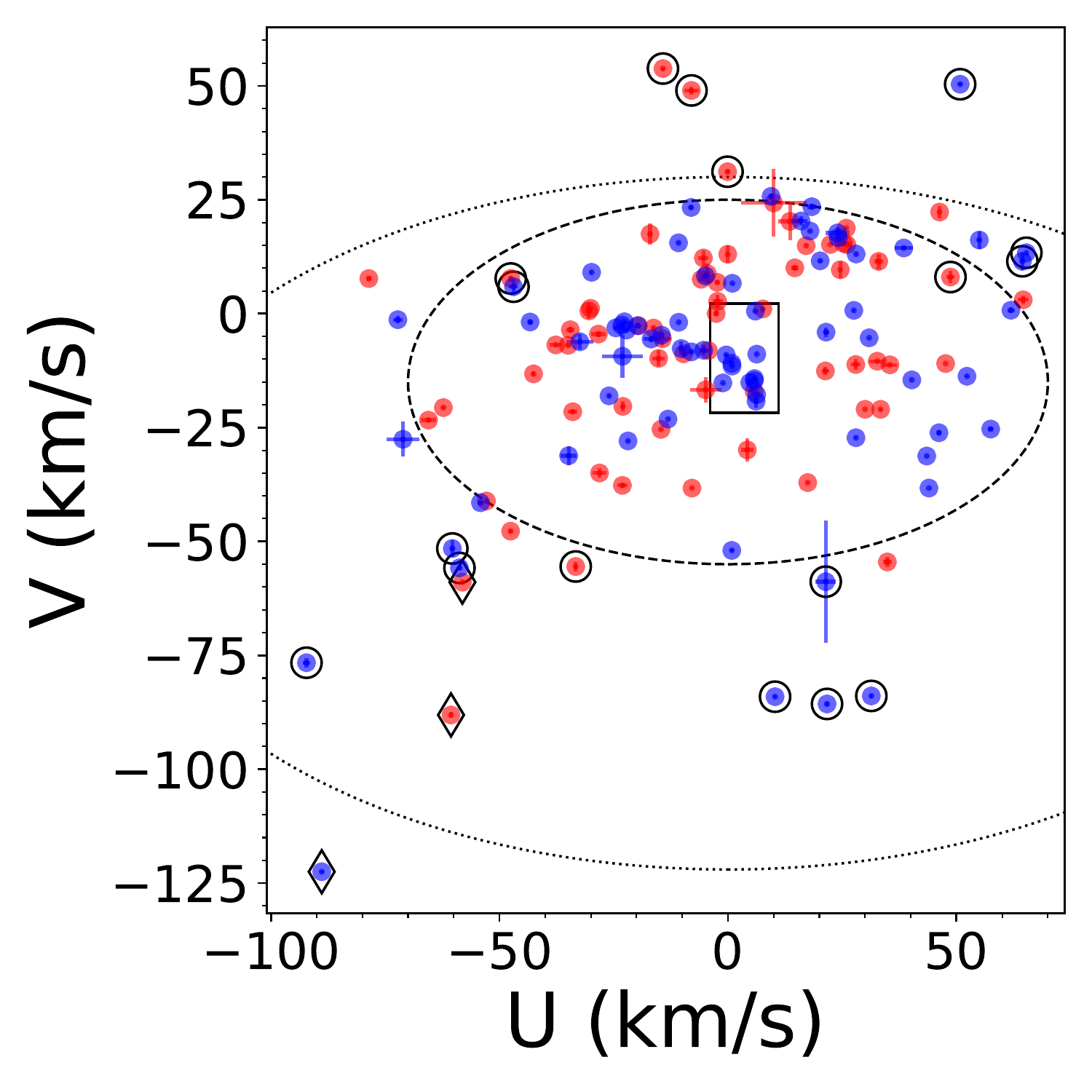}{0.45\textwidth}{}
\fig{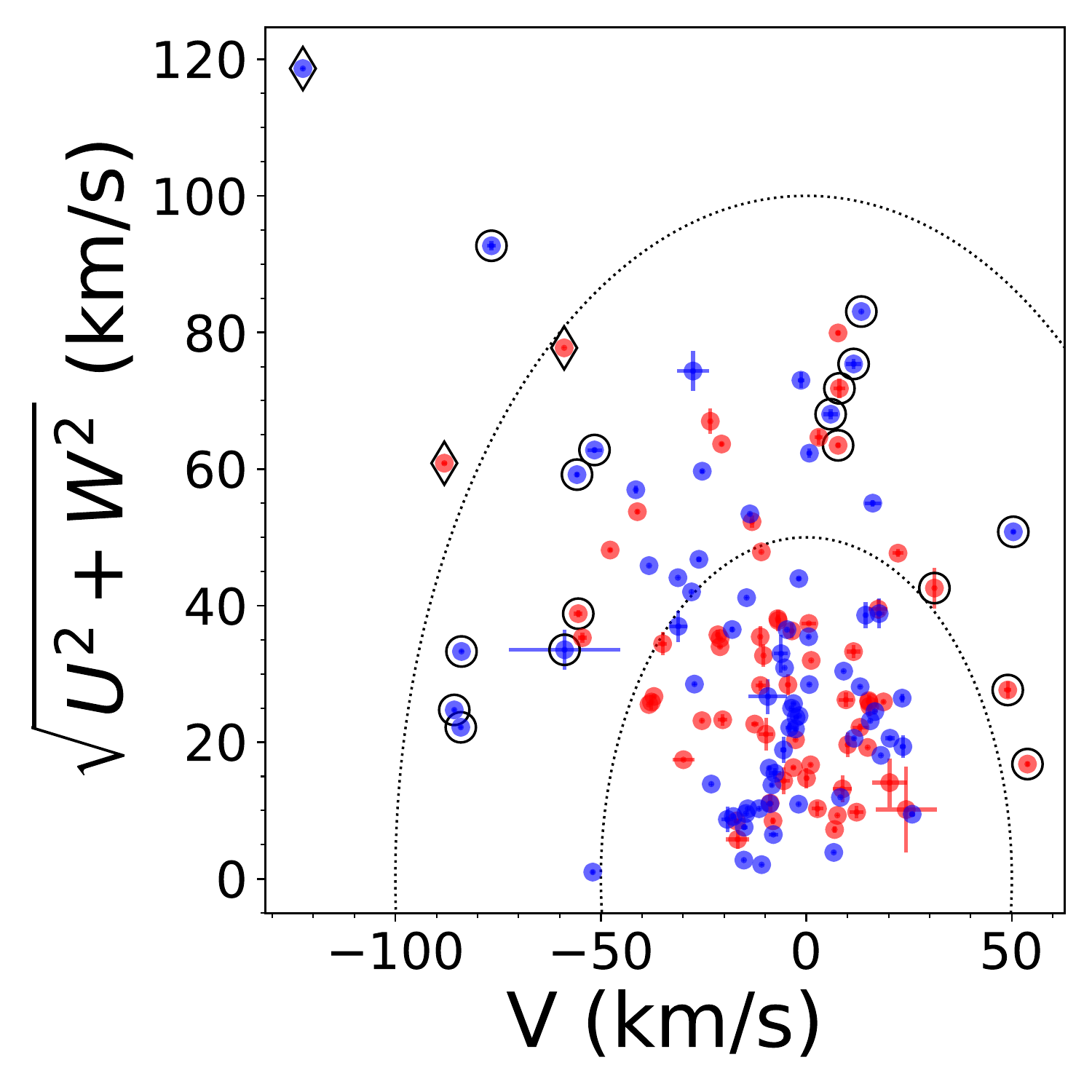}{0.45\textwidth}{}
}
\vspace{-1cm}
\gridline{\fig{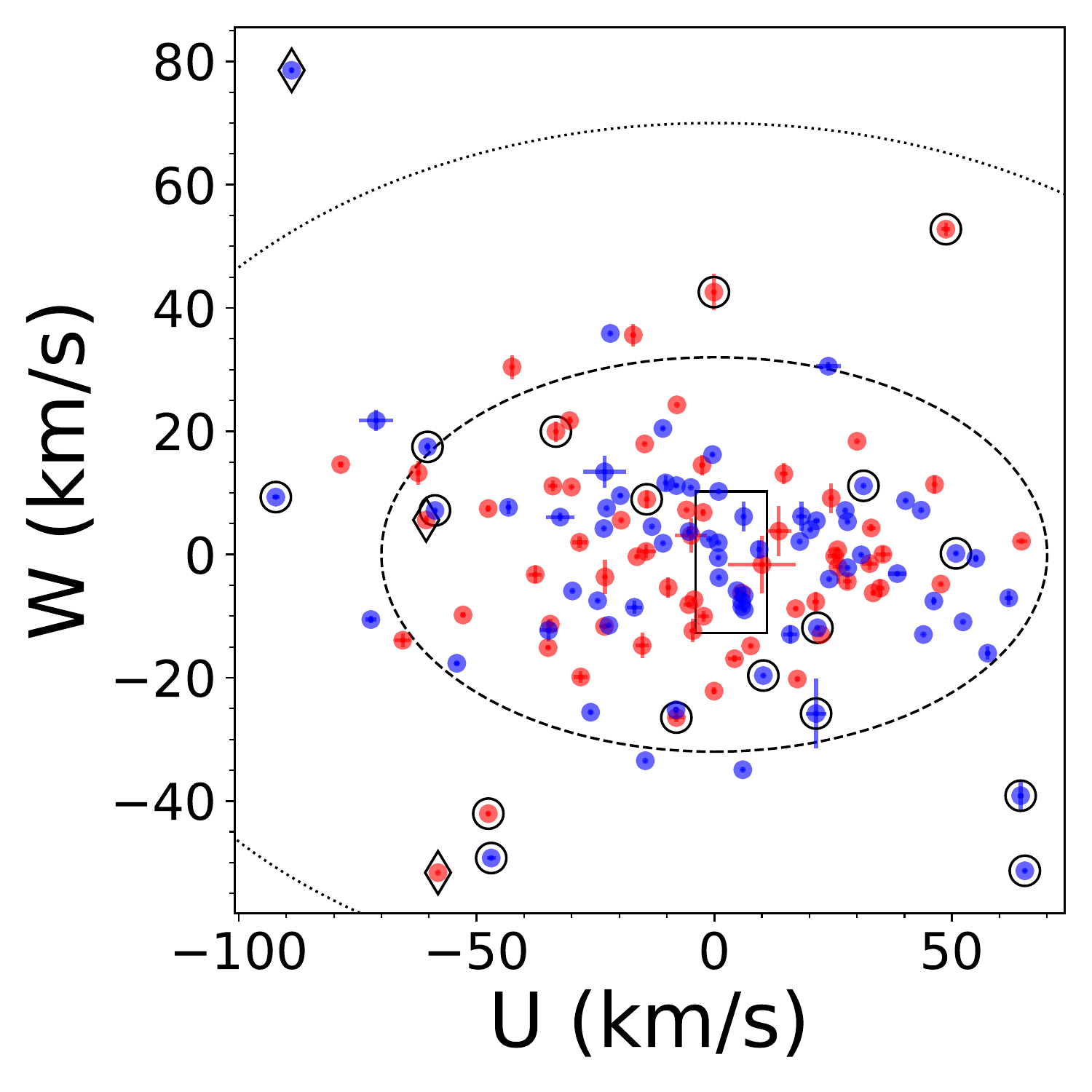}{0.45\textwidth}{}
\fig{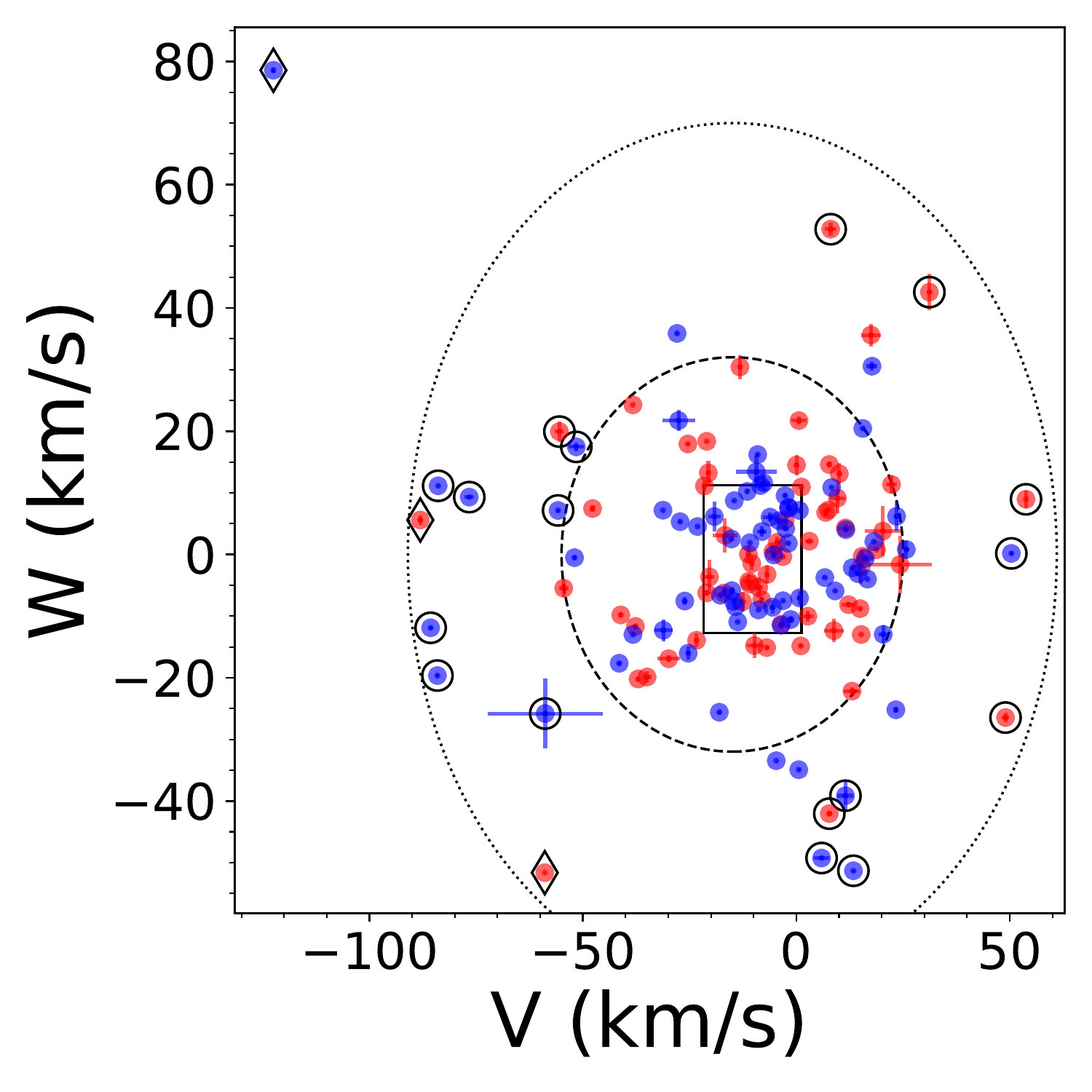}{0.45\textwidth}{}
}
\vspace{-1cm}
\caption{Same as Figure \ref{fig:figUVWvelocity} for the late-M (red) and L dwarfs (blue) in our kinematic sample. The thick disk sources ($P[\mathrm{TD}]/P[\mathrm{D}] > 10$) are highlighted as open diamonds. \label{fig:UVWvelocity_lateM_dwarf}}
\end{figure*}

Results are summarized in Table \ref{table:lateML_uvw} and Figure \ref{fig:UVWvelocity_lateM_dwarf}. 
We note that two late-M dwarfs in this sample (M7 2MASS J02530084+1652532 and M9 2MASS J03341218$-$4953322) are kinematically associated with the thick disk population, the L5.5 2MASSI J1721039+334415 is associated with the intermediate thick disk/halo population, and 17 sources (6 late-M dwarfs and 11 L dwarfs) are associated with intermediate thin/thick disk population, which includes J1331$-$0116, which we classify as a blue L6.
Evaluating the distribution of \textit{UVW} velocities, we find that the L dwarfs in this sample
exhibit significant correlations between 
\textit{UV} (\textit{R} = 0.33, p-value $<$ 0.01), 
\textit{UW} (\textit{R} = $-$\replaced{0.31}{0.32}, p-value $<$ 0.01) and 
\textit{VW} velocities (\textit{R} = $-$\replaced{0.26}{0.27}, p-value = 0.02). 
For the late-M dwarfs, we find a significant correlation between \textit{UV} velocities (\textit{R} = 0.32, p-value $<$ 0.01), but not between \textit{UW} or \textit{VW} velocities.
The \textit{UV} velocity correlation for the L dwarfs is weaker than that previously reported in \citet[$R = 0.43\pm0.03$]{Burgasser:2015ac}, but nevertheless significant. If we remove the thick disk, intermediate thick disk/halo, and intermediate thin/thick disk population members from our sample,
the \textit{UV} velocity correlation becomes less significant for the late-M dwarf sample (\textit{R} = 0.25, p-value = 0.06) and insignificant for the L dwarfs in all three velocity pairs.
This result indicates that the \textit{UVW} velocity correlations are driven by the older kinematic populations.
As noted above, the T dwarfs, which are all thin disk sources, show no significant correlations in \textit{UV}, \textit{UW}, or \textit{VW} velocity pairs. 
The average \textit{U} and \textit{W} velocities of the late-M and L dwarfs are each consistent with zero, while the negative average \textit{V} velocity (greater for the L dwarfs) can again be attributed to asymmetric drift. We note that the average \textit{U} velocity offset for the L dwarfs reported in \cite{Burgasser:2015ac} is not seen here and is likely an artifact of small sample statistics in that study.

\begin{figure*}[!htbp]
\gridline{\fig{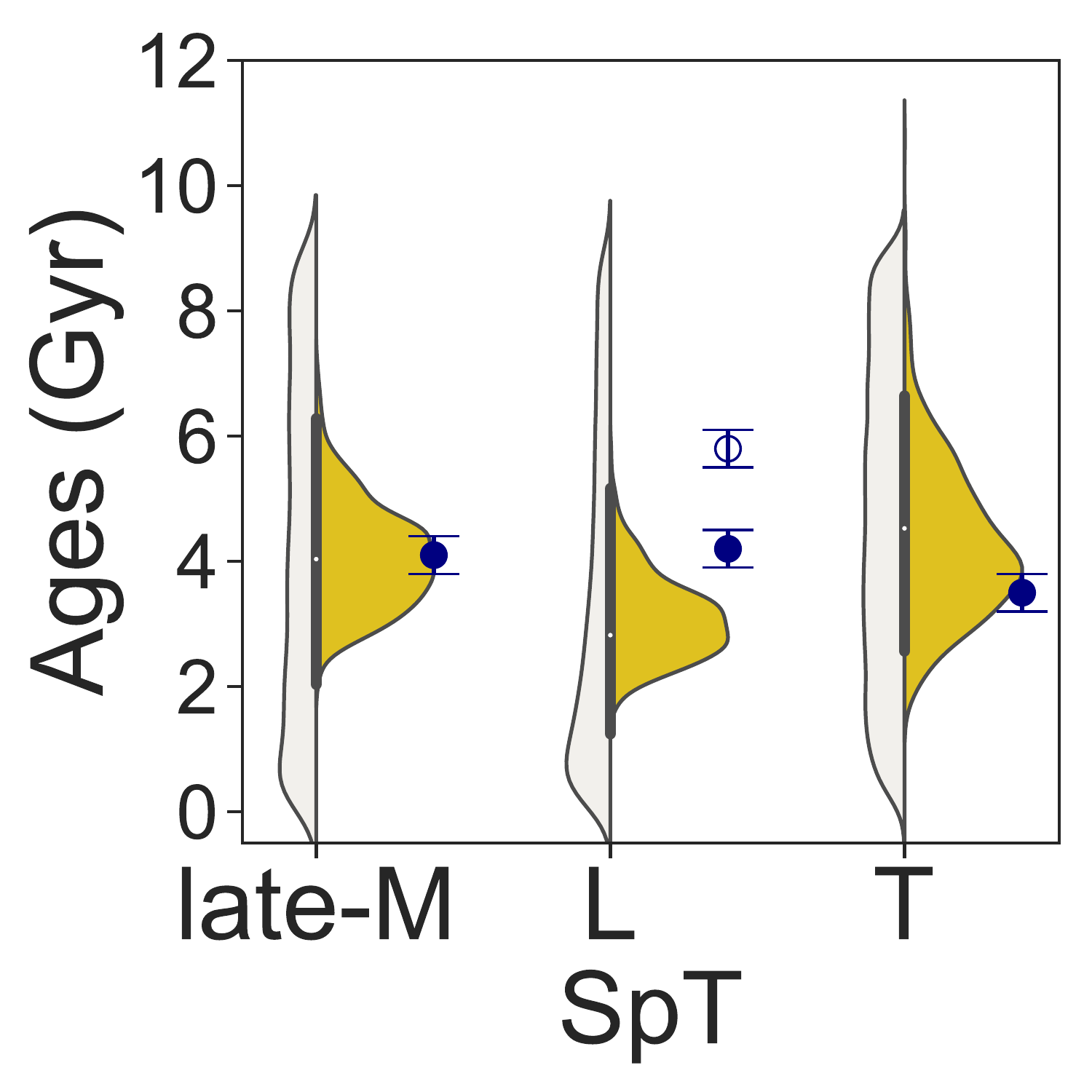}{0.5\textwidth}{}
\fig{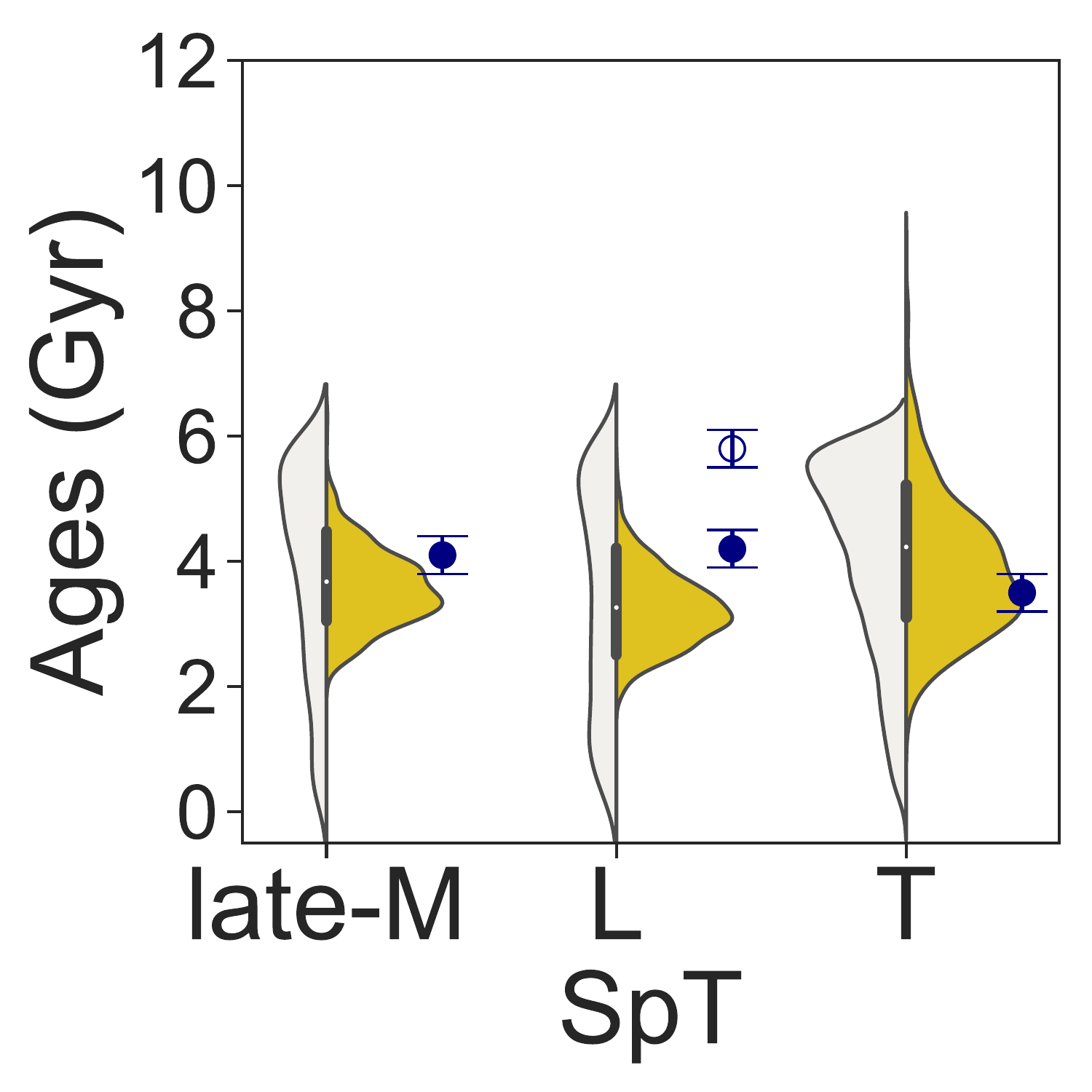}{0.5\textwidth}{}
}
\gridline{\fig{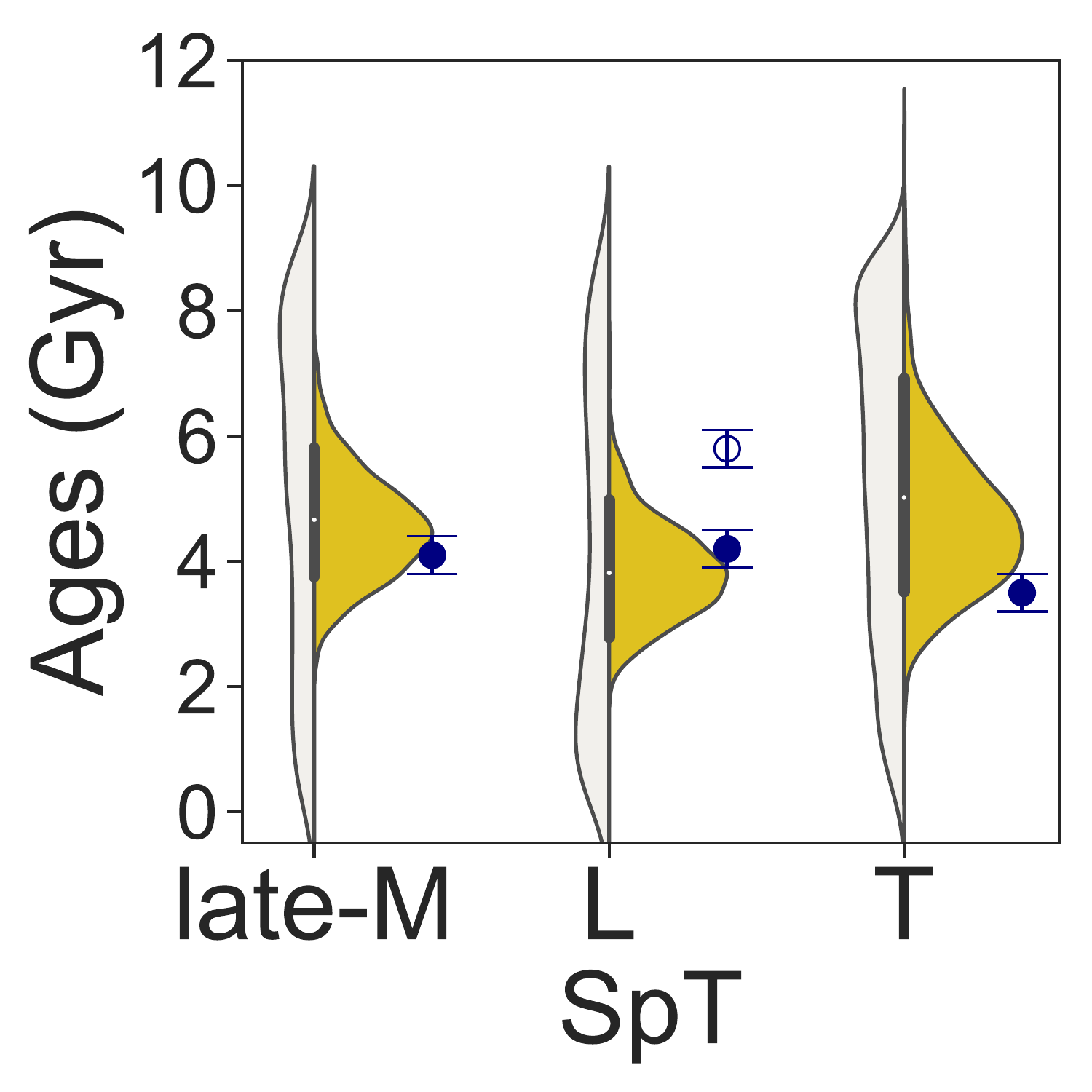}{0.5\textwidth}{}
\fig{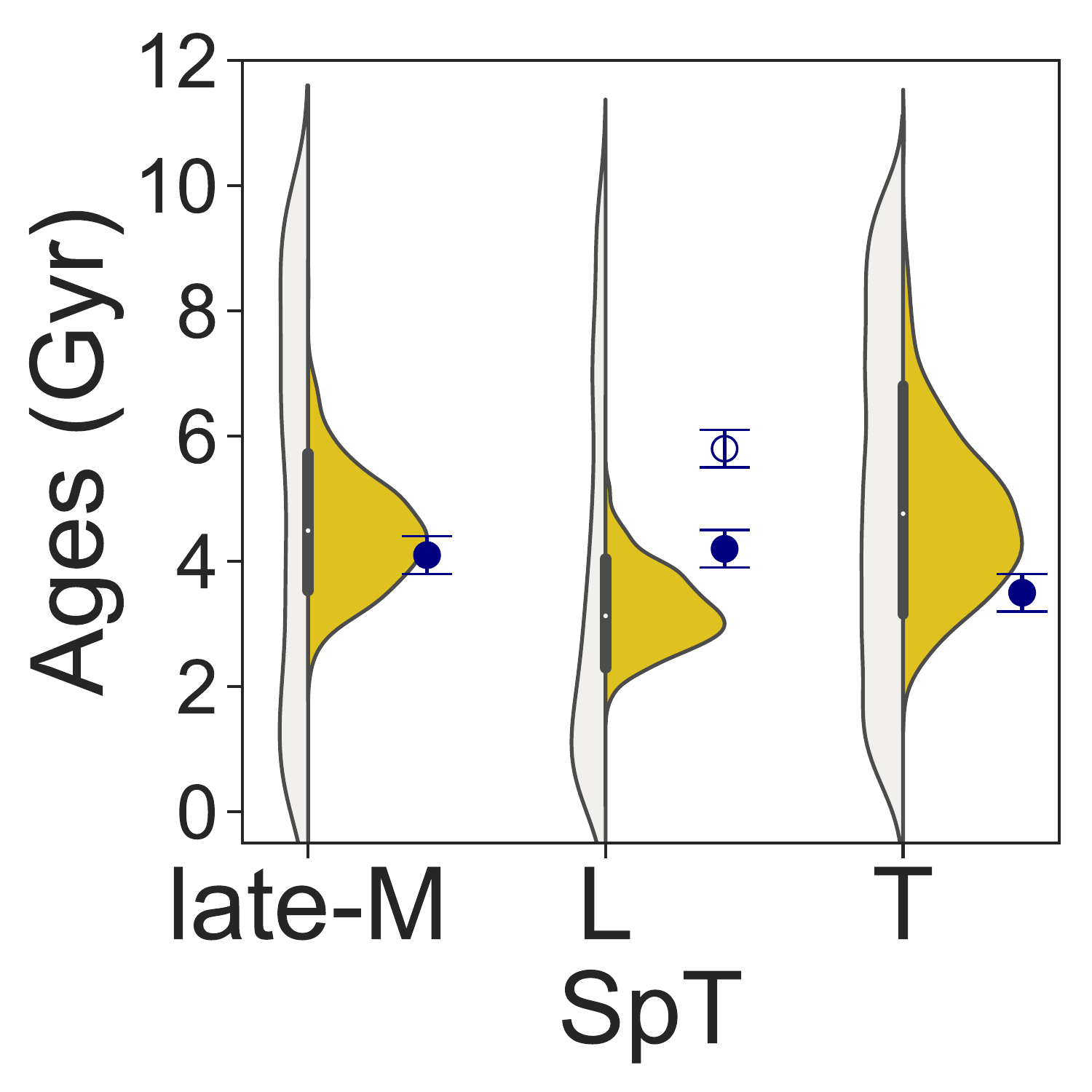}{0.5\textwidth}{}
}
\caption{Simulated age distributions (white/yellow violin plots for individual/inferred ages, respectively) and measured kinematic ages (offset blue points and sampling uncertainties indicated) for all (empty blue points) and thin disk ($P(\mathrm{TD})/P(\mathrm{D})\, \leq$ 1; solid blue point) late-M, L, and T dwarfs. 
Panels are for simulations that assume a mass function $dN/dM \propto M^{-0.5}$, \citet{Baraffe:2003aa} evolutionary models, and the following star formation rates and population ages:
\textit{Upper left}: constant birthrate over 9~Gyr (baseline simulation); 
\textit{Upper right}: \citet{Rujopakarn:2013aa} cosmic birthrate over 6~Gyr; 
\textit{Lower left}: \citet{Aumer:2009aa} exponential birthrate over 9~Gyr; 
\textit{Lower right}: constant birthrate over 12~Gyr.
\label{fig:simulated_population2}}
\end{figure*}

The corresponding kinematic ages for all of the late-M and L dwarfs in our sample using the \cite{Wielen:1977aa} and \cite{Aumer:2009aa} relations are given in Table \ref{table:lateML_kinematicage}. 
In our sample, we find highly significant correlations between $v_\text{tot}^2$ and $|W|$ for the late-M dwarfs ($N$ = 65, \textit{R} = 0.42, p-value $<$ 0.001) and L dwarfs ($N$ = \replaced{72}{71}, \textit{R} = 0.61, p-value $<$ 10$^{-4}$), and significant correlation for T dwarfs ($N$ = 37, \textit{R} = 0.33, p-value = 0.05). However, if the thin/thick disk sources are removed, the significances of these correlations are reduced: marginally significant for late-M dwarfs ($N$ = \replaced{58}{57}, \textit{R} = 0.22 and p-value = 0.09) , significant for L dwarfs ($N$ = \replaced{60}{59}, \textit{R} = 0.27 and p-value $<$ 0.05), and insignificant for T dwarfs ($N$ = 36, \textit{R} = 0.25 and p-value = 0.13).
These results suggest that the $v_\text{tot}^2$, $|W|$ correlation is dominated by \replaced{the}{a} few sources from a distinct population, so that Wielen's age-dispersion relation may not be as accurate for this sample as the Aumer and Binney relations.
The ages from the latter relations are 
$4.9 \pm 0.3$ Gyr and $\replaced{7.0}{7.1} \pm 0.4$ Gyr, respectively, for the full \added{late-M and L dwarf} samples (Figure \ref{fig:simulated_population2}, upper left panel). These values confirm the significant (\replaced{4.2}{4.4}$\sigma$) discrepancy between late-M and L dwarf kinematic ages found in previous studies.
If we remove the thick disk sources, the kinematic ages are reduced to
$4.1 \pm 0.3$ Gyr for the late-M dwarfs and $\replaced{5.7}{5.8} \pm 0.3$ Gyr for the L dwarfs. These ages are
still significantly (\replaced{3.8}{4.0}$\sigma$) discrepant, while
the late-M and T dwarfs in our sample have statistically equivalent ages.

\begin{figure*}[!htbp]
\centering
\gridline{\fig{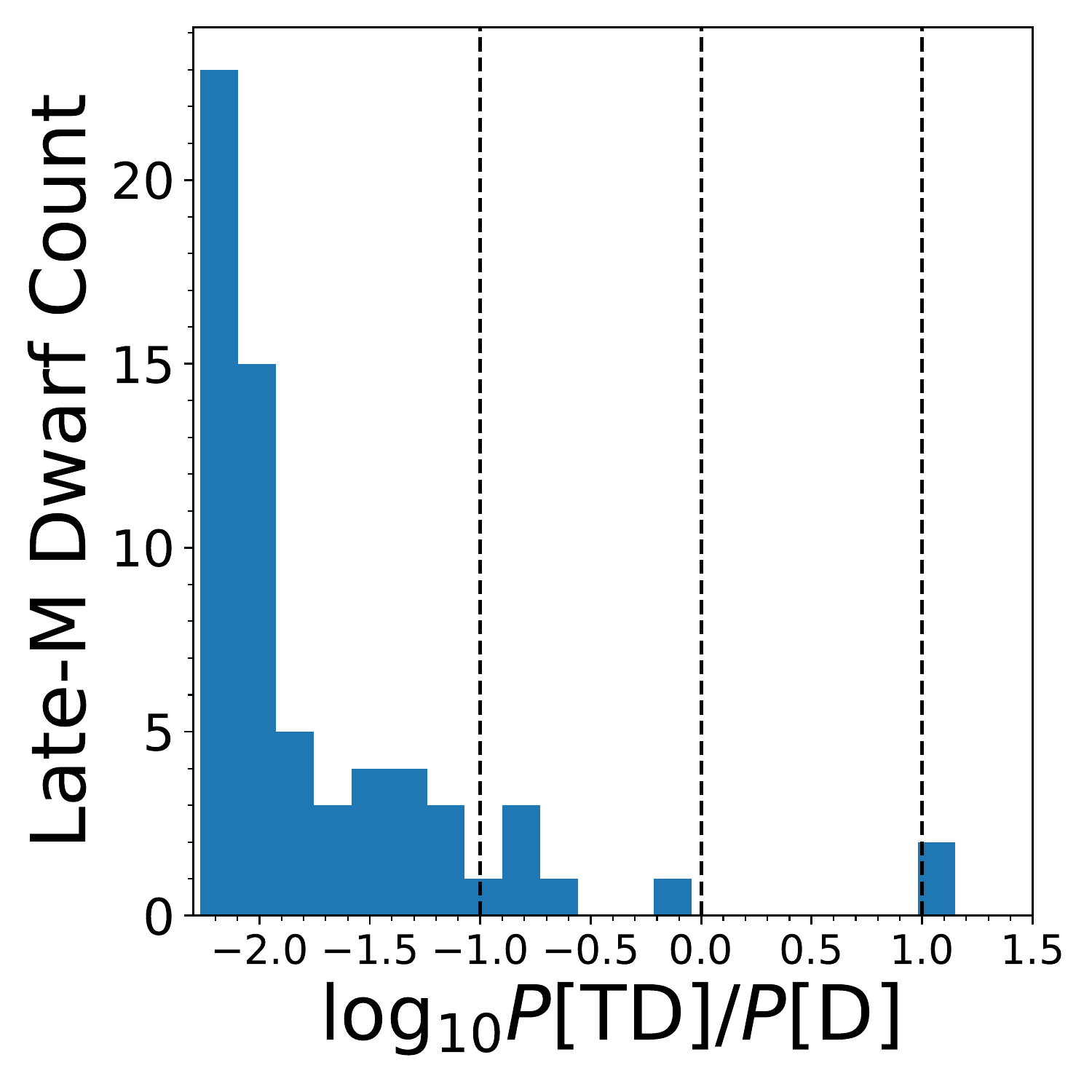}{0.33\textwidth}{}
\fig{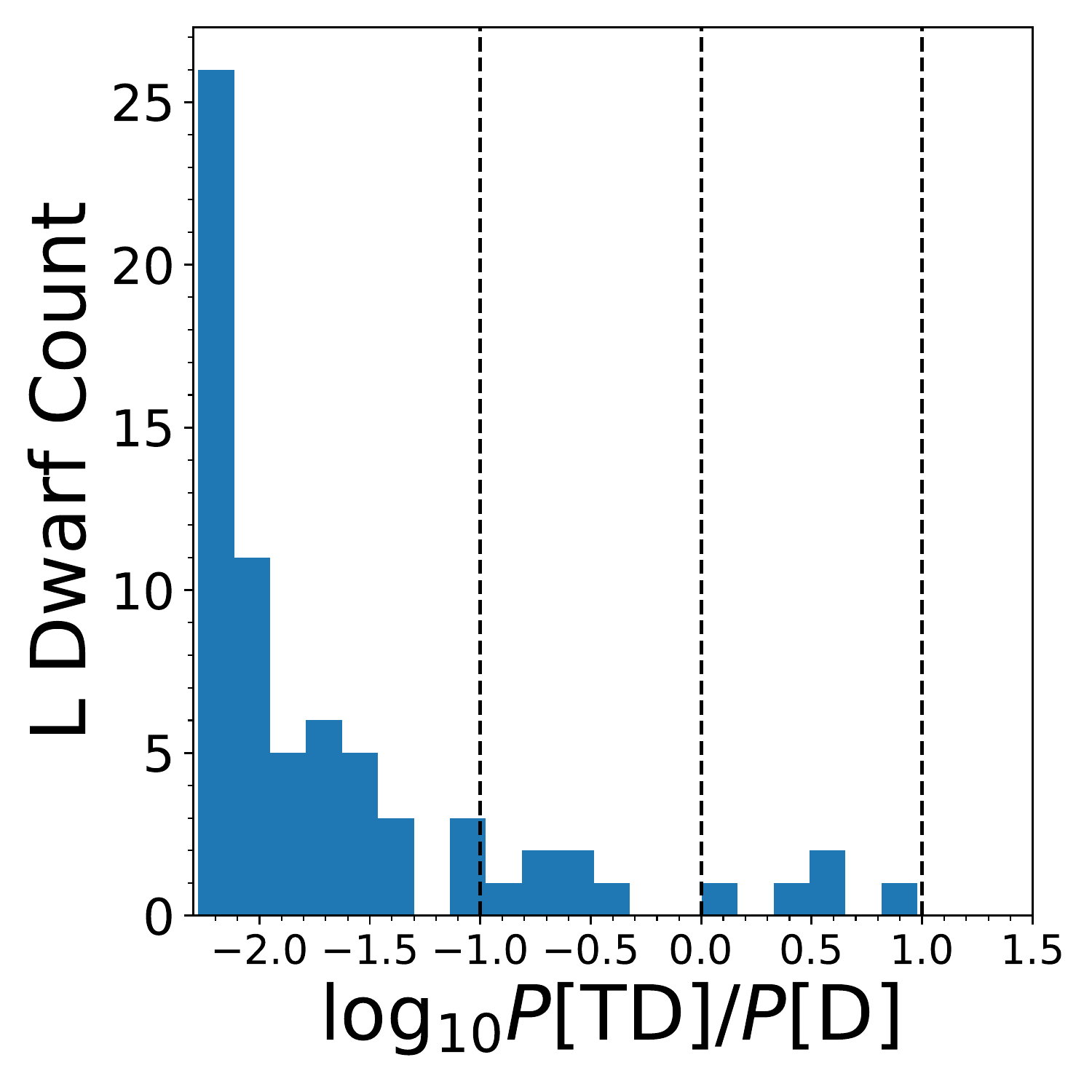}{0.33\textwidth}{}
\fig{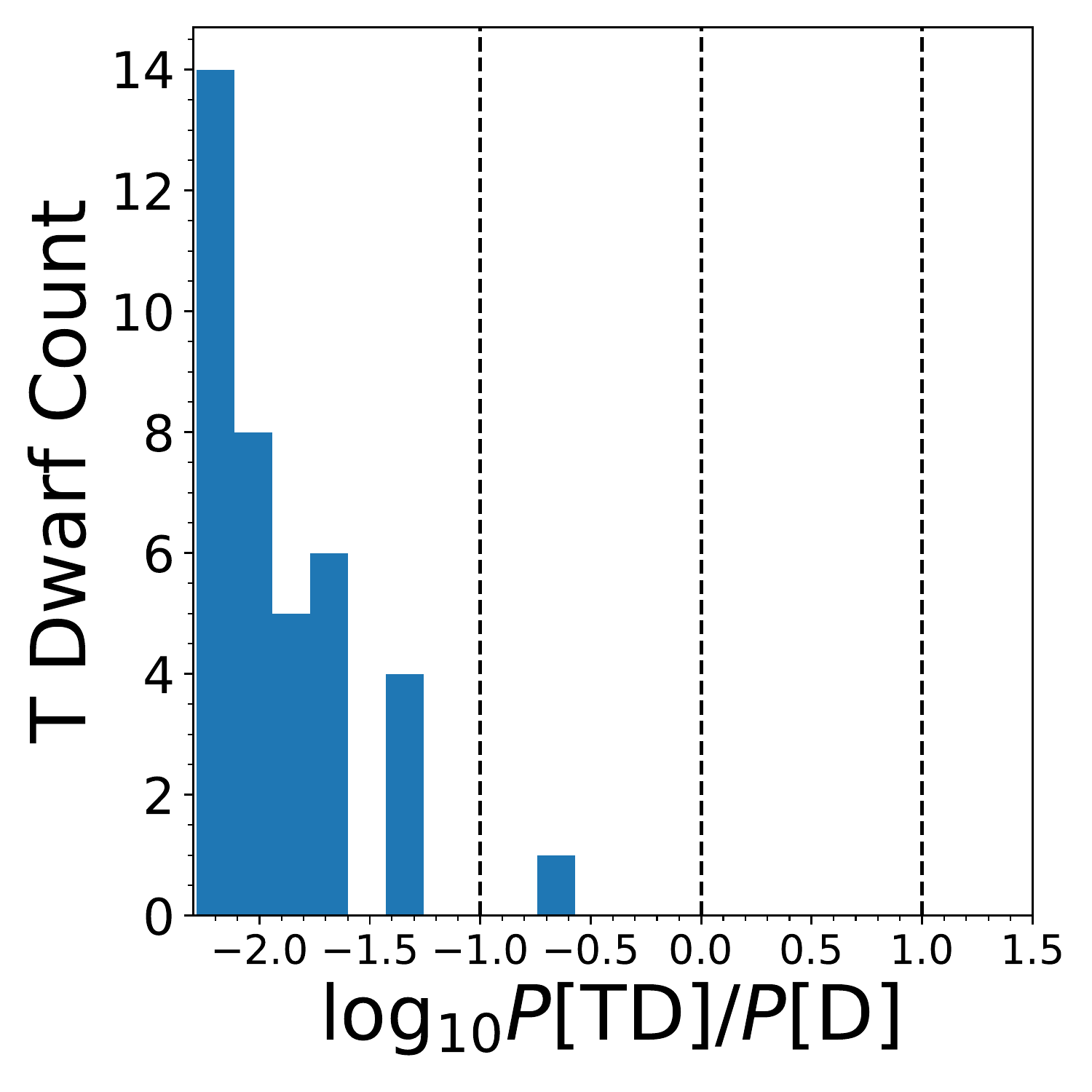}{0.33\textwidth}{}
}
\caption{Distributions of $\log$ probability ratios of \replaced{thin/thick}{thick/thin} disk \replaced{(D/TD)}{(TD/D)} sources for late-M, L, and T dwarfs, respectively. The vertical dashed lines denote the probability ratios of 0.1, 1, and 10, respectively. \label{fig:prob_dist}}
\end{figure*}

Figure \ref{fig:prob_dist} displays the log probability of thin disk to thick disk membership for late-M, L and T dwarfs, again following \citet{Bensby:2003aa}. These distributions show that there is a marginally higher proportion of intermediate thin/thick disk and thick disk sources relative to thin disk sources among the L dwarfs (8$^{+5}_{-2}$\%). as compared to the late-M dwarfs (3$^{+3}_{-1}$\%) and T dwarfs (3$^{+6}_{-2}$\%), with ratio statistics computed using binomial statistics following \citet{2003ApJ...586..512B}. 
The intermediate thin/thick disk sources in particular skew the kinematic dispersions and ages toward higher values.
To truly assess the kinematic age of the thin disk population without discarding too many old thin disk sources, 
we refined our selection requirement to $P[\mathrm{TD}]/P[\mathrm{D}]$ $\leq$ 1.0,  
since $P[\mathrm{TD}]/P[\mathrm{D}]$ = 1.0 denotes a 50\% probability as a thin or thick disk source.
This thin disk sample has \replaced{equivalent}{similar} kinematic ages 
of $4.1 \pm 0.3$ Gyr\added{ and $4.2 \pm 0.3$ Gyr} for \deleted{both }the late-M and L dwarfs\added{, respectively}, which \replaced{is}{are} slightly older but consistent with the T dwarfs.
These values imply equivalent ages across the entire \added{late-}MLT sequence.

Refining the sample to high-probability thin disk sources appears to resolve the long-standing discrepancy between late-M and L dwarf kinematics \citep{Faherty:2009aa, Seifahrt:2010aa, Burgasser:2015ac}. In Section \ref{sec:compare_simulated_ages}, we will show the observed thin disk ages of \added{late-}MLT dwarfs are also consistent with our population simulations. What remains unclear is why the local L dwarf sample has a higher fraction of thick disk sources compared to late-M and T dwarfs.
We will evaluate the properties of the thick disk L dwarfs, several of which are classified as unusually blue L dwarfs, in Section \ref{subsec:subpopulation}. 

For completeness, we evaluated the more recent age-velocity dispersion relation of \citet{Yu:2018aa}, based on the kinematics and ages of $>$3500 sub-giant and giant branch stars. The ages were estimated from empirical trends in [${\mathrm{C}}/{\mathrm{M}}$] and [${\mathrm{N}}/{\mathrm{M}}$] abundances \citep{Ho:2017aa}.
\citet{Yu:2018aa} fit power-law relations to the velocity dispersions in Galactic cylindrical coordinates, $\mathrm{age} = \added{C} \sigma_{i}^{\beta}$, where $i$ = R, $\mathrm{\phi}$, and $\mathrm{Z}$ direction\added{, and $C = 1$~Gyr (km s$^{-1}$)$^{-\beta}$ is a unit conversion factor}.
Using the $z$-coordinate relation from this study with $\beta_{z}$ = 0.56 $\pm$ 0.14, and propagating uncertainties with Monte Carlo sampling, we compute ages for 
late-M dwarfs (all/thin disk = 4.9$^{+2.4}_{-1.6}$~Gyr/4.8$^{+2.3}_{-1.5}$~Gyr), 
L dwarfs (all/thin disk = \replaced{5.2}{5.1}$^{+2.5}_{-1.7}$~Gyr/4.6$^{+2.1}_{-1.4}$~Gyr),  
and T dwarfs (4.1$^{+1.7}_{-1.2}$~Gyr) that are consistent with each other, albeit with 
higher statistical uncertainties. The $R$ and $\phi$ relations of \citet{Yu:2018aa} yield significantly younger ages (2.6--3.4~Gyr) which may be attributed to the very different spatial distributions of the sub-giant/giant sample as compared to the local ultracool dwarf sample.

\begin{longrotatetable}
\begin{deluxetable*}{lcccccccccl}
\tablecaption{Velocity Dispersions and Group Kinematic Ages \label{tablekinematicage}}
\tablewidth{700pt}
\tabletypesize{\scriptsize}
\tablehead{
\colhead{Sample} & \colhead{N} & \colhead{$\langle U \rangle$} & \colhead{$\langle V \rangle$} & \colhead{$\langle W \rangle$} & \colhead{$\sigma_U$} & \colhead{$\sigma_V$} & \colhead{$\sigma_W$} & \colhead{$\sigma_{tot}$} & \colhead{Age} & \colhead{Note} \\ 
\colhead{} & \colhead{} & \colhead{(km s$^{-1}$)} & \colhead{(km s$^{-1}$)} & \colhead{(km s$^{-1}$)} & \colhead{(km s$^{-1}$)} & \colhead{(km s$^{-1}$)} &  \colhead{(km s$^{-1}$)} &  \colhead{(km s$^{-1}$)} & \colhead{(Gyr)} & \colhead{}
} 
\startdata
All Sources & 36 & $-$1.6 $\pm$ 5.5 & $-$3.6 $\pm$ 2.7 & 2.1 $\pm$ 2.0 & 33.1 $\pm$ 0.8 & 16.5 $\pm$ 0.5 & 12.3 $\pm$ 0.3 & 39.0 $\pm$ 1.0 & 3.5 $\pm$ 0.3 & Unweighted \\ 
 & &  &  &  & 35.5 $\pm$ 0.9 & 16.4 $\pm$ 0.5 & 11.6 $\pm$ 0.3 & 40.8 $\pm$ 0.8 & 3.0 $\pm$ 0.1 & $|W|$ Weighted \\ 
Shallow\tablenotemark{a}  & 24 &  $-$2.7 $\pm$ 3.5 & $-$2.6 $\pm$ 1.7 &  2.3 $\pm$ 1.4 & 31.9 $\pm$ 2.2 & 15.8 $\pm$ 1.0 & 12.7 $\pm$ 1.0 & 37.8 $\pm$ 1.9 & 3.2 $\pm$ 0.5 & Unweighted \\ 
Wide\tablenotemark{a} & 12 & 0.9 $\pm$ 15.0 & $-$5.0 $\pm$ 7.4 & 2.3 $\pm$ 5.5 & 34.7 $\pm$ 1.0 & 17.2 $\pm$ 0.7 & 12.9 $\pm$ 0.7 & 40.8 $\pm$ 0.9 & 4.0 $\pm$ 0.3 & Unweighted \\ 
Not Young T Dwarfs & \replaced{32}{31} & $-$\replaced{1.1}{0.2} $\pm$ \replaced{6.2}{6.3} & $-$\replaced{2.9}{2.6} $\pm$ \replaced{3.0}{3.1} & 2.6 $\pm$ 2.3 & \replaced{34.9}{35.1} $\pm$ \replaced{0.9}{1.0} & \replaced{17.3}{17.4} $\pm$ 0.5 & \replaced{13.0}{13.1} $\pm$ 0.3 & \replaced{41.1}{41.3} $\pm$ 1.1 & 4.1 $\pm$ 0.4 & Unweighted \\ 
 & &  &  &  & \replaced{36.2}{36.3} $\pm$ \replaced{1.0}{1.1} & 16.8 $\pm$ \replaced{0.6}{0.5} & \replaced{11.8}{12.0} $\pm$ 0.3 & \replaced{41.7}{41.8} $\pm$ \replaced{0.9}{1.0} & 3.1 $\pm$ 0.2 & $|W|$ Weighted \\ 
Not Binary T Dwarfs & 30 & $-$0.3 $\pm$ 5.9 & $-$1.0 $\pm$ 2.7 & 2.4 $\pm$ 2.3 & 32.1 $\pm$ 1.1 & 14.6 $\pm$ 0.4 & 12.6 $\pm$ 0.3 & 37.4 $\pm$ 1.2 & 3.1 $\pm$ 0.3 & Unweighted \\ 
 & &  &  &  & 35.3 $\pm$ 1.2 & 15.9 $\pm$ 0.6 & 11.9 $\pm$ 0.3 & 40.5 $\pm$ 1.0 & 2.9 $\pm$ 0.2 & $|W|$ Weighted \\
\enddata
\tablecomments{Ages for unweighted velocities are computed from equation~(\ref{eq:AumerBinney2009}) using the parameters in \citet{Aumer:2009aa}. Ages for $|W|$-weighted velocities are computed from equation~(\ref{eq:Wielen1977}) using the parameters in \citet{Wielen:1977aa}.}
\tablenotetext{a}{Piece-wise linear fits to unweighted velocities, with separate fits for sources within $\pm$1$\sigma$ (shallow) and outside of $\pm$1$\sigma$ (wide); see Section \ref{ages}.}
\end{deluxetable*}
\end{longrotatetable}

\begin{longrotatetable}
\startlongtable
\begin{deluxetable}{lllrrrrrcl}
\tablecaption{Late-M and L Dwarf Sample at the Local 20 pc\label{table:lateML_sample}}
\tablewidth{700pt}
\tabletypesize{\scriptsize}
\tabletypesize{\tiny}
\tablehead{
\colhead{} & \colhead{} & 
\colhead{} & \colhead{} & 
\colhead{} & \colhead{} & 
\colhead{} & \colhead{} & 
\colhead{} & \colhead{} \\
\colhead{Source Name} & \colhead{Coordinates} & 
\colhead{SpT} & \colhead{2MASS $J$} & 
\colhead{$J \, - \, K$} & \colhead{$\mu_{\alpha}$} & 
\colhead{$\mu_{\delta}$} & \colhead{$d$} & 
\colhead{Published RV} & \colhead{References\tablenotemark{a}} \\ 
\colhead{} & \colhead{(J2000)} & \colhead{} & \colhead{(mag)} & 
\colhead{(mag)} & \colhead{(mas yr$^{-1}$)} & \colhead{(mas yr$^{-1}$)} &
\colhead{(pc)} & \colhead{(km s$^{-1}$)} & \colhead{}
} 
\startdata
J0004$-$4044 & 00 04 34.86 $-$40 44 06.4 & L5+L5 & 13.11 $\pm$ 0.02 & 1.71 $\pm$ 0.04 & 671.1 $\pm$ 0.4 & $-$1498.2 $\pm$ 0.5 & 12.18 $\pm$ 0.06 & 32.8 $\pm$ 0.2 & (17); (53); (53); (61)\\ 
J0004$-$2058 & 00 04 41.46 $-$20 58 29.8 & M9 & 12.4 $\pm$ 0.02 & 1.01 $\pm$ 0.03 & 758.2 $\pm$ 0.3 & 85.2 $\pm$ 0.2 & 15.08 $\pm$ 0.04 & $-$33 $\pm$ 2 & (18); (53); (53); (47)\\ 
J0015+3516 & 00 15 44.77 +35 16 02.6 & L1 & 13.88 $\pm$ 0.03 & 1.62 $\pm$ 0.04 & 55.2 $\pm$ 0.5 & $-$257.1 $\pm$ 0.3 & 17.06 $\pm$ 0.11 & $-$37.4 $\pm$ 0.2 & (33); (53); (53); (61)\\ 
J0019+5213 & 00 19 45.83 +52 13 17.5 & M9 & 12.79 $\pm$ 0.02 & 1.17 $\pm$ 0.03 & 310.7 $\pm$ 0.2 & $-$279.1 $\pm$ 0.2 & 19.96 $\pm$ 0.07 & $-$29 $\pm$ 3 & (16); (53); (53); (60)\\ 
J0024$-$0158 & 00 24 24.63 $-$01 58 19.9 & L0.5 & 11.99 $\pm$ 0.04 & 1.45 $\pm$ 0.04 & $-$77.2 $\pm$ 0.5 & 141.1 $\pm$ 0.3 & 12.51 $\pm$ 0.03 & 11.6 $\pm$ 1.6 & (33); (53); (53); (39)\\ 
J0024$-$2708B & 00 24 44.22 $-$27 08 25.0 & M8.5 & 9.25 $\pm$ 0.03 & 1.01 $\pm$ 0.05 & $-$107 $\pm$ 6 & 691 $\pm$ 8 & 7.72 $\pm$ 0.15 & $-$35 $\pm$ 3 & (2); (48); (42); (60)\\ 
J0027+2219 & 00 27 55.98 +22 19 32.8 & M8e & 10.61 $\pm$ 0.02 & 1.05 $\pm$ 0.03 & 406 $\pm$ 5 & $-$170 $\pm$ 2 & 15.3 $\pm$ 0.9 & $-$11.9 $\pm$ 1.3 & (10); (41); (46); (20)\\ 
J0036+1821 & 00 36 16.11 +18 21 10.2 & L3.5 & 12.47 $\pm$ 0.03 & 1.41 $\pm$ 0.03 & 901.6 $\pm$ 0.4 & 124.0 $\pm$ 0.2 & 8.74 $\pm$ 0.02 & 20.9 $\pm$ 0.14 & (3); (53); (53); (71)\\
\enddata
\end{deluxetable}
\end{longrotatetable}

\startlongtable
\begin{deluxetable*}{llcccccc}
\tablecaption{Radial Velocities and Heliocentric Space Motions of Late-M and L Dwarfs Within 20 pc of the Sun\label{table:lateML_uvw}}
\tablewidth{700pt}
\tabletypesize{\scriptsize}
\tablehead{
\colhead{} & \colhead{} & \colhead{} & \colhead{} & \colhead{} & \colhead{} & \colhead{} & \colhead{} \\ 
\colhead{Source Name} & \colhead{SpT} & \colhead{Published RV} & \colhead{$U$} & \colhead{$V$} & \colhead{$W$} & \colhead{P[TD]/P[D]\tablenotemark{a}} & \colhead{Population\tablenotemark{a}} \\ 
\colhead{} & \colhead{} & \colhead{(km s$^{-1}$)} & \colhead{(km s$^{-1}$)} & \colhead{(km s$^{-1}$)} & \colhead{(km s$^{-1}$)} & \colhead{} & \colhead{}
} 
\startdata
J0004$-$4044 & L5+L5 & 32.84 $\pm$ 0.17 & 21.68 $\pm$ 0.05 & $-$85.7 $\pm$ 0.4 & $-$11.90 $\pm$ 0.17 & 3.42 & D/TD \\ 
J0004$-$2058 & M9 & $-$33 $\pm$ 2 & $-$42.5 $\pm$ 0.2 & $-$13.2 $\pm$ 0.4 & 30 $\pm$ 2 & 0.04 & D \\ 
J0015+3516 & L1 & $-$37.35 $\pm$ 0.16 & 28.01 $\pm$ 0.07 & $-$27.23 $\pm$ 0.14 & 5.31 $\pm$ 0.14 & 0.01 & D \\ 
J0019+5213 & M9 & $-$29 $\pm$ 3 & 4.2 $\pm$ 1.4 & $-$30 $\pm$ 3 & $-$16.9 $\pm$ 0.6 & 0.02 & D \\ 
J0024$-$0158 & L0.5 & 11.65 $\pm$ 1.6 & 9.4 $\pm$ 0.2 & 25.8 $\pm$ 0.7 & 0.8 $\pm$ 1.4 & 0.01 & D \\ 
J0024$-$2708B & M8.5 & $-$35 $\pm$ 3 & $-$0.1 $\pm$ 0.4 & 31.2 $\pm$ 0.5 & 43 $\pm$ 3 & 0.24 & D/TD \\ 
J0027+2219 & M8e & $-$11.9 $\pm$ 1.3 & $-$5 $\pm$ 3 & $-$17 $\pm$ 3 & 3 $\pm$ 3 & 0.01 & D \\ 
J0036+1821 & L3.5 & 20.90 $\pm$ 0.14 & $-$29.83 $\pm$ 0.08 & 9.08 $\pm$ 0.09 & $-$5.9 $\pm$ 0.1 & 0.01 & D \\
\enddata
\tablenotetext{a}{Galactic thin disk (D), thick disk (TD), intermediate thin/thick disk populations (D/TD), and intermediate thick disk/halo populations (D/H) assigned in the same manner as Table~\ref{tableuvw}.}
\end{deluxetable*}

\begin{longrotatetable}
\begin{deluxetable*}{lcccccccccl}
\tablecaption{Velocity Dispersions and Group Kinematic Ages of Late-M and L Dwarfs at Local 20 pc\label{table:lateML_kinematicage}}
\tablewidth{700pt}
\tabletypesize{\scriptsize}
\tablehead{
\colhead{Sample} & \colhead{N\tablenotemark{a}} & \colhead{$\langle U \rangle$} & \colhead{$\langle V \rangle$} & \colhead{$\langle W \rangle$} & \colhead{$\sigma_U$} & \colhead{$\sigma_V$} & \colhead{$\sigma_W$} & \colhead{$\sigma_{tot}$} & \colhead{Age} & \colhead{Note} \\ 
\colhead{} & \colhead{} & \colhead{(km s$^{-1}$)} & \colhead{(km s$^{-1}$)} & \colhead{(km s$^{-1}$)} & \colhead{(km s$^{-1}$)} & \colhead{(km s$^{-1}$)} &  \colhead{(km s$^{-1}$)} &  \colhead{(km s$^{-1}$)} & \colhead{(Gyr)} & \colhead{}
} 
\startdata
late-M dwarfs & 65 & $-$4.7 $\pm$ 3.9 & $-$6.8 $\pm$ 3.0 & 0.5 $\pm$ 2.1 & 31.9 $\pm$ 0.3 & 24.8 $\pm$ 0.4 & 17.3 $\pm$ 0.3 & 43.9 $\pm$ 0.6 & 4.9 $\pm$ 0.3 & Unweighted \\ 
 & &  &  &  & 34.1 $\pm$ 0.5 & 27.7 $\pm$ 0.5 & 21.2 $\pm$ 0.7 & 48.8 $\pm$ 0.5 & 4.4 $\pm$ 0.1 & $|W|$ Weighted \\ 
\added{late-M dwarfs No Youth & 60 & $-$4.8 $\pm$ 4.2 & $-$6.9 $\pm$ 3.3 & 0.0 $\pm$ 2.3 & 32.5 $\pm$ 0.4 & 25.8 $\pm$ 0.4 & 17.9 $\pm$ 0.4 & 45.2 $\pm$ 0.7 & 5.3 $\pm$ 0.3 & Unweighted \\ 
 & &  &  &  & 33.9 $\pm$ 0.6 & 28.4 $\pm$ 0.6 & 22.2 $\pm$ 1.0 & 49.5 $\pm$ 0.7 & 4.5 $\pm$ 0.1 & $|W|$ Weighted \\ }
late-M dwarfs NTD\tablenotemark{b} & 63 & $-$3.7 $\pm$ 3.8 & $-$4.5 $\pm$ 2.8 & 1.0 $\pm$ 2.0 & 30.5 $\pm$ 0.4 & 22.1 $\pm$ 0.4 & 16.3 $\pm$ 0.3 & 41.0 $\pm$ 0.6 & 4.1 $\pm$ 0.3 & Unweighted \\ 
 & &  &  &  & 34.5 $\pm$ 0.6 & 24.5 $\pm$ 0.4 & 18.1 $\pm$ 0.4 & 46.0 $\pm$ 0.5 & 3.9 $\pm$ 0.1 & $|W|$ Weighted \\
late-M dwarfs D\tablenotemark{c} & 63 & $-$3.7 $\pm$ 3.8 & $-$4.5 $\pm$ 2.8 & 1.0 $\pm$ 2.0 & 30.5 $\pm$ 0.3 & 22.1 $\pm$ 0.3 & 16.3 $\pm$ 0.3 & 41.0 $\pm$ 0.6 & 4.1 $\pm$ 0.3 & Unweighted \\ 
 & &  &  &  & 34.5 $\pm$ 0.6 & 24.4 $\pm$ 0.4 & 18.1 $\pm$ 0.4 & 46.0 $\pm$ 0.5 & 3.9 $\pm$ 0.1 & $|W|$ Weighted \\
\replaced{L dwarfs & \replaced{72}{71} & 0.6 $\pm$ 4.2 & $-$12.8 $\pm$ 3.5 & $-$1.5 $\pm$ 2.2 & 35.5 $\pm$ 0.4 & 29.5 $\pm$ 0.6 & 18.5 $\pm$ 0.4 & 49.7 $\pm$ 0.8 & 7.0 $\pm$ 0.4 & Unweighted \\
 & &  &  &  & 46.3 $\pm$ 1.0 & 46.5 $\pm$ 2.1 & 24.2 $\pm$ 0.7 & 70.0 $\pm$ 2.2 & 8.2 $\pm$ 0.4 & $|W|$ Weighted \\}{L dwarfs & 71 & 0.6 $\pm$ 4.2 & $-$12.8 $\pm$ 3.5 & $-$1.4 $\pm$ 2.2 & 35.8 $\pm$ 0.4 & 29.7 $\pm$ 0.5 & 18.7 $\pm$ 0.3 & 50.1 $\pm$ 0.7 & 7.1 $\pm$ 0.4 & Unweighted \\ 
 & &  &  &  & 46.6 $\pm$ 0.8 & 46.8 $\pm$ 1.5 & 24.3 $\pm$ 0.5 & 70.4 $\pm$ 1.6 & 8.3 $\pm$ 0.3 & $|W|$ Weighted \\} 
\replaced{L dwarfs NTD\tablenotemark{b} & \replaced{71}{70} & 1.9 $\pm$ 4.1 & $-$11.3 $\pm$ 3.2 & $-$2.6 $\pm$ 1.9 & 34.1 $\pm$ 0.4 & 26.6 $\pm$ 0.5 & 16.1 $\pm$ 0.3 & 46.2 $\pm$ 0.7 & 5.7 $\pm$ 0.3 & Unweighted \\ 
 & &  &  &  & 39.0 $\pm$ 0.6 & 30.7 $\pm$ 1.1 & 18.9 $\pm$ 0.4 & 53.1 $\pm$ 0.8 & 5.2 $\pm$ 0.1 & $|W|$ Weighted \\}{L dwarfs NTD\tablenotemark{b} & 70 & 1.9 $\pm$ 4.1 & $-$11.2 $\pm$ 3.2 & $-$2.6 $\pm$ 1.9 & 34.4 $\pm$ 0.4 & 26.8 $\pm$ 0.5 & 16.2 $\pm$ 0.3 & 46.5 $\pm$ 0.7 & 5.8 $\pm$ 0.3 & Unweighted \\ 
 & &  &  &  & 39.1 $\pm$ 0.5 & 30.8 $\pm$ 1.1 & 18.9 $\pm$ 0.4 & 53.3 $\pm$ 0.8 & 5.2 $\pm$ 0.1 & $|W|$ Weighted \\}
\replaced{L dwarfs NB\tablenotemark{d} & 67 & 3.5 $\pm$ 4.0 & $-$10.1 $\pm$ 3.1 & $-$3.0 $\pm$ 1.9 & 32.9 $\pm$ 0.4 & 25.3 $\pm$ 0.4 & 15.9 $\pm$ 0.3 & 44.4 $\pm$ 0.6 & 5.1 $\pm$ 0.3 & Unweighted \\ 
 & &  &  &  & 38.9 $\pm$ 0.6 & 28.5 $\pm$ 0.6 & 19.1 $\pm$ 0.4 & 51.8 $\pm$ 0.6 & 4.9 $\pm$ 0.1 & $|W|$ Weighted \\}{L dwarfs NB\tablenotemark{d} & 67 & 3.8 $\pm$ 4.0 & $-$10.8 $\pm$ 3.2 & $-$3.3 $\pm$ 2.0 & 33.0 $\pm$ 0.4 & 26.0 $\pm$ 0.5 & 16.1 $\pm$ 0.3 & 45.0 $\pm$ 0.7 & 5.3 $\pm$ 0.3 & Unweighted \\ 
 & &  &  &  & 38.4 $\pm$ 0.6 & 30.2 $\pm$ 1.1 & 19.2 $\pm$ 0.3 & 52.5 $\pm$ 0.8 & 5.1 $\pm$ 0.1 & $|W|$ Weighted \\} 
\replaced{L dwarfs No Youth & 70 & 0.7 $\pm$ 4.3 & $-$13.0 $\pm$ 3.6 & $-$1.5 $\pm$ 2.2 & 36.0 $\pm$ 0.4 & 29.9 $\pm$ 0.5 & 18.8 $\pm$ 0.3 & 50.4 $\pm$ 0.7 & 7.3 $\pm$ 0.4 & Unweighted \\
 & &  &  &  & 46.7 $\pm$ 0.8 & 46.9 $\pm$ 1.6 & 24.3 $\pm$ 0.5 & 70.5 $\pm$ 1.6 & 8.3 $\pm$ 0.3 & $|W|$ Weighted \\}{L dwarfs No Youth & 57 & 0.9 $\pm$ 5.0 & $-$12.4 $\pm$ 4.3 & $-$1.7 $\pm$ 2.7 & 38.1 $\pm$ 0.5 & 32.5 $\pm$ 0.7 & 20.4 $\pm$ 0.5 & 54.0 $\pm$ 1.0 & 8.8 $\pm$ 0.6 & Unweighted \\
 & &  &  &  & 48.5 $\pm$ 1.2 & 49.8 $\pm$ 2.5 & 25.6 $\pm$ 0.8 & 74.1 $\pm$ 2.6 & 8.9 $\pm$ 0.4 & $|W|$ Weighted \\}
L dwarfs No Binary & 63 & 0.7 $\pm$ 4.6 & $-$12.7 $\pm$ 3.7 & $-$1.2 $\pm$ 2.4 & 36.7 $\pm$ 0.5 & 29.4 $\pm$ 0.7 & 19.2 $\pm$ 0.5 & 50.8 $\pm$ 1.0 & 7.4 $\pm$ 0.5 & Unweighted \\ 
 & &  &  &  & 47.7 $\pm$ 1.2 & 47.1 $\pm$ 2.8 & 25.3 $\pm$ 0.9 & 71.6 $\pm$ 2.8 & 8.5 $\pm$ 0.5 & $|W|$ Weighted \\
\added{L dwarfs No Binary D\tablenotemark{c} & 58 & 2.0 $\pm$ 4.3 & $-$7.7 $\pm$ 2.8 & $-$1.8 $\pm$ 2.0 & 33.0 $\pm$ 0.4 & 21.2 $\pm$ 0.7 & 15.4 $\pm$ 0.3 & 42.1 $\pm$ 0.9 & 4.4 $\pm$ 0.3 & Unweighted \\ 
 & &  &  &  & 37.0 $\pm$ 0.7 & 21.5 $\pm$ 1.6 & 18.3 $\pm$ 0.4 & 46.5 $\pm$ 1.0 & 4.0 $\pm$ 0.2 & $|W|$ Weighted \\ }
\replaced{L dwarfs D\tablenotemark{c} & \replaced{66}{65} & 1.5 $\pm$ 4.0 & $-$7.3 $\pm$ 2.5 & $-$1.9 $\pm$ 1.9 & 32.3 $\pm$ 0.3 & 20.7 $\pm$ 0.6 & 15.2 $\pm$ 0.3 & 41.2 $\pm$ 0.7 & 4.1 $\pm$ 0.3 & Unweighted \\ 
 & &  &  &  & 36.2 $\pm$ 0.6 & 21.9 $\pm$ 1.4 & 17.5 $\pm$ 0.4 & 45.8 $\pm$ 0.8 & 3.8 $\pm$ 0.1 & $|W|$ Weighted \\}{L dwarfs D\tablenotemark{c} & 65 & 1.5 $\pm$ 4.0 & $-$7.2 $\pm$ 2.6 & $-$1.8 $\pm$ 1.9 & 32.5 $\pm$ 0.4 & 20.9 $\pm$ 0.6 & 15.3 $\pm$ 0.3 & 41.5 $\pm$ 0.8 & 4.2 $\pm$ 0.3 & Unweighted \\ 
 & &  &  &  & 36.3 $\pm$ 0.6 & 22.0 $\pm$ 1.4 & 17.6 $\pm$ 0.4 & 46.0 $\pm$ 0.9 & 3.9 $\pm$ 0.2 & $|W|$ Weighted \\} 
L0--L5 dwarfs & 57 & 3.3 $\pm$ 5.0 & $-$14.3 $\pm$ 4.2 & $-$1.3 $\pm$ 2.6 & 37.6 $\pm$ 0.5 & 31.6 $\pm$ 0.6 & 19.9 $\pm$ 0.5 & 53.0 $\pm$ 0.9 & 8.3 $\pm$ 0.5 & Unweighted \\ 
 & &  &  &  & 49.9 $\pm$ 1.3 & 49.6 $\pm$ 2.5 & 26.0 $\pm$ 0.9 & 75.0 $\pm$ 2.7 & 9.0 $\pm$ 0.5 & $|W|$ Weighted \\
\added{L0--L5 dwarfs D\tablenotemark{c} & 51 & 4.8 $\pm$ 4.7 & $-$7.3 $\pm$ 3.0 & $-$1.8 $\pm$ 2.2 & 33.6 $\pm$ 0.5 & 21.4 $\pm$ 0.3 & 15.9 $\pm$ 0.3 & 42.9 $\pm$ 0.7 & 4.6 $\pm$ 0.3 & Unweighted \\ 
 & &  &  &  & 38.8 $\pm$ 0.8 & 20.3 $\pm$ 0.5 & 18.5 $\pm$ 0.4 & 47.6 $\pm$ 0.7 & 4.1 $\pm$ 0.1 & $|W|$ Weighted \\} 
\replaced{L6--L9 dwarfs & \replaced{15}{14} & $-$9.7 $\pm$ 6.1 & $-$7.4 $\pm$ 4.6 & $-$2.2 $\pm$ 3.1 & 23.7 $\pm$ 1.8 & 18.0 $\pm$ 3.1 & 12.2 $\pm$ 0.9 & 32.1 $\pm$ 3.7 & 2.0 $\pm$ 0.7 & Unweighted \\
 & &  &  &  & 25.9 $\pm$ 1.7 & 26.4 $\pm$ 5.8 & 14.9 $\pm$ 1.7 & 40.1 $\pm$ 4.6 & 2.9 $\pm$ 0.8 & $|W|$ Weighted \\}{L6--L9 dwarfs & 14 & $-$10.5 $\pm$ 6.5 & $-$6.7 $\pm$ 4.9 & $-$1.9 $\pm$ 3.3 & 24.0 $\pm$ 2.1 & 18.4 $\pm$ 3.3 & 12.6 $\pm$ 1.0 & 32.8 $\pm$ 4.0 & 2.2 $\pm$ 0.8 & Unweighted \\ 
 & &  &  &  & 26.0 $\pm$ 1.9 & 26.8 $\pm$ 6.0 & 15.1 $\pm$ 1.8 & 40.5 $\pm$ 4.9 & 3.0 $\pm$ 0.8 & $|W|$ Weighted \\}
L0--L1 dwarfs & 26 & 10.1 $\pm$ 6.9 & $-$12.7 $\pm$ 5.3 & $-$4.5 $\pm$ 3.7 & 34.9 $\pm$ 1.1 & 27.2 $\pm$ 1.1 & 18.8 $\pm$ 0.7 & 48.1 $\pm$ 1.7 & 6.4 $\pm$ 0.7 & Unweighted \\ 
 & &  &  &  & 45.7 $\pm$ 1.4 & 31.0 $\pm$ 1.8 & 24.9 $\pm$ 0.8 & 60.6 $\pm$ 1.3 & 6.5 $\pm$ 0.2 & $|W|$ Weighted \\ 
L1--L2 dwarfs & 28 & 8.8 $\pm$ 6.6 & $-$15.0 $\pm$ 4.7 & $-$3.6 $\pm$ 3.5 & 34.8 $\pm$ 0.9 & 25.0 $\pm$ 1.0 & 18.3 $\pm$ 0.6 & 46.6 $\pm$ 1.5 & 5.8 $\pm$ 0.6 & Unweighted \\ 
 & &  &  &  & 43.7 $\pm$ 1.3 & 31.0 $\pm$ 1.7 & 22.4 $\pm$ 0.9 & 58.1 $\pm$ 1.4 & 6.1 $\pm$ 0.3 & $|W|$ Weighted \\ 
L2--L3 dwarfs & 16 & $-$0.9 $\pm$ 8.7 & $-$13.0 $\pm$ 4.5 & $-$1.3 $\pm$ 3.7 & 34.5 $\pm$ 1.3 & 18.0 $\pm$ 0.8 & 14.6 $\pm$ 0.9 & 41.6 $\pm$ 1.8 & 4.2 $\pm$ 0.6 & Unweighted \\ 
 & &  &  &  & 35.7 $\pm$ 2.0 & 28.1 $\pm$ 1.9 & 14.4 $\pm$ 1.1 & 47.7 $\pm$ 2.1 & 4.2 $\pm$ 0.4 & $|W|$ Weighted \\ 
L3--L4 dwarfs & 13 & 8.6 $\pm$ 7.6 & $-$7.4 $\pm$ 7.2 & $-$1.1 $\pm$ 3.4 & 27.4 $\pm$ 1.3 & 25.8 $\pm$ 1.7 & 11.9 $\pm$ 1.8 & 39.5 $\pm$ 2.8 & 3.7 $\pm$ 0.8 & Unweighted \\ 
 & &  &  &  & 28.9 $\pm$ 1.7 & 28.6 $\pm$ 3.5 & 15.6 $\pm$ 3.3 & 43.7 $\pm$ 2.5 & 3.5 $\pm$ 0.4 & $|W|$ Weighted \\ 
L4--L5 dwarfs & 15 & $-$3.9 $\pm$ 11.0 & $-$18.3 $\pm$ 12.0 & 4.3 $\pm$ 6.4 & 42.2 $\pm$ 2.4 & 46.1 $\pm$ 2.6 & 24.4 $\pm$ 2.7 & 67.1 $\pm$ 4.4 & 16.2 $\pm$ 3.0 & Unweighted \\ 
 & &  &  &  & 57.2 $\pm$ 7.3 & 72.9 $\pm$ 11.0 & 39.5 $\pm$ 6.2 & 100.8 $\pm$ 14.3 & 12.9 $\pm$ 2.0 & $|W|$ Weighted \\ 
L4--L5 dwarfs NB\tablenotemark{d} & 12 & 11.1 $\pm$ 8.1 & $-$7.6 $\pm$ 10.3 & $-$3.7 $\pm$ 4.2 & 27.6 $\pm$ 2.5 & 35.5 $\pm$ 2.6 & 14.5 $\pm$ 1.4 & 47.2 $\pm$ 3.9 & 6.1 $\pm$ 1.4 & Unweighted \\ 
 & &  &  &  & 26.6 $\pm$ 3.1 & 32.1 $\pm$ 3.9 & 17.9 $\pm$ 2.6 & 45.6 $\pm$ 3.1 & 3.8 $\pm$ 0.5 & $|W|$ Weighted \\
L4--L5 dwarfs D\tablenotemark{c} & 12 & 8.4 $\pm$ 8.2 & 0.8 $\pm$ 7.9 & $-$1.0 $\pm$ 4.5 & 28.2 $\pm$ 2.1 & 27.2 $\pm$ 1.8 & 15.6 $\pm$ 1.3 & 42.2 $\pm$ 3.1 & 4.4 $\pm$ 1.0 & Unweighted \\ 
 & &  &  &  & 25.1 $\pm$ 2.4 & 20.3 $\pm$ 2.1 & 19.4 $\pm$ 2.1 & 37.7 $\pm$ 2.3 & 2.5 $\pm$ 0.4 & $|W|$ Weighted \\
L5--L6 dwarfs & 15 & $-$14.1 $\pm$ 11.1 & $-$22.0 $\pm$ 11.2 & 1.4 $\pm$ 6.9 & 42.7 $\pm$ 1.8 & 43.3 $\pm$ 2.5 & 26.7 $\pm$ 2.5 & 66.4 $\pm$ 4.0 & 15.8 $\pm$ 2.7 & Unweighted \\ 
 & &  &  &  & 50.6 $\pm$ 5.2 & 69.0 $\pm$ 8.7 & 32.4 $\pm$ 4.5 & 91.5 $\pm$ 10.8 & 11.6 $\pm$ 1.6 & $|W|$ Weighted \\ 
L5--L6 dwarfs NB\tablenotemark{d} & 11 & $-$3.7 $\pm$ 10.5 & $-$8.0 $\pm$ 9.2 & $-$5.5 $\pm$ 5.1 & 34.9 $\pm$ 2.5 & 30.0 $\pm$ 3.7 & 16.7 $\pm$ 1.4 & 48.9 $\pm$ 4.7 & 6.8 $\pm$ 1.8 & Unweighted \\ 
 & &  &  &  & 29.7 $\pm$ 2.5 & 30.0 $\pm$ 3.4 & 16.5 $\pm$ 0.9 & 45.5 $\pm$ 2.8 & 3.8 $\pm$ 0.5 & $|W|$ Weighted \\
L5--L6 dwarfs D\tablenotemark{c} & 12 & $-$4.4 $\pm$ 9.7 & $-$3.8 $\pm$ 7.0 & $-$4.5 $\pm$ 5.4 & 33.4 $\pm$ 2.3 & 24.1 $\pm$ 3.2 & 18.6 $\pm$ 1.2 & 45.2 $\pm$ 4.2 & 5.4 $\pm$ 1.4 & Unweighted \\ 
 & &  &  &  & 28.3 $\pm$ 1.8 & 28.3 $\pm$ 5.1 & 17.1 $\pm$ 0.8 & 43.7 $\pm$ 3.7 & 3.5 $\pm$ 0.6 & $|W|$ Weighted \\
L6--L7 dwarfs & 10 & $-$11.8 $\pm$ 8.2 & $-$8.7 $\pm$ 6.1 & $-$3.5 $\pm$ 4.2 & 25.7 $\pm$ 2.9 & 18.9 $\pm$ 4.5 & 13.3 $\pm$ 1.4 & 34.6 $\pm$ 5.5 & 2.6 $\pm$ 1.2 & Unweighted \\ 
 & &  &  &  & 27.2 $\pm$ 2.5 & 28.6 $\pm$ 7.5 & 16.3 $\pm$ 2.2 & 43.1 $\pm$ 6.1 & 3.4 $\pm$ 1.0 & $|W|$ Weighted \\ 
L6--L7 dwarfs NB\tablenotemark{d} & 9 & $-$15.4 $\pm$ 8.2 & $-$3.1 $\pm$ 3.5 & $-$1.0 $\pm$ 3.9 & 24.1 $\pm$ 3.7 & 10.2 $\pm$ 2.0 & 11.6 $\pm$ 1.2 & 28.6 $\pm$ 4.3 & 1.5 $\pm$ 0.7 & Unweighted \\ 
 & &  &  &  & 23.1 $\pm$ 3.8 & 14.3 $\pm$ 2.9 & 13.3 $\pm$ 2.1 & 30.6 $\pm$ 2.6 & 1.5 $\pm$ 0.3 & $|W|$ Weighted \\ 
\replaced{L7--L8 dwarfs & \replaced{8}{7} & $-$9.5 $\pm$ 6.6 & $-$3.3 $\pm$ 3.5 & 1.4 $\pm$ 3.1 & 18.5 $\pm$ 1.6 & 9.7 $\pm$ 2.1 & 8.7 $\pm$ 0.9 & 22.6 $\pm$ 2.8 & 0.7 $\pm$ 0.3 & Unweighted \\ 
 & &  &  &  & 21.4 $\pm$ 2.9 & 11.3 $\pm$ 2.4 & 7.6 $\pm$ 1.0 & 25.4 $\pm$ 3.4 & 0.9 $\pm$ 0.3 & $|W|$ Weighted \\}{L7--L8 dwarfs & 7 & $-$11.1 $\pm$ 7.4 & $-$1.4 $\pm$ 3.5 & 2.5 $\pm$ 3.4 & 19.1 $\pm$ 2.1 & 8.7 $\pm$ 2.7 & 8.7 $\pm$ 1.2 & 22.8 $\pm$ 3.6 & 0.7 $\pm$ 0.4 & Unweighted \\ 
 & &  &  &  & 21.4 $\pm$ 3.3 & 10.8 $\pm$ 3.1 & 7.6 $\pm$ 1.2 & 25.3 $\pm$ 4.0 & 0.9 $\pm$ 0.4 & $|W|$ Weighted \\} 
\replaced{L7--L8 dwarfs NB\tablenotemark{d} & \replaced{7}{6} & $-$8.1 $\pm$ 7.4 & $-$3.4 $\pm$ 4.0 & 0.2 $\pm$ 3.3 & 19.4 $\pm$ 1.7 & 10.4 $\pm$ 2.3 & 8.7 $\pm$ 1.0 & 23.7 $\pm$ 3.0 & 0.8 $\pm$ 0.3 & Unweighted \\ 
 & &  &  &  & 24.4 $\pm$ 3.1 & 12.1 $\pm$ 2.7 & 8.1 $\pm$ 1.0 & 28.5 $\pm$ 3.7 & 1.2 $\pm$ 0.4 & $|W|$ Weighted \\}{L7--L8 dwarfs NB\tablenotemark{d} & 6 & $-$9.7 $\pm$ 8.4 & $-$1.2 $\pm$ 4.0 & 1.3 $\pm$ 3.7 & 20.2 $\pm$ 2.1 & 9.3 $\pm$ 3.1 & 8.9 $\pm$ 1.3 & 24.0 $\pm$ 3.9 & 0.8 $\pm$ 0.5 & Unweighted \\ 
 & &  &  &  & 24.5 $\pm$ 4.2 & 11.3 $\pm$ 3.6 & 8.2 $\pm$ 1.3 & 28.3 $\pm$ 4.9 & 1.2 $\pm$ 0.6 & $|W|$ Weighted \\} 
\replaced{L8--L9 dwarfs & \replaced{5}{4} & $-$5.7 $\pm$ 7.9 & $-$4.8 $\pm$ 6.3 & 0.5 $\pm$ 3.9 & 16.9 $\pm$ 2.6 & 13.2 $\pm$ 3.4 & 8.3 $\pm$ 1.3 & 23.0 $\pm$ 4.4 & 0.7 $\pm$ 0.5 & Unweighted \\ 
 & &  &  &  & 18.6 $\pm$ 2.8 & 14.9 $\pm$ 2.9 & 7.7 $\pm$ 2.0 & 25.3 $\pm$ 2.5 & 0.9 $\pm$ 0.3 & $|W|$ Weighted \\}{L8--L9 dwarfs & 4 & $-$7.5 $\pm$ 9.7 & $-$1.9 $\pm$ 7.1 & 2.2 $\pm$ 4.4 & 18.1 $\pm$ 2.8 & 12.9 $\pm$ 3.9 & 7.7 $\pm$ 3.4 & 23.6 $\pm$ 5.9 & 0.9 $\pm$ 0.7 & Unweighted \\ 
 & &  &  &  & 20.3 $\pm$ 3.1 & 14.4 $\pm$ 3.7 & 7.6 $\pm$ 3.7 & 26.6 $\pm$ 2.7 & 1.0 $\pm$ 0.3 & $|W|$ Weighted \\} 
\replaced{L8--L9 dwarfs NB\tablenotemark{d} & \replaced{4}{3} & $-$2.2 $\pm$ 9.1 & $-$5.4 $\pm$ 7.8 & $-$1.8 $\pm$ 4.1 & 16.3 $\pm$ 6.0 & 14.0 $\pm$ 4.9 & 7.8 $\pm$ 1.4 & 22.8 $\pm$ 7.9 & 0.9 $\pm$ 0.9 & Unweighted \\ 
 & &  &  &  & 17.5 $\pm$ 6.1 & 14.6 $\pm$ 5.3 & 6.4 $\pm$ 1.6 & 24.9 $\pm$ 3.0 & 0.8 $\pm$ 0.3 & $|W|$ Weighted}{L8--L9 dwarfs NB\tablenotemark{d} & 3 & $-$3.4 $\pm$ 12.1 & $-$1.7 $\pm$ 9.5 & $-$0.2 $\pm$ 5.2 & 15.9 $\pm$ 8.3 & 13.3 $\pm$ 5.5 & 6.8 $\pm$ 4.1 & 21.8 $\pm$ 10.8 & 1.0 $\pm$ 1.3 & Unweighted \\ 
 & &  &  &  & 19.2 $\pm$ 8.7 & 12.6 $\pm$ 5.0 & 5.5 $\pm$ 3.6 & 25.6 $\pm$ 4.0 & 0.9 $\pm$ 0.4 & $|W|$ Weighted \\ }
\enddata
\tablecomments{Ages for unweighted velocities are computed from equation~(\ref{eq:AumerBinney2009}) using the parameters in \citet{Aumer:2009aa}. Ages for $|W|$-weighted velocities are computed from equation~(\ref{eq:Wielen1977}) using the parameters in \citet{Wielen:1977aa}.}
\tablenotetext{a}{Number of sources in sample}
\tablenotetext{b}{Excluding thick disk sources}
\tablenotetext{c}{Thin disk sources only}
\tablenotetext{d}{Excluding unusual blue L dwarfs}
\end{deluxetable*}
\end{longrotatetable}

\subsection{Vertical Action Dispersion} \label{sec:vertical_action_dispersion}

Vertical action ($J_Z$) measures the excursion of momentum perpendicular to the Galactic plane, which is invariant under orbital evolution provided that the Galactic potential is axisymmetric and perturbations are on average planar. The Galactic potential does have non-axisymmetric components, such as the spiral arms, interior bar, and giant molecular clouds above and below the Galactic disk, all of which can perturb $J_Z$ and the planar actions $J_R$ and $J_{\phi}$ \citep{Beane:2018aa, Ting:2019aa}. Nevertheless, it has been argued that vertical action may be a better physical quantity to probe statistical ages for stellar populations than \textit{UVW} velocities \citep{Kiman:2019aa, Ting:2019aa}.

Vertical actions and their dispersion can be computed directly from 6D spatial and velocity coordinates. We used \textit{galpy} \citep{Bovy:2015aa} to calculate $J_{Z}$, using a Galactocentric solar position of  $(R_{\odot}, \, Z_{\odot}) = (8.2 , 0.025)$ kpc, an LSR Galactic circular velocity at the solar radius of $\nu_{\phi}(R_{\odot}) = 240$ km s$^{-1}$ \citep{Ting:2019aa}, and the LSR solar velocity vector used above. 
The position and circular velocity used here differ from our prior assumptions in Section~\ref{galacticorbits} in order to align our analysis with that of \cite{Ting:2019aa}.

By analyzing the APOGEE-\textit{Gaia} DR2 red clump giant sample \citep{Ting:2018aa, Gaia-Collaboration:2018ab},
\citet{Ting:2019aa} derived an empirical relation to estimate kinematic age from the mean vertical action $\widehat{J}_\text{Z}$ (kpc km s$^{-1}$) as a function of mean Galactic radius $\overline{R}_\text{GC}$ (kpc) and age $\tau$ for a star:
\begin{equation}
\label{eqn:ting}
\begin{split}
\widehat{J}_Z(\overline{R}_\text{GC}, \tau)  & = \widehat{J}_{Z, 0}(\overline{R}_\text{GC}) + \Delta\widehat{J}_{\text{Z}, 1 \mbox{Gyr}} \big(\frac{\tau}{1 \mbox{Gyr}}\big)^{\gamma(\overline{R}_\text{GC})} \\
& = (0.91 + 0.18 \Delta R_\text{GC} + 0.087 \Delta R_\text{GC}^2 \\ 
& + 0.014  \Delta R_\text{GC}^3) \\
& + (1.81 + 0.050  \Delta R_\text{GC}) \tau^{1.09 + 0.060 \Delta R_\text{GC}} ,  
\end{split}
\end{equation}
where $ \Delta R_\text{GC} = \overline{R}_\text{GC} - 8$ kpc,  $\overline{R}_\text{GC} = \frac{(R_\text{GC} +  R_\text{birth})}{2}$, $R_\text{birth}$ is the birth Galactic radius, and $R_\text{GC}$ is the current Galactic radius. Since we cannot determine the birth radii for the sample, we simply use $\overline{R}_\text{GC}$ = $R_\text{GC}$.

The \citet{Ting:2019aa} relation turns out to be problematic for this sample, as the zero-age baseline vertical action $\widehat{J}_{\text{Z},0}(\overline{R}_\text{GC}) = 0.95$ kpc km s$^{-1}$ is greater than the ${J_\text{Z}}$ values of 66\% of our sample, resulting in negative ages for these sources. 
Applying equation~(\ref{eqn:ting}) to the remaining sources yields vertical action ages of $1.9 \pm 2.7$~Gyr for late-M dwarfs ($1.7 \pm 2.3$~Gyr for thin disk), $2.3 \pm 2.5$~Gyr for L dwarfs ($1.9 \pm 2.1$~Gyr for thin disk), and $0.9 \pm 0.7$~Gyr for T dwarfs.
These values are similar to the age of $2.7 \pm 2.2$ Gyr inferred from vertical action analysis of late-M and L dwarfs with SDSS spectra by \citet{Kiman:2019aa}, and both are considerably younger than the ages inferred from velocity dispersions. These age estimates are likely in error, as the \citet{Ting:2019aa} relations produce {negative} ages for a significant fraction of the stars in our sample. 
We therefore discard these age determinations as absolute measures, but note that the relative consistency of ages for thin disk late-M, L, and T dwarfs is concurrent with our velocity dispersion analysis.

\subsection{Comparison to Simulated Populations \label{sec:compare_simulated_ages}}

\subsubsection{Baseline Simulations \label{sec:sim_baseline}}

To evaluate whether the kinematic ages determined here are consistent with our understanding of the formation and evolution of UCDs, we conducted a Monte-Carlo population simulation for local thin disk late-M, L, and T dwarfs. We simulated 10$^5$ sources assuming a uniform spatial distribution and a uniform star formation rate over 0.1 Gyr $\leq \tau \leq$ 9 Gyr, with masses sampling the range  $0.01\, M_{\odot} \leq M \leq 0.15\, M_{\odot}$\added{\footnote{The choice of lowest-mass 0.01 M$_{\odot}$ is limited by evolutionary models for the field sample. \citet{Saumon:2008aa} models have the lowest-mass 0.01 M$_{\odot}$ that evolve to 10~Gyr. A 10 M$_\text{Jup}$ T dwarf is relatively young (Baraffe 2003 models for $T_\mathrm{eff}$ = 953~K and mass = 0.01 M$_{\odot}$ = 100~Myr), so it is unlikely that lower mass objects will be significant contributors to the field sample.}}
drawn according to a power-law initial mass function
\begin{equation}
\frac{dN}{dM} \propto M^{-\alpha},
\end{equation}
where $M$ is mass, $N$ is the number density of stars in the local volume, and $\alpha$ is a power-law index. We chose a baseline value $\alpha = 0.5$, which is roughly consistent with UCD populations in young clusters \citep{Bastian:2010aa} and the local Galactic environment \citep{Kirkpatrick:2019aa, Kirkpatrick:2021aa}. We used the evolutionary models of \cite{Baraffe:2003aa} to convert ages and masses into effective temperatures ($T_\mathrm{eff}$), and assigned spectral types (SpT) using the empirical SpT-$T_\mathrm{eff}$ relation of \cite{Filippazzo:2015aa}.  \textit{UVW} space velocities in the LSR were then assigned based on age, by assuming\footnote{This is consistent with our kinematic analysis for the observed sample; see Sections \ref{ages} and  \ref{ages_lateML} for more details.} \textit{UVW} = 0 at $\tau$ = 0 and drawing from normal distributions in all three components using widths based on the \cite{Aumer:2009aa} age-dispersion relations for each axis of motion. 
For the \textit{V} velocity, we also added a time-dependent asymmetric drift term \citep{Aumer:2009aa}:
\begin{equation}
V_a = \frac{ - \sigma_U^2(\tau)}{74 \, \mbox{km} \, \mbox{s}^{-1}} = 23.7 \Big( \frac{\tau}{10 \text{ Gyr}} \Big)^{0.614} \text{km s$^{-1}$}.
\end{equation}

\begin{figure*}[!htbp]
\centering
\gridline{\fig{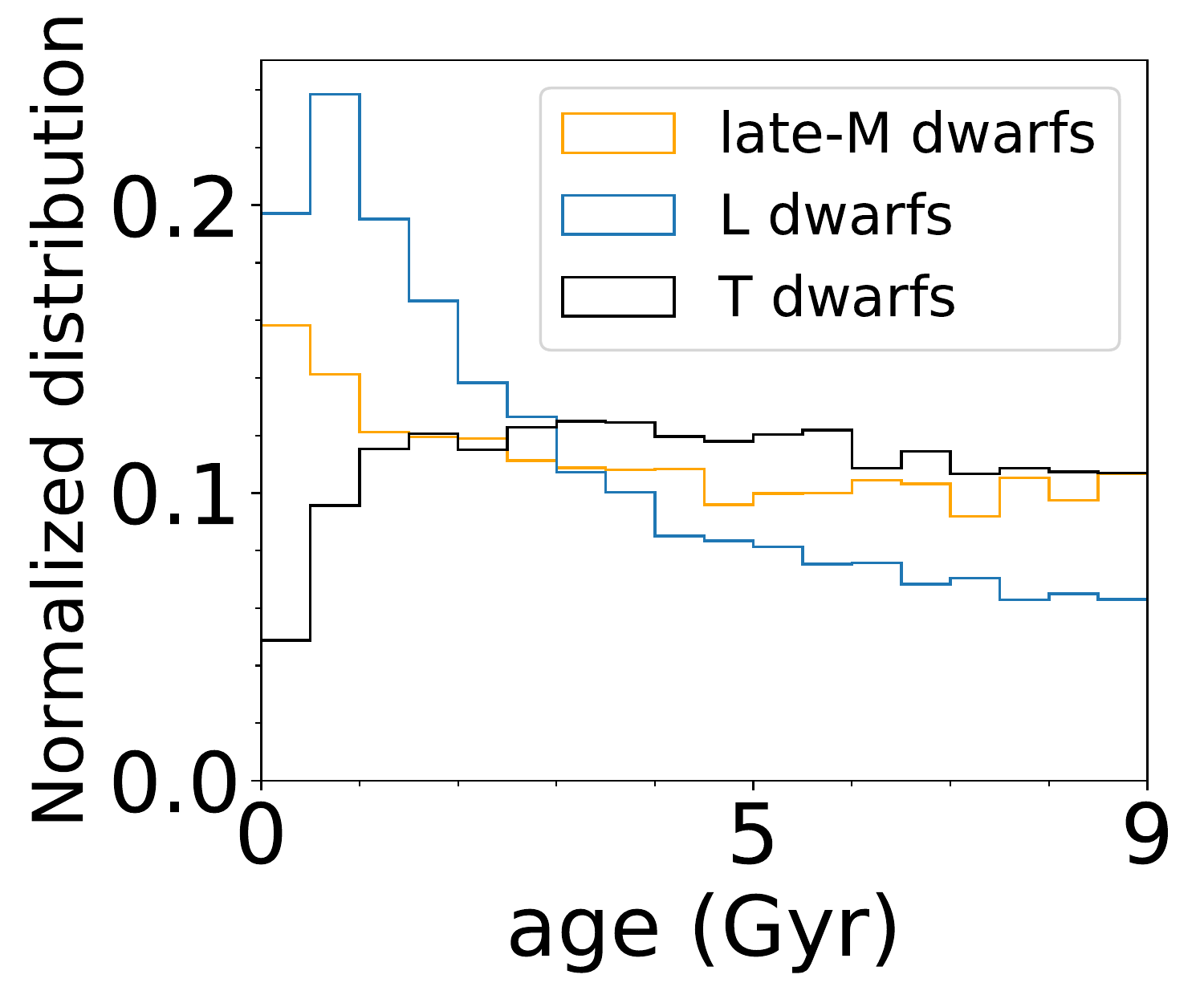}{0.5\textwidth}{}
\fig{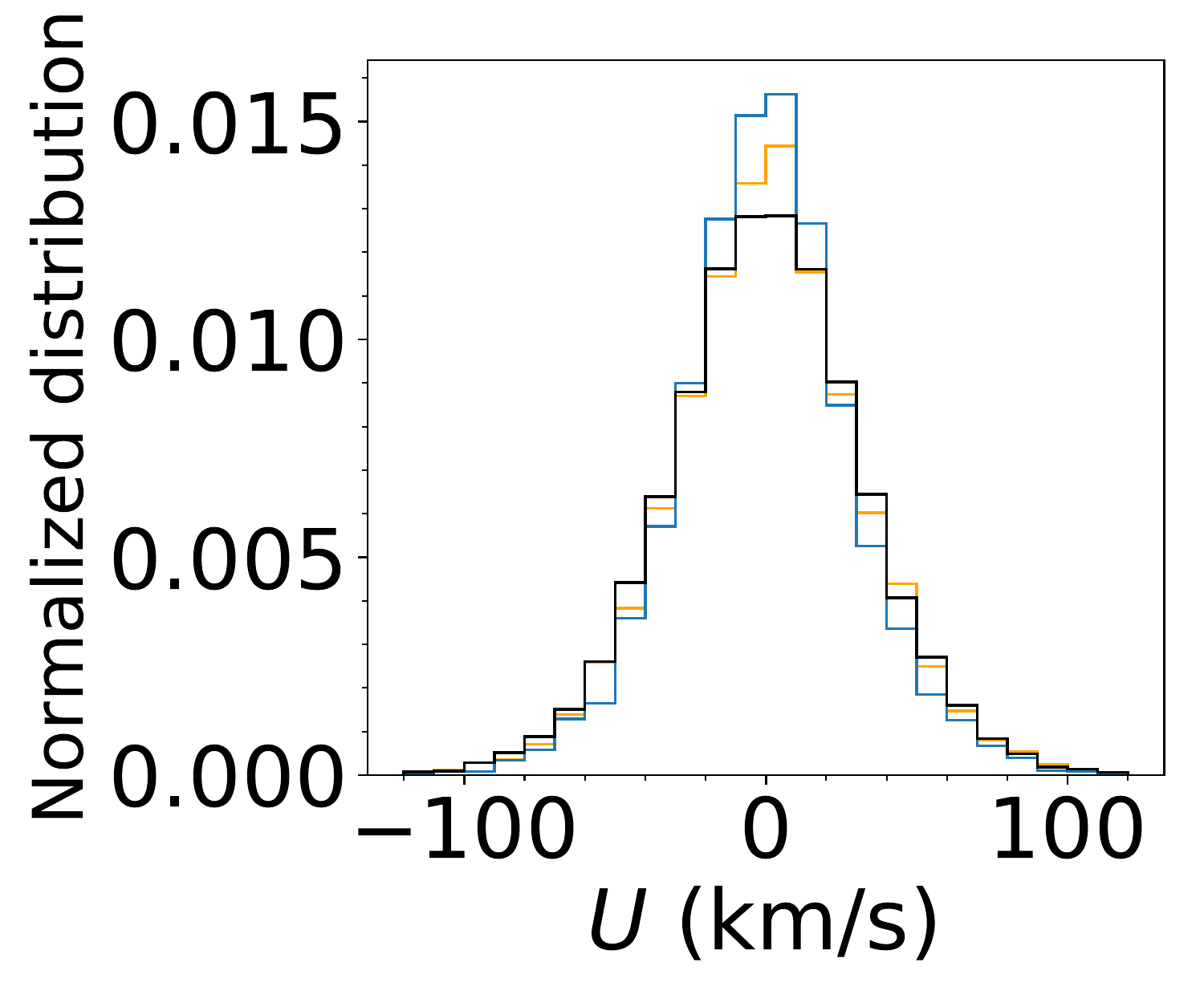}{0.5\textwidth}{}
}
\gridline{\fig{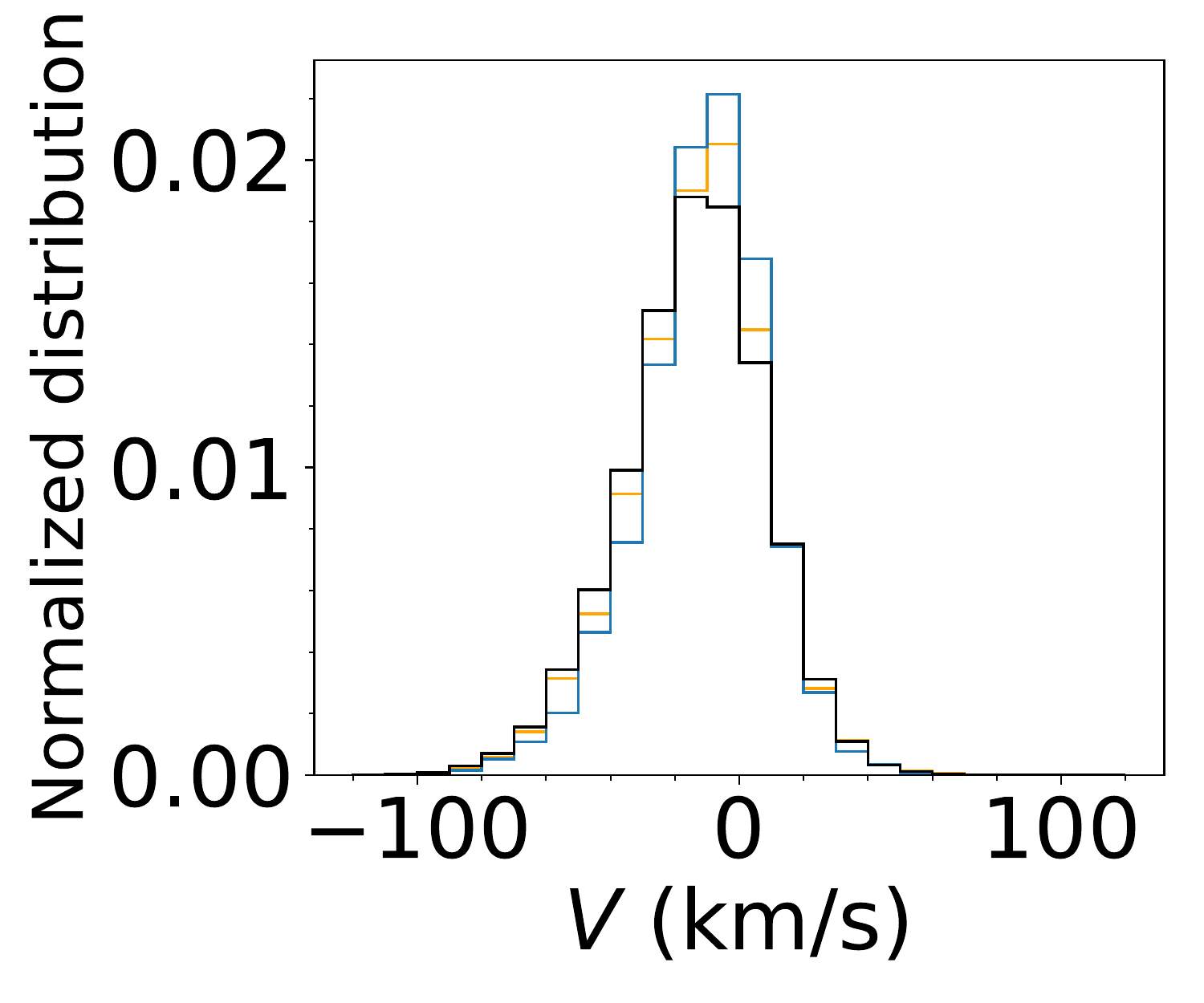}{0.5\textwidth}{}
\fig{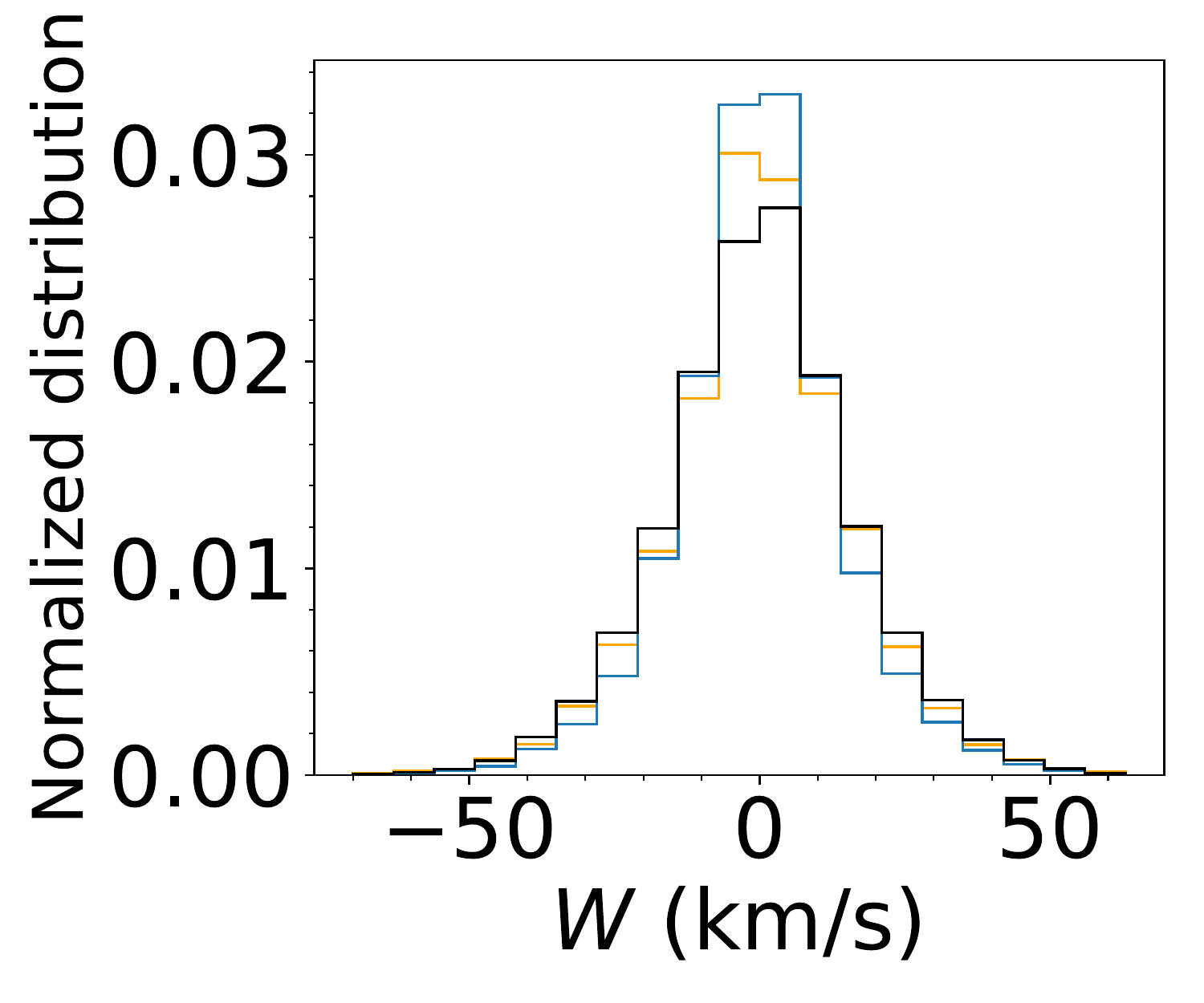}{0.5\textwidth}{}
}
\caption{Normalized distributions of ages (upper-left) and \textit{UVW} space motions for our baseline simulated population. 
Distributions are segregated between  
late-M (orange), L (blue), and T dwarfs (black). \label{fig:simulated_population}}
\end{figure*}

From this simulated (and assumed volume-complete) velocity sample, we computed the kinematic ages of UCD spectral subgroups using the same analysis as that of our observational sample.
We simulated sample selection effects by making 1,000 random draws of $N_\text{s} - 1$ sources from the simulated population, where $N_\text{s}$ corresponds to the sizes 
of our late-M, L, and T dwarf RV samples (63, \replaced{66}{65}, and \replaced{35}{36} sources, respectively)\added{ and $-1$ corresponds to the Jackknife sampling}. 
The median values and standard deviations from this sampling are summarized in Table \ref{table:simulated_population} and Figure \ref{fig:simulated_population}. 

\begin{deluxetable}{lccccc}
\tablecaption{Simulated UCD Population Parameters \label{table:simulated_population}}
\tablewidth{700pt}
\tabletypesize{\scriptsize}
\tablehead{\colhead{} & \colhead{} & \colhead{} & \colhead{} & \colhead{} & \colhead{} \\
\colhead{} & \colhead{Simulated} & \colhead{} & \colhead{} & \colhead{} & \colhead{Kinematic} \\
\colhead{Sample} & \colhead{Age\tablenotemark{a}} & \colhead{$\langle U \rangle$} & \colhead{$\langle V \rangle$} & \colhead{$\langle W \rangle$} & \colhead{Age} \\ 
\colhead{} & \colhead{(Gyr)} & \colhead{(km s$^{-1}$)} & \colhead{(km s$^{-1}$)} & \colhead{(km s$^{-1}$)} & \colhead{(Gyr)}
} 
\startdata
late-M dwarfs & 4.1  & 1 $\pm$ 4 & $-$13 $\pm$ 3 & 0 $\pm$ 2 & 4.1 $\pm$ 0.8 \\ 
L dwarfs &  2.8 & 0 $\pm$ 4 & $-$11 $\pm$ 3 & 0.1 $\pm$ 1.7 & 3.1 $\pm$ 0.7 \\ 
T dwarfs & 4.6  & 0 $\pm$ 6 & $-$14 $\pm$ 4 & 0 $\pm$ 3 & 4.3 $\pm$ 1.2
 \enddata
\tablecomments{Ages for simulated populations are computed from equation~(\ref{eq:AumerBinney2009}) using the parameters in \citet{Aumer:2009aa}, with sampling errors accounted for using a Jackknife test statistic; see Section~\ref{sec:compare_simulated_ages}. 
\tablenotetext{a}{Median simulation age $<\tau>$}}
\end{deluxetable}

The resulting kinematic ages are $4.1 \pm 0.8$ Gyr, $3.1 \pm 0.7$ Gyr, and $4.3 \pm 1.2$ Gyr for late-M, L, and T dwarfs, respectively\replaced{, which}{. These ages} are statistically equivalent, although the L dwarfs are \added{about} 1~Gyr younger on average.
The relatively large uncertainties highlight the importance of sampling effects. 
Comparing to the kinematic ages of the observed thin disk sample, we find excellent agreement 
for the late-M and T dwarfs, 
\replaced{and a modest discrepancy ($\Delta\tau$ = 0.9~Gyr, 1.3$\sigma$ deviation) for the L dwarfs. Again, by more carefully removing potentially thick disk sources from the L dwarf sample, the previously identified discrepancy between observed and simulated ages for these objects appears to be resolved.}
{but significant disagreement ($\Delta\tau$ = 4.0~Gyr, 5.0$\sigma$ deviation) compared to the full sample of L dwarfs.
However, if we exclude potential thick disk sources ($P[\mathrm{TD}]/P[\mathrm{D}]$ $>$ 1), the discrepancy for the L dwarfs is significantly reduced ($\Delta\tau$ = 1.1~Gyr) with only marginal significance of deviation (1.4$\sigma$).
Again, accounting for potential thick disk sources appears to mostly resolve the previously identified discrepancy between observed and simulated ages for L dwarfs, although our baseline simulations still predict a young L dwarf population compared to late-M and T dwarfs.}

\subsubsection{Variations on Simulated Populations} \label{sec:other_simulated_pop}

The \added{modest dis}agreement between simulated and observed kinematics \replaced{is not perfect, suggesting}{suggests} the need for some fine-tuning of simulation parameters as they relate to the kinematics of late-M, L, and T dwarfs.
To explore this, we evaluated the \replaced{individual contributions of the following parameters in comparison to the baseline parameters}{the influence of following simulation parameters on the ages of the late-M, L, and T dwarf disk population}: 

\begin{itemize}
    \item {\it Star formation rate (SFR)}: In addition to a uniform star formation rate, we examined an exponentially declining birth rate, SFR $\propto e^{\gamma\tau}$, where $\gamma$ = 0.117 \added{Gyr$^{-1}$} and $\tau$ is age \added{in Gyr} \citep{Aumer:2009aa}; and a star formation rate that reflects cosmic star formation \added{history}, SFR \replaced{$\propto (1+ z(t))^{\beta}$}{$\propto (1+ z(\tau))^{\beta}$}, where $z$ is the redshift and $\beta$ = 3.5 \citep{Rujopakarn:2010aa, Planck-Collaboration:2016aa}. 
    \item {\it Mass function}: We examined additional power-law relations with $\alpha = -1.5$, $-0.5$, and +1.5; and \deleted{evaluated} two cases of an age-dependent mass function: $\alpha$ = 0.0 $\Rightarrow$ +1.0 and $\alpha$ = +1.0 $\Rightarrow$ 0.0 \replaced{at 3~Gyr}{10~Gyr in the past}\footnote{The cosmological star formation rate peaks at $z\sim2$, corresponding to a cosmic age of $\sim$3~Gyr ($\sim$10~Gyr in the past) under the standard $\Lambda$CDM cosmology \citep{Madau:2014aa}. The star formation rate of the inner Milky Way (R $\leq$ 10~kpc) peaks at $\sim$z = 1--3 ($\sim$8--12 Gyr ago) \citep{Haywood:2016aa}.}
    (cf. \citealt{Burgasser:2015ac})\replaced{;  
    as well as}{. We also examined} a log-normal mass function from \citet{2003PASP..115..763C}. 
    \item {\it Choice of brown dwarf evolution model}: In addition to the \citet{Baraffe:2003aa} evolutionary models, we evaluated the models of \citet{Burrows:2001aa}, \citet{Saumon:2008aa}, \citet{Marley:2018aa}, and \citet{Phillips:2020aa}.
    \item {\it Maximum simulation age}: In addition to the baseline maximum age of 9~Gyr, We considered maximum ages of 6~Gyr and 12~Gyr.
    \item {\it Minimum brown dwarf mass (MBDM)}: The lowest-mass brown dwarfs are also the youngest in the L dwarf phase, which may skew simulated ages downward. We considered additional minimum masses for our simulation of 0.02~M$_{\odot}$ and 0.03~M$_{\odot}$ to explore this effect.
\end{itemize}

For computational expediency, the number of simulated sources used for these simulations was 10$^4$ versus 10$^5$ for our baseline simulations. Varying each of these parameters individually, we produced 578 additional simulations. For \citet{Marley:2018aa} and \citet{Phillips:2020aa} models, we only ran the simulations for uniform star formation rate, $\alpha$ = 0.5 and minimum brown dwarf mass of 0.01~M$_{\odot}$.
We quantified the agreement of the simulations to the observations using a $\chi^2$ statistic:
\begin{equation}
\chi^2 = \Sigma_{i} \frac{ ( \tau_{\mathrm{obs}, i} - \tau_{\mathrm{sim}, i} )^2 }{ \sigma_{ \tau_{\mathrm{obs}, i}}^2 + \sigma_{\tau_{\mathrm{sim}, i}}^2 }
\end{equation}
with $\tau_{\mathrm{obs}}$ and $\tau_{\mathrm{sim}}$ being the observed and simulated ages, $\sigma_{\mathrm{obs}}$ and $\sigma_{\mathrm{sim}}$ the observed and simulated age uncertainties, and $i$ = late-M, L, and T dwarf thin disk samples
\added{(P[TD]/P[D] $\leq$ 1)}. 
A select set of the results discussed here are summarized in Table~\ref{table:simulated_population_select}\replaced{, and a}{. A}ll simulations are fully compiled in Table~\ref{table:simulated_population_all}.

\startlongtable
\begin{deluxetable*}{cccccccccc}
\tablecaption{Select Simulated UCD Population Ages Under Different Assumptions \label{table:simulated_population_select}}
\tablewidth{700pt}
\tabletypesize{\scriptsize}
\tablehead{
 \\
\colhead{$\tau$} & \colhead{Star formation} & \colhead{$\alpha$} & \colhead{Models} & \colhead{MBDM} & \colhead{late-M dwarf age} & \colhead{L dwarf age} & \colhead{T dwarf age} & \colhead{$\chi^2$} & \colhead{Note}\\
\colhead{(Gyr)} & \colhead{(rate)} & \colhead{} & \colhead{} & \colhead{($M_{\odot}$)} & \colhead{(Gyr)} & \colhead{(Gyr)} & \colhead{(Gyr)} & 
} 
\startdata
\multicolumn{10}{c}{Observations} \\
\cline{1-10}
\multicolumn{5}{c}{ALL SOURCES\dotfill} & 4.9 $\pm$ 0.3 & \replaced{7.0}{7.1} $\pm$ 0.4 & 3.5 $\pm$ 0.3 & \nodata \\
\multicolumn{5}{c}{NOT THICK DISK ($P(\text{TD})/P(\text{D}) < 10$)\dotfill} & 4.1 $\pm$ 0.3 & \replaced{5.7}{5.8} $\pm$ 0.3 & 3.5 $\pm$ 0.3 & \nodata \\
\multicolumn{5}{c}{THIN DISK ($P(\text{TD})/P(\text{D}) < 1$)\dotfill} & 4.1 $\pm$ 0.3 & \replaced{4.1}{4.2} $\pm$ 0.3 & 3.5 $\pm$ 0.3 & \nodata \\
\cline{1-10}
\multicolumn{10}{c}{Simulations} \\
\cline{1-10}
9 & uniform & $0.5$ & B03 & 0.01 & 4.1 $\pm$ 0.8 & 3.1 $\pm$ 0.7 & 4.3 $\pm$ 1.2 & 2.1 & a \\ 
9 & uniform & $0.5$ & B01 & 0.01 & 3.5 $\pm$ 0.7 & 3.1 $\pm$ 0.6 & 4.2 $\pm$ 1.2 & 3.0 \\ 
9 & uniform & $0.5$ & S08 & 0.01 & 4.3 $\pm$ 0.8 & 3.4 $\pm$ 0.7 & 4.5 $\pm$ 1.2 & 1.7 \\ 
9 & uniform & $0.5$ & M19 & 0.01 & 1.6 $\pm$ 0.3 & 3.1 $\pm$ 0.6 & 4.0 $\pm$ 1.1 & 42.1 \\ 
9 & uniform & $0.5$ & P20C & 0.01 & 0.2 $\pm$ 0.0 & 2.6 $\pm$ 0.5 & 4.2 $\pm$ 1.2 & 172.0 \\ 
9 & uniform & $0.5$ & P20NW & 0.01 & 0.3 $\pm$ 0.0 & 2.6 $\pm$ 0.6 & 4.3 $\pm$ 1.2 & 162.3 \\ 
9 & uniform & $0.5$ & P20NS & 0.01 & 0.4 $\pm$ 0.0 & 2.7 $\pm$ 0.6 & 4.1 $\pm$ 1.2 & 154.4 \\ 
9 & exponential & $0.5$ & B03 & 0.01 & 4.5 $\pm$ 0.8 & 3.8 $\pm$ 0.8 & 4.7 $\pm$ 1.2 & 1.3 \\ 
9 & exponential & $0.5$ & B01 & 0.01 & 3.6 $\pm$ 0.6 & 3.7 $\pm$ 0.7 & 4.9 $\pm$ 1.4 & 1.7 \\ 
9 & exponential & $0.5$ & S08 & 0.01 & 4.6 $\pm$ 0.8 & 4.2 $\pm$ 0.8 & 4.8 $\pm$ 1.2 & 1.5 \\ 
9 & log-normal & $0.5$ & B03 & 0.01 & 6.8 $\pm$ 1.3 & 6.1 $\pm$ 1.0 & 6.5 $\pm$ 1.6 & 11.6 \\ 
9 & log-normal & $0.5$ & B01 & 0.01 & 7.4 $\pm$ 1.3 & 6.0 $\pm$ 1.0 & 6.2 $\pm$ 1.5 & 12.7 \\ 
9 & log-normal & $0.5$ & S08 & 0.01 & 6.4 $\pm$ 1.0 & 6.2 $\pm$ 1.1 & 6.4 $\pm$ 1.6 & 11.3 \\
9 & exponential & $0.0/1.0/3.0$ & B03 & 0.01 & 5.3 $\pm$ 1.0 & 5.0 $\pm$ 0.9 & 4.2 $\pm$ 1.2 & 2.5 & b \\
9 & exponential & $1.0/0.0/3.0$ & B03 & 0.01 & 4.2 $\pm$ 0.7 & 4.8 $\pm$ 0.8 & 5.2 $\pm$ 1.3 & 2.3 & c  \\ 
9 & uniform & 0.5 & B03 & 0.01 & 4.0 $\pm$ 0.8 & 3.0 $\pm$ 0.6 & 3.8 $\pm$ 1.1 & 2.8 & d\\
9 & uniform & 0.5 & B03* & 0.01 & 4.1 $\pm$ 0.8 & 4.1 $\pm$ 0.8 & 4.4 $\pm$ 1.2 & 0.5 & e\\
9 & uniform & 1.5/$-$0.5/3.0 & B03 & 0.01 & 3.4 $\pm$ 0.6 & 3.5 $\pm$ 0.7 & 5.0 $\pm$ 1.2 & 3.2 & c\\
9 & uniform & 1.5/$-$0.5/4.5 & B03 & 0.01 & 2.9 $\pm$ 0.6 & 3.5 $\pm$ 0.7 & 5.4 $\pm$ 1.4 & 5.6 & c\\
9 & uniform & 1.5/$-$0.5/6.0 & B03 & 0.01 & 3.2 $\pm$ 0.7 & 3.4 $\pm$ 0.7 & 4.3 $\pm$ 1.1 & 2.7 & c\\
\cline{1-10}
\enddata
\tablecomments{Kinematics ages computed using the \cite{Aumer:2009aa} relation and the procedure described in Section \ref{sec:compare_simulated_ages}.
$\tau$ is the maximum age of the sample, $\alpha$ is the mass function power law index ($\frac{dN}{dM} = M^{-\alpha}$), MBDM is the minimum brown dwarf mass. Evolving mass functions are labeled in the order of early $\alpha$, late $\alpha$, and age (in Gyr) of transition. A log-normal mass function from \citet{2003PASP..115..763C} is labeled as ``log-normal''. Star formation rates considered in our simulations: uniform, expoential \citep{Aumer:2009aa}, and cosmic star formation rate \citep{Rujopakarn:2010aa}. Brown dwarf evolution models are B03 \citep{Baraffe:2003aa}, B01 \citep{Burrows:2001aa}, S08 \citep{Saumon:2008aa}, M19 \citep{Marley:2018aa}, and P20 \citep{Phillips:2020aa}. For the last model set, C, NW, and NS stand for chemical equilibrium, weak, and strong chemical disequilibrium, respectively. Note that only substellar models are available in the P20 set. 
See Table~\ref{table:simulated_population_all} in Appendix \ref{appendix:popsim} for the full list of simulations.}
\tablenotetext{a}{Baseline simulation}
\tablenotetext{b}{\replaced{Additional s}{S}imulations with an evolving mass function from top-heavy to bottom-heavy over time using \citet{Baraffe:2003aa} evolutionary models. See Section \ref{sec:compare_simulated_ages} for details.}
\added{\tablenotetext{c}{Simulations with an evolving mass function from bottom-heavy to top-heavy over time using \citet{Baraffe:2003aa} evolutionary models. See Section \ref{sec:compare_simulated_ages} for details.}}
\tablenotetext{d}{Baseline simulation with selection within 20~pc and $J$ or $K$ $<$ 15.5. See Section \ref{subsec:sampleincomplete} for details.}
\tablenotetext{e}{\replaced{Additional s}{S}imulation with an artificial decrease in the HBMM for the \citet{Baraffe:2003aa} evolutionary models by fixing the temperatures of brown dwarfs down to masses of 0.060~M$_{\odot}$ to their 1~Gyr values. See Section \ref{sec:compare_simulated_ages} for details.}
\end{deluxetable*}

\begin{figure*}[!htbp]
\centering
\gridline{\fig{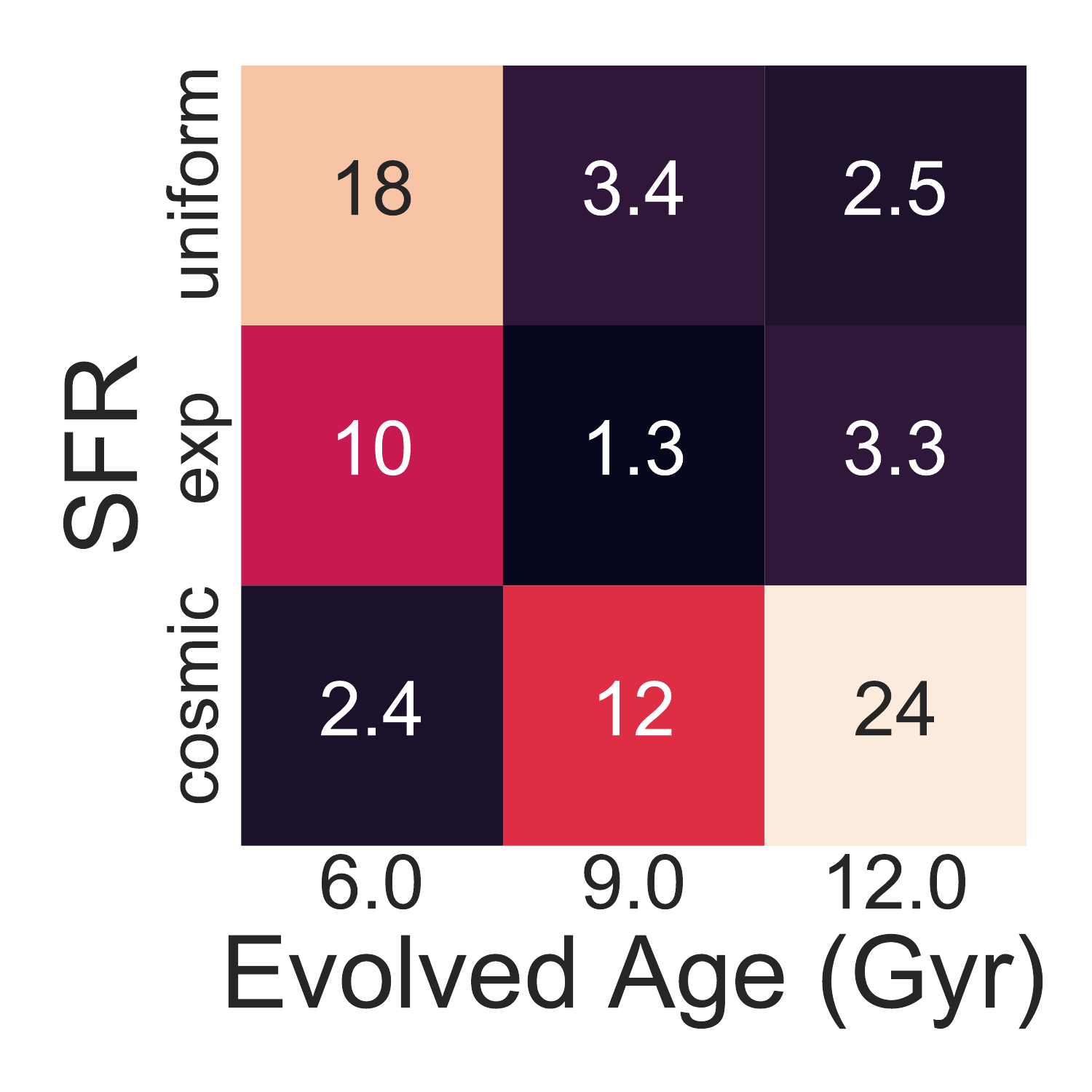}{0.33\textwidth}{}
\fig{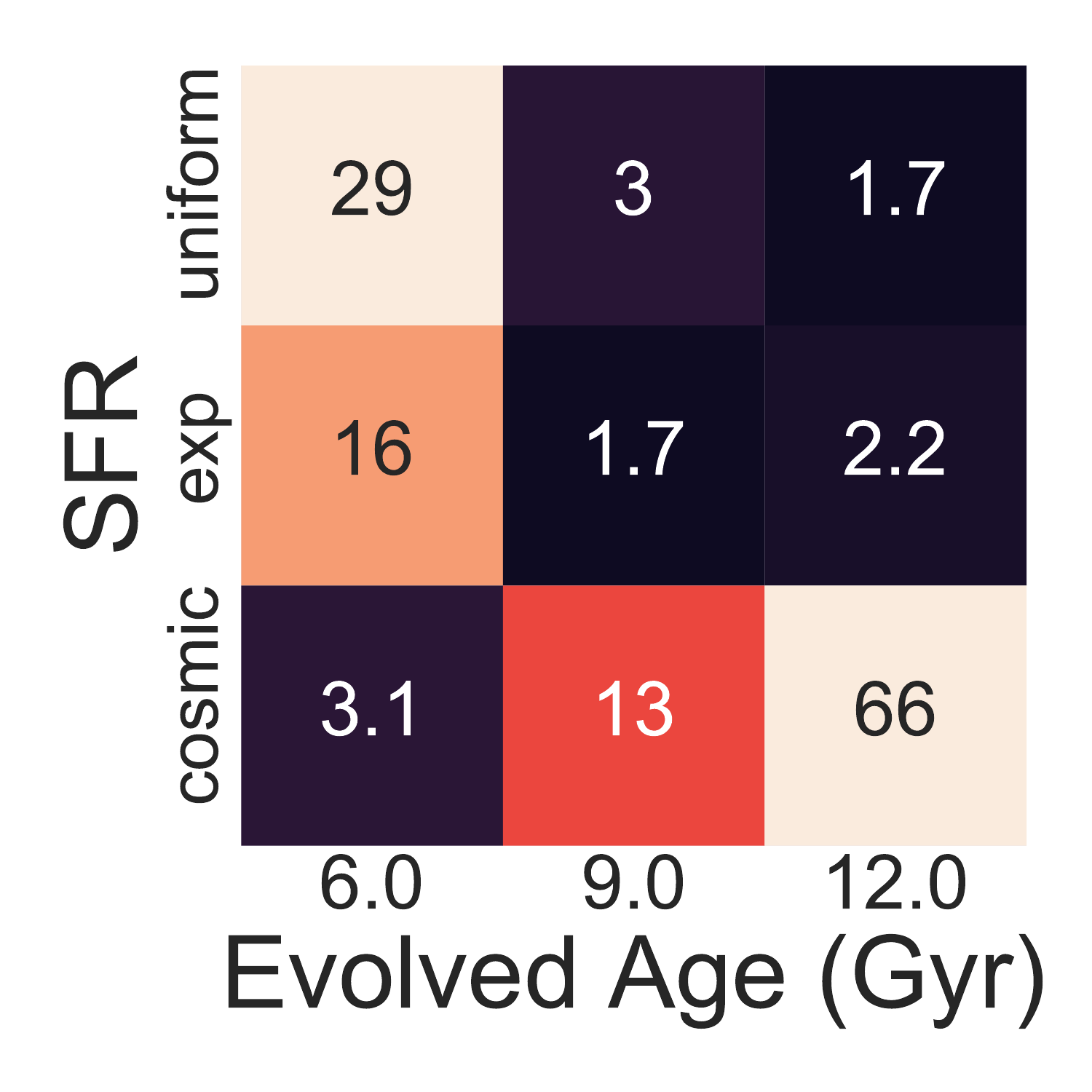}{0.33\textwidth}{}
\fig{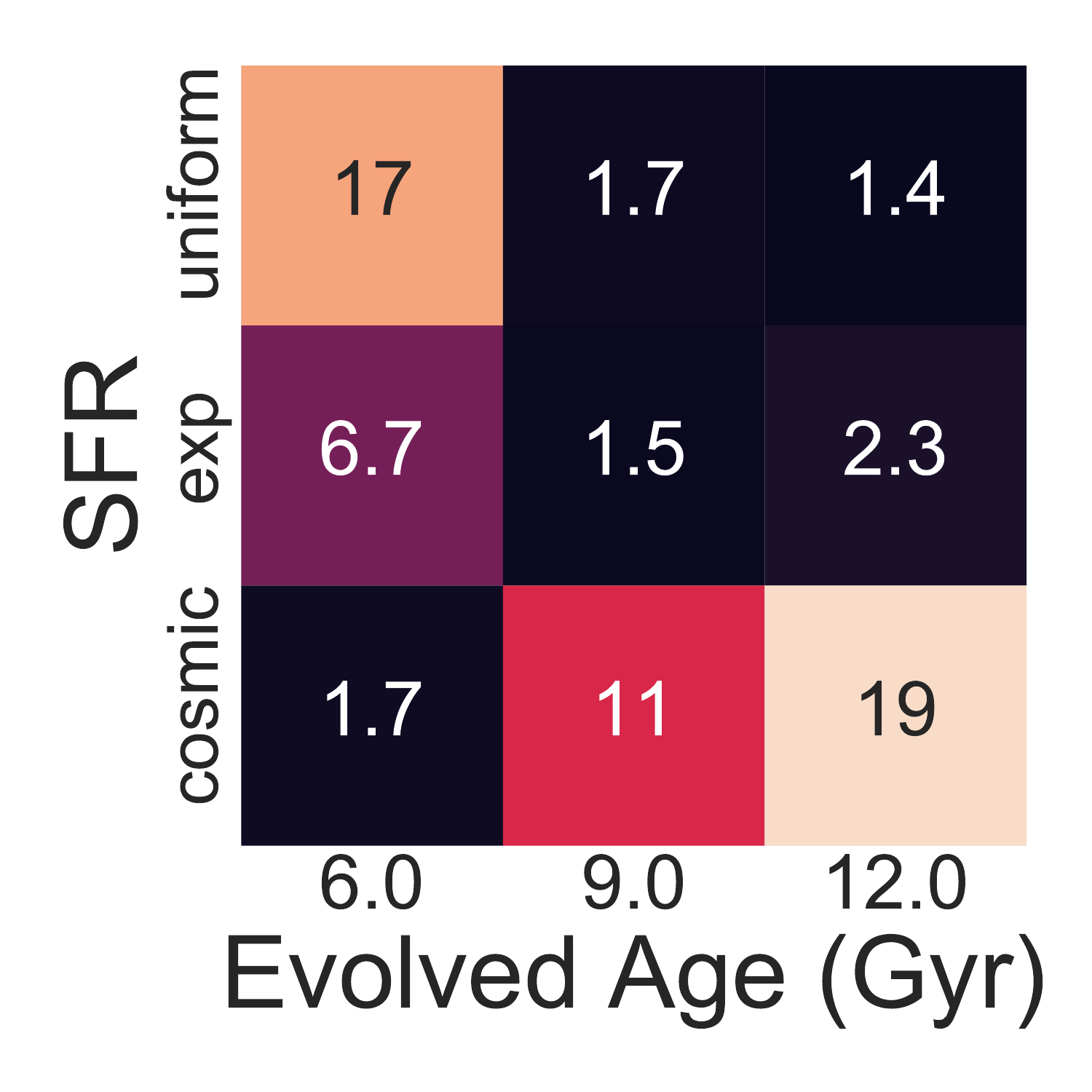}{0.33\textwidth}{}
}
\gridline{\fig{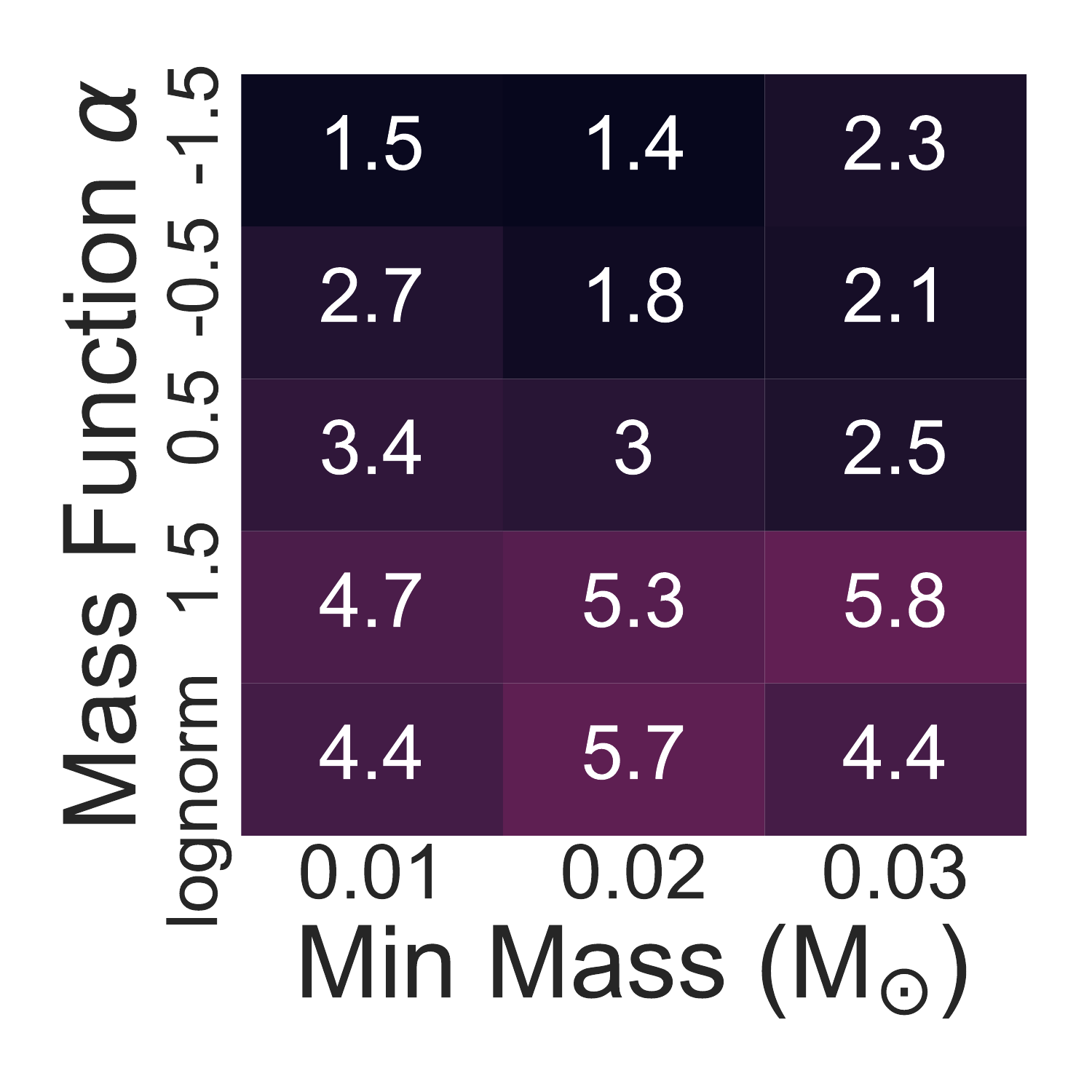}{0.33\textwidth}{}
\fig{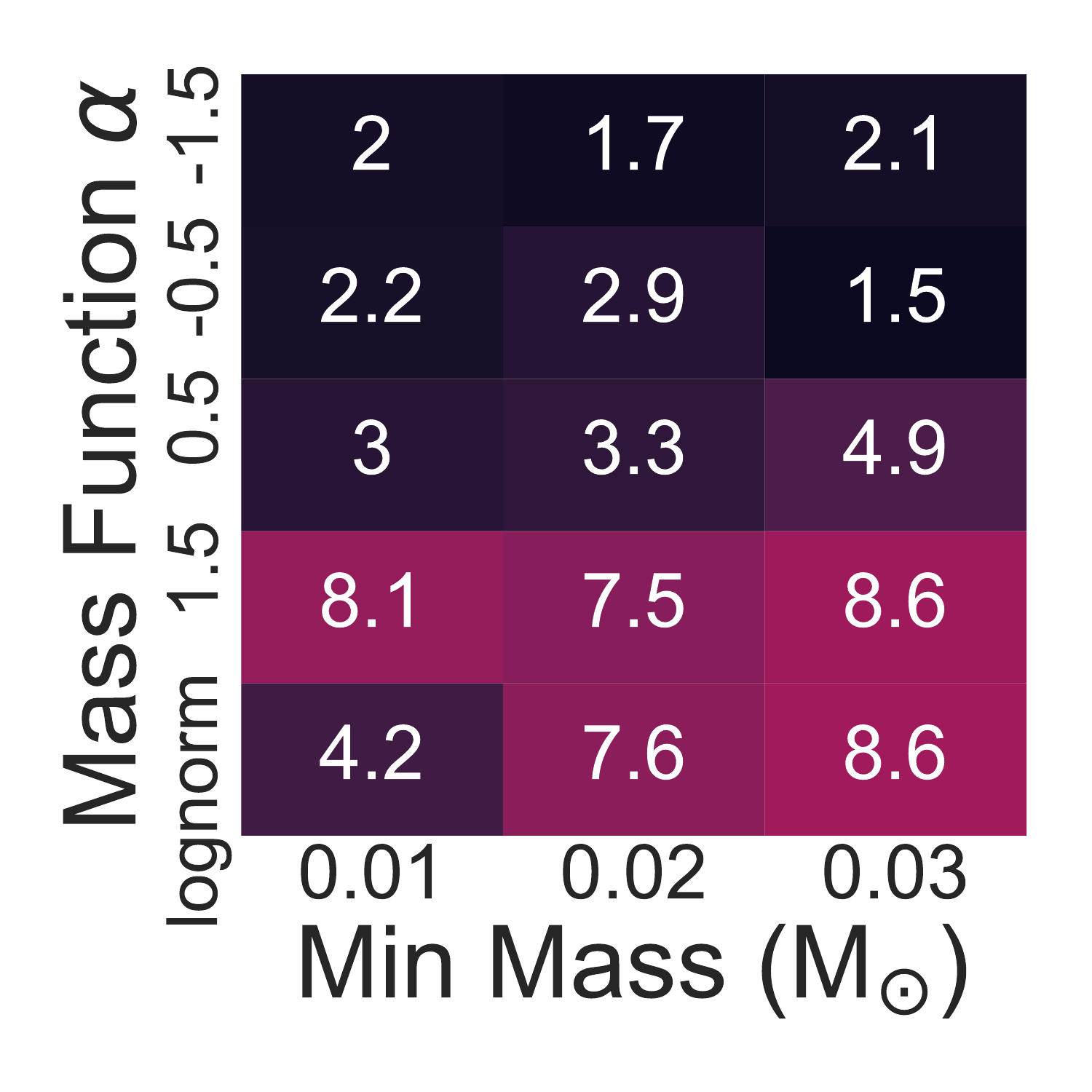}{0.33\textwidth}{}
\fig{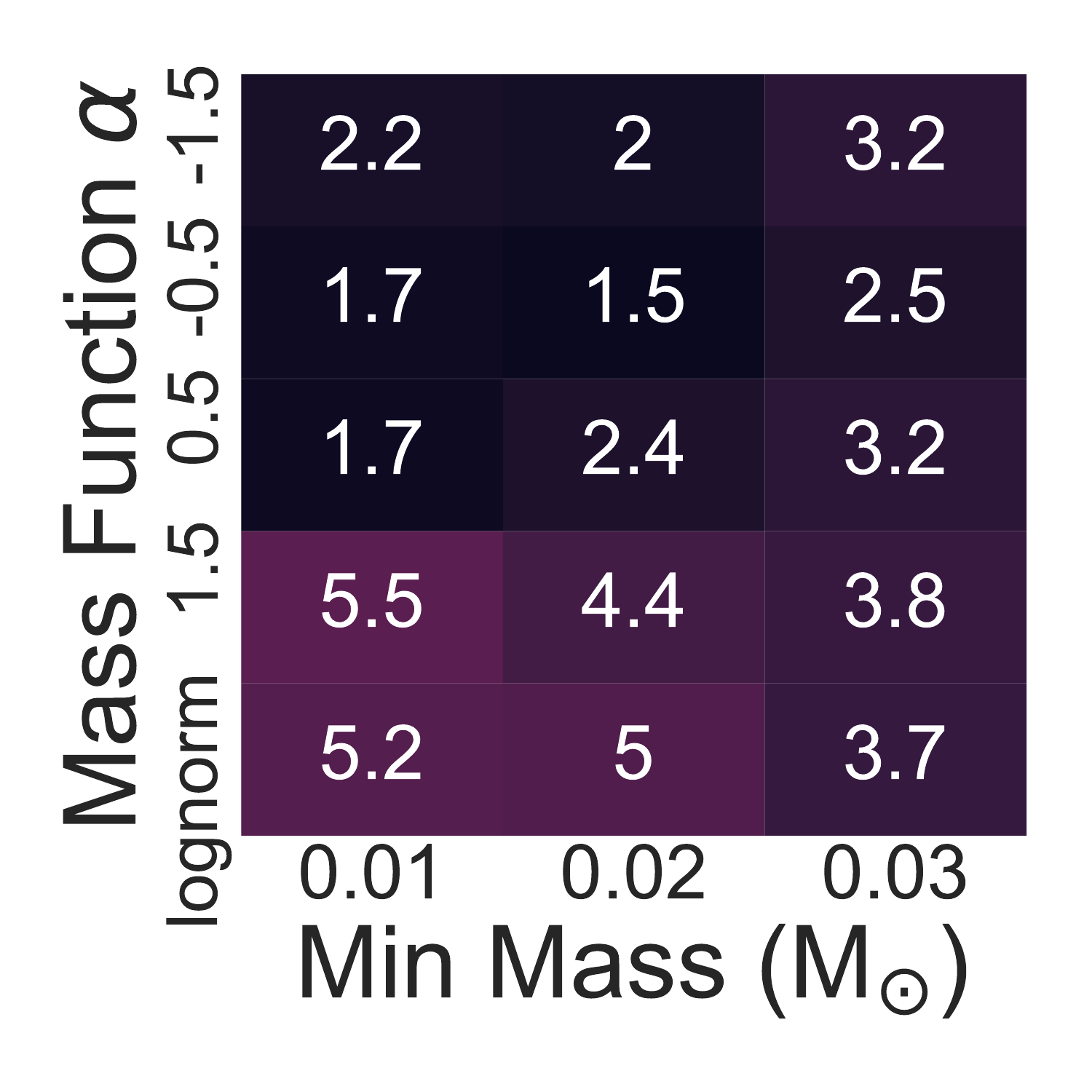}{0.33\textwidth}{}
}
\gridline{\fig{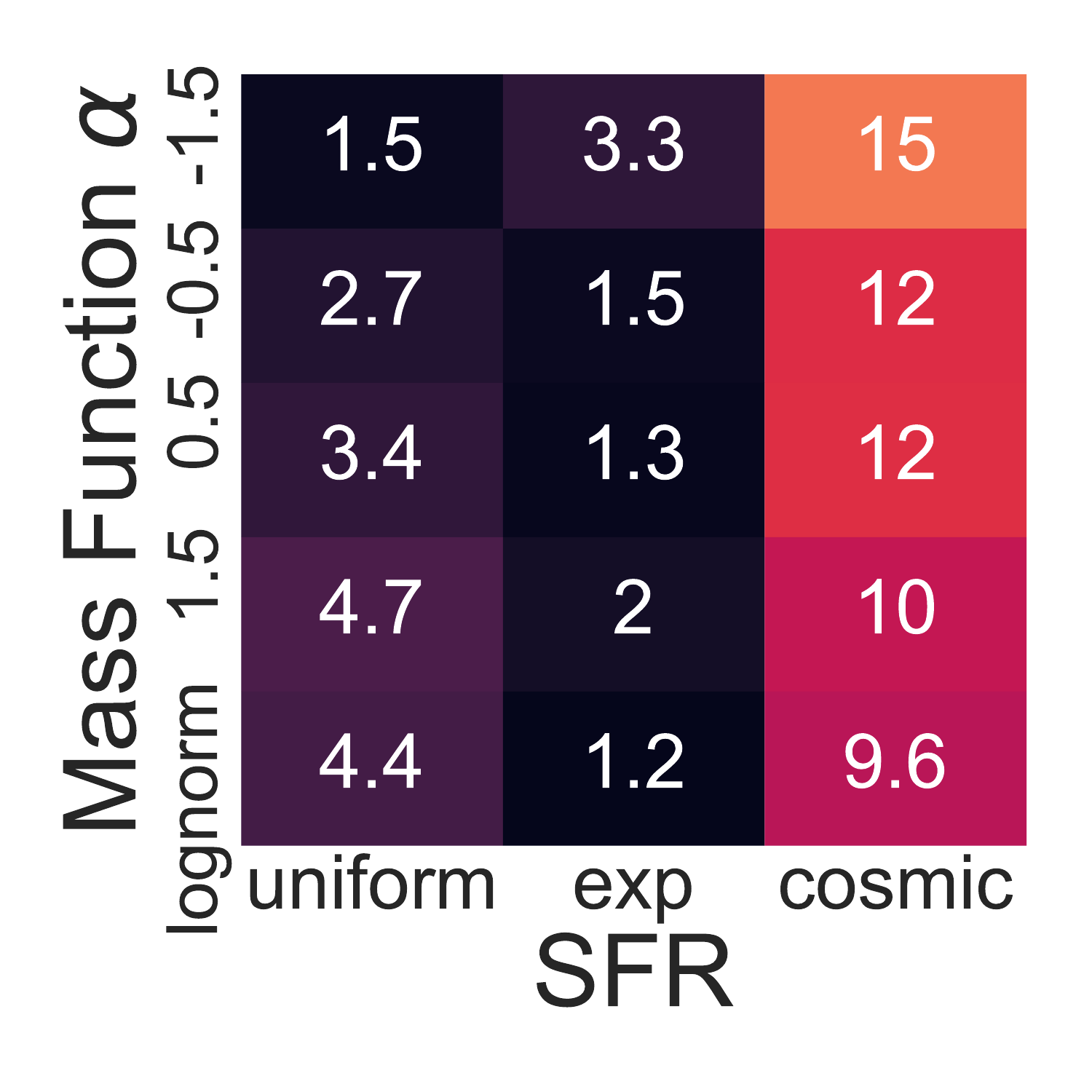}{0.33\textwidth}{}
\fig{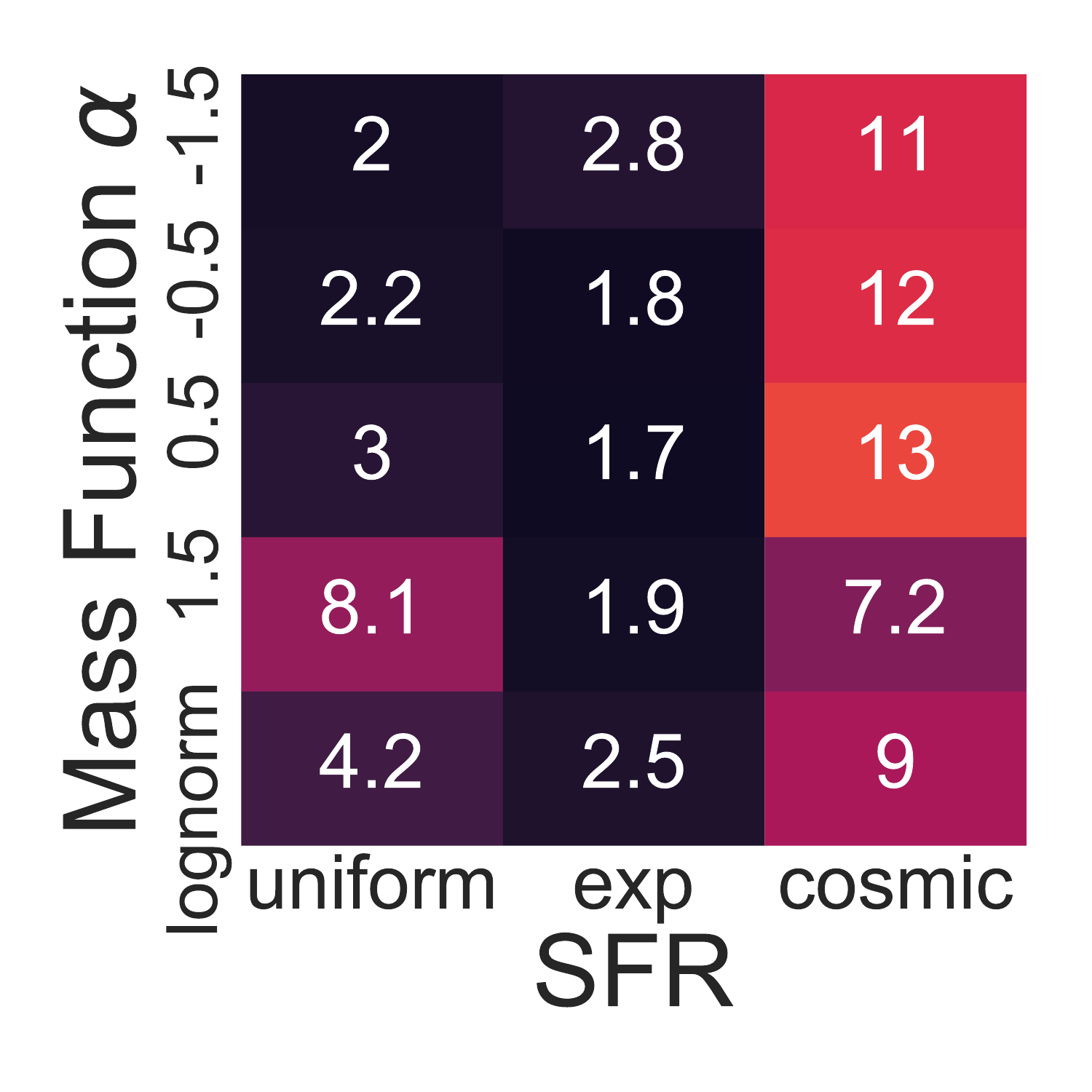}{0.33\textwidth}{}
\fig{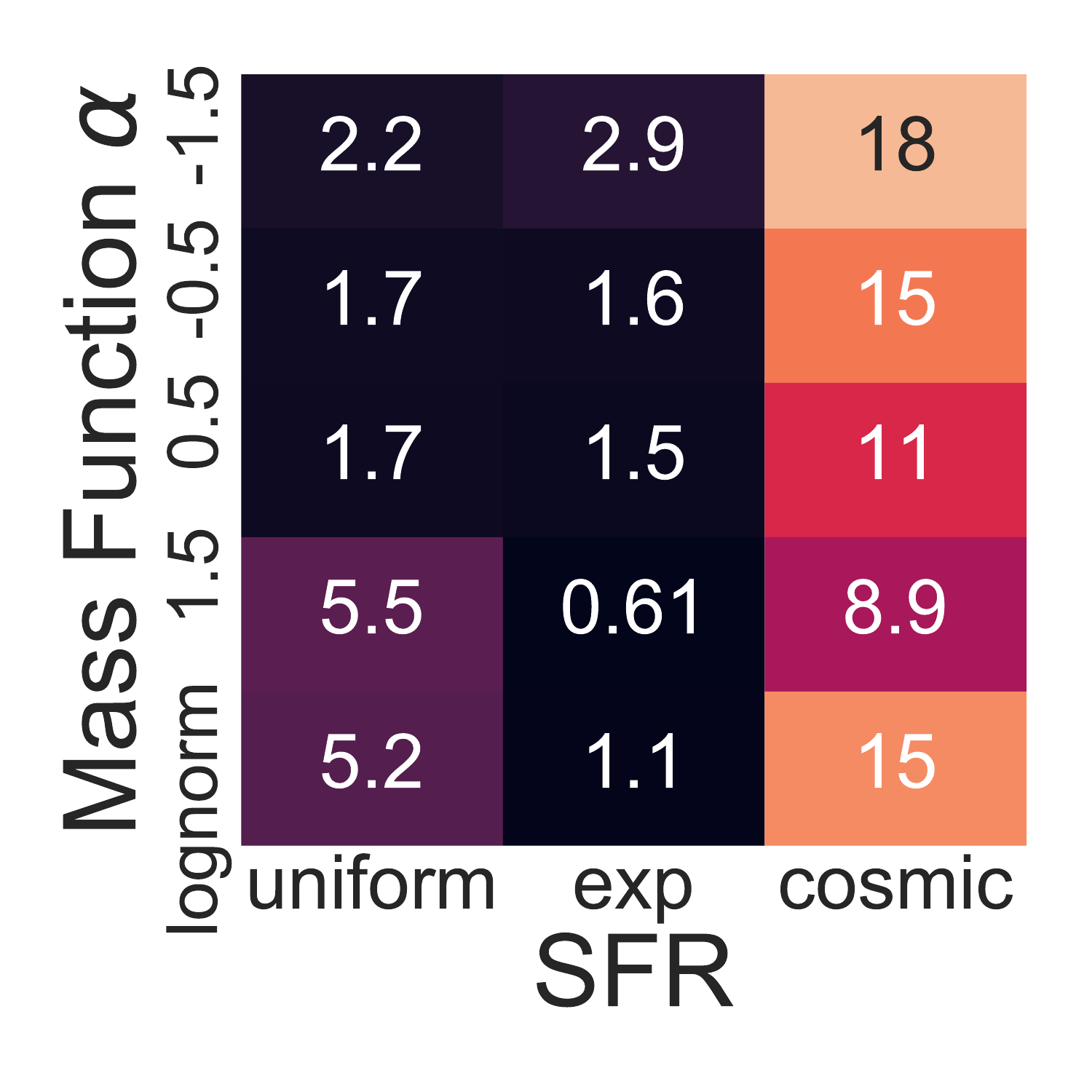}{0.33\textwidth}{}
}
\caption{$\chi^2$ distributions of simulated populations as a function of (1) \textit{top row:} star formation rate (SFR) and evolved age; (2) \textit{middle row:} minimum brown dwarf mass and mass function; (3) \textit{bottom row:} star formation rate (SFR) and mass function across brown dwarf evolution models: \textit{left column:} \citet{Baraffe:2003aa}; \textit{middle column:} \citet{Burrows:2001aa}; \textit{right column:} \citet{Saumon:2008aa}. Lower $\chi^2$ values have darker color.
\label{fig:popsim_heatmap}}
\end{figure*}

Several simulations are consistent with the observed ages, which provides some constraints on the local UCD formation history. To explore these, we first compared the results for a fixed power-law mass function with $\alpha$ = 0.5 and MBDM = 0.01~M$_{\odot}$. 
Figure \ref{fig:popsim_heatmap} shows the $\chi^2$ distribution for different SFRs and evolved ages for each of the \citet{Baraffe:2003aa}, \citet{Burrows:2001aa}, and \citet{Saumon:2008aa} models. The diagonal elements of the simulations show the best agreements with the observed ages, which are cosmic/6~Gyr, exponential/9~Gyr, and uniform/12~Gyr, with the second of these being consistently best between the models. 
These parameters generally produce similar ages for late-M, L, T populations, with L dwarfs being slightly younger and T dwarfs being slightly older. 
The agreement between these parameter sets indicates a clear degeneracy between the SFR and population age that cannot be resolved by this coarse
kinematic age comparison, although we are able to strongly rule out some combinations. For example, the cosmic/12~Gyr SFR/age combination can be ruled out to high probability using a BIC test ($\Delta$BIC $>$ 10, highly significant).

\deleted{\begin{figure*}[!htbp]
\centering
\gridline{\fig{popsim_5_1_baraffe2003.pdf}{0.33\textwidth}{}
\fig{popsim_5_1_burrows2001.pdf}{0.33\textwidth}{}
\fig{popsim_5_1_saumon2008.pdf}{0.33\textwidth}{}
}
\caption{$\chi^2$ distributions of simulated populations as a function of star formation rate (SFR) and evolved age across brown dwarf evolution models, assuming mass function $\alpha$ = 0.5: \textit{left:} \citet{Baraffe:2003aa}; \textit{middle:} \citet{Burrows:2001aa}; \textit{right:} \citet{Saumon:2008aa}. Lower $\chi^2$ values have darker color. 
\label{appendix:popsim1}}
\end{figure*}}

\begin{figure*}[!htbp]
\centering
\gridline{\fig{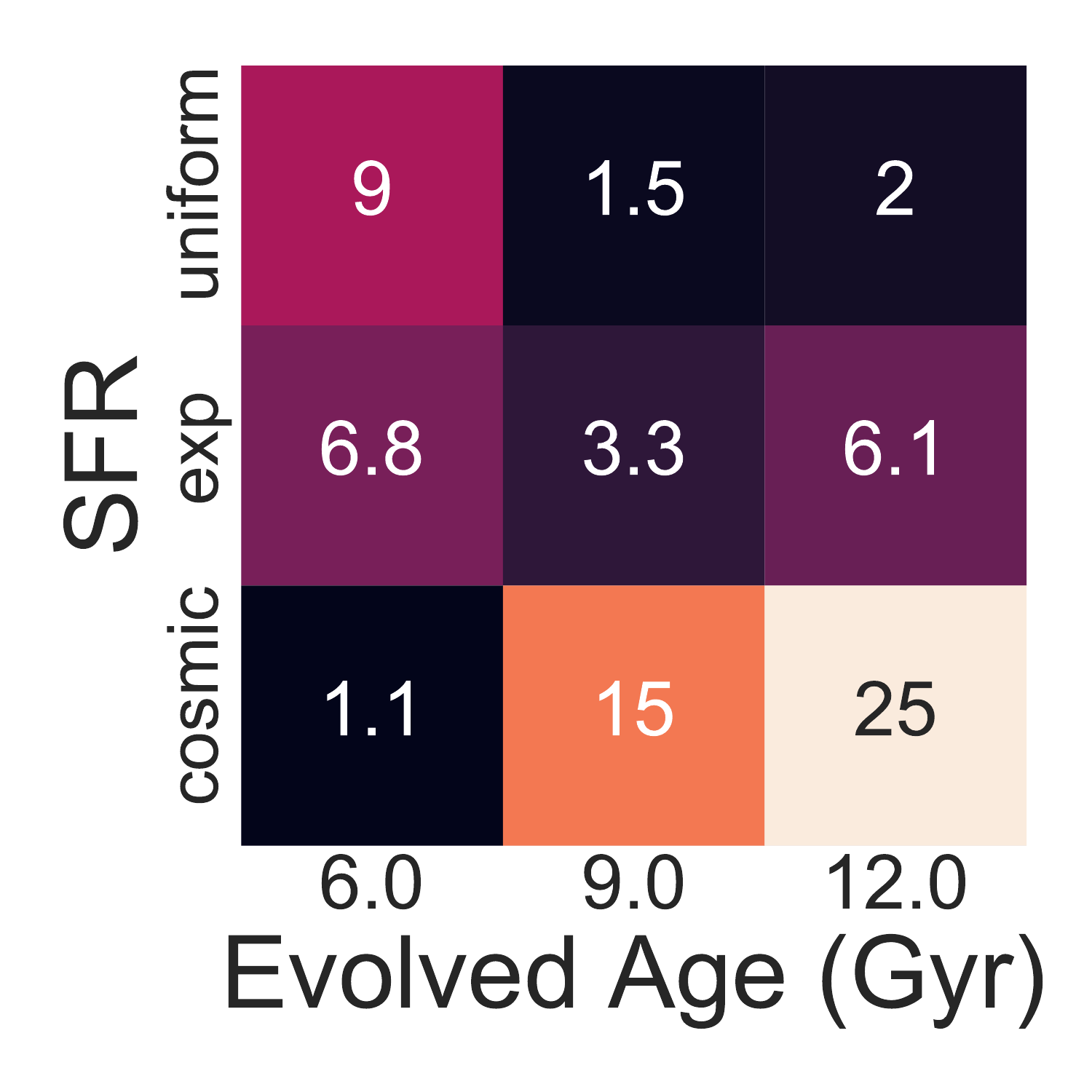}{0.33\textwidth}{}
\fig{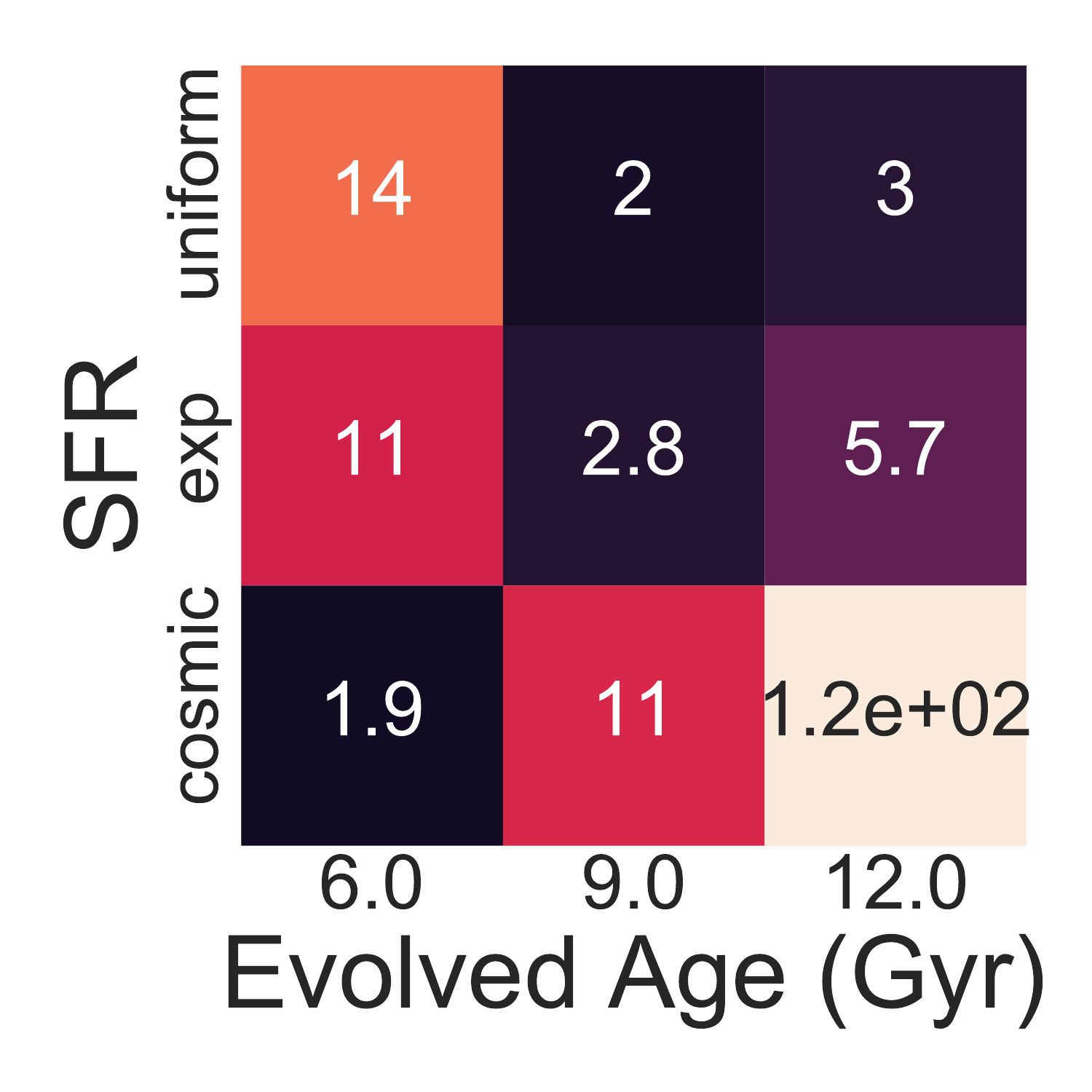}{0.33\textwidth}{}
\fig{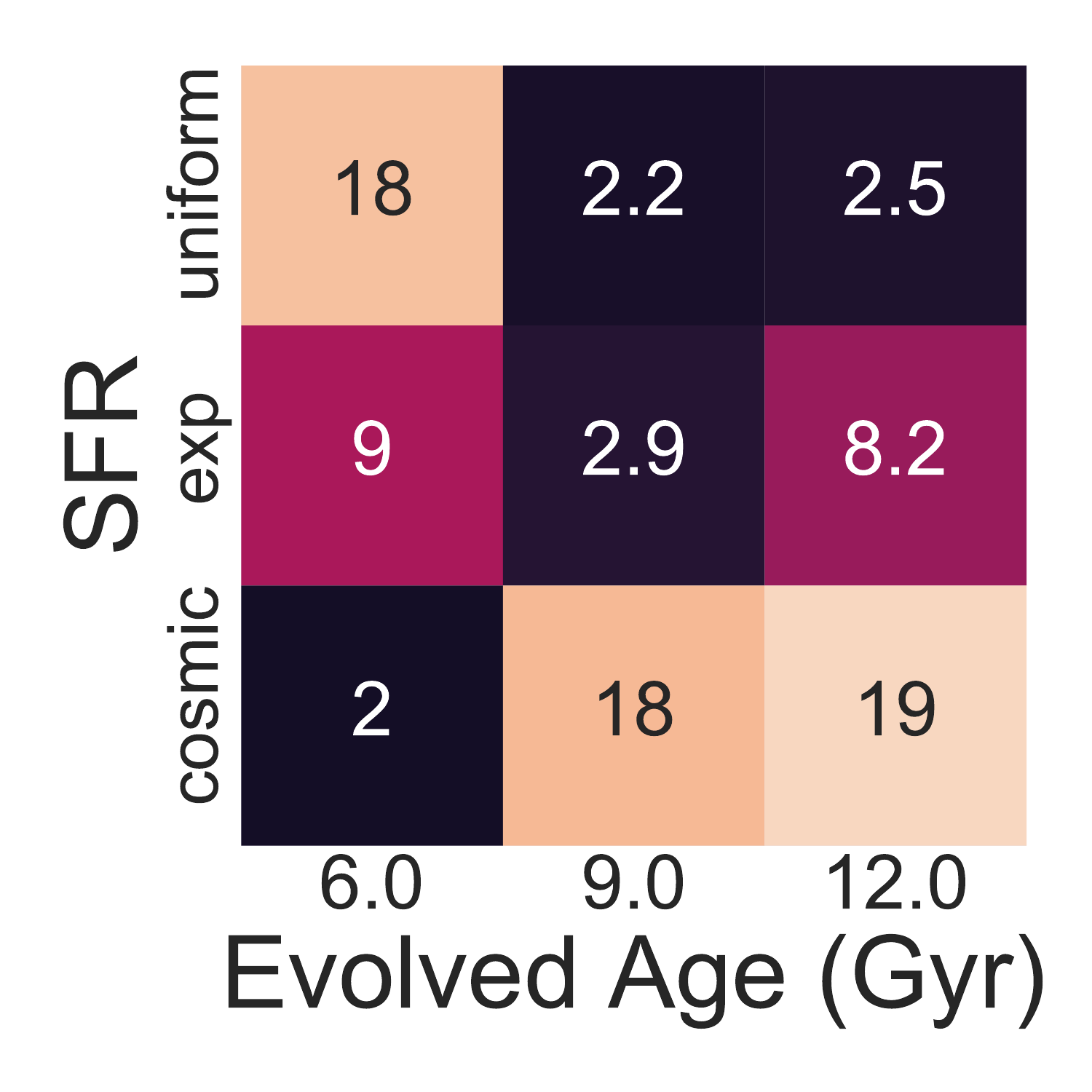}{0.33\textwidth}{}
}
\gridline{\fig{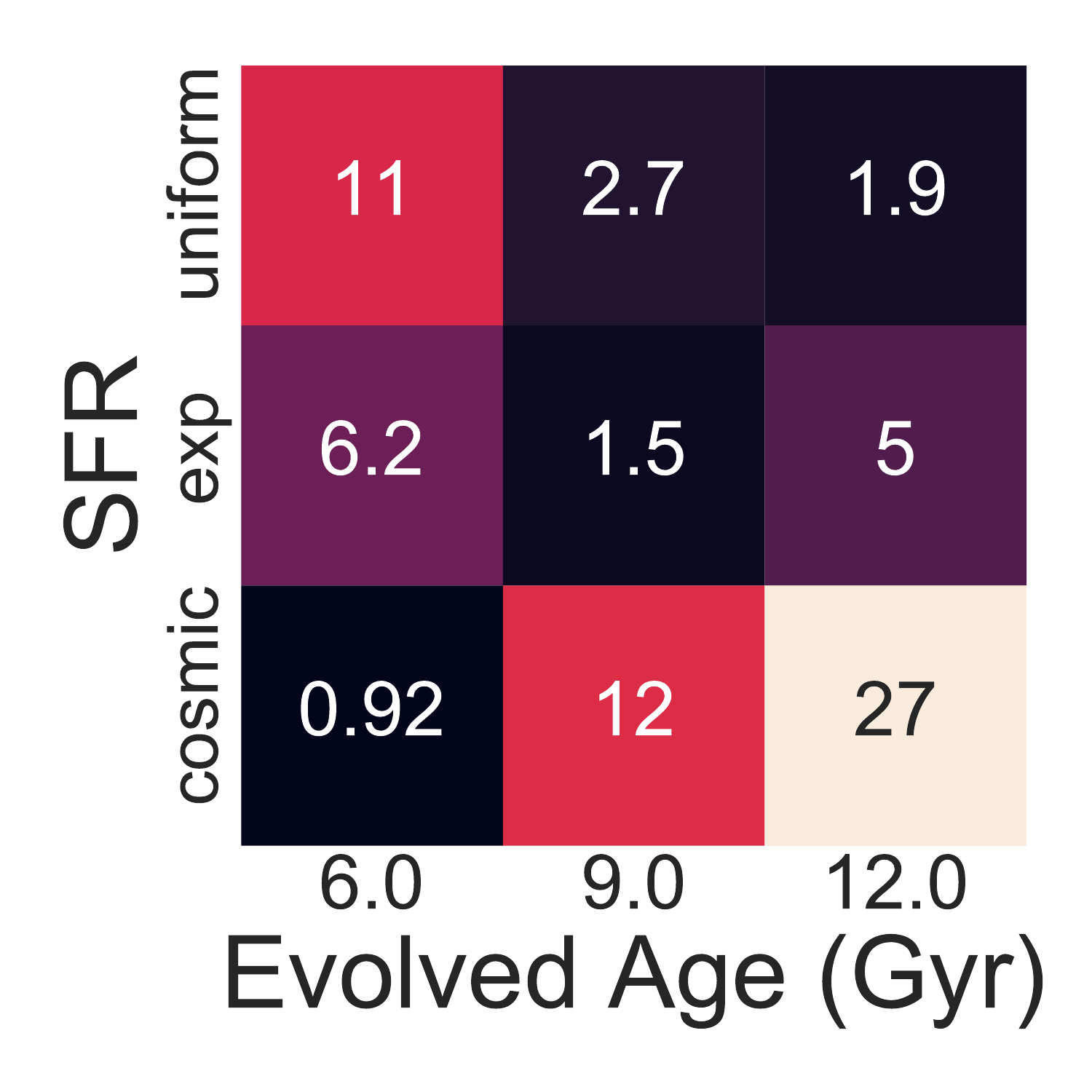}{0.33\textwidth}{}
\fig{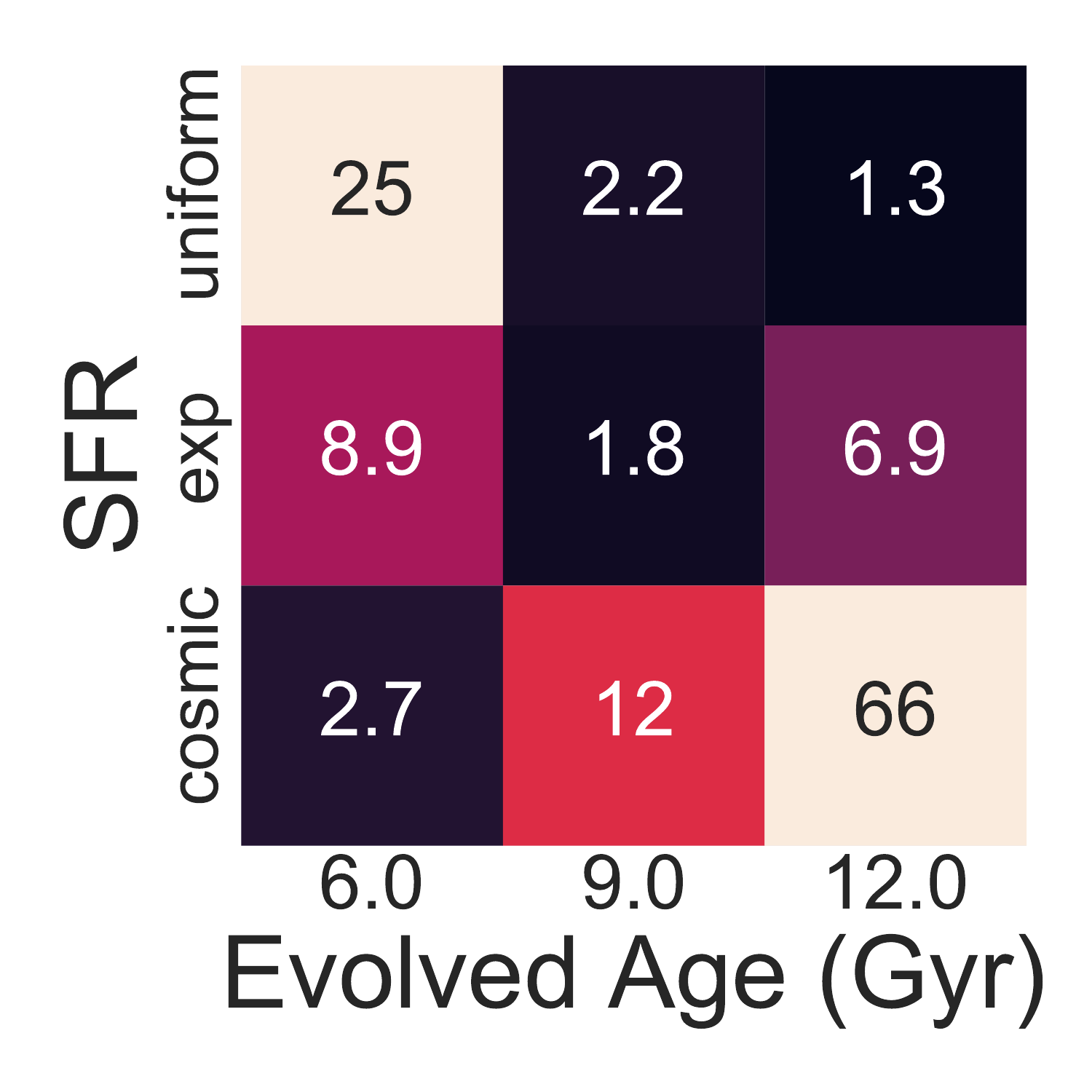}{0.33\textwidth}{}
\fig{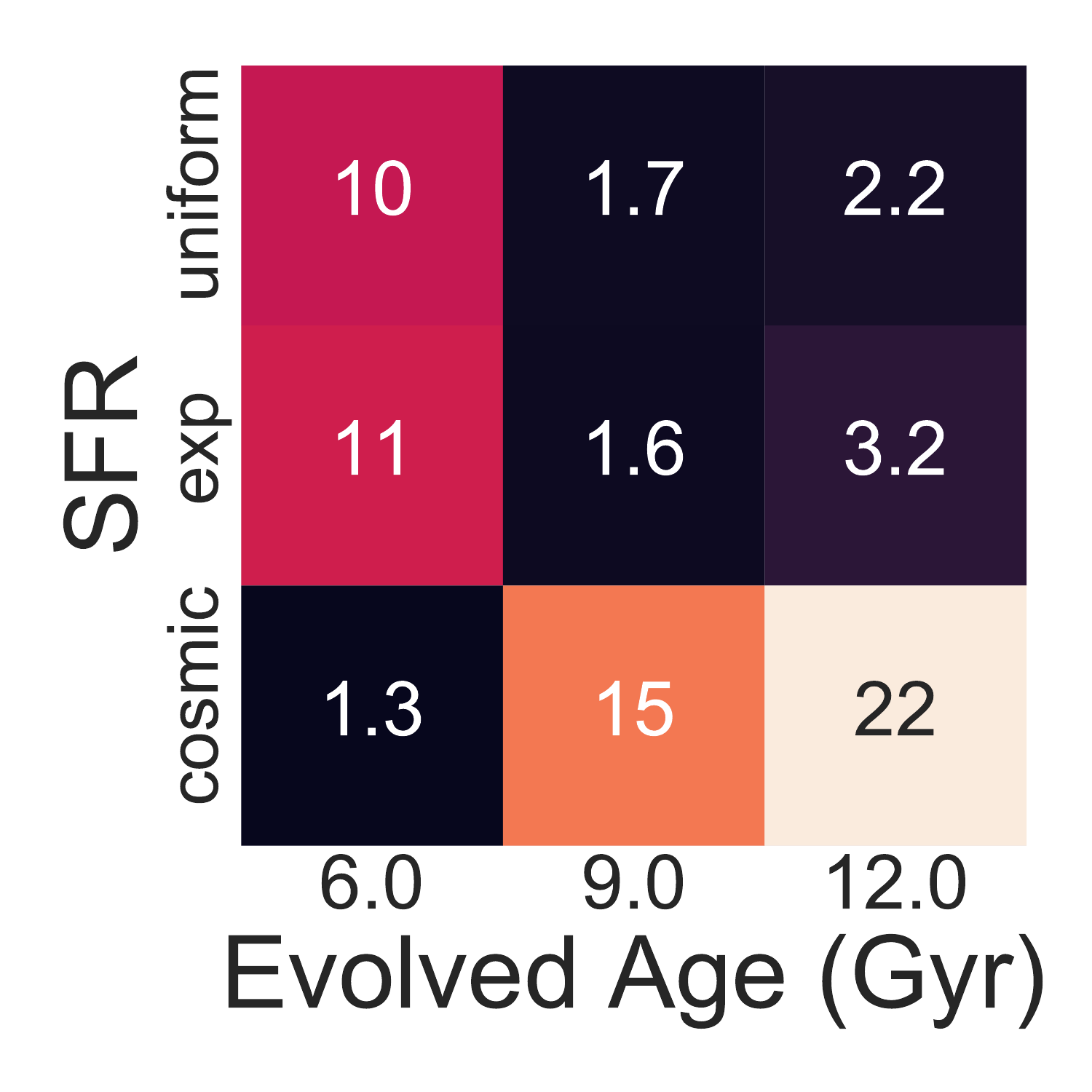}{0.33\textwidth}{}
}
\gridline{\fig{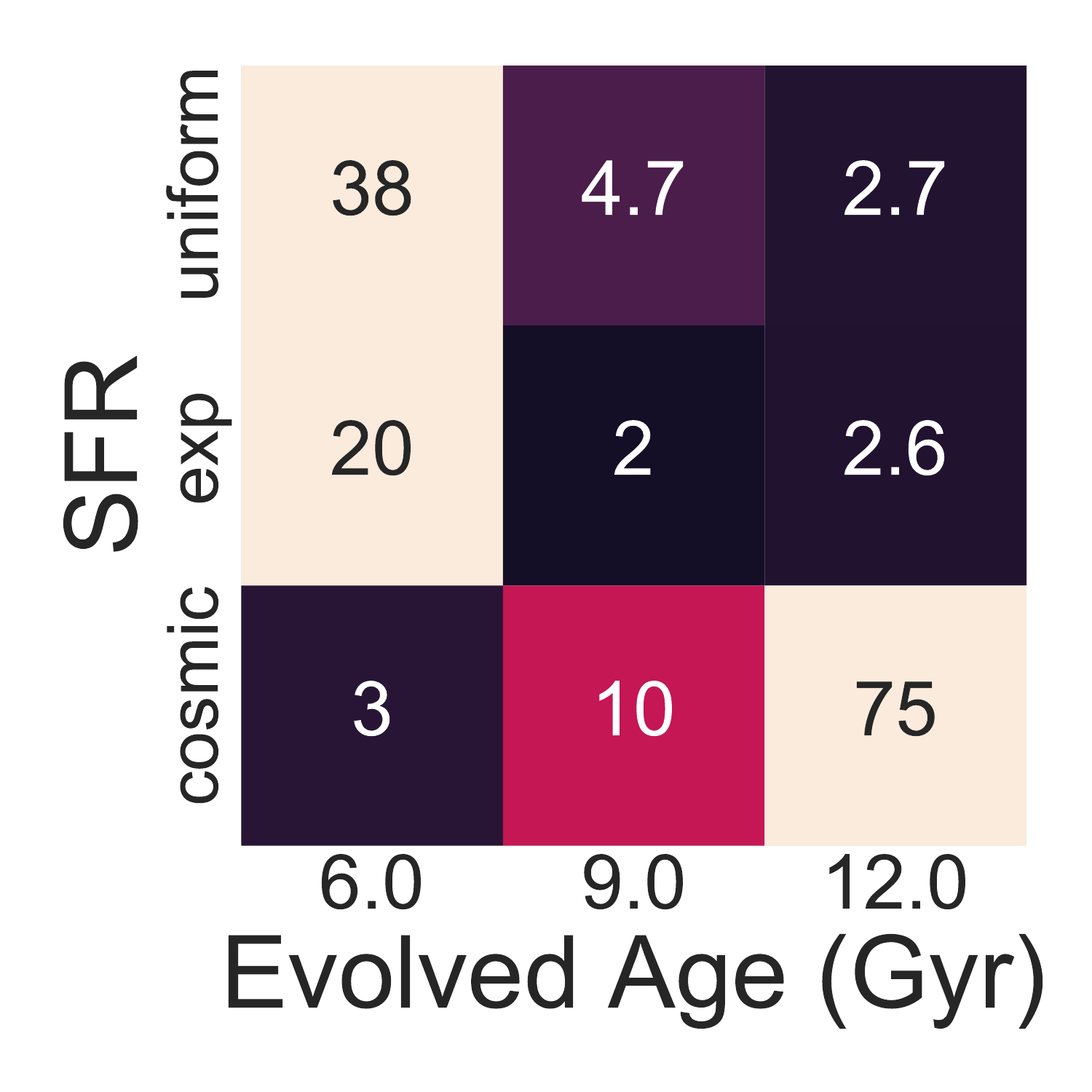}{0.33\textwidth}{}
\fig{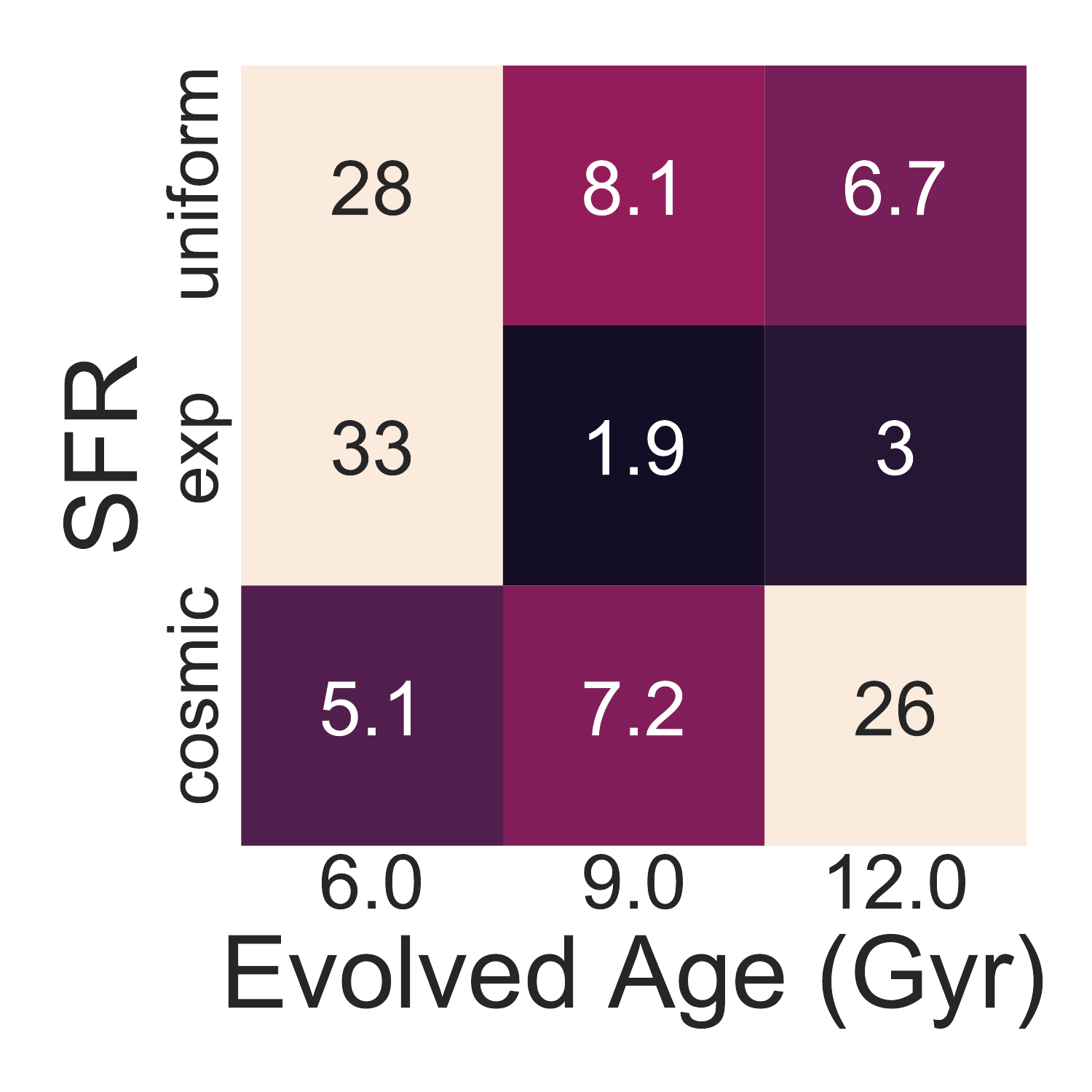}{0.33\textwidth}{}
\fig{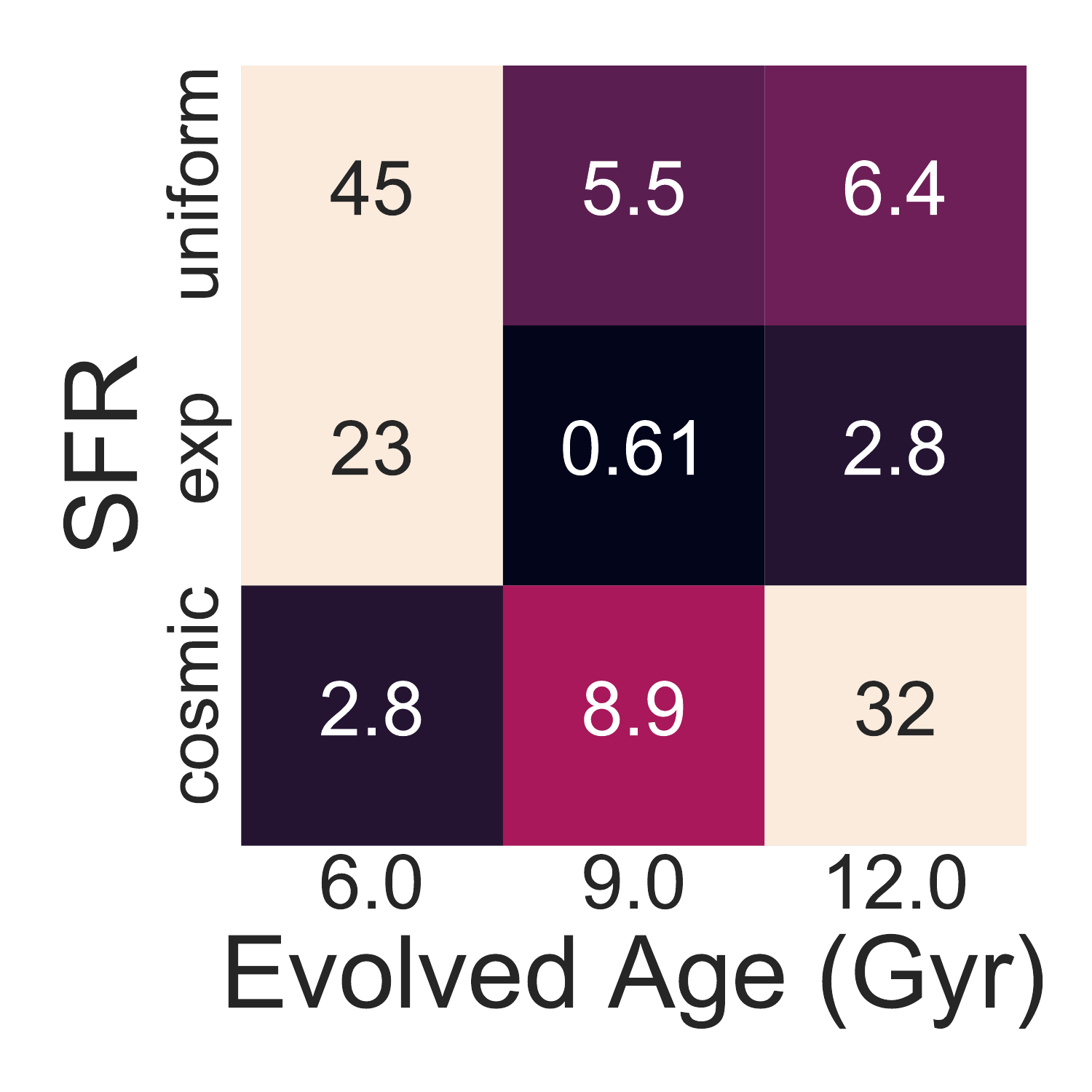}{0.33\textwidth}{}
}
\caption{Same as Figure \ref{fig:popsim_heatmap}  top panel for $\alpha$ = $-$1.5 (top), $-$0.5 (middle) and 1.5 (bottom).
\label{fig:popsim2}}
\end{figure*}

Holding the SFR and sample age fixed to our baseline parameters, we compared the mass function power-law index to the MBDM (Figure~\ref{fig:popsim_heatmap} middle panels). 
The mass functions with $\alpha$ = 0.5, $-0.5$, and $-1.5$ are statistically equivalent, while the bottom-heavy $\alpha$ = 1.5 and log-normal mass functions are not favored ($\Delta$BIC $>$ 2, positive\added{; also see Figure~\ref{fig:popsim2}}). The results are insensitive to MBDM or evolutionary models.
We also compared the SFR and mass function for a fixed age of 9~Gyr and MBDM = 0.01~M$_{\odot}$ (Figure~\ref{fig:popsim_heatmap} bottom panels).
Models using the cosmic star formation rate are significantly worse than other models. Again, the mass functions $\alpha$ = 0.5, $-0.5$, and $-1.5$ are statistically equivalent, while $\alpha$ = 1.5 and log-normal mass functions using the uniform star formation rate are not favored ($\Delta$BIC $>$ 2, positive).

\begin{figure}[!htbp]
\centering
\includegraphics[width=0.9\linewidth, trim=0 10 0 0]{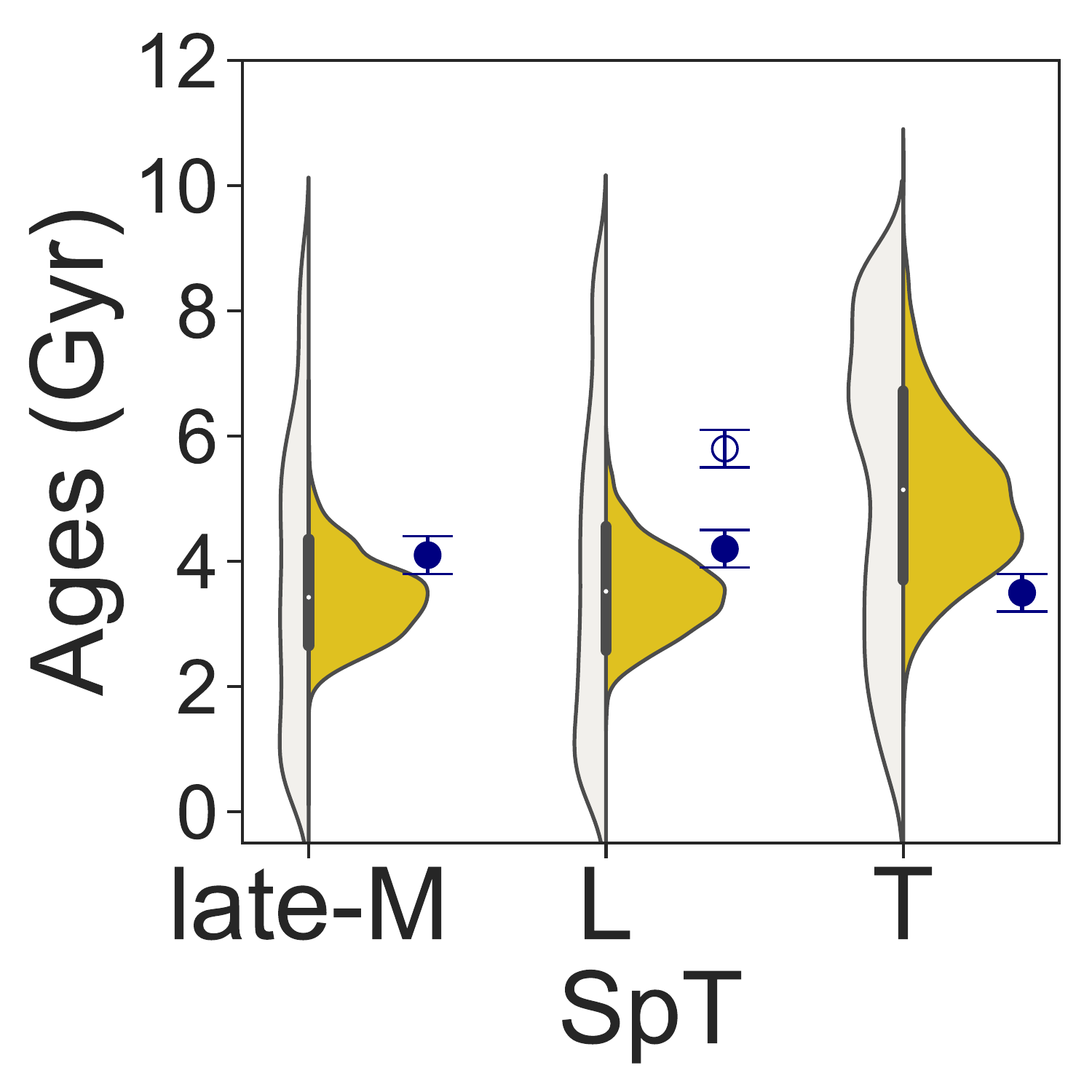}
\includegraphics[width=0.9\linewidth, trim=0 10 0 0]{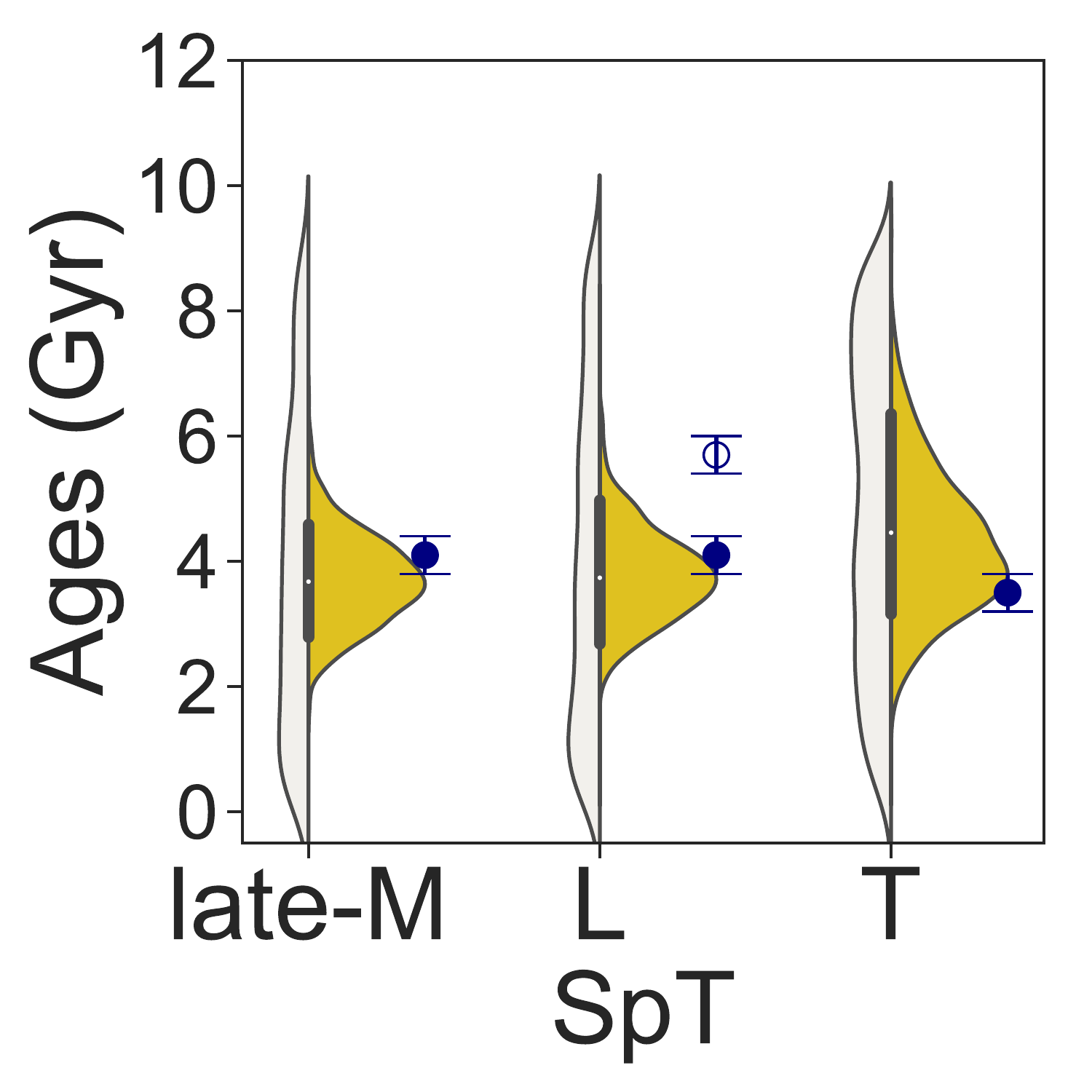}
\includegraphics[width=0.9\linewidth, trim=0 20 0 0]{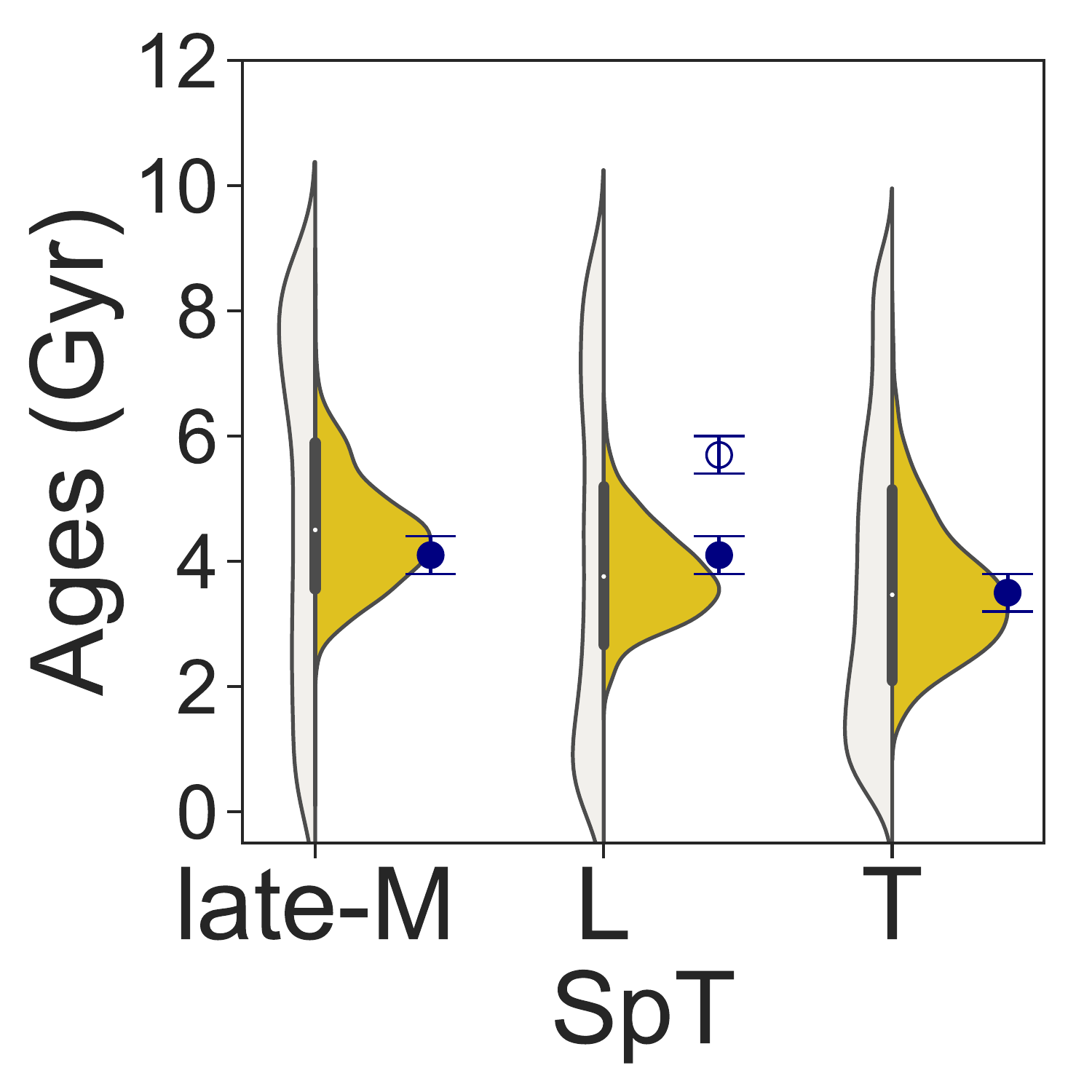}
\caption{Same as Figure~\ref{fig:simulated_population2} comparing three simulations with baseline parameters and mass functions that evolve over time: 
(Top) $\alpha$ = 1.5 $\rightarrow$ $-$0.5 (bottom-heavy to top-heavy; same model as \citealt{Burgasser:2015ac});
(Middle) $\alpha$ = 1 $\rightarrow$ 0 (bottom-heavy to top-heavy); and
(Bottom) $\alpha$ = 0 $\rightarrow$ 1 (top-heavy to bottom-heavy).
\label{fig:simulated_population_mfevolve}}
\end{figure}

\begin{figure*}[!htbp]
\centering
\gridline{\fig{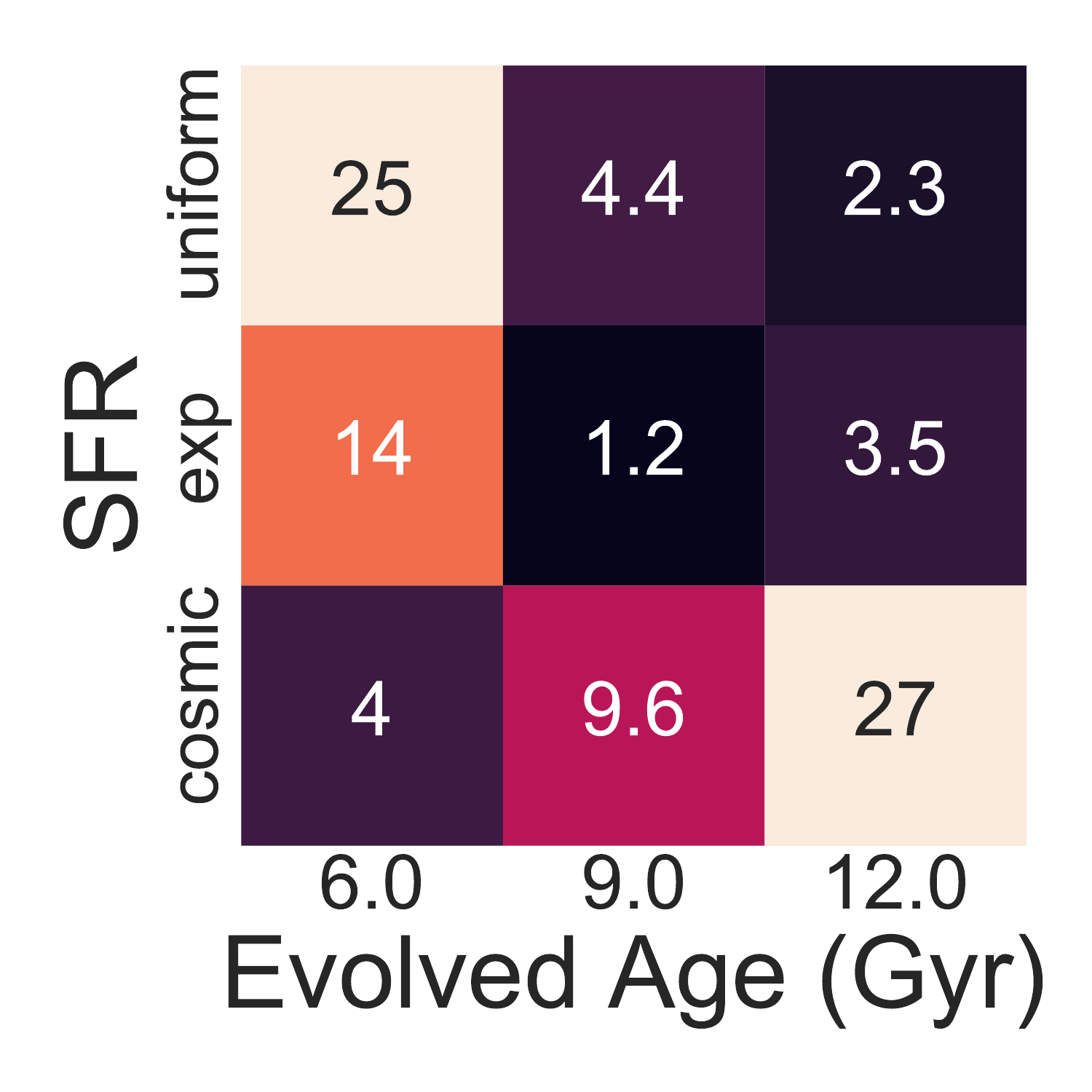}{0.33\textwidth}{}
\fig{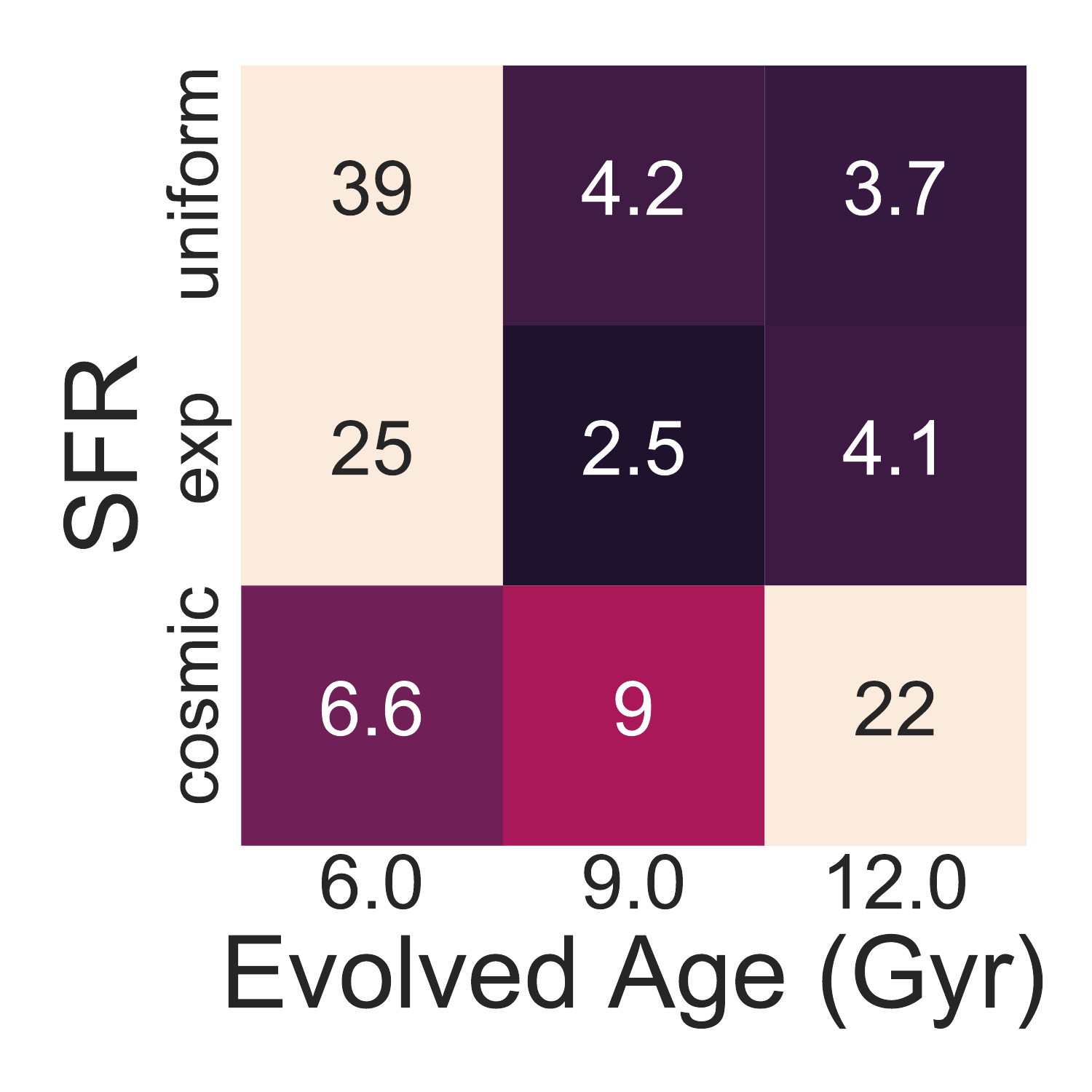}{0.33\textwidth}{}
\fig{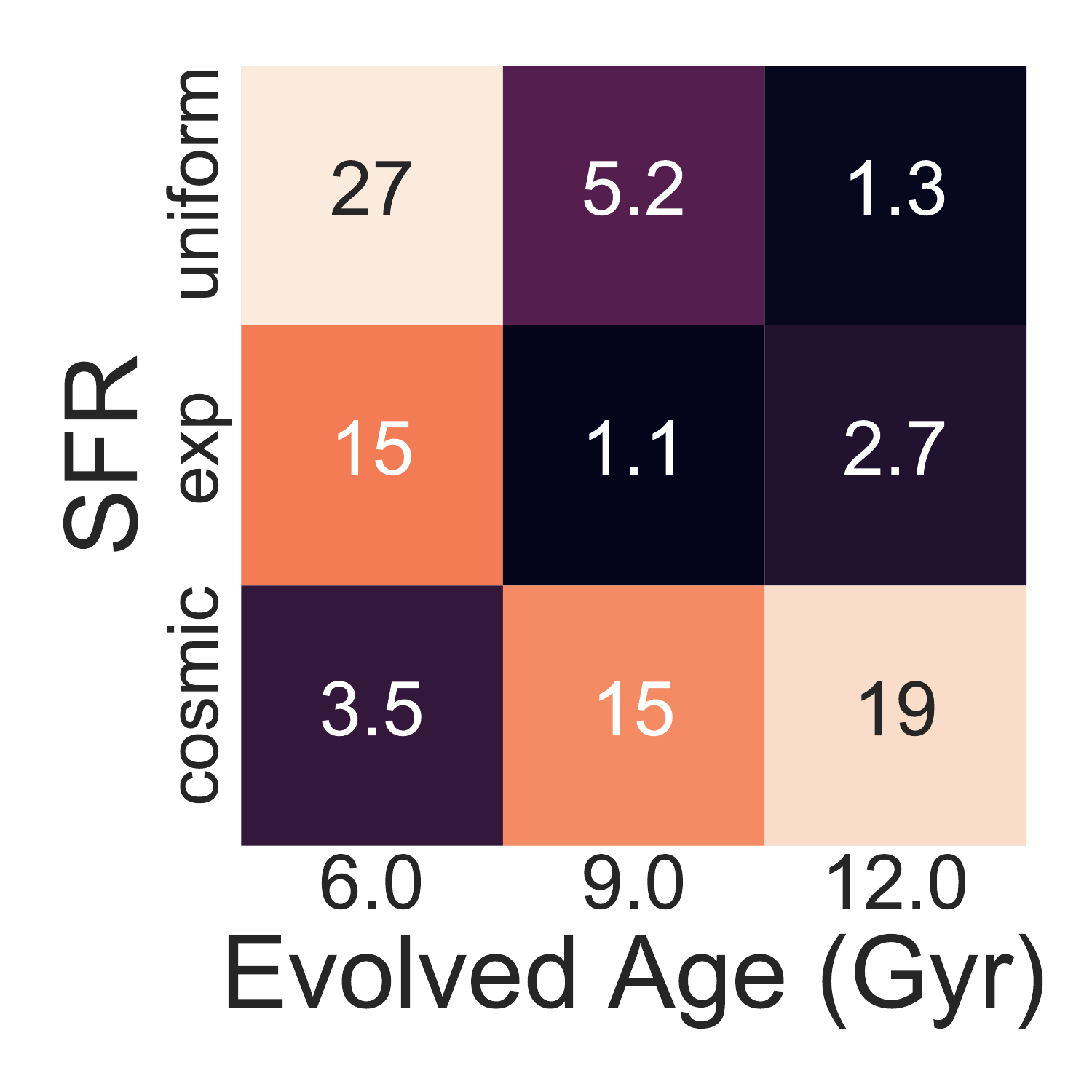}{0.33\textwidth}{}
}
\gridline{\fig{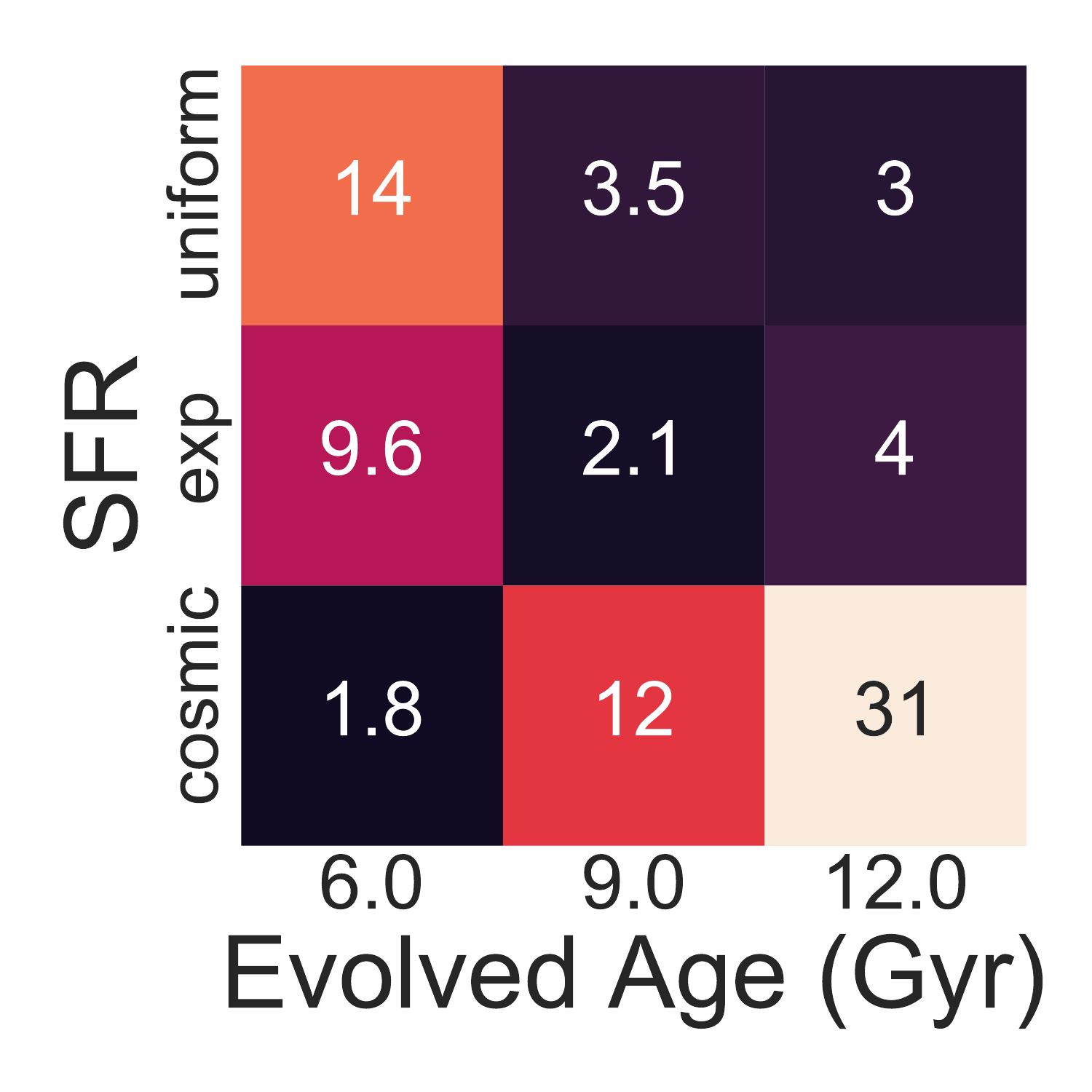}{0.33\textwidth}{}
\fig{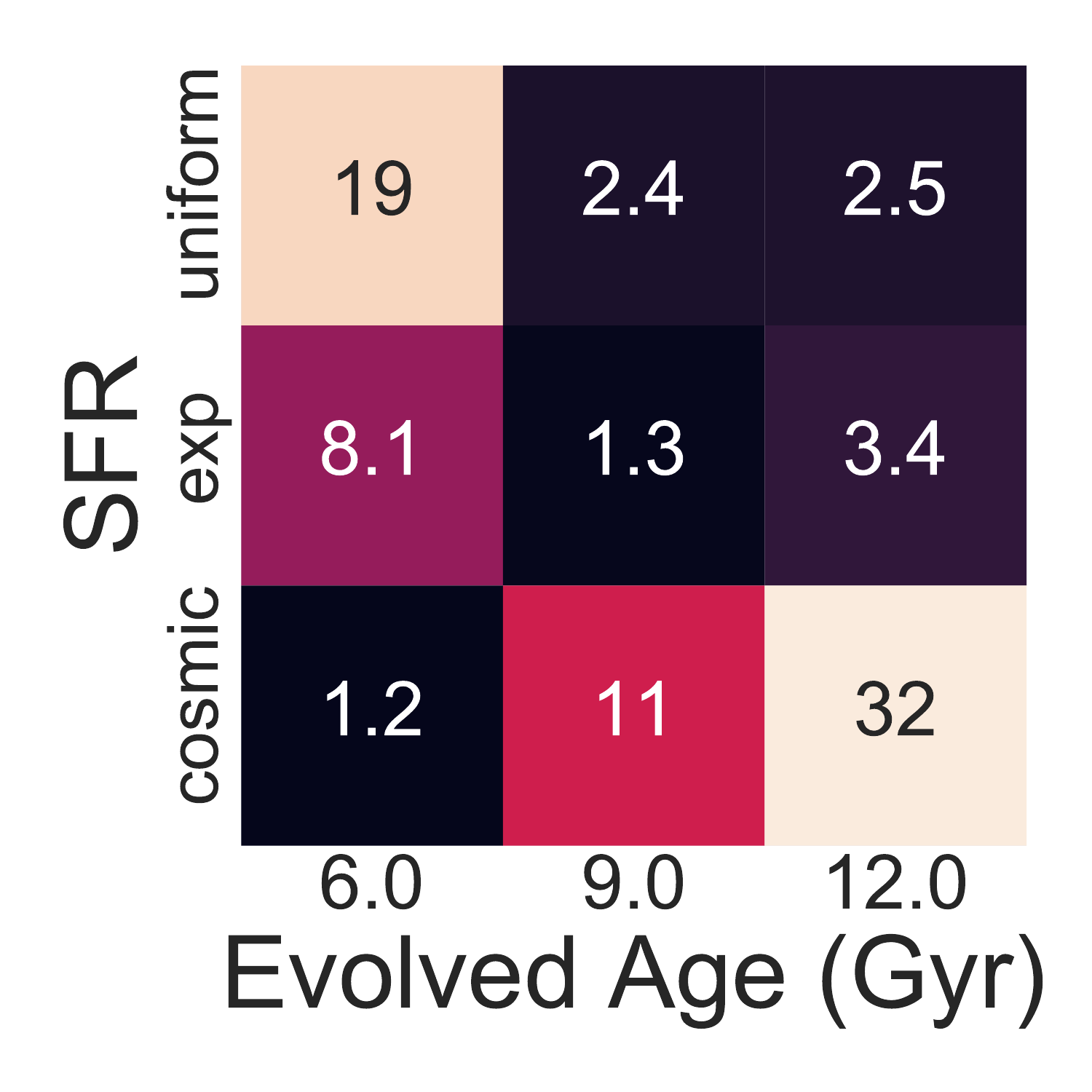}{0.33\textwidth}{}
\fig{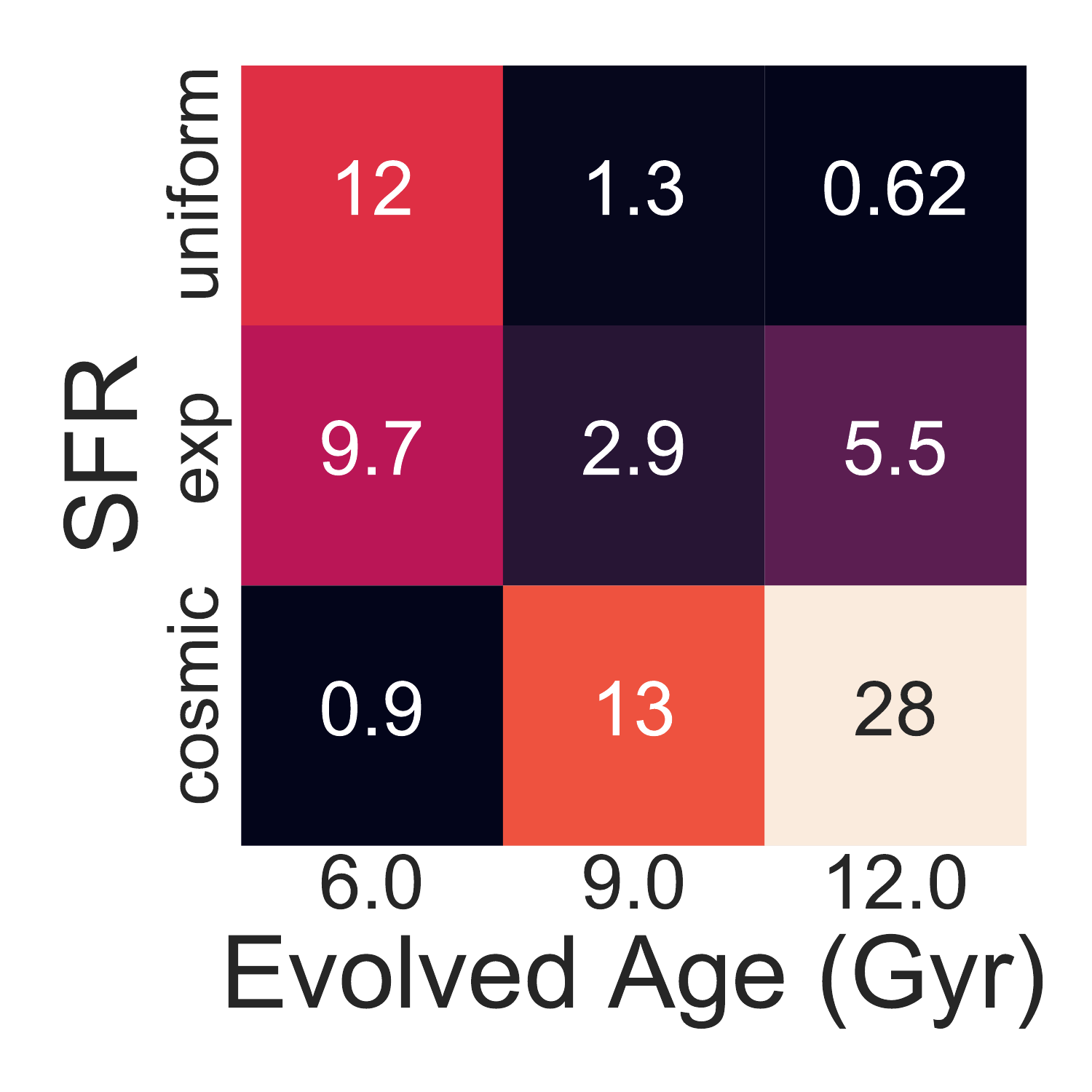}{0.33\textwidth}{}
}
\gridline{\fig{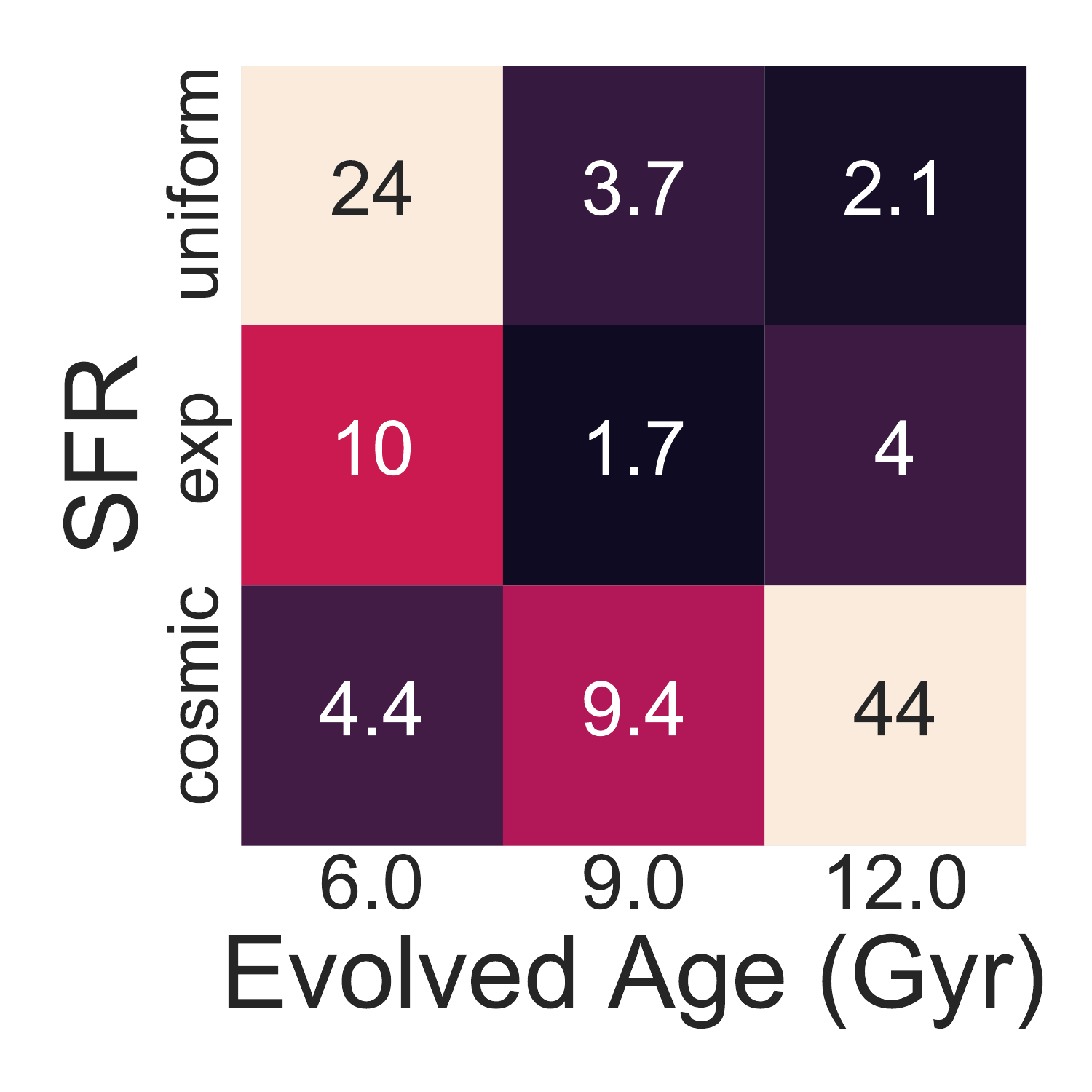}{0.33\textwidth}{}
\fig{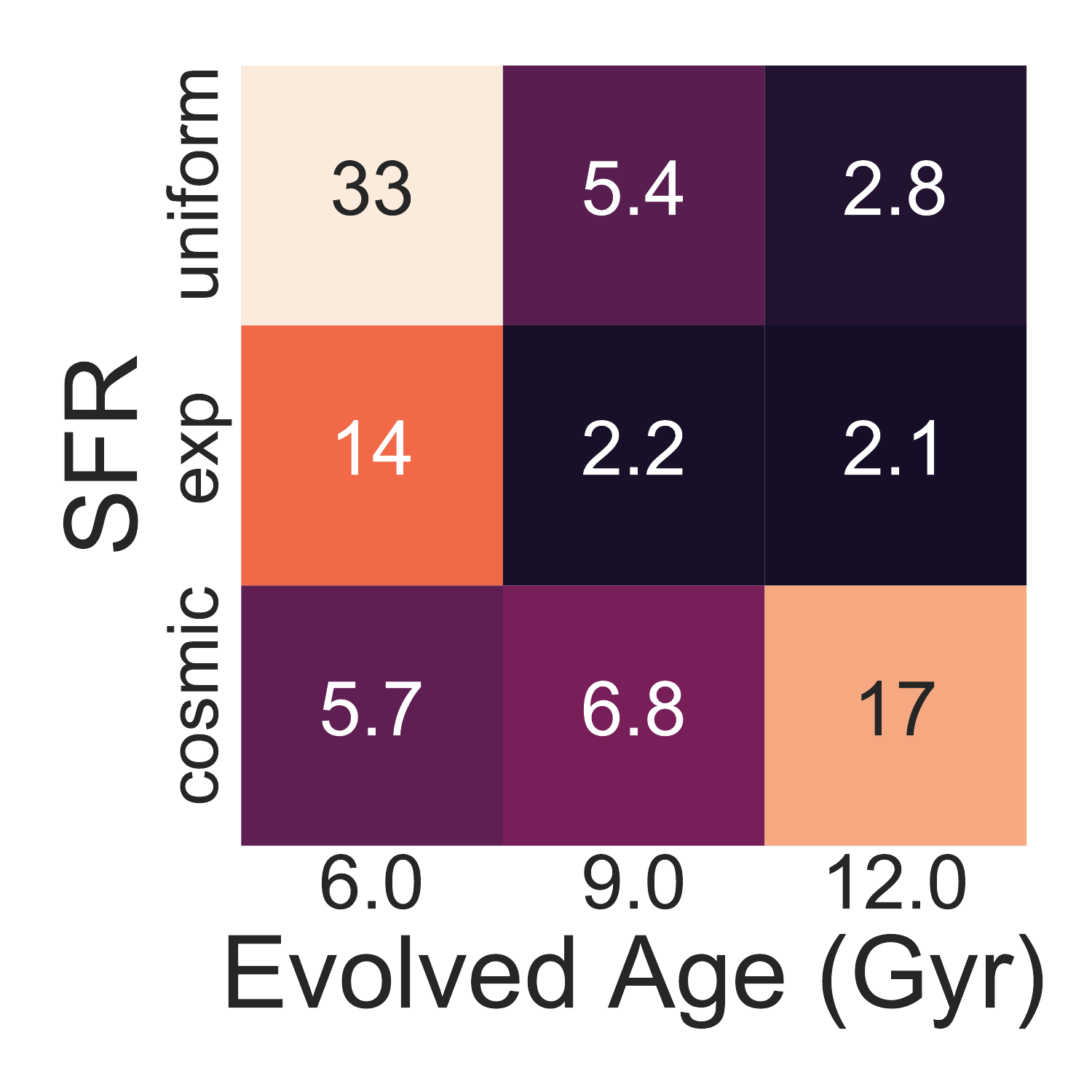}{0.33\textwidth}{}
\fig{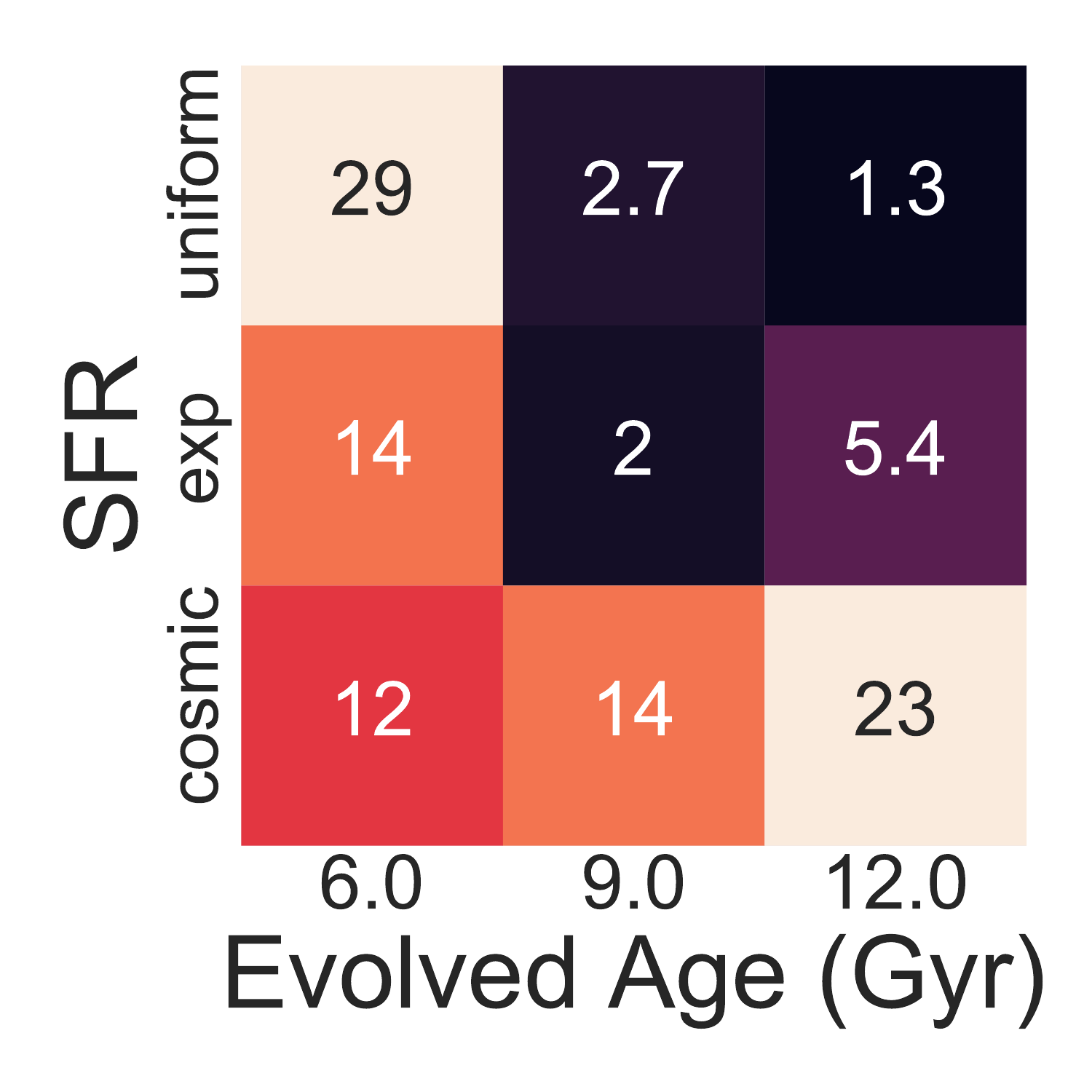}{0.33\textwidth}{}
}
\caption{Same as Figure \ref{fig:popsim_heatmap} top panel for the \citet{Chabrier:2000aa} log-normal mass function (top), and for evolving mass functions $\alpha$ = 0 $\rightarrow$ 1 at 3~Gyr (middle) and $\alpha$ = 1 $\rightarrow$ 0 at 3~Gyr (bottom).
\label{fig:popsim3}}
\end{figure*}

\citet{Burgasser:2015ac} considered whether the observed older L dwarfs in their sample could be the outcome of 
a mass function that evolves from bottom-heavy to top-heavy over time (i.e., $\alpha$ decreasing over time).
Figure~\ref{fig:simulated_population_mfevolve} shows that the original hypothesis 
$\alpha$ = 1.5 $\rightarrow$ \replaced{-0.5)}{$-$0.5 at 4.5~Gyr} is not consistent with the observed ages of late-MLT dwarfs\added{ ($\chi^2$ = 5.6; $\Delta$BIC = 3.5)}, overestimating in particular the ages of T dwarfs\added{ (5.4 $\pm$ 1.4~Gyr)}. 
\added{However, scenarios with the same transition in at different transition ages (3~Gyr and 6~Gyr) are consistent ($\Delta$BIC $\leq$ 2).}
\replaced{A}{Similarly, a} narrower range of mass function of evolution, either $\alpha$ = 0 $\rightarrow$ 1 or $\alpha$ = 1 $\rightarrow$ 0 with uniform and exponential birth rates cannot be ruled out\added{ ($\Delta$BIC $\leq$ 2; Figure~\ref{fig:popsim3})}. 
\deleted{Scenarios with different endstates for the value of $\alpha$ (+1.5 to $-$0.5) and different ages for the transition (3~Gyr, 4.5~Gyr and 6~Gyr) fail to reproduce the ages of late-M, L, and T dwarfs in our kinematic sample.}

Comparing the outcomes of different evolutionary models, we found the predictions of the \citet{Burrows:2001aa}, \citet{Saumon:2008aa}, and \citet{Baraffe:2003aa} models yield roughly identical results.
Late-M dwarf ages using the \citet{Marley:2018aa} and \citet{Phillips:2020aa} cloudless atmosphere models were too young compared to L and T dwarfs, likely due to temperature limits in the model parameter space.  
We also found that increasing the maximum age or minimum brown dwarf mass in the simulations increased the mean ages of ultracool dwarfs, but retained the relative ages for late-M, L, and T dwarfs. 

\begin{figure}[!htbp]
\centering
\includegraphics[width=0.9\linewidth, trim=0 10 0 10]{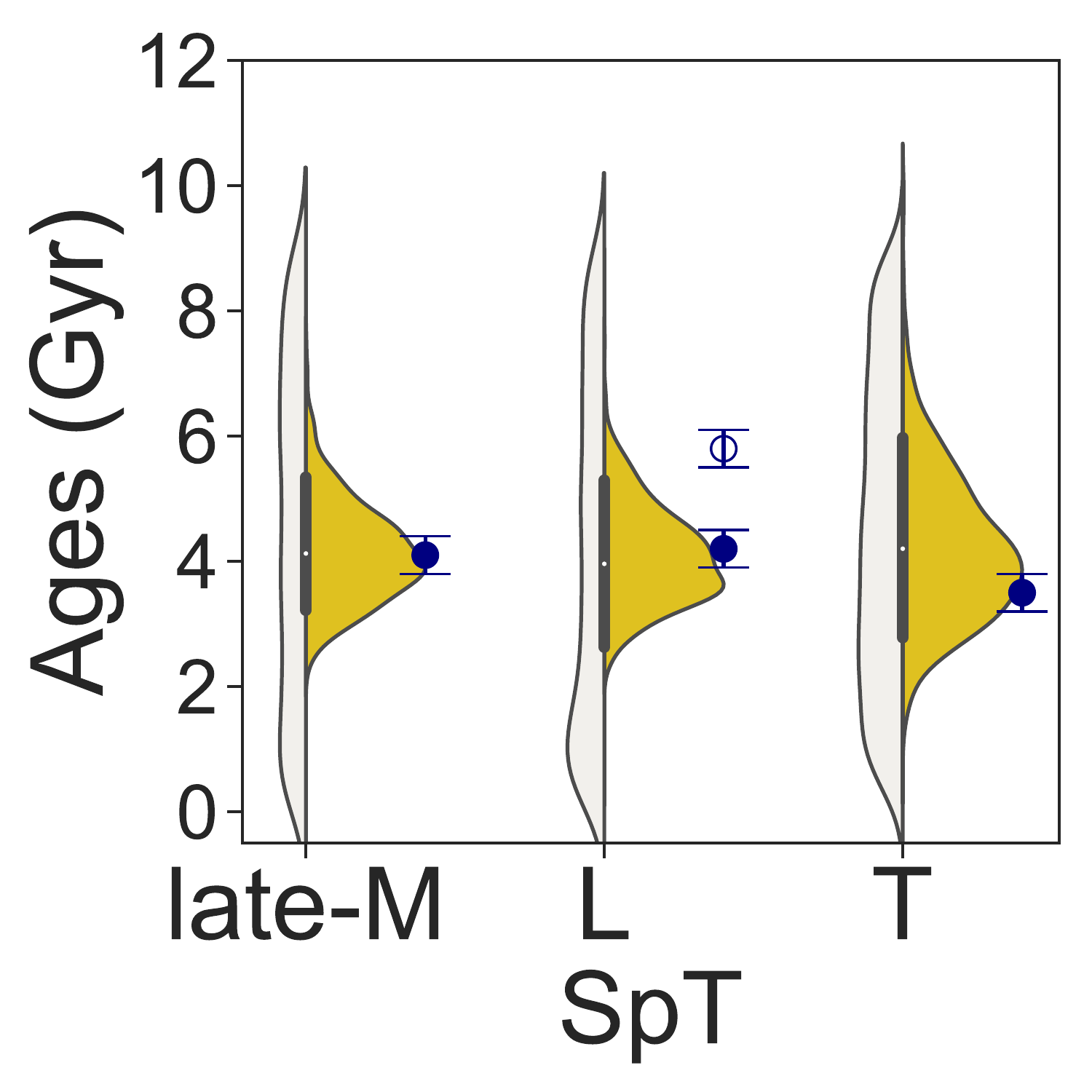}
\caption{Same as Figure~\ref{fig:simulated_population2}, comparing observed kinematic ages to a simulation assuming baseline parameters and a hydrogen burning minimum mass \replaced{artificially set to 0.06~M$_{\odot}$.}{with an artificial decrease for the \citet{Baraffe:2003aa} evolutionary models by fixing the temperatures of brown dwarfs with masses $M$ $\geq$ 0.06~M$_{\odot}$ to their 1~Gyr values.}
\label{fig:hbmm_sim}}
\end{figure}

Errors in evolutionary models could contribute to the marginal age discrepancy between simulations and observations for the L dwarfs. 
The ratio of L-type stars and brown dwarfs and the thermal evolution of the most massive brown dwarfs depends on the efficiency of hydrogen fusion reactions close to the critical temperature. An offset in the HBMM, particularly toward lower masses so that more L dwarfs are stars, could potentially increase the ages of L dwarfs in the simulations. To explore this effect, we imposed an artificial decrease in the HBMM for the \citet{Baraffe:2003aa} evolutionary models by fixing the temperatures of brown dwarfs \replaced{down to masses of 0.06~M$_{\odot}$}{with masses $M$ $\geq$ 0.06~M$_{\odot}$} to their 1~Gyr values. The resulting ages for late-M, L, and T dwarfs in these simulations 
are $4.1 \pm 0.8$~Gyr, $4.1 \pm 0.8$~Gyr, $4.4 \pm 1.2$~Gyr, respectively, fully consistent with our observed ages (Figure~\ref{fig:hbmm_sim}).
The fact that this adjustment provides the best match between the simulations and observed sources is suggestive of potential evolutionary model issues, which have also been raised with \added{mass and luminosity measurements of brown dwarf companions to age-dated stars \citep{Dupuy:2017aa} and the surprisingly high masses} \deleted{recent 
measurements of unusually massive} T-type brown dwarfs \added{in binaries} (e.g. \citealt{Dupuy:2019aa, Brandt:2020aa, Sahlmann:2020aa, Sahlmann:2021aa}). \added{These studies suggest that evolutionary model predictions of the temperatures and luminosities of objects around the HBMM may not align with the observed properties of these systems.}
However, with only a marginal age discrepancy, potential selection biases in our L dwarf kinematic sample (see below), and the degeneracies present among other simulation parameters, further work is needed 
to confirm this result.
We note that increasing the timescale of cooling could also produce older L-type brown dwarfs, but also drives up the ages of T dwarfs and is therefore an unlikely scenario.

In summary, several variations in simulation parameters were able to reproduce the observed kinematic ages of the local late-M, L, and T dwarf populations self-consistently, and highlight some degeneracies in this approach. 
Nevertheless, we are able to rule out \replaced{some}{several} parameter sets, and identify a potential indicator of a lower HBMM.

\section{Discussion\label{sec:explain}}

While our simulations are able to reproduce the kinematic ages of UCDs in the local thin disk population, the origin of the relatively high fraction of local thick disk L dwarfs remains unclear. 
As L dwarfs span the HBMM, in the local Galactic environment they consist of a mixed population of stars and (young) brown dwarfs. Accurate characterization of this population is therefore critical for validating brown dwarf evolutionary models and measuring brown dwarf formation history in the Galaxy. 
Here we explore some possible explanation as to why the local L dwarf sample studied in this paper are less well-modeled as compared to local late-M and T dwarfs.

\subsection{Sample incompleteness\label{subsec:sampleincomplete}}

The kinematic sample examined in this study, while larger than previous studies and limited to $d <$ 20~pc, is not uniformly volume complete. One primary reason for this is that fainter late-L and T dwarfs are detected at smaller distances for a given sensitivity limit. For Keck/NIRSPEC, an effective magnitude limit of 15 restricts observations of L5 dwarfs to 22~pc ($K$-band) and T5 dwarfs to 14~pc ($J$-band). 
RV measurements from the literature are also not volume- or magnitude-complete, as measurements were obtained with different optical and near-infrared spectrographs with varying sensitivity thresholds. 

The underlying target sample is itself not volume-complete, either, particularly in the Galactic plane where nearby ultracool dwarfs sample smaller scale heights \replaced{(and are hence younger)}{, and are hence younger,} but where source contamination and crowding is high.
\citet{Bardalez-Gagliuffi:2019aa} determined that the sample of late-M and early- to mid-L dwarfs is 62$^{+8}_{-7}$\% and 83$^{+10}_{-9}$\% complete within 20~pc, and new nearby sources are still being uncovered in wide-field surveys such as PanSTARRS \citep{2020AJ....159..257B}, {\em Gaia} \citep{2018ApJ...868...44F,2018A&A...619L...8R,2020A&A...637A..45S}, \added{and {\em WISE} \citep{Meisner:2020aa, Kirkpatrick:2021aa}}.
Figure \ref{fig:SpT_lateML_sample_with_simulated_population} compares our simulated population to the 20 pc samples of \citet[hereafter BG19]{Bardalez-Gagliuffi:2019aa}, \citet[hereafter B20]{2020AJ....159..257B}, and \citet[hereafter K21]{Kirkpatrick:2021aa}. While \replaced{the latter}{these} UCD samples are the most volume-complete \deleted{samples} constructed to date, there remain gaps particularly among the late-M, late-L, and T dwarfs.
Our kinematic sample contains 48\%--52\% of the 20 pc UCDs in the BG19 and B20 samples,
and based on these studies' completeness estimates, only 32\%--42\% of M7--T8 dwarfs within 20~pc.
Compared to K21, our sample contains only 27\% of L0--T8 dwarfs within 20~pc.

\begin{figure*}[!htbp]
\includegraphics[width=\textwidth]{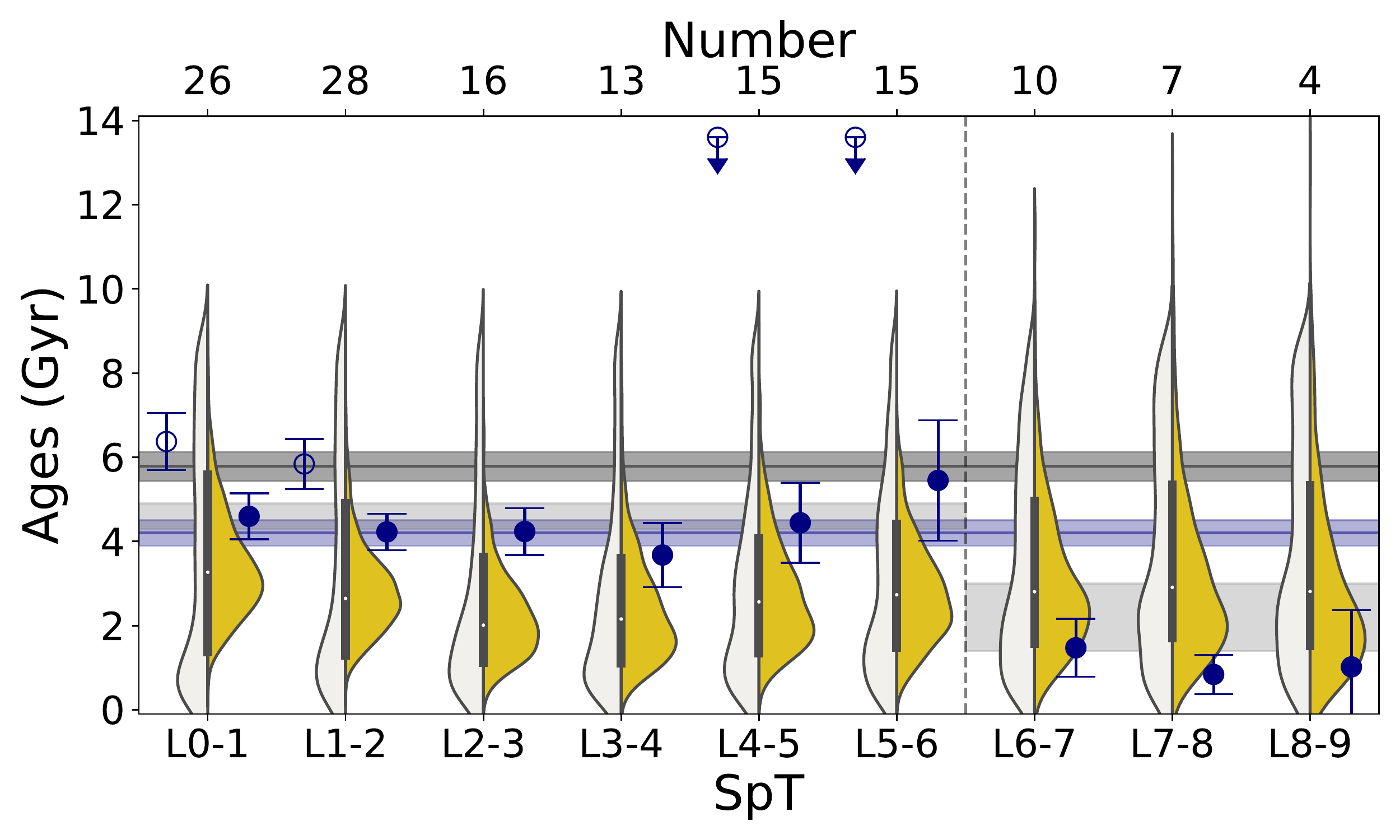}
\caption{
Age distributions of L dwarfs from the baseline simulation binned in groupings of two subtypes compared to measured kinematic ages with similar binning.
Simulated age distributions are shown as individual source ages (white violin plots) and derived kinematic ages (yellow violin plots).
Observed kinematic ages are show\added{n} with (open blue points) and without (solid blue points) \deleted{unusually blue }L dwarfs\added{ for $P[\mathrm{TD}]/P[\mathrm{D}]$ $>$ 1}. 
Our derived kinematic ages and uncertainties (\replaced{5.7}{5.8}$\pm$0.3~Gyr, \replaced{4.1}{4.2}$\pm$0.3~Gyr) 
for 
\replaced{the whole L dwarf sample are indicated by the black line 
for P(D/TD) $<$ 10 and blue line for P(D/TD) $<$ 1.}{our nominal L dwarf sample ($P[\mathrm{TD}]/P[\mathrm{D}]$ $<$ 10) are indicated by the black line, and for our constrained thin disk sample ($P[\mathrm{TD}]/P[\mathrm{D}]$ $<$ 1) by the blue line.}
We also show the average kinematic age\added{s} for \deleted{and}early- and late-L dwarfs \added{separately} as grey bands. 
The number of sources in each subtype bin is labeled at top.
The ages of the discrepant L4--L5 and L5--L6 subtype bins are due to four unusually blue L dwarfs within these groups. \added{Note that the very old tails of inferred ages for L6--L9 are due to small sampling effect.}
\label{fig:boxplot_L_dwarf}}
\end{figure*}

One way this incompleteness can produce an age discrepancy is if the fraction of early-type and late-type L dwarfs in our sample is imbalanced. 
Early-type L dwarfs are predominantly low-mass stars of all ages and very young brown dwarfs, while late-type L dwarfs are predominantly young brown dwarfs, according to our simulations and the data. 
Figure~\ref{fig:boxplot_L_dwarf} illustrates this trend in our kinematic sample, which shows a clear decline in velocity dispersion and inferred age as a function of L dwarf spectral subclass, with the exception of the L4--L6 subtypes (discussed in further detail below).
More coarsely, while the overall kinematic age of thin disk L dwarfs is $\replaced{4.1}{4.2}\pm0.3$~Gyr, the kinematic age of \replaced{15}{14} L6--L9 dwarfs is only \replaced{$2.0 \pm 0.7$}{$2.2 \pm 0.8$}~Gyr. 
We computed the ratio of L0--L5 to L5--L9 dwarfs in our baseline simulation sample, our kinematic sample, the samples of B20 and K21, and the combined sample of BG19 and B20. 
These ratios were found to be 1.3 (simulated), \replaced{3.8}{4.1} (kinematic sample, L0--L5:L5--L9 = \replaced{57:15}{57:14}), 1.0 (B20, L0--L5:L5--L9 = 21:22), 1.3 (K21, L0--L5:L5--L9 = 96:75), and 3.3 (B19+B20, L0--L5:L5--L9 = 105:32), respectively. 
Our kinematic sample is clearly biased toward early L dwarfs compared \added{to }the simulated, B20, and K21 samples, but \deleted{notably} not \added{compared to} the combined B19 and B20 sample. 
However, if we try to reproduce \replaced{this}{the kinematic sample} asymmetry through forced random draws from our baseline simulation, we find an average age for thin disk L dwarfs (\replaced{$3.0\pm0.7$}{$3.0\pm0.6$}~Gyr) that is still younger than \replaced{the observed kinematic sample}{observed}.
Similarly, sampling with replacement the thin disk L dwarfs from our kinematic sample to force a ratio of early-to-late L dwarfs of 1.3 (simulations) or 1.0 (B20) results a nearly identical age
\replaced{($4.1 \pm 0.1$~Gyr)}{($4.1 \pm 0.8$~Gyr)} as the original sample.

Our sample is also biased toward early T dwarfs due to the intrinsic faintness of later subtypes, with an early-to-late T dwarf ratio (T0--T4/T5--T8) of  \replaced{1.0 that is}{1.0, }
much higher than the most recent T dwarf sample from K21 (0.14) and our baseline population simulation (0.22). 
Again,\added{ if} we\deleted{ if} \added{randomly} draw from the simulation to match the spectral type ratio of the \added{observed} kinematic sample, we find \replaced{and}{an} older but statistically equivalent age of 4.3 $\pm$ 1.2~Gyr\replaced{, while d}{. D}rawing from the kinematic sample to match the simulation \added{T dwarf} spectral type ratio yields a nearby identical age of \replaced{3.45 $\pm$ 0.23}{3.5 $\pm$ 0.2}~Gyr.

Another selection bias is the magnitude limit of the observed kinematic sample, which 
can similarly skew the number of early-type and later-type L dwarfs. 
We modeled this in our baseline simulation by assigning distances up to 20~pc over a uniform-density volume, assigning apparent magnitudes using the \citet{Dupuy:2012aa} absolute magnitude-spectral type relations, and constraining the simulated sample to be brighter than $J$ or $K$ $<$ 15.
The resulting magnitude-limited simulated late-M, L, and T dwarf kinematic ages are $4.0\pm 0.8$~Gyr, $3.0\pm0.6$~Gyr, and $3.8\pm1.1$~Gyr, respectively, fully consistent with our baseline simulation. 
Hence, a magnitude limit does not explain the older age for L dwarfs in the kinematic sample.
A magnitude-limited sample can preferentially select younger sources, which haven't fully contracted to their fully degenerate radii. This would affect all sources in our sample, but particularly those subgroups whose limiting magnitudes place them within the volume limit. This bias may explain the slightly younger kinematic age of T dwarfs compared to the simulations, but does not explain the presence of older L dwarfs in the kinematic sample.

\subsection{Contamination by Distinct Sub-populations\label{subsec:subpopulation}}

Our kinematic L dwarf sample contains \replaced{nine}{eight} L-type binaries, \replaced{two}{fourteen} young L dwarfs and four unusually blue L dwarfs, the latter based on the 
color outlier criteria of \citet{Faherty:2009aa}\footnote{The blue L dwarfs in our sample are 2MASS J05395200$-$0059019 \citep{Geballe:2002aa}, 
DENIS J112639.9$-$500355 \citep{Phan-Bao:2008aa}, 
2MASSI J1721039+334415 \citep{Bardalez-Gagliuffi:2014aa}, and 
2MASS J05395200$-$0059019 \citep{2000AJ....119..928F}.}.
Like the thick disk population, these distinct sub-populations have the potential of skewing the kinematic dispersions and ages of the overall sample, so we re-evaluated the velocity dispersions and kinematic ages of the L dwarfs after removing each of these subgroups
(Table~\ref{table:lateML_kinematicage}).
Removal of young L dwarfs and binaries \deleted{marginally }increases the age of the total population to \replaced{$7.3 \pm 0.4$}{$8.8 \pm 0.6$}~Gyr and $7.4 \pm 0.5$~Gyr\replaced{,}{.} \added{The former is expected as removal of young sources makes the population older while the latter is} consistent with no change (when thick disk stars are left in).
Removal of blue L dwarfs reduces the kinematic age to \replaced{$5.1 \pm 0.3$}{$5.3 \pm 0.3$}~Gyr, consistent with removal of thick disk stars.
This result confirms the interpretation that unusually blue features are associated with higher surface gravities and/or lower metallicities, both of which correlate with older ages. 
We also examined whether removing L dwarfs with ages younger than 500~Myr in our baseline simulation would elevate the ages of the remaining sources; only a marginal shift to $3.4 \pm 0.7$~Gyr was found. Finally, we considered the age without the binaries (8 sources)\added{ for our thin disk L dwarf sample}, which results in a slightly older but statistically consistent age of $4.4 \pm 0.3$~Gyr.

\subsection{Evidence of a Kinematic Indicator of the Main Sequence Terminus} \label{sec:break}

Kinematic\deleted{s} \replaced{differences in subtype, rather than}{variations as a function of spectral type, rather that across} whole subclasses, provides a finer examination of UCD ages. Prior simulation work has 
predicted subtype age structure particularly among the L dwarfs due to brown dwarf evolutionary effects and the changing mixture of stars and brown dwarfs with spectral type (cf.\ \citealt{Burgasser:2004aa}). 
At late-M and early L subtypes, only the youngest and most massive brown dwarfs will have temperatures consistent with these types, restricting their representation among the overall sample and resulting in a relatively low brown dwarf-to-star ratio. As we proceed to later types, the mass range of stars with the appropriate temperatures declines, while both the age and mass range of allowed brown dwarfs expands, increasing the brown dwarf-to-star ratio. Since these brown dwarfs are preferentially young, this changing ratio drives down the average age of the population. At late enough spectral types, temperatures become sufficiently low that stars are not present, resulting in a ``pure'' brown dwarf sample that is relatively young but increases in mean age through the T and Y dwarf sequences.

As a preliminary assessment of these effects, we examined a more refined breakdown of L dwarfs by subtype. 
Figure~\ref{fig:boxplot_L_dwarf} displays the observed kinematic ages and simulation predictions of thin disk L dwarfs broken down in bins of two subtypes. Both observations and simulations confirm an overall downward trend of age with later spectral type, declining from $5.8 \pm 0.6$~Gyr at L1--L2 to \replaced{$0.9 \pm 0.3$}{$0.9 \pm 0.7$}~Gyr for L8--L9. 
However, in the L4--L5 and L5--L6 subtypes this downward trend briefly reverses, with the latter having an average age of $5.4 \pm 1.4$~Gyr. 
We note that this increase is present even with the removal of both thick-disk and unusually blue L dwarfs, which are clustered among these mid-L subtypes. The simulations show a concurrent reversal in average kinematic age, skewed by an \deleted{increasing }older population of stars near the HBMM. By spectral type L6--L7, the observed kinematic ages drop back to the downward trend line, while the simulations show a more modest decrease in mean age and a broadened distribution\deleted{s} of ages overall.

We interpret this newly-discerned 
``kinematic break'' around spectral type\added{s} L5--L6 \replaced{as}{to be} an observable of the terminus of the stellar Main Sequence.
The effective temperature range of L5--L6 dwarfs, 1500~K $\lesssim$ $T_{\mathrm{eff}}$ $\lesssim$ 1600~K \citep{Filippazzo:2015aa}, corresponds to evolutionary model predictions for the HBMM at ages of 4--5~Gyr \citep{Baraffe:2003aa}, in rough agreement with the average age of the local UCD population. \citet{Dupuy:2017aa} identify a similar HBMM boundary at slightly earlier spectral types of L3--L5 based on the distribution of 38 dynamical mass measurements from binaries; while \citet{Dieterich:2014aa} anchor the HBMM at $\sim$2075~K, corresponding to L1--L2 subtypes, based on an inferred radius minimum. 
The differences among these empirical indicators of the HBMM may reflect both sample variations and sensitivity to specific brown dwarf indicators. For example, the L1--L2 range may represent a threshold in the brown dwarf-to-star ratio in the field population, while the L5-L6 range represents the disappearance of stars entirely. 
We emphasize that all of these subtype samples are small and need to be expanded 
to confirm and quantify these empirical indicators of the transition between stars and brown dwarfs in the Galactic field population.

\subsection{Refining Constraints on UCD Population Parameters} \label{sec:popsim_subtype}

\begin{figure}[!htbp]
\includegraphics[width=\linewidth]{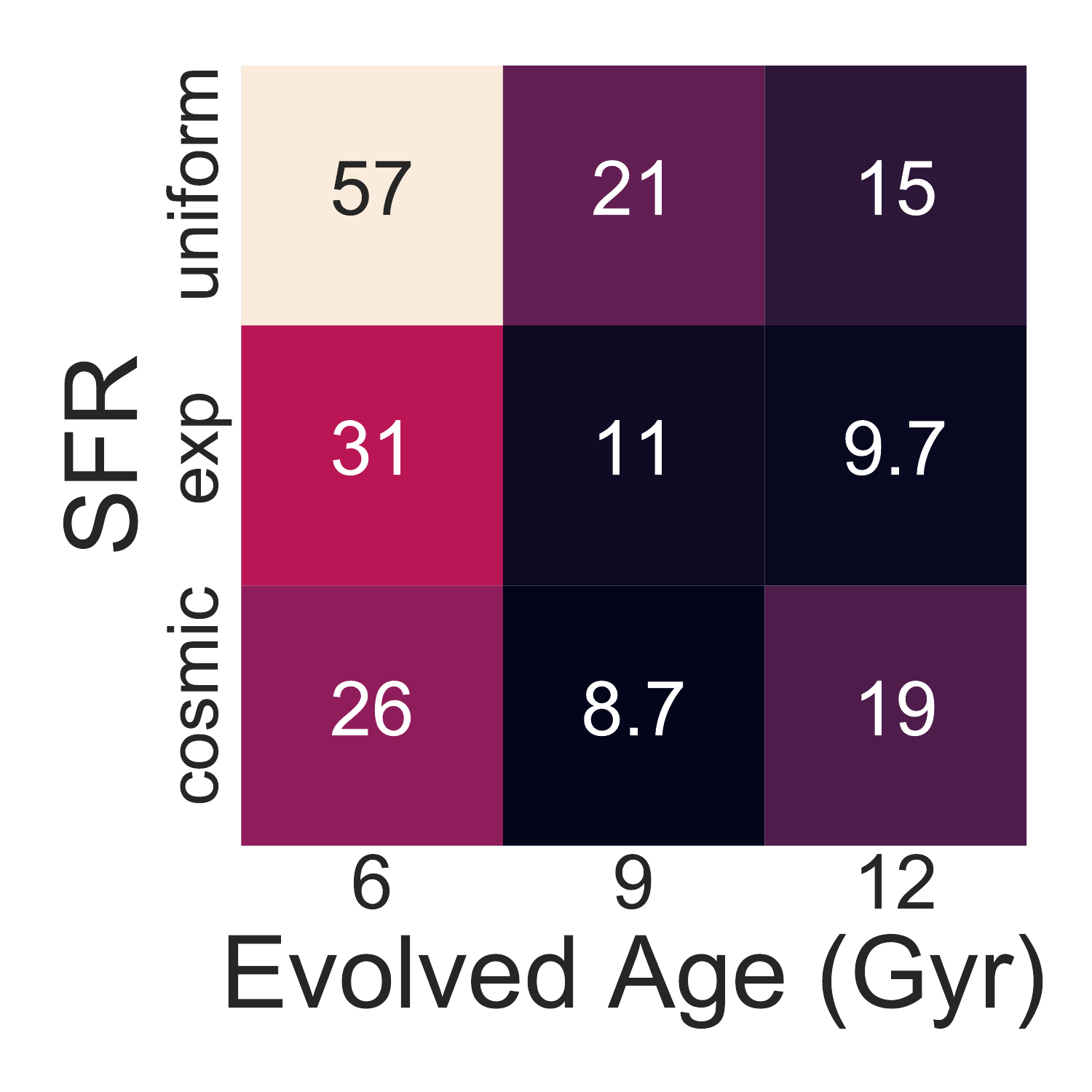}\par 
\includegraphics[width=\linewidth]{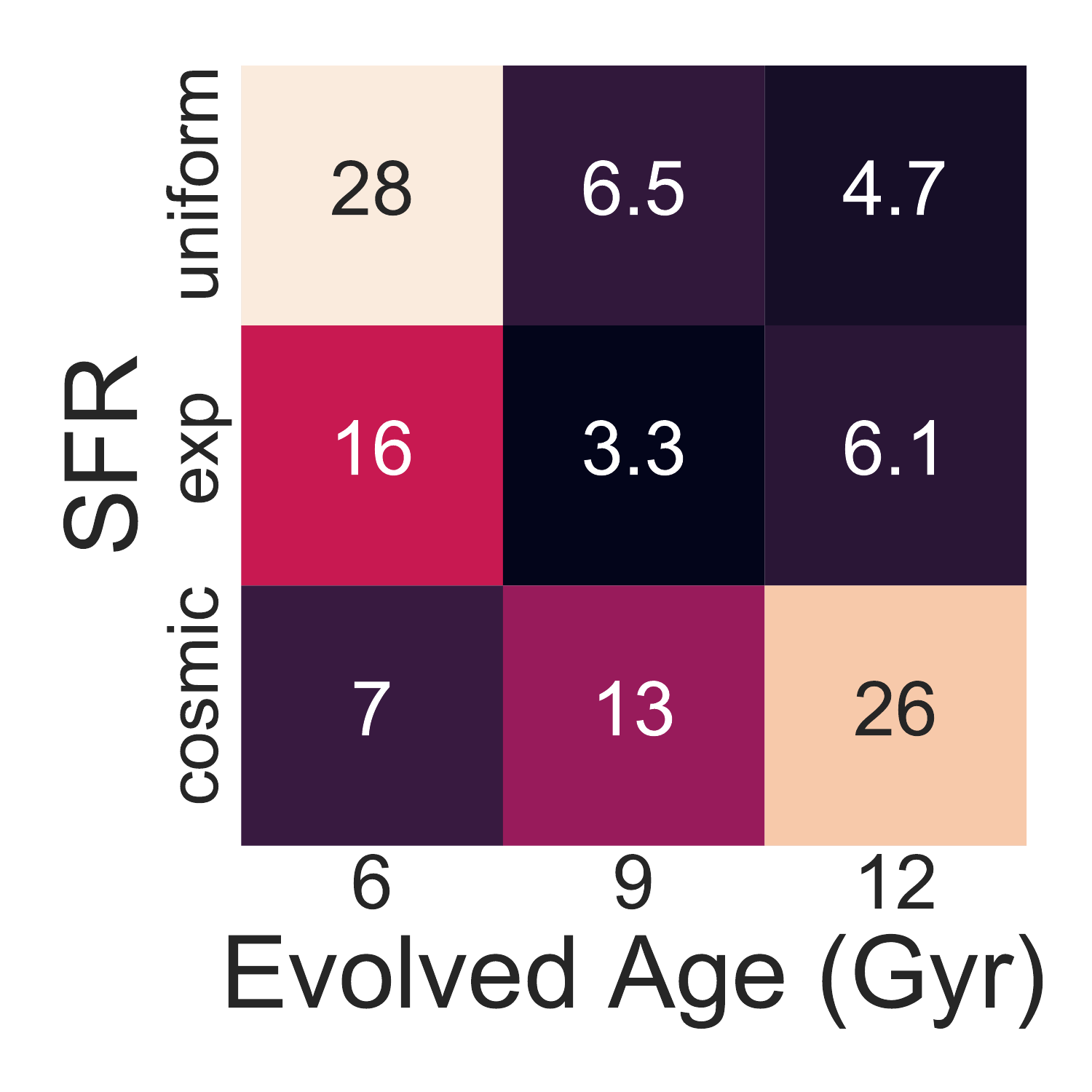}\par
\caption{$\chi^2$ distributions of simulated populations as a function of star formation rate (SFR) and evolved age for the L dwarf subtype (top) and UCD subtype samples (bottom). All simulations assume a power-law mass function with $\alpha$ = 0.5, \citet{Baraffe:2003aa} evolutionary models, and minimum mass of 0.01~M$_{\odot}$.
\label{fig:popsim_breakdown_heatmap}
}
\end{figure}

\begin{figure*}[!htbp]
\includegraphics[width=\textwidth, trim=0 10 0 10]{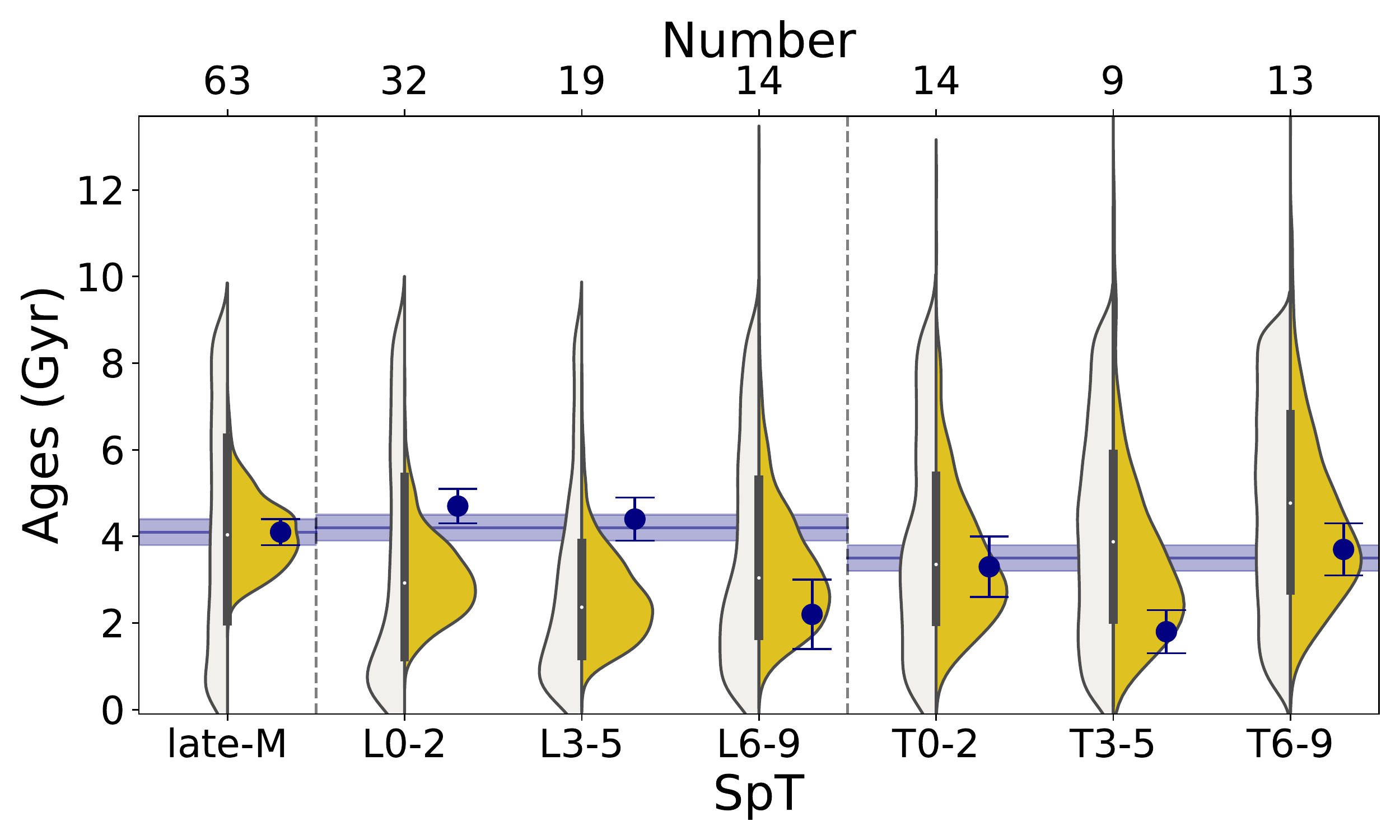}
\caption{Observed kinematic ages of thin-disk UCDs in our kinematic sample grouped into bins of three subtypes (solid blue points) compared to similarly-binned best-fit simulation predictions using a power-law mass function with $\alpha$ = 0.5, \citet{Baraffe:2003aa} evolutionary models, minimum mass of 0.01~M$_{\odot}$, the \citet{Aumer:2009aa} exponential SFR, and a population age of 9~Gyr (white/yellow violin plots for simulated ages/inferred population kinematic ages). The derived kinematic ages and uncertainties for the late-M, L, and T dwarf subgroups are indicated by blue bands. The number of sources in each subtype bin is labeled at top.
\label{fig:boxplot_UCD_best}}
\end{figure*}

The segregation of kinematic ages by subtype also provides an opportunity to more finely constrain population parameters, albeit with lower statistical accuracy. In particular, the age distribution of the underlying population plays a specific role in setting the relative balance of stars and brown dwarfs, and the terminus of the Main Sequence, within the L dwarf class. To explore these effects, we compared a subset of our simulations to two subtype breakdowns of our kinematic sample.  Following Section~\ref{sec:other_simulated_pop}, we evaluated variations in SFR and population age with the MF, minimum mass, and evolutionary model fixed to our baseline assumptions. Here we compare these to our thin disk L dwarf sample in groups of two subtypes, and our overall sample in groups of three subtypes to account for the small sample for T dwarfs. Figures~\ref{fig:popsim_breakdown_heatmap} and \ref{fig:boxplot_UCD_best} shows the $\chi^2$ distributions and best-fit distributions of simulated kinematic ages for these comparisons. 
For the L dwarf sample, we find the cosmic/9~Gyr SFR/age combination provides the best overall fit, exceeding our baseline model ($\Delta$BIC $>$ 10, highly significant) but consistent with the exponential/9~Gyr ($\Delta$BIC = 2.3, positive) and exponential/12~Gyr ($\Delta$BIC = 1.0, not significant). We can now rule out the cosmic/6~Gyr SFR/age combination by a BIC test ($\Delta$BIC $>$ 10, highly significant). 
For the UCD sample, the exponential/9~Gyr SFR/age combination provides the best overall fit, exceeding our baseline model ($\Delta$BIC = 3.2, positive) but consistent with the uniform/12~Gyr model ($\Delta$BIC = 1.4, not significant). Therefore, exponential/9~Gyr combination gives consistently the best fit for both forms of sample binning.

The refinement of the population parameters from this analysis is clearly limited, a consequence of the small sample statistics and the necessity to average over spectral types. Nevertheless, these outcomes show that a larger and more complete kinematic sample broken down by subtype could break some of the simulation parameter degeneracies and lead to a well-constrained assessment of the local UCD population properties, particularly in conjunction with other observable distributions such as the luminosity function and independent age or mass diagnostics.

\section{Summary} \label{sec:summary}
{\normalfont
The paper has presented new and refined multi-epoch RV and $v\sin{i}$ measurements for a total of 37 T dwarfs. In addition to 23 sources without previously reported RV or $v\sin{i}$ measurements, we improved the measurements of 14 T dwarfs by reducing RV uncertainties by a factor of 5 and $v\sin{i}$ uncertainties by a factor of 3 using a forward-modeling approach. 

Our key scientific results are summarized as follows:

\begin{enumerate}
    \item Most of the local T dwarfs are fast rotators, with a median $v\sin{i}$ of 27 km s$^{-1}$ independent of T dwarf spectral type, larger than the median rotational velocities of late-M dwarfs (12 km s$^{-1}$) and L dwarfs (20 km s$^{-1}$). This trend supports prior work indicating that the angular momenta of brown dwarfs are not lost effectively to magnetic winds. In addition, T dwarfs with larger space velocities, which are likely older and more massive, have larger $v\sin{i}$ values, which may reflect their larger moments of inertia (resistance to angular momentum loss) or more compact radii (greater spin-up). 
    \item Combining our RVs with published and \textit{Gaia} astrometry, we calculated the Galactic \textit{UVW} velocities and orbits for our sample. We found that all of our T dwarfs are in the thin disk population; the one exception in our sample, 2MASS J1331$-$0116, is an unusually blue L dwarf.  
    \item We kinematically confirmed two previously-identified, planetary-mass, young moving group members SIMP J0136+0933 (Carina-Near; \citealt{Gagne:2017aa}) and 2MASS J1324+6358 (AB Doradus; \citealt{Gagne:2018aa}). The T4 dwarf 2MASS J0819$-$0335 and the T6.5+T7.5 binary J1553+1532AB are identified as candidate kinematic members of $\beta$ Pictoris and Carina-Near moving groups, respectively, but the absence of spectral indicators of youth suggest that these are coincident field brown dwarfs.
    \item Among 5 T dwarfs with multiple-epoch RV measurements, we found that two objects exhibited statistically significant RV variability consistent with binary orbital motion: the T0.0+T4.5 spectral binary 2MASS J1106+2754, which has an estimated orbital period of $3.92^{+0.07}_{-0.09}$~yr and semi-amplitude of $6.30\pm0.05$~km s$^{-1}$; and the L7+T3.5 spectral binary 2MASS J2126+7617 with an orbit period of \deleted{roughly }12\added{$^{+1.5}_{-1.2}$~yr}\deleted{years} and semi-amplitude of 3.0$^{+0.7}_{-0.6}$ km s$^{-1}$. In addition, 2MASS~J0559$-$1404, a suspected overluminous T4.5, shows evidence of RV variability over the course of 20 years, but no clear periodic signal; future observations are needed to confirm and assess the origin of these variations.
    \item Using empirical age-velocity dispersion relations, we determined the kinematic age of local T dwarfs to be $3.5 \pm 0.3$~Gyr, which is consistent with the kinematic age of local late-M dwarfs ($4.1 \pm 0.3$~Gyr), but considerably younger than the kinematic age of local L dwarfs ($\replaced{5.7}{5.8} \pm 0.3$~Gyr). By excluding likely thick disk population members ($P(\mathrm{TD})/P(\mathrm{D})$ $>$ 1), the kinematic age of local L dwarfs is lowered to $\replaced{4.1}{4.2} \pm 0.3$~Gyr, in line with late-M and T dwarfs. This analysis appears to \added{partly} resolve the long-standing kinematic anomaly of local L dwarfs.
    \item Population simulations 
    reproduce the measured ages of local thin-disk late-M, L, and T dwarfs, although L dwarf ages are predicted to be younger than observed. Varying the star formation history, mass function, evolutionary models, maximum age, and minimum mass of these simulations and comparing to our kinematic sample allows us to constrain some of these population parameters, but several degeneracies remain. We are able to rule out mass function evolution (bottom-heavy to top-heavy) as an explanation for old L dwarfs due to disagreement with the T dwarf velocity dispersion, but find tentative evidence of a lower hydrogen burning minimum mass (HBMM). 
    A more refined breakdown in spectral type can improve this analysis, but will require larger samples to make robust constraints.
    \item A detailed evaluation of kinematic age as a function of spectral type for L dwarfs reveals a linear trend of decreasing mean age with spectral type, as predicted by population simulations. We also identify an age upturn and sudden break in ages at subtypes L4--L6 which likely reflects the terminus of the stellar Main Sequence. This spectral type range aligns with evolutionary model predictions for the HBMM at an age of 4--5~Gyr. 
\end{enumerate}

This study provides a significant expansion in the number and spectral type range of UCDs with precise 6D coordinates, and the inclusion of T dwarfs in particular allows us to constrain many of the underlying population parameters. It has also allowed us to resolve the L dwarf age anomaly and identify empirical diagnostics of brown dwarf evolution and the Main Sequence terminus. However, the size of the current sample remains the primary limitation in precisely quantifying key aspects of this analysis, notably the contaminating fraction of thick disk/blue L dwarfs in the local population, the location and sharpness of the stellar Main Sequence, and degeneracies among simulation parameters. There is considerable capacity to increase the size of the kinematic sample even in the local volume, which is necessary to properly address issues related to completeness, resolution and accuracy in per-spectral-type analyses, as well as the sample bias induced by distinct subpopulations such as blue L dwarfs. Moreover, even a volume-complete 20~pc sample represents a tiny fraction of the Milky Way environment, and it is possible that the immediate local volume around the Sun is not representative of the Galactic disk, a form of cosmic scatter. This motivates a deeper kinematic survey, to 50~pc or 100~pc for example, which meay be feasible with future spectroscopic survey facilities or through a more restricted analysis of 2D kinematics (cf.\ \citealt{Faherty:2009aa}).
Improvements can also be made to the characterization of the UCD kinematic sample. While we specifically evaluated the contributions of young, binary, and unusually blue UCDs, we did not explicitly evaluate or model metallicity or cloud variations\added{ or inclination angle}, which are of particular importance for the L dwarfs and can influence both empirical calibrations and the underlying evolution of brown dwarfs. 
There are also improvements to made to the population simulations. We have not explicitly taken into account thick disk or halo populations in the simulation, which are likely well-mixed and not easily separable from the thin disk sample based on simple probability thresholds. There are also spatial-temporal correlations to consider, since sources found closer to the Sun (smaller scaleheights) will be preferentially younger than the broader Milky Way population (cf.\ \citealt{Ryan:2017aa}). All of these considerations are open for exploration in future studies.

\facility{Keck: II (NIRSPEC)}
\software{ BANYAN $\Sigma$ \citep{Gagne:2018ab}, Texmaker, Sublime Text, Python, \textit{NumPy} \citep{van-der-Walt:2011aa}, \textit{SciPy} \citep{Virtanen:2019aa}, \textit{Matplotlib} \citep{Hunter:2007aa}, \textit{pandas} \citep{mckinney-proc-scipy-2010}, \textit{seaborn} \citep{waskom2020seaborn}, \textit{Astropy} \citep{Astropy-Collaboration:2013aa}, \textit{wavelets}, \textit{SPLAT} \citep{Burgasser:2017ac}, \textit{emcee} \citep{Foreman-Mackey:2013aa}, \textit{galpy} \citep{Bovy:2015aa}}

\acknowledgments
The authors thank Maria Rosa Zapatero Osorio and Emily Rice for providing their NIRSPEC data obtained on 2000 June 15 and on 2005 July 19, respectively. The authors also thank Gregory Doppmann, Percy Gomez, Carlos Alvarez, and other Keck Observatory staff and support astronomers for their assistance in obtaining Keck/NIRSPEC observations. 
C-CH and AJB
acknowledge funding support from the National Science Foundation under award No.\ AST-1517177.
The material presented in this paper is based upon work supported by the National Aeronautics and Space Administration under Grant No.\ NNX15AI75G.
Support for this work was provided by NASA through the NASA Hubble Fellowship grant HST-HF2-51447.001-A awarded by the Space Telescope Science Institute, which is operated by the Association of Universities for Research in Astronomy, Inc., for NASA, under contract NAS5-26555.
JB acknowledges support from National Science Foundation Graduate Research Fellowship Program grant DGE-1762114.
This work utilizes the measurements from 2MASS catalogs and \textit{Gaia} Data Release 2. The authors acknowledge the usefulness of the SIMBAD database and VizieR service. \added{The authors thank the anonymous referee for his/her/their useful review that has improved the original manuscript.}

The data presented herein were obtained at the W.\ M.\ Keck Observatory, which is operated as a scientific partnership among the California Institute of Technology, the University of California and the National Aeronautics and Space Administration. The Observatory was made possible by the generous financial support of the W.\ M.\ Keck Foundation. \added{This research has made use of the Keck Observatory Archive (KOA), which is operated by the W.\ M.\ Keck Observatory and the NASA Exoplanet Science Institute (NExScI), under contract with the National Aeronautics and Space Administration.} 
The authors recognize and acknowledge the very significant cultural role and reverence that the summit of Maunakea has with the indigenous Hawaiian community, and that the W.\ M.\ Keck Observatory stands on Crown and Government Lands that the State of Hawai'i is obligated to protect and preserve for future generations of indigenous Hawaiians. 
Portions of this work were conducted at the University of California, San Diego, which was built on the unceded territory of the Kumeyaay Nation, whose people continue to maintain their political sovereignty and cultural traditions as vital members of the San Diego community.  
}

\clearpage
\appendix

\section{Minimum $v\sin{i}$ Determination}
\label{appendix:minvsini}

In Section \ref{sec:min_vsini}, we described our method to determine the minimum $v\sin{i}$. Here we provide the corresponding diagnostic plots on the minimum $v\sin{i}$ values as a function of $T_{\text{eff}}$ and S/N.

\restartappendixnumbering 

\begin{figure*}[!htbp]
\includegraphics[width=\textwidth, trim=10 30 30 20, clip=true]{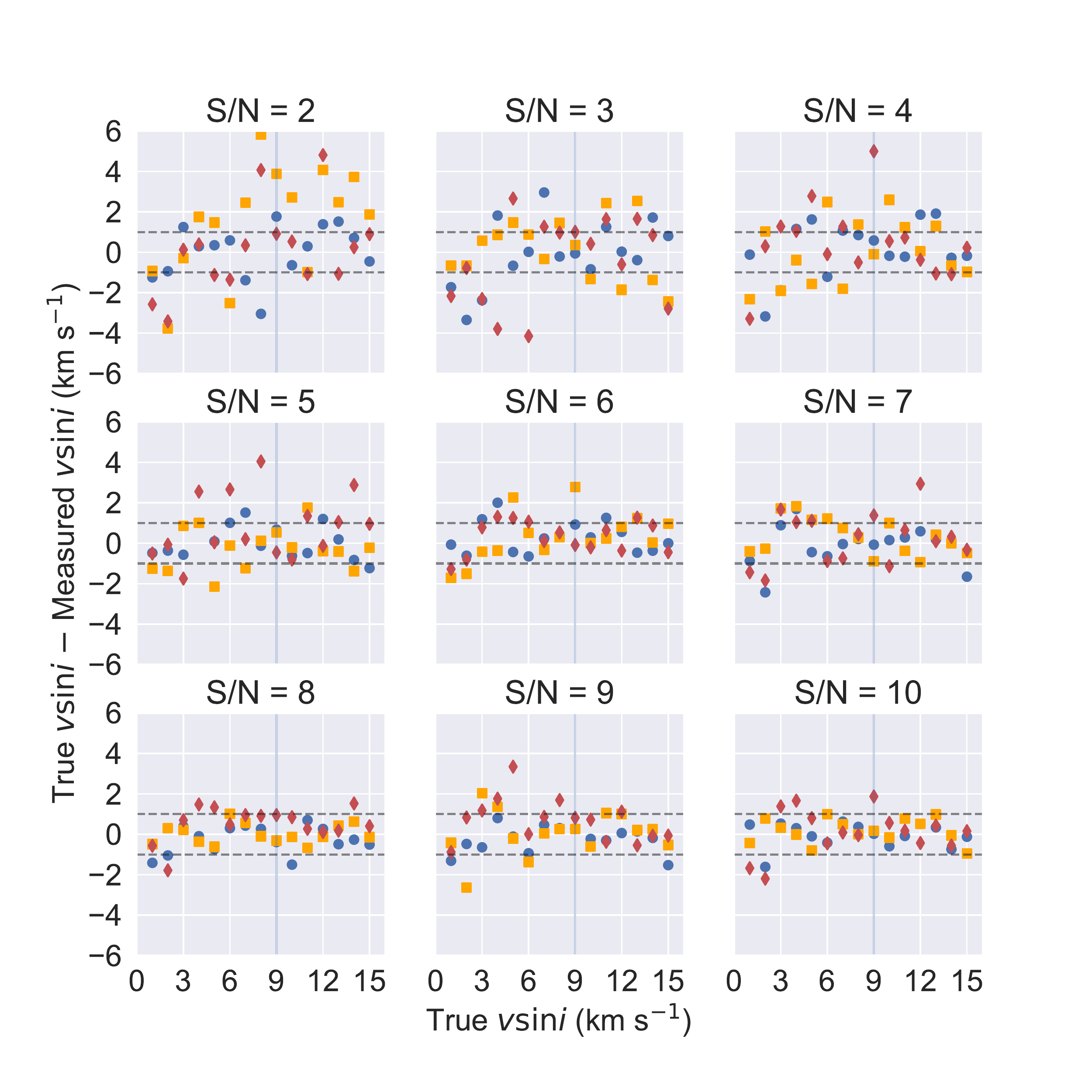}
\caption{The difference between true and measured $v\sin{i}$ compared to true $v\sin{i}$ values as a function of S/N and $T_{\text{eff}}$ for BT-Settl models. The $T_{\text{eff}}$ grid points are 900~K (blue), 1200~K (orange), and 1500~K (red). The grey horizontal dash lines represent $v\sin{i}$ difference = 1 km s$^{-1}$. The vertical blue line indicates the $v\sin{i}$ = 9 km s$^{-1}$
\label{fig:min_vsini_multipanel_btsettl}}
\end{figure*}

\begin{figure*}[!htbp]
\includegraphics[width=\textwidth]{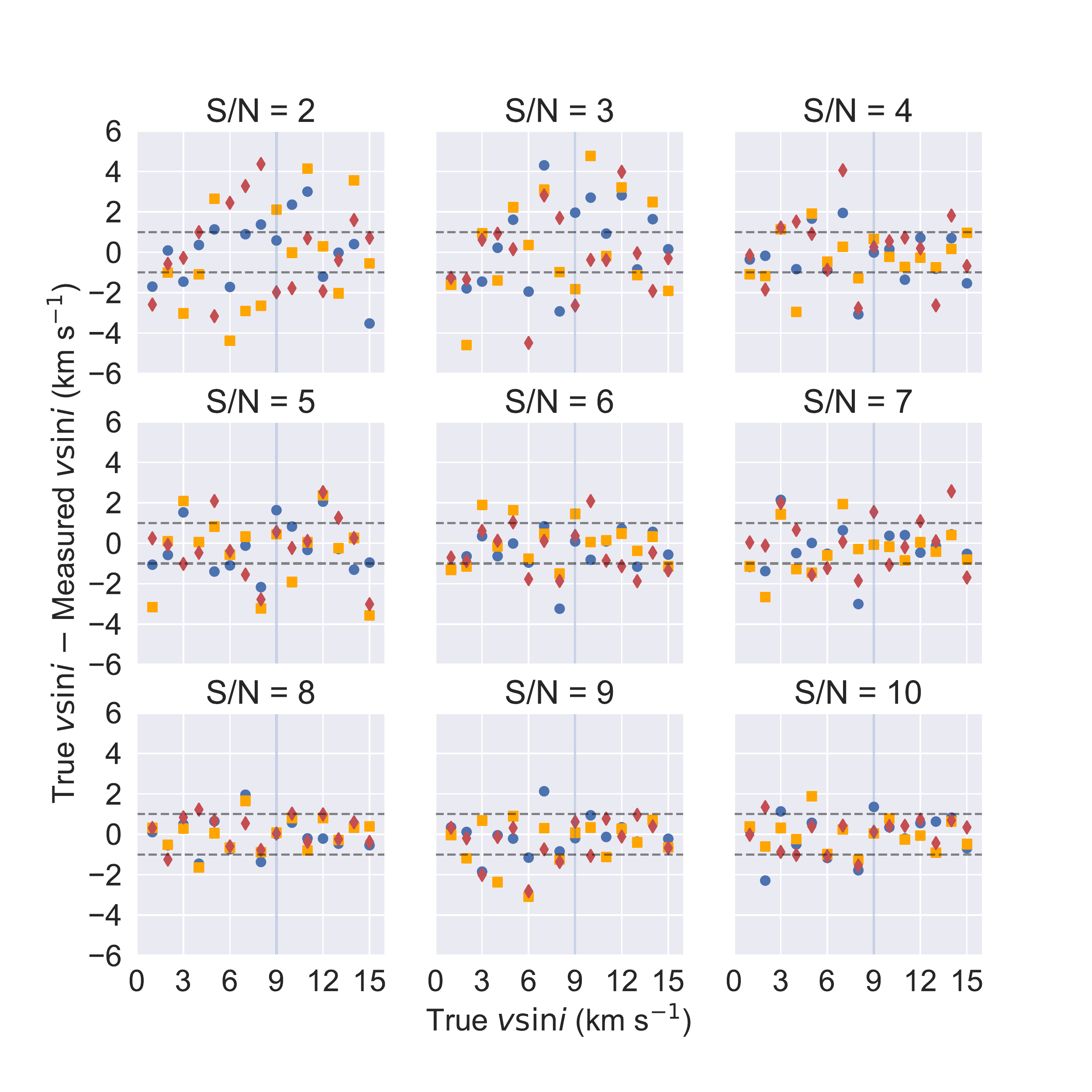}
\caption{Same as Figure \ref{fig:min_vsini_multipanel_btsettl} for Sonora models.
\label{fig:min_vsini_multipanel_sonora}}
\end{figure*}

\clearpage

\section{Simulated UCD Population Ages Under Different Assumptions}
\label{appendix:popsim}

This Appendix supplements the analysis in Sectoin~\ref{sec:other_simulated_pop} by providing the full list of parameters and corresponding $\chi^2$ fit values for the population simulation examined. Select visualizations of the age distributions and comparative $\chi^2$ plots are provided in the main text.

\startlongtable
\begin{deluxetable*}{ccccccccc}
\tablecaption{Simulated UCD Population Ages Under Different Assumptions \label{table:simulated_population_all}
}
\tablewidth{700pt}
\tabletypesize{\scriptsize}
\tablehead{
 \\
\colhead{$\tau$} & \colhead{Star formation} & \colhead{$\alpha$} & \colhead{Models} & \colhead{MBDM} & \colhead{late-M dwarf age} & \colhead{L dwarf age} & \colhead{T dwarf age} & \colhead{$\chi^2$} \\
\colhead{(Gyr)} & \colhead{(rate)} & \colhead{} & \colhead{} & \colhead{($M_{\odot}$)} & \colhead{(Gyr)} & \colhead{(Gyr)} & \colhead{(Gyr)}  & \colhead{}
} 
\startdata
\multicolumn{9}{c}{Observations} \\
\cline{1-9}
\multicolumn{5}{c}{ALL SOURCES\dotfill} & 4.9 $\pm$ 0.3 & \replaced{7.0}{7.1} $\pm$ 0.4 & 3.5 $\pm$ 0.3 & \nodata \\
\multicolumn{5}{c}{NOT THICK DISK ($P(\text{TD})/P(\text{D}) < 10$)\dotfill} & 4.1 $\pm$ 0.3 & \replaced{5.7}{5.8} $\pm$ 0.3 & 3.5 $\pm$ 0.3 & \nodata \\
\multicolumn{5}{c}{THIN DISK ($P(\text{TD})/P(\text{D}) < 1$)\dotfill} & 4.1 $\pm$ 0.3 & \replaced{4.1}{4.2} $\pm$ 0.3 & 3.5 $\pm$ 0.3 & \nodata \\
\cline{1-9}
\multicolumn{9}{c}{Simulations} \\
\cline{1-9}
9\tablenotemark{a} & uniform & $0.5$ & B03 & 0.01 & 4.1 $\pm$ 0.8 & 3.1 $\pm$ 0.7 & 4.3 $\pm$ 1.2 & 2.1 \\ 
\cline{1-9}
9 & uniform & $0.5$ & B01 & 0.01 & 3.5 $\pm$ 0.7 & 3.1 $\pm$ 0.6 & 4.2 $\pm$ 1.2 & 3.0 \\ 
9 & uniform & $0.5$ & S08 & 0.01 & 4.3 $\pm$ 0.8 & 3.4 $\pm$ 0.7 & 4.5 $\pm$ 1.2 & 1.7 \\
\enddata
\end{deluxetable*}

\bibliographystyle{aasjournal}
\bibliography{nirspec_paper_reference}

\end{document}